\documentclass[]{imsart}

\RequirePackage{amsthm,amsmath,amsfonts,amssymb}
\RequirePackage[authoryear]{natbib}
\usepackage{pdfpages}
\RequirePackage[colorlinks,citecolor=blue,urlcolor=blue]{hyperref}
\RequirePackage{graphicx}
\usepackage{amsmath}
\usepackage{subfig}
\usepackage{bbm}
\usepackage{algorithm}
\usepackage{enumerate}
\usepackage{mathtools}
\usepackage[body={15.5cm,21cm}, top=3cm]{geometry}
\startlocaldefs

\def\card{\mathrm{card}}

\newcommand{\K}{J}
\newcommand\mystretch{\rule[-2mm]{0mm}{5mm} } 
\newcommand{\vertiii}[1]{{\left\vert\kern-0.25ex\left\vert\kern-0.25ex\left\vert #1 
		\right\vert\kern-0.25ex\right\vert\kern-0.25ex\right\vert}}

\newcommand\asp{\hspace{4mm}}

\theoremstyle{plain}

\newtheorem{theorem}{Theorem}[section]

\theoremstyle{remark}

\newtheorem{assumption}[theorem]{Assumption}

\endlocaldefs

\allowdisplaybreaks

\usepackage{xr}
\begin{document}

\begin{frontmatter}
\title{VT-MRF-SPF: Variable Target Markov Random Field Scalable Particle Filter}
\runtitle{VT-MRF-SPF}

\begin{aug}
\author[]{\fnms{Ning}~\snm{Ning}\ead[label=e1]{patning@tamu.edu}}
\address[]{Department of Statistics, 
	Texas A\&M University,
	College Station\printead[presep={,\ }]{e1}}

\end{aug}
\begin{abstract}
Markov random fields (MRFs) are invaluable tools across diverse fields, and spatiotemporal MRFs (STMRFs) amplify their effectiveness by integrating spatial and temporal dimensions. However, modeling spatiotemporal data introduces additional hurdles, including dynamic spatial dimensions and partial observations, prevalent in scenarios like disease spread analysis and environmental monitoring. Tracking high-dimensional targets with complex spatiotemporal interactions over extended periods poses significant challenges in accuracy, efficiency, and computational feasibility. To tackle these obstacles, we introduce the variable target MRF scalable particle filter (VT-MRF-SPF), a fully online learning algorithm designed for high-dimensional target tracking over STMRFs with varying dimensions under partial observation. We rigorously guarantee algorithm performance, explicitly indicating overcoming the curse of dimensionality. Additionally, we provide practical guidelines for tuning graphical parameters, leading to superior performance in extensive examinations.
\end{abstract}

\begin{keyword}[class=MSC]
\kwd[Primary ]{65C05}
\kwd[; secondary ]{65C35}
\kwd{65C40}
\kwd{94C15}
\end{keyword}

\begin{keyword}
\kwd{Markov random fields}
\kwd{spatiotemporal analysis}
\kwd{latent variable}
\kwd{sequential Monte Carlo}
\kwd{curse of dimensionality}
\kwd{time-varying graphs}
\end{keyword}

\end{frontmatter}

\section{Introduction}
We start by presenting the background and motivation in Subsection \ref{sec:Background}, then outline our contributions in Subsection \ref{sec:contributions}, followed by the paper's organization in Subsection \ref{sec:organization}.

\subsection{Background and motivation} 
\label{sec:Background}	

Markov random fields (MRFs) have found wide-ranging applications spanning disciplines such as physics, computer vision, machine learning, computational biology, and materials science \citep{li2009markov}. Spatiotemporal data differs from high-dimensional data in its representation and characteristics, as it incorporates both spatial and temporal dimensions. Extending MRFs to spatiotemporal MRFs (STMRFs) offers numerous advantages in modeling dynamic systems over space and time \citep{christakos2017spatiotemporal}.  By incorporating temporal information, STMRFs enable the modeling of dependencies and interactions over both spatial and temporal dimensions, making them suitable for tasks such as video analysis, motion estimation, and tracking. This extension enhances the capability of MRFs to handle spatiotemporal data, making them a valuable tool for capturing and analyzing dynamic phenomena in various applications, including video processing, medical imaging, and environmental monitoring \citep{descombes1998fmri, prates2022non}. 

\begin{figure}[t!]
	\centering
	\subfloat{\includegraphics[width = 2.5in]{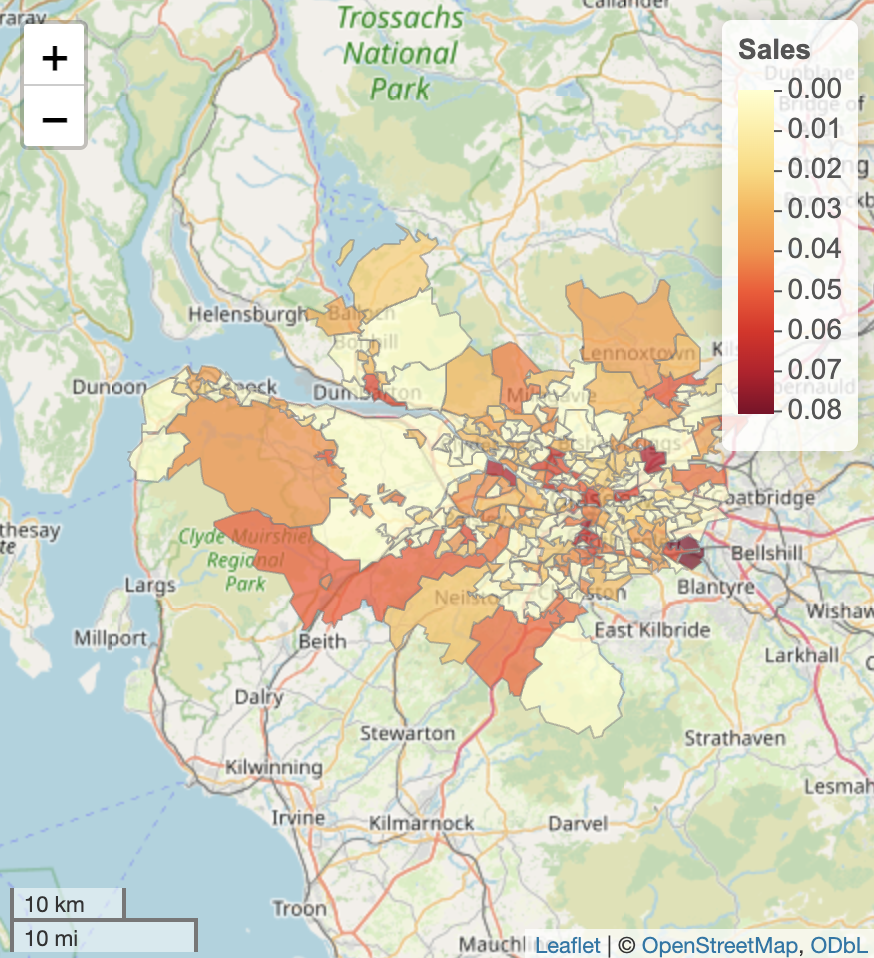}}\hfil
	\subfloat{\includegraphics[width =2.5in]{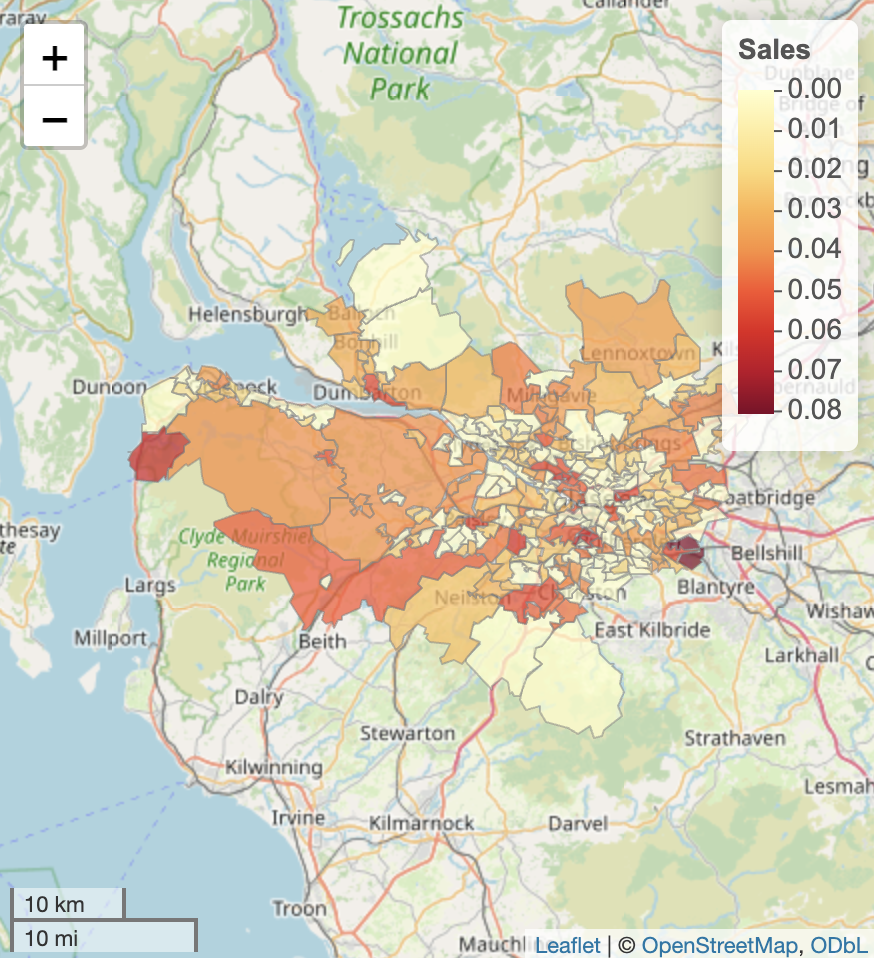}}
	\caption{Illustration of missing or irregular spatiotemporal data at two time points}
	\label{fig:graph_illustration2}
\end{figure}

Spatiotemporal data exhibits time-varying spatial dimensions due to a multitude of factors, including the presence of missing or irregular data in specific spatial areas. This is exemplified in Fig \ref{fig:graph_illustration2}, which showcases the 271 Intermediate Zones within the Greater Glasgow and Clyde health board in Scotland. The left figure demonstrates the absence of data in certain locations at one time point, while the right figure illustrates the absence of data in different spatial locations at another time point. This missing data can stem from various reasons such as sensor malfunctions, occlusions in visual data, or incomplete data collection processes \citep{jiang2019data}. Furthermore, the dynamic nature of spatiotemporal data also arises from the inherent variability in spatial interactions over time, changes in environmental conditions, as well as the continuous evolution of phenomena being observed \citep{lin2023statistical}. Additionally, the presence of artifacts and uncertainties in the collected spatial data further contributes to the time-varying spatial dimension of spatiotemporal data, posing challenges for analysis and modeling (see, e.g. \cite{lin2024understanding} and the references therein).

What made spatiotemporal data analysis more challenging is the partial observation. In epidemiology, partial observation of spatiotemporal data often arise due to several factors: 1) Incomplete reporting leads to underestimation or incomplete representation of disease spread, as not all cases are reported. 2) Spatial heterogeneity occurs due to differences in healthcare infrastructure and reporting practices across geographical regions, resulting in uneven data coverage. 3) Temporal dynamics introduce fluctuations in reporting frequency and accuracy over time due to changes in public health policies or resource availability. 4) Diagnostic uncertainty arises from misdiagnosis or delayed diagnosis, particularly for diseases with nonspecific symptoms. 5) Sampling bias may skew data towards certain population groups or regions, affecting estimates of disease prevalence. 6) Additionally, measurement errors in data collection, recording, or transmission further impact the reliability of epidemiological analyses and interpretations. Therefore, epidemiological modeling with partial observation is widely employed \citep{li2020substantial, subramanian2021quantifying, li2023machine}.

An influential study by \cite{khan2005mcmc} analyzed  hidden STMRF of varying dimension (HSTMRF-VD) and introduced an online learning algorithm based on the particle filter (PF). Also known as the sequential Monte Carlo method, PFs are especially well-suited for analyzing spatiotemporal data because of their capacity to effectively capture the dynamic nature of the data across both space and time \citep{doucet2001introduction, chopin2020introduction}. This method represents the posterior distribution of the state vector by a set of random samples called particles, which are recursively propagated through the dynamic model as new measurements become available \citep{del2006sequential}. The online fashion of particle filters is crucial for real-time estimation and inference in dynamic systems, providing the capability for continual adaptation to changing system dynamics and uncertainties \citep{chopin2023computational}. Analyzing spatiotemporal data in an online fashion is crucial as it enables real-time monitoring and decision-making, such as disease outbreaks or weather phenomena. By analyzing data online, insights can be obtained promptly, enabling timely interventions and responses.  

Despite the advantageous features of the PF for handling spatiotemporal data, the challenge of curse of dimensionality (COD) significantly hampers its effectiveness, particularly when dealing with high-dimensional datasets. However, the proliferation of sensor networks, satellite imagery, IoT devices, and social media platforms has led to an explosion in spatiotemporal data of both high spatial and long temporal dimensions. PF encounters diminishing performance with increasing model dimensionality \citep{bengtsson2008curse, snyder2008obstacles}. In mathematical terms, the algorithmic error of PF exhibits an exponential increase with the dimension of the state space of the underlying model \citep{rebeschini2015can}. 
 Numerous endeavors have been made to adapt the PF to diverse high-dimensional model types and scenarios, such as \cite{finke2017approximate, singh2017blocking, guarniero2017iterated, goldman2021spatiotemporal, rimella2022exploiting, ning2023iterated, finke2023conditional, ionides2024iterated}. However, a scalable (fully) online learning algorithm that is generically applicable to HSTMRF models, with or without time-varying dimensions, remains an open challenge, and this paper aims to address this gap.

\subsection{Our contributions}
\label{sec:contributions}

In this paper, we work on general HSTMRF-VD models, allowing the latent state and observations at each spatial unit to take continuous or discrete values (demonstrated in Fig \ref{fig:continus_discrete}), real values or complex values, and importantly being high-dimensional or infinite-dimensional thus incorporating functional spaces. This expansion distinguishes our model from the HSTMRF-VD model proposed by \cite{khan2005mcmc}, where each spatial unit's random variable is constrained to a single value. Furthermore, our model features time-inhomogeneous transition densities and measurement densities, allowing for variations over time, in contrast to the time-homogeneous nature of \cite{khan2005mcmc}. This extension is crucial; for instance, in the context of COVID-19, the precision of measurements may vary during different stages of transmission, reflecting changes in measurement equipment availability. Additionally, we incorporate general neighborhood interactions, as opposed to the pairwise neighborhood interaction considered by \cite{khan2005mcmc}, which is a special case of ours. This distinction is illustrated in Section \ref{sec:mrf} using the theory of Gibbs measure. Importantly, we permit spatial interactions to evolve over time while accommodating non-overlapping regional dependencies, as visualized in Fig \ref{fig:g}. This feature is motivated by within-state interactions and state-specific policies during the COVID-19 pandemic.

We propose the variable target MRF scalable PF (VT-MRF-SPF) in Algorithm \ref{Algorithm_VT-MRF-SPF} for inferring the latent states of general high-dimensional HSTMRF-VD models. Given the model's time-evolving graph structure, we employ time-evolving cluster partitions for these graphs, as illustrated in Fig \ref{fig:time-evolvingcluster}. Unlike the PF and its variants, which predict the next state using the model's dynamics and calculate weights using the observation density, the VT-MRF-SPF takes a different approach. At each time $t$, the VT-MRF-SPF predicts the latent state of all available targets at time $t+1$, without considering spatial interactions. Subsequently, within each cluster in the time-evolving partition $\mathcal{B}(k_t)$, weights are computed as the product of the measurement density and the spatial interaction density of the latent state. Resampling is then performed cluster-wise based on these cluster-specific weights. These particles are recursively propagated through the dynamic model as new measurements become available, enabling the filter to dynamically adjust its estimation based on the most recent observations. The fully online learning scheme enables continuous updating of estimates based on incoming data, eliminating the necessity to revisit past observations. This characteristic effectively circumvents the common storage challenge in spatiotemporal analysis, minimizing the need for large storage capacity.

Mathematically understanding the VT-MRF-SPF's mechanism and rigorously bounding its algorithmic error pose significant challenges, especially considering the state space of the latent state ($\mathbb{X}^v$) associated with each vertex $v$ is a Polish space. Given $\mathbb{N}$ is countable, under the product topology, $\mathcal{X}=(\mathbb{X}^v)^\mathbb{N}$ is also Polish spaces. Even in cases where $\mathbb{X}^v$ is a Banach space, $\mathcal{X}$ is a Polish space not a Banach space; thus, techniques only applicable to Banach spaces are not  applicable here. To overcome this, we borrowed concepts such as decay of correlation alongside the Dobrushin comparison theorem (Theorem 8.20 in \cite{georgii2011gibbs}) from statistical physics, both suitable for Polish spaces and high-dimensional graphs. The algorithmic bias and variance are bounded in Theorems \ref{thm:main_theorem1} and \ref{thm:main_theorem2} respectively, each followed by sketches outlining distinct proof techniques employed. The upper bound of the overall algorithmic error, provided in equation \eqref{eqn:error}, solely relies on local quantities, thus beating the COD. We allow all graphical quantities to vary with time, but if they are fixed over time, the resulting upper bound becomes uniform in time, which is achieved by delicate stability analysis in  spatiotemporal framework. 

Furthermore, the algorithmic error bounds explict reveal the importance of each graphical quantity, offering practical insights to understand and mitigate errors inherent in the VT-MRF-SPF. Following the guideline of favoring small cluster sizes, we conducted extensive numerical analyses to assess the performance of the VT-MRF-SPF across diverse scenarios. We introduced a variant of the widely used conditional autoregressive (CAR) model proposed in \cite{leroux2000estimation}, incorporating varying spatial dimensions, time-evolving network interactions, and partial observations. Our evaluations encompassed both discrete and continuous observation models, with equal and unequal target entering and staying probabilities, and utilizing both complete graph structures and real spatial structures as depicted in Fig \ref{fig:graph_illustration2}. Comparing the performance of the VT-MRF-SPF with the Variable Target Joint MRF PF (VT-MRF-PF) proposed in \cite{khan2005mcmc}, which to date is the only fully online learning algorithm applicable to general HSTMRF-VD models, we observed that the VT-MRF-SPF consistently demonstrates stability and scalability, outperforming the VT-MRF-PF across all scenarios.

\subsection{Organization of the paper}
\label{sec:organization}
The organization of the paper unfolds as follows: In Section \ref{sec:Model}, we introduce HSTMRF-VD models. In Section \ref{sec:Main_results}, we present the main results of this paper: proposing the VT-MRF-SPF in Algorithm \ref{Algorithm_VT-MRF-SPF} and establishing its algorithmic error bounds in Theorems \ref{thm:main_theorem1} and \ref{thm:main_theorem2}. In Section \ref{sec:numerical}, we present numerical analyses of the VT-MRF-SPF's performance, in comparison with the VT-MRF-PF. The supplementary materials provide preliminary proofs and proofs of the main theoretical results.

 \section{Model description}
\label{sec:Model}

In this section, we begin by reviewing the MRF model in Section \ref{sec:mrf}, followed by an exploration of the STMRF model incorporating time-varying spatial dimensions in Section \ref{sec:stmrf}. Finally, we present our HSTMRF-VD model in Section \ref{sec:Hidden_stmrf}.

\subsection{Markov random field}
\label{sec:mrf}

In this subsection, we review MRFs and their connection to the Gibbs distribution, which is essential for understanding our model.
The elements within a finite set $V$ are considered being interconnected through a neighborhood system. For any vertex $v \in V$, its neighbor set is determined as the collection of nearby vertices within a specified radius $r$:
$$
N(v)=\Big\{v' \in V \,\big|\, d (v,v')\leq r,\, v' \neq v\Big\},
$$
where $d(v, v')$ represents the Euclidean distance between $v$ and $v'$, and $r$ is a positive integer-valued parameter. A neighborhood system for the set $V$ is characterized by the collection of all such neighbor sets:
$$
\mathcal{N}=\Big\{N(v) \,\big|\, \forall v \in V\Big\}.
$$
It has the following properties:
a vertex is not neighboring to itself, i.e., $v \notin N(v)$;
the neighboring relationship is mutual, i.e.,  $v \in N(v^{\prime}) \Leftrightarrow v^{\prime} \in N(v)$.

\begin{figure}[t!]
	\centering
	\label{fig:real-score-fn}
	\includegraphics[width=0.9\textwidth]{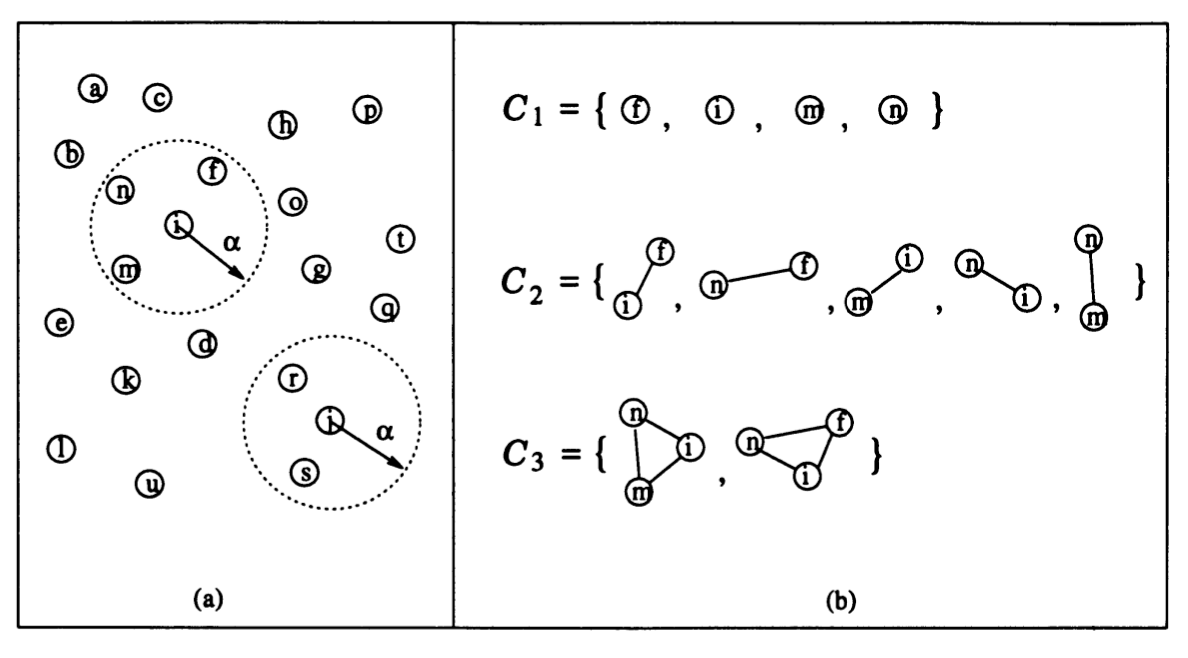}
	\caption{Neighborhood and cliques on a set of irregular vertices.  (Source: Figure 1.3  in \cite{li2012markov})
	}
\end{figure}
The pair $(V, \mathcal{N}) \triangleq \mathcal{G}$ forms a graph in the conventional sense, where $V$ represents the vertices and $\mathcal{N}$ dictates the connections between vertices based on neighboring relationships. A clique for $\mathcal{G}$ is defined as a subset of vertices in $V$. It can either be a single vertex $\{v\}$, or a pair of neighboring vertices $\left\{v, v^{\prime}\right\}$, or a triple of neighboring vertices $\left\{v, v^{\prime}, v^{\prime \prime}\right\}$, and so forth.  The collections of single-vertex, pair-vertex, and triple-vertex cliques are denoted by $\mathcal{C}_1$, $\mathcal{C}_2$, and $\mathcal{C}_3$, respectively, where
$$
\begin{gathered}
	\mathcal{C}_1=\big\{v \,\big|\, v \in V\big\}, \\
	\mathcal{C}_2=\Big\{\left\{v, v^{\prime}\right\} \,\Big|\, v^{\prime} \in N(v),\, v \in V\Big\},\\
	\text{and}\quad	\mathcal{C}_3=\Big\{\left\{v, v^{\prime}, v^{\prime \prime}\right\} \,\Big|\, v, v^{\prime}, v^{\prime \prime} \in V \text { are neighbors to one another }\Big\}.
\end{gathered}
$$
The collection of all cliques for $(V, \mathcal{N})$ is represented as
\begin{align}
	\label{eqn:cliques}
	\mathcal{C}=\mathcal{C}_1 \cup \mathcal{C}_2 \cup \mathcal{C}_3 \ldots
\end{align}
where $\ldots$ denotes possible sets of larger cliques.

\begin{figure}[t!]
	\centering
	\includegraphics[width=0.9\textwidth]{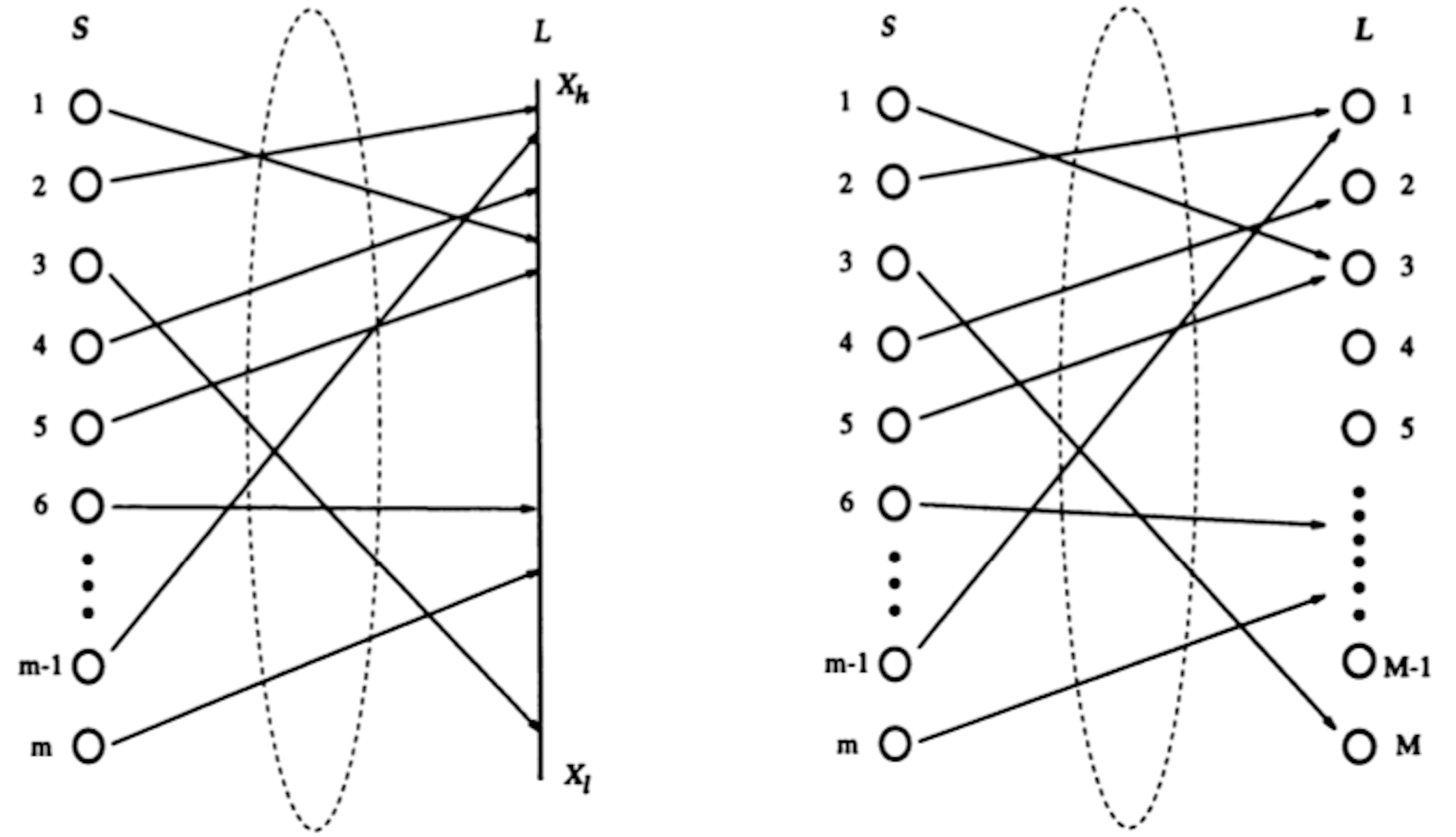}
	\caption{The above shows mappings with continuous label set (left) and discrete label set (right).  (Source: Figure 1.1  in \cite{li2012markov})
	}
	\label{fig:continus_discrete}
\end{figure}

Consider a family of random variables $X=\left\{X^v\right\}_{v\in V}$ defined on the set $V$, where each $X^v$ assumes values/lables $x^v$ in a label set $\mathbb{X}^v$. 
We borrow Figure 1.1 of \cite{li2012markov} to visualize the mappings with continuous label set and discrete label set in Fig \ref{fig:continus_discrete}. 
Supposing the cardinality $\card(V)=m$, this family $X$ is termed a $m$-dimensional random field. The notation $X^v=x^v$ denotes the event that $X^v$ takes the value $x^v$, and $\left(X^1=x^1, \ldots, X^m=x^m\right)$ represents the joint event. For simplicity, a joint event is abbreviated as $X=x$, where $x=\left\{x^1, \ldots, x^m\right\}$ forms a configuration of $X$, corresponding to a realization of the random field. In the case of a discrete label set $\mathbb{X}^v$, the probability that the random variable $X^v$ equals $x^v$ is denoted as $P(X^v=x^v)$, and the joint probability is denoted as $P(X=x)=P(\{X^v=x^v\}_{v\in V})$, abbreviated as $P(x^v)$ and $P(x)$ unless there is a need to elaborate the expressions. In the case of a continuous $\mathbb{X}^v$, probability density functions are denoted as $f^v\left(X^v=x^v\right)$ and $f(X=x)$, abbreviated as $f^v(x^v)$ and $f(x)$. Since the discrete-valued random variable has a probability density function with respect to (w.r.t.) the counting measure, we consistently use probability density functions throughout the paper.

The family $X$ is termed a MRF on $V$ w.r.t. a neighborhood system $\mathcal{N}$ if and only if $f^v(x)>0$ and the following Markovian condition holds:
$$
f^v\big(x^v \,\big|\, x^{V\backslash \{v\}}\big)=f^v\big(x^v \,\big|\,  x^{N(v)}\big),
$$
for any $x \in \mathbb{X}$, 
where $V\backslash \{v\}$ denotes the set difference, $x^{V\backslash {v}}$ represents the set of labels at the vertices in $V\backslash \{v\}$, and
$$
x^{N(v)}=\Big\{x^{v^{\prime}} \,\big|\,  v^{\prime} \in N(v)\Big\}
$$
stands for the set of labels at the vertices neighboring $v$. The Markovianity characterizes the local properties of $X$, indicating that the label at a vertex depends solely on those at the neighboring vertices, i.e., only labels from neighboring vertices directly influence each other. However, it is always feasible to choose a sufficiently large $r$ to uphold the Markovian condition, and the largest neighborhood encompasses all other vertices. Thus, any $X$ is an MRF w.r.t. such a neighborhood system.

There are two approaches to specify a MRF: one involves conditional probabilities $f^v(x^v \mid x^{N(v)})$ and the other is based on the joint probability $f(x)$. The Hammersley-Clifford theorem  establishing the equivalence between MRFs and Gibbs distributions provides a mathematically tractable way to specify the joint probability of an MRF;
 see \cite{li2012markov} for further details. A set of random variables $X$ is said to be a Gibbs random field on $V$ w.r.t. $\mathcal{N}$ if and only if its configurations follow a Gibbs distribution. The density of the Gibbs distribution is expressed as
$$
f(x)=Z^{-1} \times \mathrm{e}^{-\frac{1}{T} U(x)},
$$
where $Z$ is a normalizing constant known as the partition function, $T$ is a constant referred to as the temperature (assumed to be 1 unless stated otherwise), and $U(x)$ is the energy function. The energy function is defined as the sum of clique potentials $V_{\card(c)}(x)$ over all possible cliques $\mathcal{C}$ defined in equation \eqref{eqn:cliques}, i.e.,
$$
U(x)=\sum_{c \in \mathcal{C}} V_{\card(c)}(x),
$$
where $V_{\card(c)}(x)$ depends on the local configuration within the clique $c$, with $\card(c)$ being the cardinality of $c$. It may be convenient to express the energy of a Gibbs distribution as the sum of several terms, each ascribed to cliques of a certain size, i.e.,
\begin{align}
	\label{eqn:generalU}
U(x)=\sum_{\{v\} \in \mathcal{C}_1} V_1\left(x^v\right)+\sum_{\left\{v, v^{\prime}\right\} \in \mathcal{C}_2} V_2(x^v, x^{v^{\prime}})+\sum_{\left\{v, v^{\prime}, v^{\prime \prime}\right\} \in \mathcal{C}_3} V_3(x^v, x^{v^{\prime}}, x^{v^{\prime \prime}})+\cdots.
\end{align}
An important special case is when only cliques of size up to two are considered:
\begin{align}
	\label{eqn:autonormal}
	U(x)=\sum_{v \in V} V_1\left(x^v\right)+\sum_{v \in V} \sum_{v^{\prime} \in N(v)} V_2(x^v, x^{v^{\prime}}).
\end{align}

Clearly, the Gaussian distribution is a special case  of the Gibbs distribution. The Gaussian MRF (GMRF), also known as the auto-model and the auto-normal model, is characterized by the conditional probability density: for any $v \in V$,
\begin{align}
	\label{eqn:GMRF_pdf}
f^v(x^v\, \big|\, x^{N(v)})=\frac{1}{\sqrt{2 \pi \sigma_v ^2}} \mathrm{e}^{-\frac{1}{2 \sigma_v ^2}\left[x^v-\mu_v-\sum_{v^{\prime} \in N(v)} \beta_{vv^{\prime}}(x^{v^{\prime}}-\mu_{v^{\prime}})\right]^2}.
\end{align}
This is the normal distribution with conditional mean
$$
E(x^v \, \big|\,  x^{N(v)})=\mu_v + \sum_{v^{\prime} \in N(v)} \beta_{v v^{\prime}}(x^{v^{\prime}}-\mu_{v^{\prime}}),
$$
and conditional variance
$$
\operatorname{Var}(x^v \, \big|\, x^{N(v)})=\sigma_v ^2.
$$
By Theorem 2.1 of \cite{mardia1988multi}, the joint probability density is the density of a Gibbs distribution
$$
f(x)=\frac{\sqrt{\operatorname{det}(B)}}{(2 \pi)^{m/2}} \exp\Bigg(-\frac{(x-\mu)^{\mathrm{T}} B(x-\mu)}{2 }\Bigg),
$$
where, for the cardinality $\card(V)=m$, the vectors $x$ and $\mu$ are $m$-dimensional, and $B=\left[b_{v, v}\right]$ is the $m \times m$-dimensional matrix such that
$$b_{vv^{\prime}}=(\delta_{vv^{\prime}}-\beta_{v v^{\prime}})/\sigma_v^2$$ 
with $\delta$ being the Dirac delta function and $\beta_{v v}=0$.  The corresponding single-vertex and pair-vertex clique potential functions in equation \eqref{eqn:autonormal} are given respectively as
\begin{align}
	\label{eqn:autonormal1}
	V_1\left(x^v\right)&=\left(x^v-\mu_v\right)^2 / 2 \sigma_v^2,\nonumber\\
	\text{and}\quad V_2(x^v, x^{v^{\prime}})&=-\beta_{v v^{\prime}}\left(x^v-\mu_v\right)(x^{v^{\prime}}-\mu_{v^{\prime}}) / 2 \sigma_v^2.
\end{align}
\begin{figure}[t!]
	\centering
	\includegraphics[width=.33\textwidth]{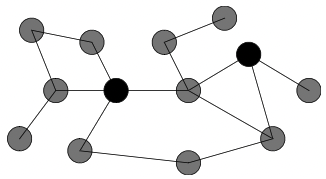}\hfill
	\includegraphics[width=.33\textwidth]{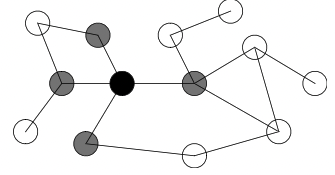}\hfill
	\includegraphics[width=.33\textwidth]{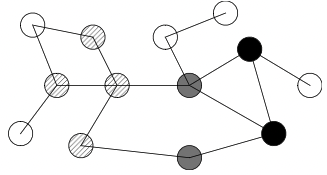}
	\caption{Left: The pairwise Markov property; the two black nodes are conditionally independent given the gray nodes (one-to-one). Middle: The local Markov property; the black and white nodes are conditionally independent given the gray nodes (one-to-many). Right: The global Markov property; the black and striped nodes are conditionally independent given the gray nodes (many-to-many).  (Source: Figure 2.3 in  \cite{rue2005gaussian})
	}
	\label{fig:GMRF_MP}
\end{figure}
Note that there are in fact three types of Markov properties describing conditionally independence in the MRF: the pairwise Markov property, the local Markov property, and the global Markov property. An illustration is provided in Fig \ref{fig:GMRF_MP} and further information can be seen from \cite{rue2005gaussian}. The global Markov property is stronger than the local Markov property, which in turn is stronger than the pairwise one. We note that the GMRF is the most frequently employed type of MRF, with all three properties being equivalent for the GMRF.

\subsection{Spatiotemporal Markov random field of varying dimension}
\label{sec:stmrf}

\begin{figure}[t!]
	\centering
	\includegraphics[width=0.43\textwidth]{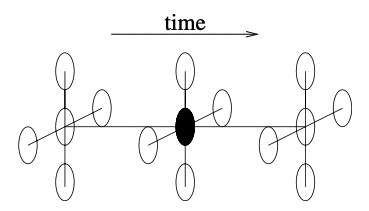}
	\caption{A common neighborhood structure in STMRF models. In addition to spatial neighbors, also the same node in next and previous time-step can be neighbors. (Source: Figure 2.12 in \cite{rue2005gaussian})
	}
	\label{STMRF}
\end{figure}

STMRF models extend MRF models to incorporate additional temporal variations. An illustrative scenario is depicted in Fig \ref{STMRF}. Consider a sequence of graphs over time $[T]:=\{1,2,\ldots,T\}$, akin to the one shown in Fig \ref{STMRF}. If we focus on a specific slice at time $t\in [T]$ and a particular node $v\in V$ within it—let's call it $x_t^v$ and represent it as the black node—its spatial neighbors consist of nearby nodes as schematically indicated with four neighbors in this depiction. A common enhancement in spatiotemporal MRF models involves considering additional neighbors in time, which includes nodes from the preceding and succeeding slices, $x_{t-1}^v$ and $x_{t+1}^v$. That is, we consider $(X_t)_{t\in [T]\cup \{0\}}$ as a Markov chain in a Polish state space $\mathbb{X}$. Recall that a Polish space is defined as a separable completely metrizable topological space. This definition encompasses a wide range, accommodating both discrete and continuous values, including real and complex numbers. Furthermore, it can be high-dimensional or infinite-dimensional spaces, making it versatile to include functional spaces.
Define the reference measure of $X_t$ on $\mathbb{X}$ as $\psi$. The state of $X_t$ at each time $t$ is a random field $(X_t^v)_{v\in V}$ indexed by a finite undirected graph with vertex set $V$. Define the reference measure of $X_t^v$ on its state space $\mathbb{X}^v$ as $\psi^v$, where $\mathbb{X}^v$ can also be a Polish space. Based on the network structure, 
$$X_t:=(X_t^v)_{v\in V},\qquad\mathbb{X}=\prod_{v\in V}\mathbb{X}^v\qquad\text{and}\qquad \psi=\prod_{v\in V}\psi^v.$$
Furthermore, for any set $W\subset V$, define
\begin{equation}
	\label{eqn:cluster_measure}
	X_t^W:=(X_t^v)_{v\in W},\quad\mathbb{X}^W:=(\mathbb{X}^v)_{v\in W}	\quad\text{and}\quad \psi^W(dx_t^W):=\prod_{v\in W}\psi^v (dx_t^v).
\end{equation}

The MRF and STMRF models described above have fixed graph structures. However, in reality, it is common to encounter graphs with time-varying dimensions, such as social networks. 
We refer to the STMRF model with varying dimensions as the STMRF-VD model. 
As in the influential work \cite{khan2005mcmc}, 
a new identifier random vector $K_t\subseteq \mathbb{N}$  at each time $t$ is introduced to indicate targets of interest. It is clear that there are many such hypotheses, and each of these distinct hypotheses corresponds to a joint state variable $X_t^{K_t}$ in the space $\mathcal{X}=(\mathbb{X}^v)^\mathbb{N}$.  Its dimensionality depends on the number of non-zero entries in $K_t$ and thus evolves over time. For example, if the dimension of state space $\mathbb{X}^v$ of a single target $v$ is $2$, $K_t=\{1,3,4\}$ corresponds to a joint state $X_t^{\{1,3,4\}}$ of dimension $6$.  Recall that we consider the state space $\mathbb{X}^v$ as a Polish space. With $\mathbb{N}$ being countable, under the product topology, $\mathcal{X}$ is also a Polish space. Given that Polish spaces can include discrete or continuous values or both, to make notation visually compact, we define the transition density of $X_t$ w.r.t. $\psi$ as $f_t$ and that of $X_t^v$ w.r.t. $\psi^v$ as $f_t^v$. When $X_t$ takes discrete values, $\psi$ is understood as the counting measure. 
Similarly, we define the transition density of $K_t$ as $p_t$.

\begin{figure}[t!]
	\centering
	\subfloat{\includegraphics[width = 2in]{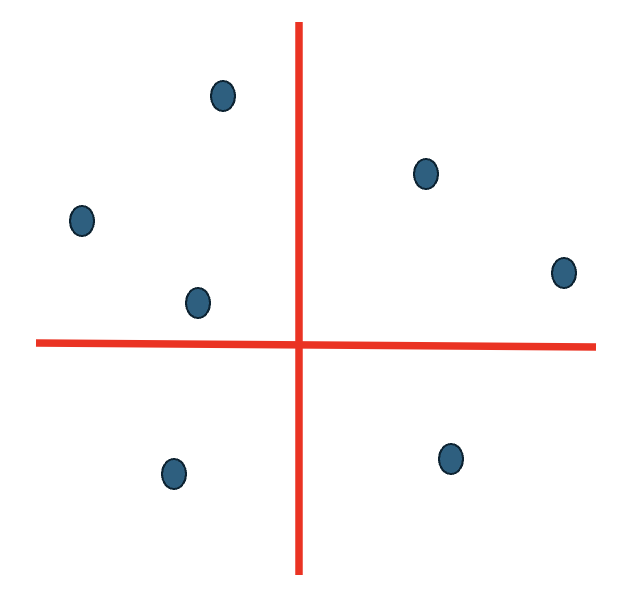}}\hfil
	\subfloat{\includegraphics[width =2.5in]{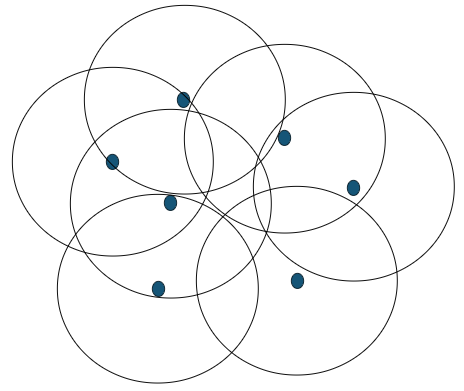}}
	\caption{The left figure is an illustration of the nonoverlapping regional patition. The right figure is an illustration of overlapping neighborhood interactions.}
	\label{fig:g}
\end{figure}

We allow $p_t$ to potentially depend on regional information. In the context of COVID-19, public health policies vary across different states in the US. Each state implements its own distinct set of policies and regulations concerning mask mandates, vaccination requirements, social distancing measures, and lockdown protocols. These policies are often customized to address the particular needs and circumstances of each state, consequently impacting the level of tracking efforts within each region. Consider $\mathcal{R}$ as a regional partion, which is a set of nonoverlapping spatial regions that will not change over time, i.e., 
$$V=\bigcup_{R\in \mathcal{R}}R, \quad\quad R\cap R'=\emptyset\; \text{for any}\; R\neq R'\;\text{and}\; R,R'\in \mathcal{R}.$$
Fig \ref{fig:g} provides illustrations of the nonoverlapping regional patition and the overlapping neighborhood interactions. 
The transition density of $K_t$ is defined as follows based on $\mathcal{R}$:
\begin{align}
	\label{eqn:transition_product_form}
	p_t(k_t\mid k_{t-1},x_{t-1}^{k_{t-1}} )=\prod_{R\in \mathcal{R}}p_t^R(k_t^{R}\mid k_{t-1}, x_{t-1}^{k_{t-1} \cap R} ).
\end{align}
That is, the value of $K_t$ in region $R\in \mathcal{R}$, denoted as $K_t^R$, is generated by the global value $K_{t-1}$ and the latent state in that region $x_{t-1}^{k_{t-1} \cap R}$. In this manner, users could adjust the tracking intensity based on the monitoring status of a given region. For instance, if $x_{t-1}^{k_{t-1} \cap R}$ is close to zero, a smaller set of $K_t$ may be more appropriate. In practical terms, when the COVID-infected population in a state is low, the state might adopt less stringent policies and track fewer patients.

In \cite{khan2005mcmc}, they firstly considered the model over a fixed graph, whose transition density takes the following form: 
\begin{align}
	\label{eqn:pairwise_interaction}
	f(x_t\mid x_{t-1})
	=\prod_{v\in V}f^v(x_t^v\mid x_{t-1}^v)\prod_{v'\in N(v)}\exp(-\overline{f}^v(x_t^v,x_t^{v'})),
\end{align}
where $\overline{f}^v$ is a function of two variables.
They then considered the variable target case, wherein $(K_t,X_t^{K_t})$ was divided into $(K_t^E,X_t^{K_t^E})$ for modeling targets entering and $(K_t^S,X_t^{K_t^S})$ for modeling targets staying. Hence, there is no transition probability of $X_t^{K_t^E}$, simply its own probability. The transition probability of $X_t^{K_t^S}$ depends on $X_{t-1}^{K_t^S}$ in a product form, i.e., $$P(X_t^{K_t^S}\mid X_{t-1}^{K_t^S})=\prod_{i\in K_t^S}P(X_t^i\mid X_{t-1}^i).$$ 
In this paper, we combine and generalize their transition densities into a unified form.

Our transition density, for each time $t\in [T]$, is given by
\begin{align}
	\label{eqn:transition_product_form2}
	f_t(x_t^{k_t}\mid k_t,  x_{t-1}^{k_{t-1}})=\prod_{v\in k_t}f_t^v(x_t^{v}\mid k_t^{v(\mathcal{R})}, x_{t-1}^{ k_{t-1}\cap \{v\}})\widetilde{f}_t^v(x_t^v,x_t^{N_t(v)}),
\end{align}
where $\widetilde{f}_t^v$ is a function of the variable $x_t^v$ and his neighborhood vector $x_t^{N_t(v)}$. Here, 
$v(\mathcal{R})$ gives the region that vertex $v$ belongs to, and $N_t(v)$ is the $r$-range neighborhood defined as 
\begin{align}
	\label{eqn:def_neighborhood}
	N_t(v):=\big\{v'\in k_t: d(v,v')\leq r,\, v' \neq v\big\},
\end{align}
with $r$ being a positive interger and $d(v,v')$ being the Euclidean distance between two vertices $v$ and $v'$.
Now, we outline the advantages of our modeling approach in equation \eqref{eqn:transition_product_form2} compared to the STMRF-VD model of \cite{khan2005mcmc}:
\begin{enumerate}[(i)]
	\item We properly formulate the probability transition densities for target entering and staying into a single equation \eqref{eqn:transition_product_form2}. To the best of our knowledge, this is the first instance of incorporating dimension changes. Specifically, for each $v$ in the identifier $K_t=k_t$ at time $t$, $X_t^{v}$ depends on $X_{t-1}^{v}$ only if $v\in k_{t-1}$ (i.e., $v$ stays).
	
	\item We introduce the feature that $X_t^{v}$ could depend on $k_t^{v(\mathcal{R})}$, the identifier set in the region to which $v$ belongs. This feature is necessary; for example, in a region with a high number of active cars, the movements of each car could be restricted.
	
	\item The function $\widetilde{f}_t^v$ serves as a generalization of the pairwise interaction described in equation \eqref{eqn:pairwise_interaction}. This can be easily illustrated in the context of the Gibbs distribution, as explained in Subsection \ref{sec:mrf}.  The STMRF-VD model of \cite{khan2005mcmc} corresponds to a special case with the first and second-order terms as described in equation \eqref{eqn:autonormal}, whereas our model could include higher order terms described in equation \eqref{eqn:generalU}.

	\item At last, a crucial difference is that our transition density is time-inhomogeneous that can vary over time, whereas those in equation \eqref{eqn:pairwise_interaction} are time-homogeneous and remain the same across all time points.
\end{enumerate}

\subsection{Hidden spatiotemporal Markov random field of varying dimension}
\label{sec:Hidden_stmrf}
In epidemiology, $X_t^v$ could represent the true disease status of an individual (e.g., susceptible, exposed, infected, recovered), while the observable data may include noisy measurements. By incorporating partial and noisy observations into the model, one can estimate the true disease dynamics more accurately, allowing for better predictions of future outbreaks and informing public health interventions. We refer to the partially observed STMRF-VD model as the hidden STMRF-VD (HSTMRF-VD) model. Specifically, the state of the STMRF-VD model, $X_t^{K_t}$, is not directly observable and hence is called the latent state; its partial and noisy observations are represented by the observation $Y_t^{K_t}$. 
At each time $t$, the state of $Y_t^{K_t}$ is a random field that is considered conditionally independent given $(X_t^{K_t})_{t\in [T]}$. As $X_t^{v}$ takes values in a Polish state space, we consider the state space of $Y_t^{v}$ to be a Polish space as well, thus allowing it to be discrete or continuous, finite-dimensional or infinite-dimensional. Moreover, since $\mathbb{N}$ is countable, under the product topology, the state space of $Y_t^{K_t}$, defined as $\mathcal{Y}=(\mathbb{Y}^v)^\mathbb{N}$, is also a Polish space. We define the reference measure of $Y_t^{K_t}$ on its state space $\mathcal{Y}$ as $\phi$, and the reference measure of $Y_t^v$ on its state space $\mathbb{Y}^v$ as $\phi^v$, such that $\phi=\prod_{v}\phi^v$.

We define the measurement density of $Y_t$ w.r.t. $\phi$ as $g_t(y_t^{k_t}\mid k_t,x_t^{k_t})$, and define the measurement density of $Y_t^v$  w.r.t. $\phi^v$ as $g_t^v$. Based on the network structure, we have the following product-formed expression:
\begin{align}
	\label{eqn:emission_product_form}
	g_t(y_t^{k_t}\mid k_t, x_t^{k_t})=\prod_{v\in k_t}g_t^v(y_t^{v}\mid k_t^{v(\mathcal{R})},x_t^v).
\end{align}
Our measurement density generalized that of \cite{khan2005mcmc} in the following two  aspects:
\begin{enumerate}[(i)]
	\item We allow the density of $Y_t^{v}$ to depend on $k_t^{v(\mathcal{R})}$, the identifier set within the region to which $v$ belongs. This feature is essential. When the number of measurements to be done is large, the precision of some measurements may be compromised. In practice, healthcare workers may experience fatigue or burnout when conducting a high volume of tests over an extended period. Conversely, the measurement errors are low when only a few tests are required.
	
	\item Our measurement density is time-inhomogeneous, allowing it to change over time $t$, whereas that of \cite{khan2005mcmc} is time-homogeneous. This extension is crucial. Using the example of COVID-19, during the initial stages of transmission, the precision of COVID-19 measurements may be less accurate due to the scarcity of measurement equipment compared to later stages.
\end{enumerate}

\section{Main results}
\label{sec:Main_results}
In this section, we present our main results. In Section \ref{sec:notations_distances}, definitions of graphical notations and distances are presented. In Section \ref{sec:Problemformulation}, we formulate of the high-dimensional latent target tracking problem. Section \ref{sec:Algorithm} outlines the details of our the VT-MRF-SPF algorithm. Section \ref{sec:assumption} elucidates the assumptions required to establish the algorithmic error bound. In Section \ref{sec:bias_bound}, we provide an upper bound for the algorithmic bias. Lastly, Section \ref{sec:variance_bound} offers an upper bound for the algorithmic variance.

\subsection{Graphical notations and distances}
\label{sec:notations_distances}
Define the distance between two vertex sets $W$ and $W'$ as
\begin{equation}
	\label{eqn:distance_clusters_def}
	d(W,W'):=\min_{v\in W}\min_{v'\in W'} d(v,v');
\end{equation}
denote the $r$-inner boundary of $W_t\in k_t$ as the subset of vertices in $W_t$ that can interact with vertices outside $W_t$, i.e.,
\begin{equation}
	\label{eqn:partial_definition}
	\partial W_t:=\big\{v\in W_t: N_t(v)\nsubseteq W_t \big\};
\end{equation}
denote the maximal size of one single cluster up to time $T$ as
\begin{equation}
	\label{eqn:maxsize_cluster}
	|\mathcal{B}|_T^{\infty}:=\max_{s\in [T]\cup \{0\}} \max_{B_s\in\mathcal{B}(k_s)}\card(B_s);
\end{equation}
denote the maximal number of vertices that interact with any vertex in its $r$-neighborhood up to time $T$ as 
\begin{equation}
	\label{eqn:maxn_inter_vertices}
	\Delta_T:=\max_{s\in [T]\cup \{0\}}\max_{v\in k_s}\card\big\{v'\in k_s:d(v,v')\leq r\big\};
\end{equation}
denote the maximal number of vertices in one single region up to time $T$ as 
\begin{equation}
	\label{eqn:maxn_inter_clusters}
	\Delta_T^\mathcal{R}:=\max_{s\in [T]\cup \{0\}}\max_{v\in k_s}\card\big\{v(\mathcal{R})\big\};
\end{equation}
and denote the maximal distance in one single region up to time $T$ as  
\begin{equation}
	\label{eqn:maxn_inter_clusters}
	r_T^\mathcal{R}:=\max_{s\in [T]\cup \{0\}}\max_{v,v'\in k_s, v'\in v(\mathcal{R})}d(v,v').
\end{equation}
We assume that $r$, $\Delta_T$, $\Delta_T^\mathcal{R}$, and $r_T^\mathcal{R}$ are greater than and equal to one throughout the paper.

Between two random measures $\rho$ and $\rho'$ on space $\mathbb{S}$, we define the distance
\begin{equation}
	\label{eqn:vertiii_definition}
	\vertiii{\rho- \rho'}:=\sup_{h\in \mathcal{S}:|h|\leq 1}\left[\mathbb{E}|\rho(h)-\rho'(h)|^2\right]^{1/2},
\end{equation}
where $\mathcal{S}$ denotes the class of measurable
functions $h: \mathbb{S}\rightarrow \mathbb{R}$ and $$\rho(h):=\int h(x) d\rho(x)=\int h(x) \rho(dx).$$
Between two random measures $\rho$ and $\rho'$ on space $\mathbb{S}$, we define the local distance, for $\K\subseteq k_s$ with $s\in [T]\cup \{0\}$,
\begin{equation}
	\label{eqn:vertiii_K_definition}
	\vertiii{\rho- \rho'}_{\K}:=\sup_{h\in \mathcal{S}^\K:|h|\leq 1}\left[\mathbb{E}|\rho(h)-\rho'(h)|^2\right]^{1/2},
\end{equation}
where $\mathcal{S}^\K$ denotes the class of measurable
functions $h: \mathbb{S}\rightarrow \mathbb{R}$ such that $h(x)=h(\overline{x})$ whenever $x^{\K}=\overline{x}^{\K}$. 
Between two probability measures $\rho$ and $\rho'$ on $\mathbb{S}$, we define the total variation distance 
\begin{equation}
	\label{eqn:tvd}
	\|\rho-\rho'\|:=\sup_{h\in \mathcal{S}:|h|\leq 1}|\rho(h)-\rho'(h)|,
\end{equation}
and define the local total
variation distance, for $\K\subseteq k_s$ with $s\in [T]\cup \{0\}$,
\begin{equation}
	\label{eqn:ltvd}
	\|\rho-\rho'\|_{\K}:=\sup_{h\in \mathcal{S}^{\K}:|h|\leq 1}|\rho(h)-\rho'(h)|.
\end{equation}

\subsection{Problem formulation}
\label{sec:Problemformulation}
We assume that the triple of processes $(X,Y, K)$ is realized on one canonical probability space, and denote $\mathbb{P}$ and $\mathbb{E}$ as the probability measure and expectation on that space respectively. Given the observations $\{Y_1^{K_1},\ldots,Y_T^{K_T}\}$, we aim to approximate the nonlinear filter
\begin{align*}
	\pi_T(A)=\mathbb{P}\Big[X_T^{K_T} \in A \,\big | \,Y_1^{K_1},\ldots,Y_T^{K_T}\Big].
\end{align*}
We consider $\pi_0=\delta_x$, the Dirac measure centered on $x$. Then the nonlinear filter could be expressed recursively as
\begin{align}
	\label{eqn:pi_recursive} 
	\pi_s= \mathsf{F}_s \pi_{s-1}, \qquad s\in [T],
\end{align}
where, for any probability measure $\mu_{s-1}$ on $\mathcal{X}$ at time $s-1$ and any set $A \subseteq \mathcal{X}$, 
the operator $\mathsf{F}_s$ is given by
\begin{align}
	&(\mathsf{F}_s\mu_{s-1})(A)\\
	&=\dfrac{\splitdfrac{
			\int\mathbbm{1}_A(x_s^{k_s})\prod_{\omega\in k_s}f_s^{\omega}(x_s^{\omega}\mid k_s^{\omega(\mathcal{R})}, x_{s-1}^{k_{s-1}\cap \{\omega\}})g_s^{\omega}(y_s^{\omega}\mid k_s^{\omega(\mathcal{R})}, x_s^{\omega}) \widetilde{f}_s^{\omega}(x_s^{\omega},x_s^{N_s(\omega)})}{\times \prod_{R\in \mathcal{R}}p_s^R(k_s^{R}\mid k_{s-1},  x_{s-1}^{k_{s-1} \cap R} ) \mu_{s-1}(dx_{s-1}^{k_{s-1}})\psi(dx_s^{k_s})}}
	{\splitdfrac{
			\int\prod_{\omega\in k_s}f_s^{\omega}(x_s^{\omega}\mid k_s^{\omega(\mathcal{R})}, x_{s-1}^{k_{s-1}\cap \{\omega\}})g_s^{\omega}(y_s^{\omega}\mid k_s^{\omega(\mathcal{R})}, x_s^{\omega}) \widetilde{f}_s^{\omega}(x_s^{\omega},x_s^{N_s(\omega)})}{\times \prod_{R\in \mathcal{R}}p_s^R(k_s^{R}\mid k_{s-1},  x_{s-1}^{k_{s-1} \cap R} ) \mu_{s-1}(dx_{s-1}^{k_{s-1}})\psi(dx_s^{k_s})}}.\nonumber
\end{align}
It is instructive to write the recursion  $\mathsf{F}_s=\mathsf{C}_s\mathsf{P}_s$ in two steps
\begin{align*}
	\pi_{s-1} \xrightarrow[]{\text{prediction}} \pi_{s|s-1}=\mathsf{P}_s\pi_{s-1}
	\xrightarrow[]{\text{correction}} \pi_s=\mathsf{C}_s\pi_{s|s-1}
\end{align*}
where the prediction operator $\mathsf{P}_s$ is defined as 
\begin{align}
	\label{eqn:prediction_operator_definition}
	&(\mathsf{P}_s\rho)(h)\\
	&=\int  h(x_s^{k_s})\prod_{\omega\in k_s}f_s^{\omega}(x_s^{\omega}\mid k_s^{\omega(\mathcal{R})}, x_{s-1}^{k_{s-1}\cap \{\omega\}})\prod_{R\in \mathcal{R}}p_s^R(k_s^{R}\mid k_{s-1},  x_{s-1}^{k_{s-1} \cap R} )\psi(dx_s^{k_s})\rho(dx_{s-1}^{k_{s-1}}),\nonumber
\end{align}
and the correction operator $\mathsf{C}_s$ is defined as 
\begin{equation}
	\begin{split}
		\label{eqn:correction_operator_definition}
		(\mathsf{C}_s\rho)(h)
		&=\dfrac{
			\int h(x_s^{k_s})\prod_{\omega\in k_s}g_s^{\omega}(y_s^{\omega}\mid k_s^{\omega(\mathcal{R})}, x_s^{\omega}) \widetilde{f}_s^{\omega}(x_s^{\omega},x_s^{N_s(\omega)}) \rho(dx_{s}^{k_{s}})}
		{
			\int \prod_{\omega\in k_s}g_s^{\omega}(y_s^{\omega}\mid k_s^{\omega(\mathcal{R})}, x_s^{\omega}) \widetilde{f}_s^{\omega}(x_s^{\omega},x_s^{N_s(\omega)}) \rho(dx_{s}^{k_{s}})},
	\end{split}
\end{equation}
for any probability measure $\rho$ on $\mathcal{X}$.

\subsection{Algorithm}
\label{sec:Algorithm} 
\begin{algorithm}[t!]
	{	\small
		\noindent	Notations: $[T]$ is the time index set, $[N]$ is the Monte Carlo index set, and $\mathcal{B}(k_t)$ is the cluster partition for indentifier $k_t$ at time $t$.
		\smallskip
		
		\noindent\begin{tabular}{l}
			Iterate for $t\in [T]$:	\mystretch\\
			1. Sample i.i.d. $(k_{t-1}^{(n)},x_{t-1}^{k_{t-1}^{(n)},(n)})$ with probability $\widehat{\pi}_{t-1}$ for $n\in [N]$.
			\mystretch\\
			2. Sample identifier according to $p_t(k_t^{(n)}\mid k_{t-1}^{(n)}, x_{t-1}^{k_{t-1}^{(n)},(n)})$ for $n\in [N]$. 
			\mystretch\\
			3. Sample staying targets from $f_t^v(x_t^{v,(n)}\mid k_t^{v(\mathcal{R}),(n)}, x_{t-1}^{v,(n)})$ to each target $v\in k_t^{S,(n)}$ for $n\in [N]$.
			\mystretch\\
			4. Sample new targets from $f_t^v(x_t^{v,(n)}\mid k_t^{v(\mathcal{R}),(n)})$ to each target $v\in k_t^{E,(n)}$ for $n\in [N]$. \\
			\hspace{0.45cm}Then one obtains new states $(k_t^{(n)},x_t^{k_t^{(n)},(n)})$ for $n\in [N]$.
			\mystretch\\
			5. For $B_t\in\mathcal{B}(k_t)$\\
			6. \asp\; Compute $w_t^{B_t,(n)} =  \prod\limits_{v\in B_t}g_t^v(y_t^{v}\mid k_t^{v(\mathcal{R}),(n)},x_t^{v,(n)})\widetilde{f}_t^v(x_t^{v,(n)},x_t^{N_t(v),(n)})$ for $n\in [N]$  \mystretch\\
			7. \asp\; Compute $\widetilde{w}_t^{B_t,(n)} = w_t^{B_t,(n)}\Big\slash \sum_{n=1}^{N}w_t^{B_t,(n)}$.  \mystretch\\
			8. End For\\
			9. Compute $\widehat{\pi}_t =\bigotimes_{B_t\in\mathcal{B}(k_t)}\sum_{n=1}^N \widetilde{w}_t^{B_t,(n)}\delta(k_t^{(n)},x_t^{k_t^{(n)},(n)})$.
	\end{tabular}}
	\caption{(The the VT-MRF-SPF algorithm)}
	\label{Algorithm_VT-MRF-SPF}
\end{algorithm}

\begin{figure}[t!]
	\centering
	\subfloat{\includegraphics[width = 2in]{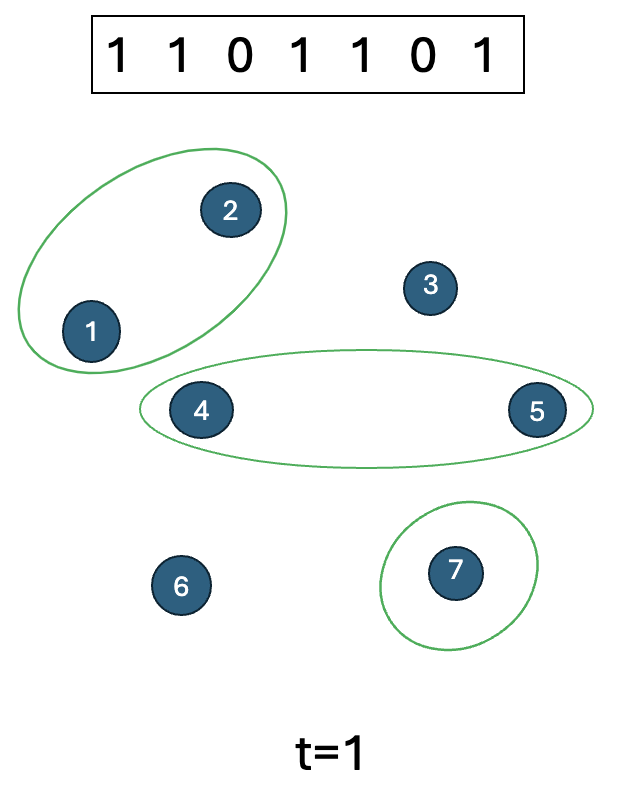}}\hspace{1cm}\hfil
	\subfloat{\includegraphics[width =2in]{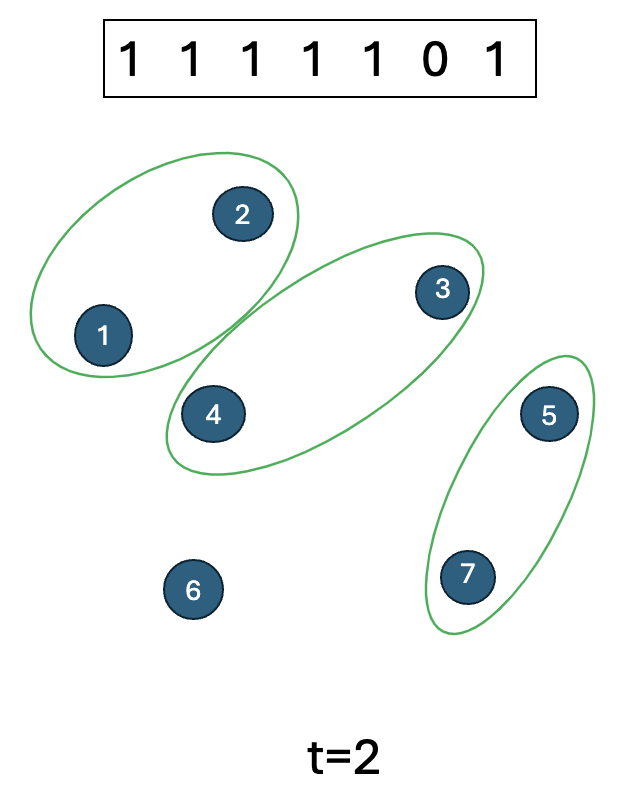}}
	\caption{Illustration of the time-evolving cluster partition based on the time-evolving identifier. Consider a fixed cluster size of 2, the identifier at time 1 as $k_1=\{1,2,4,5,7\}$, and the identifier at time 2 as   $k_2=\{1,2,3,4,5,7\}$.}
	\label{fig:time-evolvingcluster}
\end{figure}

We propose the VT-MRF-SPF in Algorithm \ref{Algorithm_VT-MRF-SPF}. Upon reviewing the pseudocode, we can see that a key distinction is in the update procedure, which utilizes time-evolving clusters. To be precise, we consider a time-evolving cluster partition $\mathcal{B}(k_t)$ that divides $k_t$ into nonoverlapping clusters, i.e.,
\begin{align}
	\label{eqn:cluster_partition}
	k_t=\bigcup_{B_t\in \mathcal{B}(k_t)}B_t, \quad\quad B_t\cap B_t'=\emptyset\; \text{for any}\; B_t\neq B_t'\;\text{and}\; B_t,B_t'\in \mathcal{B}(k_t).
\end{align}
Using the partition, we can express the joint conditional density of $x_t^{k_t}$ and $y_t^{k_t}$ as follows:
\begin{align*}
	&\hspace{-1cm}q_t(x_t^{k_t},\, y_t^{k_t} \mid k_t,  x_{t-1}^{k_{t-1}})\\
	&=\prod_{B_t\in \mathcal{B}(k_t)}\prod_{{\omega}\in B_t}f_t^{\omega}(x_t^{\omega}\mid k_t^{\omega(\mathcal{R})}, x_{t-1}^{k_{t-1}\cap \{\omega\}})\widetilde{f}_t^{\omega}(x_t^{\omega},x_t^{N_t(\omega)})g_t^{\omega}(y_t^{\omega}\mid k_t^{\omega(\mathcal{R})}, x_t^{\omega}).
\end{align*}
Fig \ref{fig:time-evolvingcluster} illustrates the time-evolving cluster partition based on the time-evolving identifier. For a fixed cluster size of 2,  the cluster partition $\mathcal{B}(k_1)=\{\{1,2\}, \{4,5\}, \{7\}\}$ at time $t=1$ when $k_1=\{1,2,4,5,7\}$, while the cluster partition $\mathcal{B}(k_2)=\{\{1,2\}, \{3,4\}, \{5,7\}\}$ at time $t=2$ when $k_2=\{1,2,3,4,5,7\}$.

The the VT-MRF-SPF approximates the true filter $\pi_s$, whose recursion is provided in \eqref{eqn:pi_recursive}, with two primary features: Monte Carlo sampling and cluster-based updates.  These can be mathematically quantified using two operators, $\mathsf{S}^{N}$ and $\mathsf{B}_s$ for $s\in [T]$,  where $\mathsf{S}^{N}$ represents the sampling operator for any probability measure $\rho$ 
\begin{equation}
	\label{eqn:samplingoperator}
	\mathsf{S}^{N}\rho=\frac{1}{N}\sum_{j=1}^N\delta_{x_j},
\end{equation}
with $\{x_j\}_{j\in [J]}$ being i.i.d. samples distributed according to $\rho$,
and $\mathsf{B}_s$ is defined as the clustering operator 
\begin{equation}
	\label{eqn:clusteringoperator}
	\mathsf{B}_s\rho:=\bigotimes_{B_s\in \mathcal{B}(k_s)}\mathsf{B}^{B_s} \rho,
\end{equation}
with $\mathsf{B}^{B_s} \rho$ being the marginal of $\rho$ on $\mathbb{X}^{B_{s}}$. Using these two operators, we can recursively formulate the approximate filter $\widehat{\pi}_t$ for the VT-MRF-SPF as
\begin{align}
	\label{eqn:widehat_pi_recursive} \widehat{\pi}_s=\widehat{\mathsf{F}}_s \widehat{\pi}_{s-1}, \qquad s\in [T],
\end{align}
where $\widehat{\pi}_0=\delta_x$. Here,
the opertator $\widehat{\mathsf{F}}_s$ is given by
\begin{equation}
	\label{eqn:widehat_pi_def} 
	\widehat{\mathsf{F}}_s=\mathsf{C}_s \mathsf{B}_s\mathsf{S}^{N}\mathsf{P}_s
\end{equation}
evolving as follows:
\begin{align*}
	\widehat{\pi}_{s-1} \xrightarrow[\text{sampling}]{\text{prediction}} \widehat{\pi}_{s|s-1}=\mathsf{S}^{N}\mathsf{P}_s\widehat{\pi}_{s-1}
	\xrightarrow[\text{correction}]{\text{clustering}} \widehat{\pi}_s=\mathsf{C}_s \mathsf{B}_s \widehat{\pi}_{s|s-1},
\end{align*}
where the operators $\mathsf{P}_s$ and $\mathsf{C}_s$ are those utilized in the recursion of the true filter and are defined in equations \eqref{eqn:prediction_operator_definition} and \eqref{eqn:correction_operator_definition}, respectively.

\subsection{Assumption}
\label{sec:assumption}

Theoretical results of this paper rely on the following assumption: 
\begin{assumption} 
	\label{assumption}
	For any $v\in k_t$, $x_{t-1}^{ k_{t-1}}\in \mathcal{X}$, $x_t^v\in \mathbb{X}^v$, $y_t^v\in \mathbb{Y}^v$, and $t\in [T]$, we impose the following local conditions:
	\begin{enumerate}[(1)]
		\item there exist $\epsilon_u>0$ and $\epsilon_d>0$ such that
		$$\epsilon_d\leq f_t^v(x_t^{v}\mid k_t^{v(\mathcal{R})}, x_{t-1}^{ k_{t-1}\cap \{v\}})\leq \epsilon_u;$$
		
		\item there exist $\epsilon'_u>0$ and $\epsilon'_d>0$ such that
		$$\epsilon'_d\leq\widetilde{f}_t^{v}(x_t^{v},x_t^{N_t(v)})\leq \epsilon'_u;$$
		
		\item there exist $\gamma_u>0$ and $\gamma_d>0$ such that
		$$\gamma_d \leq g_t^v(y_t^{v}\mid k_t^{v(\mathcal{R})},x_t^v)\leq \gamma_u.$$
	\end{enumerate}
	For any $R\in \mathcal{R}$, $x_{t-1}^{ k_{t-1}}\in \mathcal{X}$, $t\in [T]$, and $k_t, k_{t-1}\subseteq \mathbb{N}$, we impose the regional condition: there exist $\kappa_u>0$ and $\kappa_d>0$ such that
	$$\kappa_d\leq p_t^R(k_t^{R}\mid k_{t-1},  x_{t-1}^{k_{t-1} \cap R} )\leq  \kappa_u.$$
	At last, we suppose
	\begin{align}
		\label{eqn:main_thm_condition}
		\frac{\epsilon_d}{\epsilon_u}\frac{\epsilon_d'}{\epsilon_u'}\frac{\kappa_d}{\kappa_u}>\left(1-\frac{1}{6(\Delta_T+\Delta_T^\mathcal{R})|\mathcal{B}|_T^{\infty}} \right).
	\end{align}
\end{assumption}

In Assumption \ref{assumption}, we assume that the local and regional densities are bounded on both sides, thereby endowing the underlying Markov chain with strong ergodicity.  Such assumptions are routinely assumed in PF literatures, where they are typically considered from global perspectives. However, we interpret them here in local and regional contexts, respectively.  The last condition \eqref{eqn:main_thm_condition} and the fact that $|\mathcal{B}|_T^{\infty}\geq 1$ ensure the positivity of $\beta_T$ defined as 
\begin{align}
	\label{eqn:beta_definition}
	\beta_T=-\frac{1}{r+r_T^\mathcal{R}}\log\left(6\left(1-\frac{\epsilon_d}{\epsilon_u}\frac{\epsilon_d'}{\epsilon_u'}\frac{\kappa_d}{\kappa_u}\right)(\Delta_T+\Delta_T^\mathcal{R}) \right),
\end{align}
which will be used throughout the paper.
In contrast to  the local and regional assumptions, this condition stems from the neighborhood interaction modeling, which necessitates the aggregation of factors influencing transition dynamics.

\subsection{Algorithmic bias bound}
\label{sec:bias_bound}
Since the VT-MRF-SPF utilizes a cluster-based update scheme, in Theorem \ref{thm:main_theorem1} below, we  examine the bias arising from the time-evolving cluster partition that is mathematically described in equation \eqref{eqn:cluster_partition}. For this purpose, we define a cluster filter $\widetilde{\pi}_t$ with $\widetilde{\pi}_0=\delta_x$, which can be recursively expressed as: 
\begin{align}
	\label{eqn:widetilde_pi_recursive} \widetilde{\pi}_s=\widetilde{\mathsf{F}}_s \widetilde{\pi}_{s-1}, \qquad s\in [T],
\end{align}
such that for any probability measure $\mu_{s-1}$ on $\mathcal{X}$ at time $s-1$ and any set $A \subseteq \mathcal{X}$, 
\begin{align}
	\label{eqn:widetilde_pi_def} 
	(\widetilde{\mathsf{F}}_s\mu_{s-1})(A)
	=\dfrac{\splitdfrac{\splitdfrac{
				\int\mathbbm{1}_A(x_s^{k_s})\prod_{B_s\in \mathcal{B}(k_s)} \int\prod_{{\omega}\in B_s}f_s^{\omega}(x_s^{\omega}\mid k_s^{\omega(\mathcal{R})}, x_{s-1}^{k_{s-1}\cap \{\omega\}})}{\hspace{4cm}\times g_s^{\omega}(y_s^{\omega}\mid k_s^{\omega(\mathcal{R})}, x_s^{\omega})\widetilde{f}_s^{\omega}(x_s^{\omega},x_s^{N_s(\omega)}) }}{\times \prod_{R\in \mathcal{R}}p_s^R(k_s^{R}\mid k_{s-1},  x_{s-1}^{k_{s-1} \cap R} ) \mu_{s-1}(dx_{s-1}^{k_{s-1}})\psi(dx_s^{k_s})}}
	{\splitdfrac{\splitdfrac{
				\int \prod_{B_s\in \mathcal{B}(k_s)} \int\prod_{{\omega}\in B_s}f_s^{\omega}(x_s^{\omega}\mid k_s^{\omega(\mathcal{R})}, x_{s-1}^{k_{s-1}\cap \{\omega\}})}{\hspace{4cm}\times g_s^{\omega}(y_s^{\omega}\mid k_s^{\omega(\mathcal{R})}, x_s^{\omega})\widetilde{f}_s^{\omega}(x_s^{\omega},x_s^{N_s(\omega)}) }}{\times \prod_{R\in \mathcal{R}}p_s^R(k_s^{R}\mid k_{s-1},  x_{s-1}^{k_{s-1} \cap R} ) \mu_{s-1}(dx_{s-1}^{k_{s-1}})\psi(dx_s^{k_s})}}.
\end{align}	
That is, $\widetilde{\mathsf{F}}_s=\mathsf{C}_s \mathsf{B}_s\mathsf{P}_s$ evolves as follows:
\begin{align*}
	\widetilde{\pi}_{s-1} \xrightarrow[]{\text{prediction}} \widetilde{\pi}_{s|s-1}=\mathsf{P}_s\widetilde{\pi}_{s-1}
	\xrightarrow[\text{correction}]{\text{clustering}}  \widetilde{\pi}_s=\mathsf{C}_s \mathsf{B}_s \widetilde{\pi}_{s|s-1}.
\end{align*}

We can see that the difference between the evolutions of $\widehat{\pi}_{T}$ and $\widetilde{\pi}_{T}$ lies in the sampling process. Therefore, $\widetilde{\pi}_T$ represents a theoretical filter generated by cluster updates based on the time-evolving cluster partition. Subsequently, the bias introduced by the VT-MRF-SPF  can be quantified mathematically as $\|\widetilde{\pi}_T-\pi_T\|_{\K}$ for any set $\K$, utilizing the local total variation distance defined in equation \eqref{eqn:ltvd}, rather than the $\vertiii{\,\cdot\,}_{\K}$ distance defined in equation \eqref{eqn:vertiii_K_definition} for two random measures. The following theorem demonstrates that the bias can be upper bounded by local quantities alone. A rigorous proof is provided in the Supplement.
\begin{theorem}
	\label{thm:main_theorem1}
	Under Assumption \ref{assumption}, 
	for $B_T\in \mathcal{B}(k_T)$ and $\K \subseteq B_T$, we have
	\begin{align*}
		\|\widetilde{\pi}_T-\pi_T\|_{\K}<  \frac{8e^{-\beta_T}}{1-e^{-\beta_T}}\left(1-\frac{\epsilon_d}{\epsilon_u}\right)\card(\K) \left[\max_{s\in [T]}\max_{B_s'\in \mathcal{B}(k_s)} e^{-\beta_T d(\K,\partial B_s')}\right].
	\end{align*}
\end{theorem}

The definition of $\beta_T$ in equation \eqref{eqn:beta_definition} involves only constants, with $T$ used to quantify these constants up to a specific time point of interest. If one takes fixed  $r_T^\mathcal{R}$, $\Delta_T$, and $\Delta_T^\mathcal{R}$ across time, then $\beta_T$ would be irrelevant of $T$. Additionally, the term $e^{-\beta_T d(\K,\partial B_s')}$ by definitions \eqref{eqn:distance_clusters_def} and \eqref{eqn:partial_definition}, says that as the distance of $\K$ to a graph partition increase, the less impact that graph patition would be on it, which conforms to the common sense. 
Given that the constant $\beta_T$ is positive and the distance $d(\K,\partial B_s')$ is positive, this term lies in the range $(0,1)$. In sum, we can see that the upper bound in Theorem \ref{thm:main_theorem1} has no error accumulated over the time dimension. Next, the upper bound only involves local graphical quantities. Notably, it only has the cardinality of a set  $\card(\K)$, not the cardinality of a whole identifier set $\card(k_t)$, thus overcoming the COD. As the upper bound increases monotonically with the cardinality of $J$, the algorithm favors smaller cluster sizes. The term $\max_{s\in [T]}\max_{B_s'\in \mathcal{B}(k_s)} e^{-\beta_T d(\K,\partial B_s')}$ further supports this preference. It suggests that with a larger size of $J$, the distance of $J$ to the boundary of any cluster becomes smaller, leading to a more pronounced bias. 

The proof of Theorem \ref{thm:main_theorem1} relies on the Dobrushin comparison theorem to control the error accumulation over the time dimension and to quantify the impact induced by the graph partition within a cluster, which is described below to have this paper self-contained.
\begin{theorem}[Dobrushin comparison theorem, Theorem $8.20$ in \cite{georgii2011gibbs}]
	\label{thm:Dobrushin}
	\hfill \break
	Let $I$ be a finite set. Let $\mathbb{S}=\prod_{i\in I}\mathbb{S}^i$, where $\mathbb{S}^i$ is a Polish space for each $i\in I$. Define the coordinate projections $X^i: x \rightarrow x^i$ for $x\in \mathbb{S}$ and $i\in I$. For probability measures $\rho$ and $\overline{\rho}$ on $\mathbb{S}$, define
	$$\rho_x^i(A)=\rho(X^i\in A\mid X^{I\backslash \{i\}}=x^{I\backslash \{i\}}),$$
	$$\rho_{\overline{x}}^i(A)=\rho(X^i\in A\mid X^{I\backslash \{i\}}={\overline{x}}^{I\backslash \{i\}}),$$
	$$\overline{\rho}_x^i(A)=\overline{\rho}(X^i\in A\mid X^{I\backslash \{i\}}=x^{I\backslash \{i\}}),$$
	$$C_{ij}=\frac{1}{2}\sup_{x,\overline{x}\in \mathbb{S}:\atop x^{I\backslash \{j\}}=\overline{x}^{I\backslash \{j\}}}\| \rho_{x}^i- \rho_{\overline{x}}^i\| \quad\text{and}\quad b_j=\sup_{x\in \mathbb{S}}\|\rho_{x}^j-\overline{\rho}_{x}^j\|.$$
	If the Dobrushin condition 
	$\max_{i\in I}\sum_{j\in I}C_{ij}<1$
	holds, then for every $J \subseteq I$,
	$$\|\rho-\overline{\rho}\|_J\leq \sum_{i\in J}\sum_{j\in I}D_{ij}b_j,$$
	where $D:=\sum_{n\in \mathbb{N}}C^n<\infty.$
\end{theorem}

\subsection{Algorithmic variance bound}
\label{sec:variance_bound}
After addressing the algorithmic bias in Theorem \ref{thm:main_theorem1}, our next objective is to investigate the variance generated by the VT-MRF-SPF. Recalling that the bias is generated by the graph partition, the variance is merely produced by the Monte Carlo samplings. In Theorem \ref{thm:main_theorem2}, we quantify the variance using the local metric $\vertiii{\,\cdot\,}_{\K}$ defined in equation \eqref{eqn:vertiii_K_definition} for random measures. A rigorous proof is provided in the Supplement.
\begin{theorem}
	\label{thm:main_theorem2}
	Under Assumption \ref{assumption}, for $B_T\in \mathcal{B}(k_T)$ and $\K \subseteq B_T$, we have
	\begin{align*}
		\vertiii{\widetilde{\pi}_T-\widehat{\pi}_T}_{\K}
		< \frac{64}{\sqrt{N}}\left(\frac{\epsilon_u^2\kappa_u}{\epsilon_d^2 \epsilon_d'\kappa_d}\right)^{|\mathcal{B}|_T^{\infty}}\left(\frac{\gamma_u}{\gamma_d}\frac{\epsilon_u'}{\epsilon_d'}\right)^{ |\mathcal{B}|_T^{\infty}+(|\mathcal{B}|_T^{\infty})^2} \frac{|\mathcal{B}|_T^{\infty}
			\card(\K)}{1-\exp\Big(-\beta_T+\log(|\mathcal{B}|_T^{\infty})\Big)}.
	\end{align*}
\end{theorem}
Given that the variance generated by the VT-MRF-SPF is due to the Monte Carlo sampling, routine Monte Carlo analysis provides the $\frac{1}{\sqrt{N}}$ factor.
Considering $\beta_T$ only involves local constants up to time $T$, the upper bound of the variance term  is uniform in the time dimension if time uniform graphical quantities (e.g. $\Delta_{t_1}=\Delta_{t_2}$ for any $t_1\neq t_2$) are used. Furthermore, the upper bound only involves local constants and the cardinality of $J$, thus overcoming the COD.   It is worth noting that  $(\epsilon_u^2\kappa_u)/(\epsilon_d^2 \epsilon_d'\kappa_d)\geq 1$ and $(\gamma_u\epsilon_u')/(\gamma_d\epsilon_d')\geq 1$, the upper bound grows exponentially with $|\mathcal{B}|_T^{\infty}$. Recall that $|\mathcal{B}|_T^{\infty}$ defined in equation \eqref{eqn:maxsize_cluster} stands for the maximal size of one single cluster up to time $T$. This is as expected for our clusterwised update scheme, in the same way as the variance  of the PF growing exponentially in terms of the graph dimension. As the bias error bound, the variance error bound grows monotonically with the cardinality of $J$. However, if we let $|\mathcal{B}|_T^{\infty}$ be small, $J$ as a subset of $B_T$ would be small.

Finally, using the triangle inequality and then noting the absence of random sampling in $\|\widetilde{\pi}_T-\pi_T\|_{\K}$, we have
\begin{align}
	\label{eqn:triangle_inequality_CCOD}
	\vertiii{\widehat{\pi}_T-\pi_T}_{\K}
	\leq  \vertiii{\widetilde{\pi}_T-\pi_T}_{\K}+\vertiii{\widetilde{\pi}_T-\widehat{\pi}_T}_{\K}=\|\widetilde{\pi}_T-\pi_T\|_{\K}+\vertiii{\widetilde{\pi}_T-\widehat{\pi}_T}_{\K}.
\end{align} 
Then, Theorems \ref{thm:main_theorem1} and \ref{thm:main_theorem2} yield that under Assumption \ref{assumption}, 
for every $B_T\in \mathcal{B}(k_T)$ and $\K \subseteq B_T$, the algorithmic error of the VT-MRF-SPF
\begin{align}
	\label{eqn:error}
	&\hspace{-0.5cm}\vertiii{\widehat{\pi}_T-\pi_T}_{\K}\\
	< &
\frac{8e^{-\beta_T}}{1-e^{-\beta_T}}\left(1-\frac{\epsilon_d}{\epsilon_u}\right)\card(\K) \left[\max_{s\in [T]}\max_{B_s'\in \mathcal{B}(k_s)} e^{-\beta_T d(\K,\partial B_s')}\right]\nonumber\\
	&+\frac{64}{\sqrt{N}}\left(\frac{\epsilon_u^2\kappa_u}{\epsilon_d^2 \epsilon_d'\kappa_d}\right)^{|\mathcal{B}|_T^{\infty}}\left(\frac{\gamma_u}{\gamma_d}\frac{\epsilon_u'}{\epsilon_d'}\right)^{ |\mathcal{B}|_T^{\infty}+(|\mathcal{B}|_T^{\infty})^2} \frac{|\mathcal{B}|_T^{\infty}
		\card(\K)}{1-\exp\Big(-\beta_T+\log(|\mathcal{B}|_T^{\infty})\Big)}.\nonumber
\end{align}

\section{Numerical analysis}
\label{sec:numerical}
In this section, we conduct numerical analysis to examine the performance of the VT-MRF-SPF. In Section \ref{sec:numerical_model}, we introduce a variant of the widely used CAR model proposed in \cite{leroux2000estimation} that incorporates time-evolving spatial dimensions and time-evolving network interactions, as well as partial observations. In Section \ref{sec:Performance_analysis}, we demonstrate the algorithmic performance of the proposed the VT-MRF-SPF compared to the VT-MRF-PF in \cite{khan2005mcmc}, which are both online learning algorithms applicable to general HSTMRF-VD models. In Section \ref{sec:numerical_real}, further numerical analysis results are provided using the real adjacency matrix, generated by the Greater Glasgow and Clyde health board in Scotland as visualized in Fig \ref{fig:graph_illustration2}.

\subsection{HSTMRF-VD model with CAR latent states}
\label{sec:numerical_model}
In fact, hidden CAR models with fixed spatial dimension and network interaction have been used for example in \cite{napier2016model} and \cite{lee2018spatio}. Specifically, their models' framework contains two components: an overall temporal trend and distinct spatial surfaces for each time period.
However, we allow each spatial location has its own temporal trend. That is, their models can be seen as a special case of ours when all spatial locations' temporal trends take the same value. 

We assess the performance of the VT-MRF-SPF using both continuous and discrete observation models. At each time $t\in [T]$ and location $i\in k_t$, the observation follows a normal distribution in the continuous case:
\begin{align}
	\label{eqn:Gaussian_observation}
	Y_{t}^i \sim \operatorname{Normal}\left(\psi_{t}^i,\nu^2\right),
\end{align}
and the observation follows a Poisson distribution in the discrete case:
\begin{align}
	\label{eqn:Poisson_observation}
	Y_{t}^i \sim \operatorname{Poisson}\left(\mu_{t}^i\right),\qquad\ln \left(\mu_{t}^i\right)=\psi_{t}^i.
\end{align}
The latent state of our HSTMRF-VD model has a temporal component $\varphi=(\varphi_1^{k_1}, \ldots, \varphi_T^{k_T})$ and a spatial component $\phi=(\phi_1^{k_1}, \ldots, \phi_T^{k_T})$. For each $i \in k_t$ and $t\in [T]$,
\begin{align}
	\psi_{t}^i & =\phi_{t}^i+\varphi_t^i, \nonumber\\
	\phi_{t}^i \mid \phi_{t}^{-i}, W(t)& \sim \operatorname{Normal}\left(\frac{\vartheta\sum_{i'\in k_t} w_{i i'}(t) \phi_{t}^{i'}}{\vartheta\sum_{i'\in k_t} w_{i i'}(t)+1-\vartheta},\; \frac{\widetilde{\sigma}_t^2}{\vartheta\sum_{i'=1}^{k_t} w_{i i'}(t)+1-\vartheta}\right),  \label{eqn:phi_dynamics}\\
	\varphi_t^i \mid \varphi_{-t}^i, D & \sim \operatorname{Normal}\left(\frac{\overline{\vartheta} \sum_{t'=1}^T d_{t t'} \varphi_{t'}^{\{i\}\cap k_{t'}}}{\overline{\vartheta}^2 \sum_{t'=1}^T d_{t t'}+1-\overline{\vartheta}^2},\; \frac{\sigma^2}{\overline{\vartheta}^2 \sum_{t'=1}^T d_{t t'}+1-\overline{\vartheta}^2}\right),\label{eqn:varphi_dynamics}
\end{align}
where $$\phi_{t}^{-i}=\left(\phi_{t}^1, \ldots, \phi_{t}^{i-1}, \phi_{t}^{i+1}, \ldots, \phi_{t}^{k_t}\right),$$ $$\varphi_{-t}^i=(\varphi_{1}^{\{i\}\cap k_1},\ldots,\varphi_{t-1}^{\{i\}\cap k_{t-1}}, \varphi_{t+1}^{\{i\}\cap k_{t+1}}\ldots,\varphi_{T}^{\{i\}\cap k_T}).$$ That is, both the temporal and spatial components are modeled by the CAR model, while the latter has a temporally-varying variance parameter $\widetilde{\sigma}_t^2$. 

In the above equations, spatial correlations are controlled by the temporal sequence of symmetric $k_t \times k_t$-dimensional adjacency matrices $\{W(t)\}_{t\in [T]} = \{(w_{i i'}(t))_{i,i'\in k_t}\}_{t\in [T]}$,  where $w_{i i'}(t)$ denotes the spatial proximity between areal units $(S_i, S_{i'})$ and the data availability of these two units. This matrix is assumed to be binary, with $w_{i i'}(t)=1$ if the areal units $(S_i, S_{i'})$ share a common border and both have data available at time $t$, and $w_{i i'}(t)=0$ otherwise. Moreover, $w_{ii}=0$ for all areal units $i \in k_t$. 
Temporal autocorrelation is controlled by a $T \times T$-dimensional  tridiagonal neighborhood matrix $D = (d_{tt'} )$, where $d_{tt'}=1$ if $|t - t'| = 1$ and $d_{tt'}=0$ otherwise. For example, when $k_1=6$ and $T=7$, the two matrices $W(1)$ and $D$ could be given respectively as 
$$ W(1)=\begin{pmatrix}0\;\,&1\;&0\;\,&0\;\,&1\;\,&0\\1&0&1&0&1&0\\0&1&0&1&0&0\\0&0&1&0&1&1\\1&1&0&1&0&0\\0&0&0&1&0&0\end{pmatrix}_{6\times 6}
\qquad\text{and}\qquad
D = \begin{pmatrix}
	0 \;\;& 1 \;\;& 0 \;\;& 0 \;\;& 0 \;\;& 0 \;\;& 0 \\
	1 & 0 & 1 & 0 & 0 & 0 & 0 \\
	0 & 1 & 0 & 1 & 0 & 0 & 0 \\
	0 & 0 & 1 & 0 & 1 & 0 & 0 \\
	0 & 0 & 0 & 1 & 0 & 1 & 0 \\
	0 & 0 & 0 & 0 & 1 & 0 & 1 \\
	0 & 0 & 0 & 0 & 0 & 1 & 0
\end{pmatrix}_{7\times 7}.
$$

To illustrate the earlier-described theory, the conditional probability density $\phi_{t}^i$ can be expressed by equation \eqref{eqn:GMRF_pdf} as
$$
f_t^i(\phi_{t}^i\, \big|\, \phi_{t}^{-i}, W(t))=\frac{1}{\sqrt{2 \pi \sigma_i ^2}} \mathrm{e}^{-\frac{1}{2 \sigma_i ^2}\left[\phi_{t}^i-\mu_i-\sum_{i^{\prime}=1}^{k_t} \beta_{i i^{\prime}}(\phi_{t}^{i^{\prime}}-\mu_{i^{\prime}})\right]^2},
$$
where $\mu_i=0$ for all $i$,
$$\beta_{i i'}=\frac{\vartheta  w_{i i'}(t)}{\vartheta\sum_{i^{\prime}=1}^{k_t} w_{i i'}(t)+1-\vartheta}\qquad\text{and}\qquad \sigma_{i}^2=\frac{\widetilde{\sigma}_t^2}{\vartheta\sum_{i^{\prime}=1}^{k_t} w_{i i'}(t)+1-\vartheta}.$$
By \cite{rue2005gaussian} pages 1-3 therein, $\varphi_t^i$ follows the classical autoregressive process of order 1:
$$
\varphi_t^i=\overline{\vartheta} \varphi_{t-1}^{\{i\}\cap k_{t-1}}+\epsilon_t^i, \qquad \epsilon_t^i \stackrel{\text { iid }}{\sim} \operatorname{Normal}(0,\sigma^2), \qquad|\overline{\vartheta}|<1,
$$
where the index $t$ represents time. That is, for $t=2, \ldots, n$,
$$
\varphi_t^i \;\Big | \; \varphi_1^{\{i\}\cap k_{1}}, \ldots, \varphi_{t-1}^{\{i\}\cap k_{t-1}} \sim \operatorname{Normal}\left(\overline{\vartheta} \varphi_{t-1}^{\{i\}\cap k_{t-1}} , \sigma^2\right).
$$

\subsection{Performance analysis using full adjacency matrix}
\label{sec:Performance_analysis}
In all the experiments conducted, the initial values $\varphi_0=(\varphi_0^1,\ldots, \varphi_0^{k_0})$ were drawn from a uniform distribution in the range $[1,2]$.  The variance parameter $\sigma^2$ in equation \eqref{eqn:varphi_dynamics} was set as $0.1$. The variance parameters $(\widetilde{\sigma}_1^2,\ldots,\widetilde{\sigma}_T^2)$ in equation \eqref{eqn:phi_dynamics} were drawn from the uniform distribution in the range $[1,2]$.
Two parameters $\vartheta$ and $\overline{\vartheta}$ were drawn from the uniform distribution in the range $[0,1]$. 
We now clarify the three notations used in Algorithm \ref{Algorithm_VT-MRF-SPF}: for all experiments conducted, we set the time dimension $T=400$, set the Monte Carlo count $N=800$, and generated time-evolving clusters with equal cluster size $2$ as illustrated in Fig \ref{fig:time-evolvingcluster}. The spatial dimension sizes $(50, 100, 150, 200, 250, 300)$ were tested, providing sufficient range to visualize the spatial scalability of the VT-MRF-SPF for long time series. In order to measure the spatial scalability and fit the results within the same plots, we reported the spatial-scaled log-likelihood values. Specifically, we divided each log-likelihhod by the associated spatial dimension. For fair comparison, we demonstrated the results using the log-likelihood calculation methods in both the VT-MRF-PF and the VT-MRF-SPF, which we named as PF log-likelihood and SPF log-likelihood, respectively. Notably, from spatial dimension $150$, the VT-MRF-SPF consistently outperforms the VT-MRF-PF in both log-likelihood calculation methods.

\begin{figure}[t!]
	\centering
	\subfloat{\includegraphics[width = 2.5in]{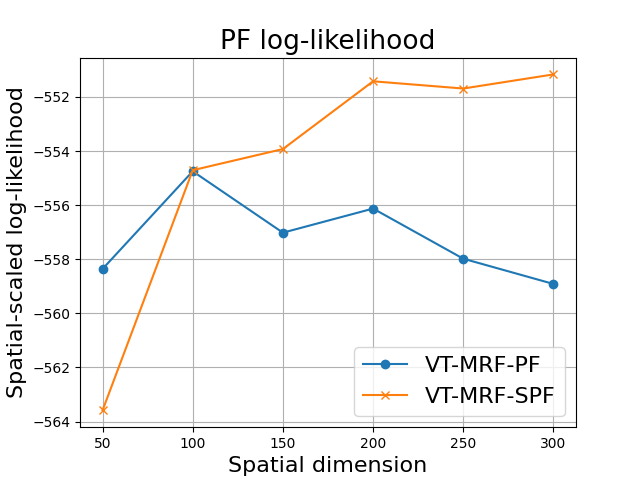}}\hfil
	\subfloat{\includegraphics[width =2.5in]{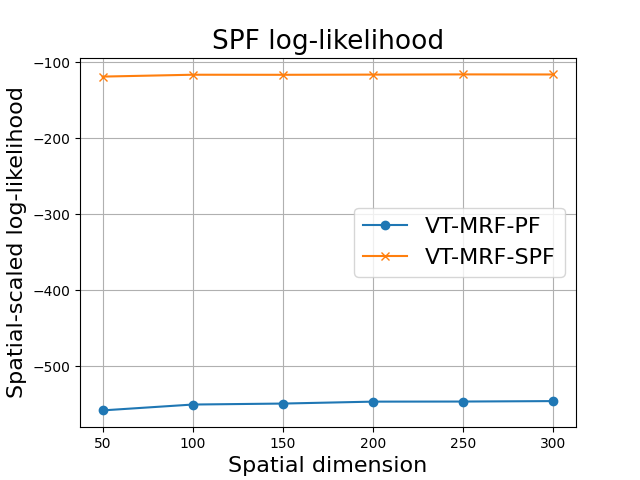}}
	\caption{Comparison of spatial-scaled log-likelihood results for latent state inference, using the HSTMRF-VD model with equal target entering and staying probabilities under a complete spatial graph, assuming normal distributed observation errors. The left figure displays results obtained using the PF log-likelihood employed in the VT-MRF-PF, while the right figure shows results obtained using the SPF log-likelihood utilized in the VT-MRF-SPF.}
	\label{fig:Gaussian_full}
\end{figure}

\begin{figure}[htbp!]
	\centering
	\subfloat{\includegraphics[width = 2.5in]{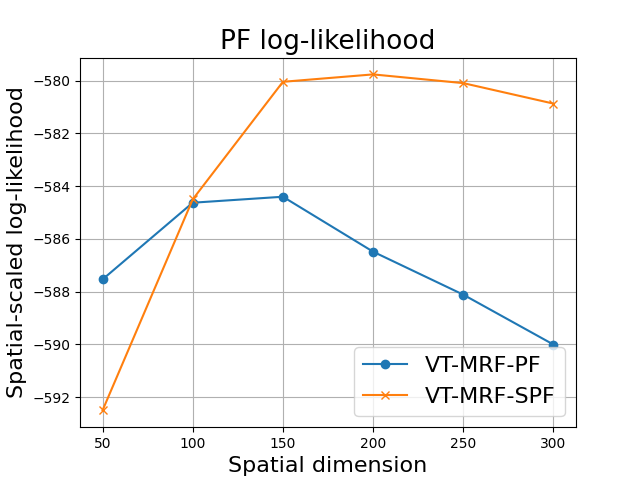}}\hfil
	\subfloat{\includegraphics[width =2.5in]{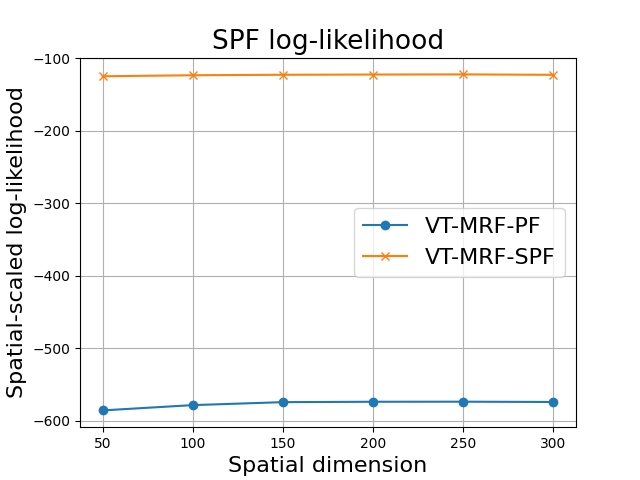}}
	\caption{Comparison of spatial-scaled log-likelihood results for latent state inference, using the HSTMRF-VD model with unequal target entering and staying probabilities under a complete spatial graph, assuming normal distributed observation errors.}
	\label{fig:Gaussian_full_un}
\end{figure}

The experimental results reported in Fig \ref{fig:Gaussian_full} and Fig \ref{fig:Gaussian_full_un} were obtained using the full adjacency matrix for a complete graph where each vertex is connected to all others, for the observation model following the normal distribution as described in equation \eqref{eqn:Gaussian_observation}. In Fig \ref{fig:Gaussian_full}, we considered the target entering probability and staying probability being the same as $0.9$, with the target leaving probability set to $0.1$. We can see that when considering the PF log-likelihood, the VT-MRF-PF is better at spatial dimension $50$, they exhibit comparable performance at spatial dimension $100$, and the VT-MRF-SPF consistently surpasses thereafter. 
Regarding the SPF log-likelihood, the spatial-scaled log-likelihood remains relatively consistent across spatial dimensions, and the VT-MRF-SPF consistently exhibits significantly much better performance compared to the VT-MRF-PF.
 In Fig \ref{fig:Gaussian_full_un}, we explored scenarios where the target entering probability was set to $0.85$ and the target staying probability was set to $0.95$, with the target leaving probability at $0.05$. We noted a similar pattern to that observed in Fig \ref{fig:Gaussian_full}.

\begin{figure}[t!]
	\centering
	\subfloat{\includegraphics[width = 2.5in]{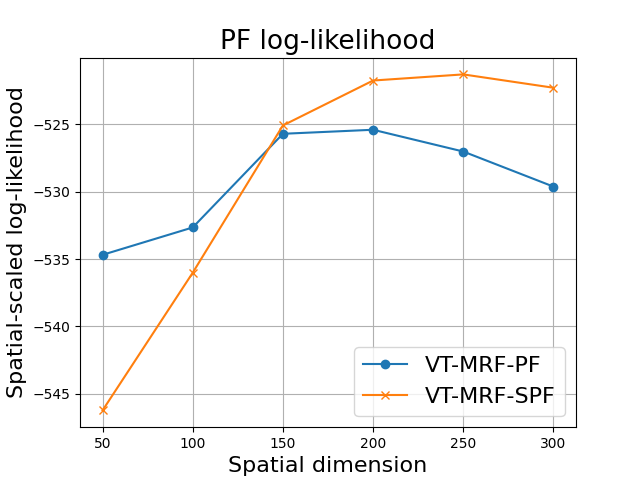}}\hfil
	\subfloat{\includegraphics[width =2.5in]{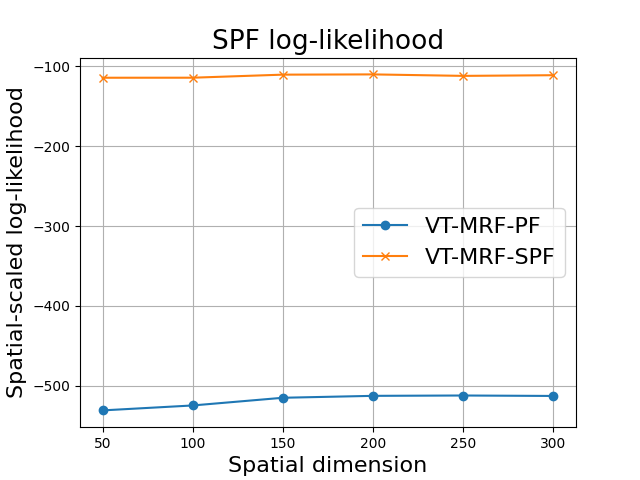}}
	\caption{Comparison of spatial-scaled log-likelihood results for latent state inference, using the HSTMRF-VD model with equal target entering and staying probabilities under a complete spatial graph, assuming Poisson distributed observation errors.}
	\label{fig:Poisson_full}
\end{figure}

\begin{figure}[t!]
	\centering
	\subfloat{\includegraphics[width = 2.5in]{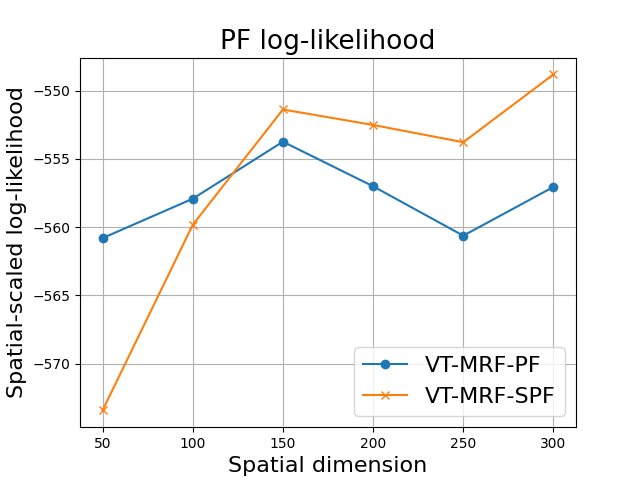}}\hfil
	\subfloat{\includegraphics[width =2.5in]{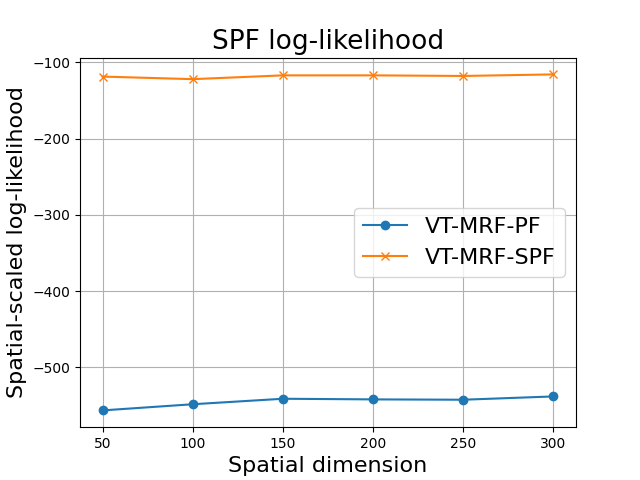}}
	\caption{Comparison of spatial-scaled log-likelihood results for latent state inference, using the HSTMRF-VD model with unequal target entering and staying probabilities under a complete spatial graph, assuming Poisson distributed observation errors. }
	\label{fig:Poisson_full_un}
\end{figure}

The experimental results presented in Fig \ref{fig:Poisson_full} and Fig \ref{fig:Poisson_full_un} were also generated using the adjacency matrix for a complete graph, but with the observation model following a Poisson distribution as described in equation \eqref{eqn:Poisson_observation}.
In Fig \ref{fig:Poisson_full}, we maintained equal entering and staying probabilities. Notably, when considering the PF log-likelihood, the VT-MRF-PF outperforms at spatial dimensions $50$ and $100$, while the VT-MRF-SPF consistently surpassing from spatial dimension $150$.
Regarding the SPF log-likelihood, the spatial-scaled log-likelihood values remain relatively consistent across spatial dimensions, while the VT-MRF-SPF exhibits significantly better performance compared to the VT-MRF-PF.
In Fig \ref{fig:Poisson_full_un}, we explored scenarios with unequal entering and staying probabilities, observing a similar pattern to that observed in Fig \ref{fig:Poisson_full}.

\subsection{Performance analysis using real adjacency matrix}
\label{sec:numerical_real}

In this section, we employ the same setup as described in Section \ref{sec:Performance_analysis}. However, instead of utilizing the full adjacency matrix, we employ the real adjacency matrix generated by the Greater Glasgow and Clyde health board in Scotland, as illustrated in Figure \ref{fig:graph_illustration2}. This health board is one of the 14 regional health boards in Scotland and encompasses the city of Glasgow along with a population of approximately 1.2 million individuals. It is partitioned into M = 271 Intermediate Zones (IZs), which serve as a key geographical unit for the dissemination of small-area statistics in Scotland. Consequently, we conducted tests across spatial dimensions ranging from $(50, 100, 150, 200, 250, 271)$.
\begin{figure}[t!]
	\centering
	\subfloat{\includegraphics[width = 2.5in]{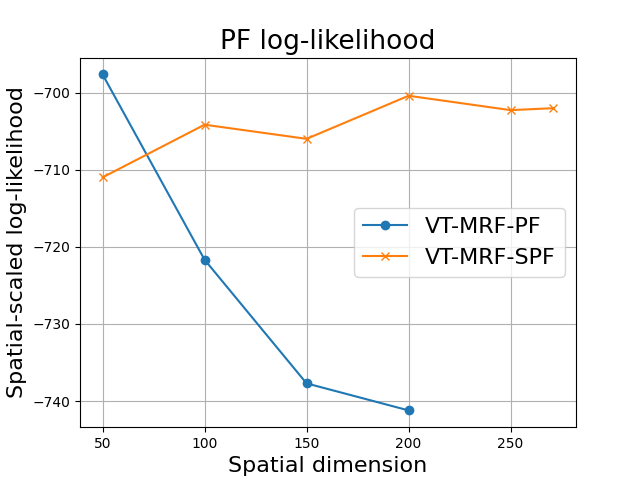}}\hfil
	\subfloat{\includegraphics[width =2.5in]{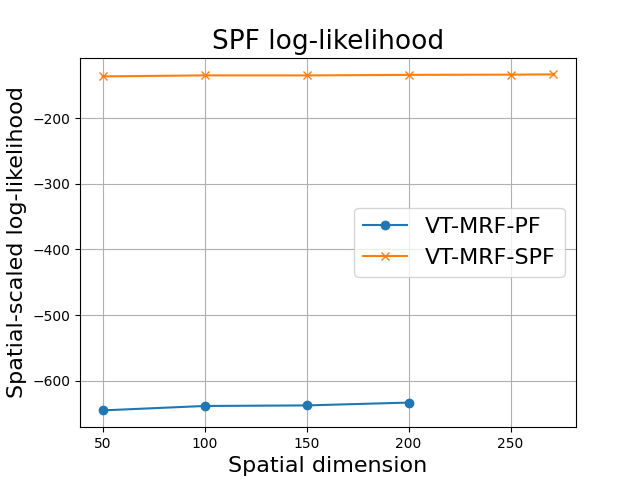}}
	\caption{Comparison of spatial-scaled log-likelihood results for latent state inference, using the HSTMRF-VD model with equal target entering and staying probabilities under a real  spatial graph, assuming normal distributed observation errors. }
	\label{fig:Gaussian_real}
\end{figure}

\begin{figure}[t!]
	\centering
	\subfloat{\includegraphics[width = 2.5in]{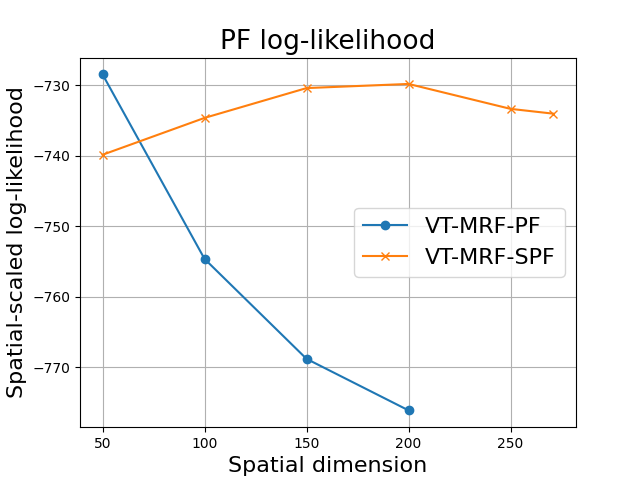}}\hfil
	\subfloat{\includegraphics[width =2.5in]{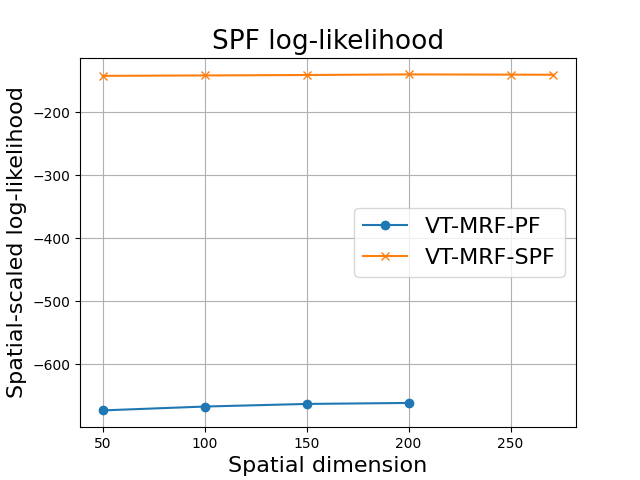}}
	\caption{Comparison of spatial-scaled log-likelihood results for latent state inference, using the HSTMRF-VD model with unequal target entering and staying probabilities under a real  spatial graph, assuming normal distributed observation errors. }
	\label{fig:Gaussian_real_un}
\end{figure}

The experimental results reported in Fig \ref{fig:Gaussian_real} (resp. Fig \ref{fig:Gaussian_real_un}) were obtained for the observation model following a normal distribution as described in equation \eqref{eqn:Gaussian_observation}, with equal (resp. unequal) target entering and staying probabilities. In both Fig \ref{fig:Gaussian_real} and Fig \ref{fig:Gaussian_real_un}, we observe that the VT-MRF-PF demonstrates effective performance only up to a spatial dimension of 200, with optimal performance observed only at spatial dimension 50 under the PF log-likelihood. Conversely, the VT-MRF-SPF exhibits stable performance across all spatial dimensions and consistently outperforms the VT-MRF-PF starting from a spatial dimension of 100, under both log-likelihood calculation methods.

\begin{figure}[t!]
	\centering
	\subfloat{\includegraphics[width = 2.5in]{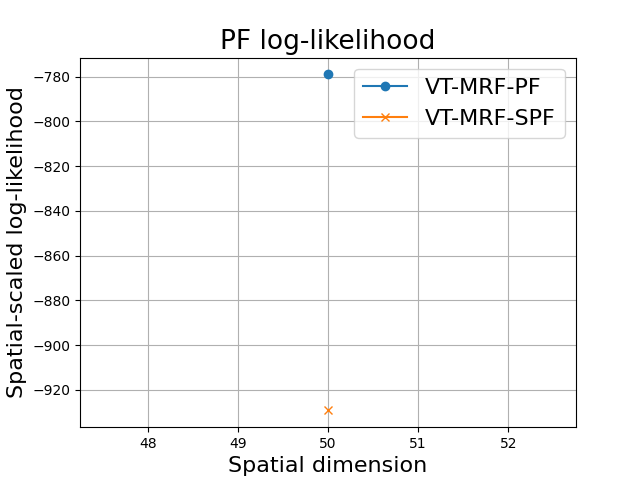}}\hfil
	\subfloat{\includegraphics[width =2.5in]{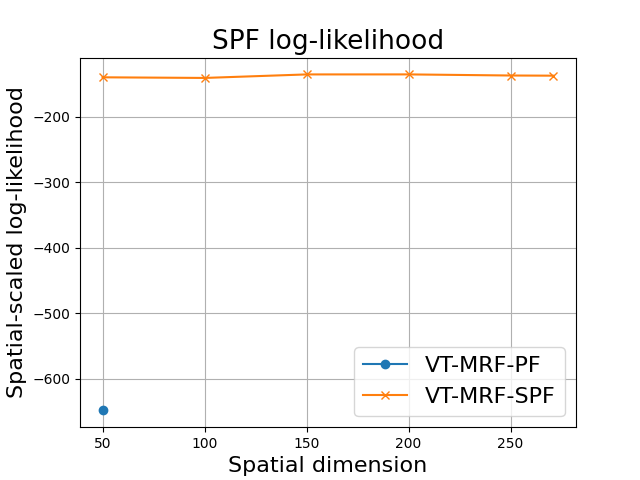}}
	\caption{Comparison of spatial-scaled log-likelihood results for latent state inference, using the HSTMRF-VD model with equal target entering and staying probabilities under a real  spatial graph, assuming Poisson distributed observation errors. }
	\label{fig:Poisson_real}
\end{figure}

\begin{figure}[t!]
	\centering
	\subfloat{\includegraphics[width = 2.5in]{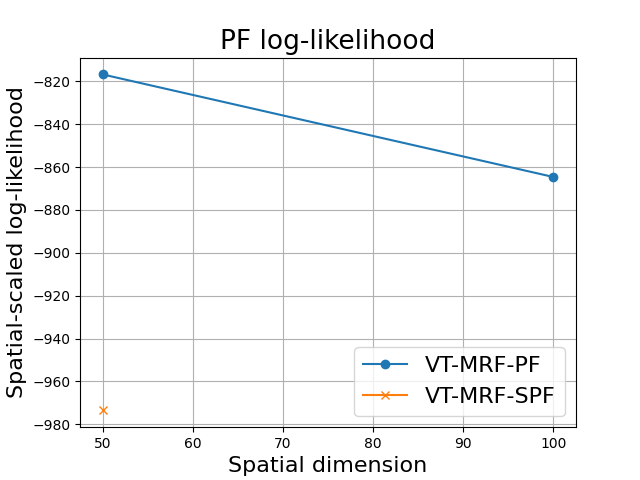}}\hfil
	\subfloat{\includegraphics[width =2.5in]{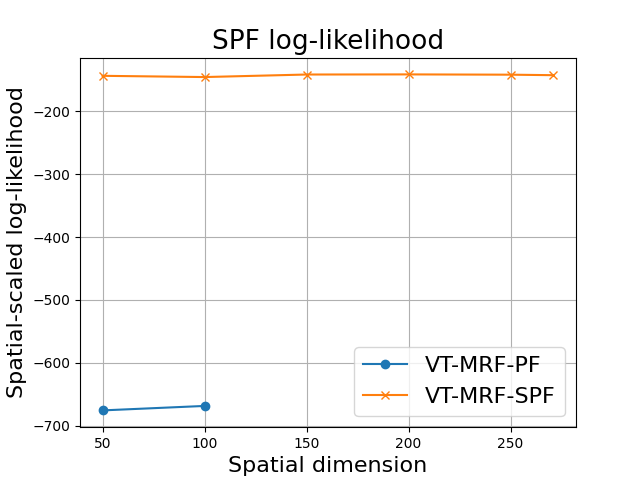}}
	\caption{Comparison of spatial-scaled log-likelihood results for latent state inference, using the HSTMRF-VD model with unequal target entering and staying probabilities under a real spatial graph, assuming Poisson distributed observation errors.}
	\label{fig:Poisson_real_un}
\end{figure}

The experimental results presented in Fig \ref{fig:Poisson_real} (resp. Fig \ref{fig:Poisson_real_un}) were obtained for the observation model following a Poisson distribution as described in equation  \eqref{eqn:Poisson_observation}. These results were obtained under scenarios of equal (resp. unequal) target entering and staying probabilities. We observe that VT-MRF-P demonstrates functionality solely at a spatial dimension of $50$ under the scenario of equal target entering and staying probabilities, and at spatial dimensions $50$ and $100$ under the unequal case. In contrast, the VT-MRF-SPF exhibits functionality across all these spatial dimensions, albeit with inferred results only available at a spatial dimension of $50$ when employing the PF log-likelihood. This outcome is unsurprising, considering the curse of dimensionality associated with the PF and the PF log-likelihood. Upon employing the appropriate SPF log-likelihood,  the VT-MRF-SPF displays stable and scalable performance, surpassing  the VT-MRF-PF in its applicable dimensions.

\begin{funding}
This research project was partially supported by NIH grant 1R21AI180492-01 and the Individual Research Grant at Texas A\&M University. 
\end{funding}


\bibliography{sample}

\begin{thebibliography}{37}
\providecommand{\natexlab}[1]{#1}
\providecommand{\url}[1]{\texttt{#1}}
\expandafter\ifx\csname urlstyle\endcsname\relax
  \providecommand{\doi}[1]{doi: #1}\else
  \providecommand{\doi}{doi: \begingroup \urlstyle{rm}\Url}\fi

\bibitem[Bengtsson et~al.(2008)Bengtsson, Bickel, and Li]{bengtsson2008curse}
Thomas Bengtsson, Peter Bickel, and Bo~Li.
\newblock Curse-of-dimensionality revisited: {C}ollapse of the particle filter
  in very large scale systems.
\newblock In \emph{Probability and statistics: Essays in honor of David A.
  Freedman}, pages 316--334. Institute of Mathematical Statistics, 2008.

\bibitem[Chopin et~al.(2020)Chopin, Papaspiliopoulos,
  et~al.]{chopin2020introduction}
Nicolas Chopin, Omiros Papaspiliopoulos, et~al.
\newblock \emph{An introduction to sequential {Monte Carlo}}, volume~4.
\newblock Springer, 2020.

\bibitem[Chopin et~al.(2023)Chopin, Fulop, Heng, and
  Thiery]{chopin2023computational}
Nicolas Chopin, Andras Fulop, Jeremy Heng, and Alexandre~H Thiery.
\newblock Computational {D}oob h-transforms for online filtering of discretely
  observed diffusions.
\newblock In \emph{International Conference on Machine Learning}, pages
  5904--5923. PMLR, 2023.

\bibitem[Christakos(2017)]{christakos2017spatiotemporal}
George Christakos.
\newblock \emph{Spatiotemporal random fields: theory and applications}.
\newblock Elsevier, 2017.

\bibitem[Del~Moral et~al.(2006)Del~Moral, Doucet, and Jasra]{del2006sequential}
Pierre Del~Moral, Arnaud Doucet, and Ajay Jasra.
\newblock Sequential {Monte Carlo} samplers.
\newblock \emph{Journal of the Royal Statistical Society Series B: Statistical
  Methodology}, 68\penalty0 (3):\penalty0 411--436, 2006.

\bibitem[Descombes et~al.(1998)Descombes, Kruggel, and von
  Cramon]{descombes1998fmri}
Xavier Descombes, Frithjof Kruggel, and D~Yves von Cramon.
\newblock f{MRI} signal restoration using a spatio-temporal {M}arkov random
  field preserving transitions.
\newblock \emph{NeuroImage}, 8\penalty0 (4):\penalty0 340--349, 1998.

\bibitem[Doucet et~al.(2001)Doucet, De~Freitas, and
  Gordon]{doucet2001introduction}
Arnaud Doucet, Nando De~Freitas, and Neil Gordon.
\newblock An introduction to sequential {Monte Carlo} methods.
\newblock \emph{Sequential {Monte Carlo} methods in practice}, pages 3--14,
  2001.

\bibitem[Finke and Singh(2017)]{finke2017approximate}
Axel Finke and Sumeetpal~S Singh.
\newblock Approximate smoothing and parameter estimation in high-dimensional
  state-space models.
\newblock \emph{IEEE Transactions on Signal Processing}, 65\penalty0
  (22):\penalty0 5982--5994, 2017.

\bibitem[Finke and Thiery(2023)]{finke2023conditional}
Axel Finke and Alexandre~H Thiery.
\newblock Conditional sequential {Monte Carlo} in high dimensions.
\newblock \emph{The Annals of Statistics}, 51\penalty0 (2):\penalty0 437--463,
  2023.

\bibitem[Georgii(2011)]{georgii2011gibbs}
Hans-Otto Georgii.
\newblock \emph{{G}ibbs measures and phase transitions}, volume~9.
\newblock Walter de Gruyter, 2011.

\bibitem[Goldman and Singh(2021)]{goldman2021spatiotemporal}
Jacob~Vorstrup Goldman and Sumeetpal~Sidhu Singh.
\newblock Spatiotemporal blocking of the bouncy particle sampler for efficient
  inference in state space models.
\newblock \emph{arXiv preprint arXiv:2101.03079}, 2021.

\bibitem[Guarniero et~al.(2017)Guarniero, Johansen, and
  Lee]{guarniero2017iterated}
Pieralberto Guarniero, Adam~M Johansen, and Anthony Lee.
\newblock The iterated auxiliary particle filter.
\newblock \emph{Journal of the American Statistical Association}, 112\penalty0
  (520):\penalty0 1636--1647, 2017.

\bibitem[Ionides et~al.(2024)Ionides, Ning, and Wheeler]{ionides2024iterated}
Edward~L Ionides, Ning Ning, and Jesse Wheeler.
\newblock An iterated block particle filter for inference on coupled dynamic
  systems with shared and unit-specific parameters.
\newblock \emph{Statistica Sinica}, 34:\penalty0 1--22, 2024.

\bibitem[Jiang and Srivastava(2019)]{jiang2019data}
Yazhou Jiang and Anurag~K Srivastava.
\newblock Data-driven event diagnosis in transmission systems with incomplete
  and conflicting alarms given sensor malfunctions.
\newblock \emph{IEEE Transactions on Power Delivery}, 35\penalty0 (1):\penalty0
  214--225, 2019.

\bibitem[Khan et~al.(2005)Khan, Balch, and Dellaert]{khan2005mcmc}
Zia Khan, Tucker Balch, and Frank Dellaert.
\newblock {MCMC}-based particle filtering for tracking a variable number of
  interacting targets.
\newblock \emph{IEEE transactions on pattern analysis and machine
  intelligence}, 27\penalty0 (11):\penalty0 1805--1819, 2005.

\bibitem[Lee et~al.(2018)Lee, Rushworth, and Napier]{lee2018spatio}
Duncan Lee, Alastair Rushworth, and Gary Napier.
\newblock Spatio-temporal areal unit modeling in {R} with conditional
  autoregressive priors using the {CARB}ayes{ST} package.
\newblock \emph{Journal of Statistical Software}, 84:\penalty0 1--39, 2018.

\bibitem[Leroux et~al.(2000)Leroux, Lei, and Breslow]{leroux2000estimation}
Brian~G Leroux, Xingye Lei, and Norman Breslow.
\newblock Estimation of disease rates in small areas: a new mixed model for
  spatial dependence.
\newblock In \emph{Statistical models in epidemiology, the environment, and
  clinical trials}, pages 179--191. Springer, 2000.

\bibitem[Li et~al.(2023)Li, Ionides, King, Pascual, and Ning]{li2023machine}
Jifan Li, Edward~L Ionides, Aaron~A King, Mercedes Pascual, and Ning Ning.
\newblock Inference on spatiotemporal dynamics for networks of biological
  populations.
\newblock \emph{arXiv preprint arXiv:2311.06702}, 2023.

\bibitem[Li et~al.(2020)Li, Pei, Chen, Song, Zhang, Yang, and
  Shaman]{li2020substantial}
Ruiyun Li, Sen Pei, Bin Chen, Yimeng Song, Tao Zhang, Wan Yang, and Jeffrey
  Shaman.
\newblock Substantial undocumented infection facilitates the rapid
  dissemination of novel coronavirus ({SARS-CoV-2}).
\newblock \emph{Science}, 368\penalty0 (6490):\penalty0 489--493, 2020.

\bibitem[Li(2009)]{li2009markov}
Stan~Z Li.
\newblock \emph{Markov random field modeling in image analysis}.
\newblock Springer Science \& Business Media, 2009.

\bibitem[Li(2012)]{li2012markov}
Stan~Z Li.
\newblock \emph{Markov random field modeling in computer vision}.
\newblock Springer Science \& Business Media, 2012.

\bibitem[Lin et~al.(2023)Lin, Dai, Zou, and Ning]{lin2023statistical}
Binbin Lin, Yimin Dai, Lei Zou, and Ning Ning.
\newblock Statistical machine learning meets high-dimensional spatiotemporal
  challenges--a case study of {COVID}-19 modeling.
\newblock \emph{arXiv preprint arXiv:2312.14161}, 2023.

\bibitem[Lin et~al.(2024)Lin, Zou, Yang, Zhou, Mandal, Abedin, Cai, and
  Ning]{lin2024understanding}
Binbin Lin, Lei Zou, Mingzheng Yang, Bing Zhou, Debayan Mandal, Joynal Abedin,
  Heng Cai, and Ning Ning.
\newblock Understanding human-{COVID}-19 dynamics using geospatial big data: A
  systematic literature review.
\newblock \emph{arXiv preprint arXiv:2404.10013}, 2024.

\bibitem[Liu and Ning(2019{\natexlab{a}})]{liu2019information}
Wenjian Liu and Ning Ning.
\newblock Information reconstruction on an infinite tree for a $4\times
  4$-state asymmetric model with community effects.
\newblock \emph{Journal of Statistical Physics}, 177\penalty0 (3):\penalty0
  438--467, 2019{\natexlab{a}}.

\bibitem[Liu and Ning(2019{\natexlab{b}})]{liu2019large}
Wenjian Liu and Ning Ning.
\newblock Large degree asymptotics and the reconstruction threshold of the
  asymmetric binary channels.
\newblock \emph{Journal of Statistical Physics}, 174\penalty0 (6):\penalty0
  1161--1188, 2019{\natexlab{b}}.

\bibitem[Liu and Ning(2021)]{liu2021phase}
Wenjian Liu and Ning Ning.
\newblock Phase transition of the reconstructability of a general model with
  different in-community and out-community mutations on an infinite tree.
\newblock \emph{SIAM Journal on Discrete Mathematics}, 35\penalty0
  (2):\penalty0 1381--1417, 2021.

\bibitem[Liu et~al.(2018)Liu, Jammalamadaka, and Ning]{liu2018tightness}
Wenjian Liu, Sreenivasa~Rao Jammalamadaka, and Ning Ning.
\newblock The tightness of the {K}esten--{S}tigum reconstruction bound of
  symmetric model with multiple mutations.
\newblock \emph{Journal of Statistical Physics}, 170\penalty0 (3):\penalty0
  617--641, 2018.

\bibitem[Mardia(1988)]{mardia1988multi}
KV~Mardia.
\newblock Multi-dimensional multivariate {G}aussian {M}arkov random fields with
  application to image processing.
\newblock \emph{Journal of Multivariate Analysis}, 24\penalty0 (2):\penalty0
  265--284, 1988.

\bibitem[Napier et~al.(2016)Napier, Lee, Robertson, Lawson, and
  Pollock]{napier2016model}
Gary Napier, Duncan Lee, Chris Robertson, Andrew Lawson, and Kevin~G Pollock.
\newblock A model to estimate the impact of changes in {MMR} vaccine uptake on
  inequalities in measles susceptibility in {S}cotland.
\newblock \emph{Statistical Methods in Medical Research}, 25\penalty0
  (4):\penalty0 1185--1200, 2016.

\bibitem[Ning and Ionides(2023)]{ning2023iterated}
Ning Ning and Edward~L Ionides.
\newblock Iterated block particle filter for high-dimensional parameter
  learning: Beating the curse of dimensionality.
\newblock \emph{Journal of Machine Learning Research}, 24\penalty0
  (82):\penalty0 1--76, 2023.

\bibitem[Prates et~al.(2022)Prates, Azevedo, MacNab, and Willig]{prates2022non}
Marcos~O Prates, Douglas~RM Azevedo, Ying~C MacNab, and Michael~R Willig.
\newblock Non-separable spatio-temporal models via transformed multivariate
  {G}aussian {M}arkov random fields.
\newblock \emph{Journal of the Royal Statistical Society Series C: Applied
  Statistics}, 71\penalty0 (5):\penalty0 1116--1136, 2022.

\bibitem[Rebeschini and Van~Handel(2015)]{rebeschini2015can}
Patrick Rebeschini and Ramon Van~Handel.
\newblock Can local particle filters beat the curse of dimensionality?
\newblock \emph{The Annals of Applied Probability}, 25\penalty0 (5):\penalty0
  2809--2866, 2015.

\bibitem[Rimella and Whiteley(2022)]{rimella2022exploiting}
Lorenzo Rimella and Nick Whiteley.
\newblock Exploiting locality in high-dimensional factorial hidden {M}arkov
  models.
\newblock \emph{Journal of Machine Learning Research}, 23\penalty0
  (4):\penalty0 1--34, 2022.

\bibitem[Rue and Held(2005)]{rue2005gaussian}
Havard Rue and Leonhard Held.
\newblock \emph{Gaussian {M}arkov random fields: theory and applications}.
\newblock CRC press, 2005.

\bibitem[Singh et~al.(2017)Singh, Lindsten, and Moulines]{singh2017blocking}
Sumeetpal~S Singh, Fredrik Lindsten, and Eric Moulines.
\newblock Blocking strategies and stability of particle {G}ibbs samplers.
\newblock \emph{Biometrika}, 104\penalty0 (4):\penalty0 953--969, 2017.

\bibitem[Snyder et~al.(2008)Snyder, Bengtsson, Bickel, and
  Anderson]{snyder2008obstacles}
Chris Snyder, Thomas Bengtsson, Peter Bickel, and Jeff Anderson.
\newblock Obstacles to high-dimensional particle filtering.
\newblock \emph{Monthly Weather Review}, 136\penalty0 (12):\penalty0
  4629--4640, 2008.

\bibitem[Subramanian et~al.(2021)Subramanian, He, and
  Pascual]{subramanian2021quantifying}
Rahul Subramanian, Qixin He, and Mercedes Pascual.
\newblock Quantifying asymptomatic infection and transmission of {COVID}-19 in
  {New York City} using observed cases, serology, and testing capacity.
\newblock \emph{Proceedings of the National Academy of Sciences}, 118\penalty0
  (9):\penalty0 e2019716118, 2021.

\end{thebibliography}



\section*{Supplementary material}

# Supplement to "VT-MRF-SPF: Variable Target Markov Random Field Scalable Particle Filter"

## Ning Ning

*Department of Statistics, Texas A&M University, College Station, e-mail:* **patning@tamu.edu**

**Contents**

## S1. Preliminary proofs

For any probability measure $\mu_{s-1}$ on $\mathcal{X}$ at time $s-1$ for any integer $s \in [T]$, any vertex $v \in k_s$, and any set $A^v \subseteq \mathbb{X}^v$, we define

$$\mu_{\chi_s}^v(A^v) := \mathbb{P}^{\mu_s}\left[X_s^v \in A^v \mid X_s^{K_s\setminus\{v\}} = x_s^{k_s\setminus\{v\}}\right], \tag{S1}$$

$$\mu_{\chi_s,\chi_{s+1}}^v(A^v) := \mathbb{P}^{\mu_s}\left[X_s^v \in A^v \mid X_s^{K_s\setminus\{v\}} = x_s^{k_s\setminus\{v\}}, X_{s+1}^{K_{s+1}} = x_{s+1}^{k_{s+1}}\right], \tag{S2}$$

$$\mu_{\overline{\chi}_s,\chi_{s+1}}^v(A^v) := \mathbb{P}^{\mu_s}\left[X_s^v \in A^v \mid X_s^{K_s\setminus\{v\}} = \overline{x}_s^{k_s\setminus\{v\}}, X_{s+1}^{K_{s+1}} = x_{s+1}^{k_{s+1}}\right]. \tag{S3}$$

If $v \notin k_{s+1}$, clearly $\mu_{\chi_s,\chi_{s+1}}^v$ and $\mu_{\overline{\chi}_s,\chi_{s+1}}^v$ degenerate to $\mu_{\chi_s}^v$ and $\mu_{\overline{\chi}_s}^v$, respectively. With $v' \in k_s$ we define

$$C_{vv'}^{\mu_s} := \frac{1}{2} \sup_{x_{s+1}^{k_{s+1}} \in \mathcal{X}} \sup_{\substack{x_s^{k_s},\overline{x}_s^{k_s} \in \mathcal{X}:\\ x_s^{k_s\setminus\{v'\}} = \overline{x}_s^{k_s\setminus\{v'\}}}} \left\| \mu_{\chi_s,\chi_{s+1}}^v - \mu_{\overline{\chi}_s,\chi_{s+1}}^v \right\|. \tag{S4}$$

The following proposition is our first preliminary result, which will be used to simplify several proofs of the rest preliminary results. It considers the situation where two vertices $v, v'$ are in the same index set $k_t$ at time $t$. It states that the degree to which the perturbations of all $v'$ directly affect any $v$ under the distribution $\widetilde{\pi}_t$ can be bounded by a positive constant that is less than one.

**Proposition S1.1.** *Under Assumption 3.1, for any $t \in [T] \cup \{0\}$, we have*

$$\max_{v \in k_t} \sum_{v' \in k_t} e^{\beta_T d(v,v')} C_{vv'}^{\widetilde{\pi}_t} \leq \frac{1}{3},$$

*where $\beta_T$ is the finite positive constant defined in equation* (32).

Before we provide the proof of the above proposition, we present the following two existing results to ensure the paper is self-contained.

**Theorem S1.2** (Lemma 4.1 of Rebeschini and Van Handel [2015]). *Let probability measures $\mu, \mu', \mathsf{F}, \mathsf{F}'$ and constant $\epsilon > 0$ be such that $\mu(A) \geq \epsilon \mathsf{F}(A)$ and $\mu'(A) \geq \epsilon \mathsf{F}'(A)$ for every measurable set $A$. Then*

$$\|\mu - \mu'\| \leq 2(1 - \epsilon) + \epsilon \|\mathsf{F} - \mathsf{F}'\|.$$

**Theorem S1.3** (Lemma 4.3 of Rebeschini and Van Handel [2015]). *Let $I$ be a finite set and let $m$ be a pseudometric on $I$. Let $C = (C_{ij})_{i,j \in I}$ be a matrix with nonnegative entries. Suppose that*

$$\max_{i \in I} \sum_{j \in I} e^{m(i,j)} C_{ij} \leq c < 1.$$

*Then the matrix $D = \sum_{n \in \mathbb{N}} C^n$ satisfies*

$$\max_{i \in I} \sum_{j \in I} e^{m(i,j)} D_{ij} \leq \frac{1}{1-c}.$$

*Proof of Proposition S1.1.* For $t = 0$, the proof is trivial; since $\widetilde{\pi}_0 = \pi_0 = \delta_x$ and then we have $C_{vv'}^{\widetilde{\pi}_0} = 0$. For any $t \in [T]$, we prove by the method of induction and assume that

$$\max_{v \in k_{t-1}} \sum_{v' \in k_{t-1}} e^{\beta_T d(v,v')} C_{vv'}^{\widetilde{\pi}_{t-1}} \leq \frac{1}{3}. \tag{S5}$$

Let $v \in B_t \subseteq k_t$, $v' \in k_t$, and $v \neq v'$. Let $x_t^{k_t}, \overline{x}_t^{k_t} \in \mathcal{X}$ be such that $x_t^{k_t \setminus \{v'\}} = \overline{x}_t^{k_t \setminus \{v'\}}$. Define

$$I = (\{t-1\} \times k_{t-1}) \cup (t,v) \quad \text{and} \quad \mathbb{S} = \mathcal{X} \times \mathbb{X}^v,$$

and the probability measures on $\mathbb{S}$ as follows:

$$\rho(A) = \frac{\begin{aligned}&\int \mathbb{1}_A(x_{t-1}^{k_{t-1}}, x_t^v) \prod_{\omega \in B_t} f_t^\omega(x_t^\omega \mid k_t^{\omega(\mathcal{R})}, x_{t-1}^{k_{t-1} \cap \{\omega\}}) \\ &\quad \times g_t^v(y_t^v \mid k_t^{v(\mathcal{R})}, x_t^v) \widetilde{f}_t^v(x_t^v, x_t^{N_t(v)}) \\ &\quad \times p_{t+1}^{v(\mathcal{R})}(k_{t+1}^{v(\mathcal{R})} \mid k_t, x_t^{k_t \cap v(\mathcal{R})}) f_{t+1}^v(x_{t+1}^{k_{t+1} \cap \{v\}} \mid k_{t+1}^{v(\mathcal{R})}, x_t^v) \\ &\quad \times \prod_{R \in \mathcal{R}} p_t^R(k_t^R \mid k_{t-1}, x_{t-1}^{k_{t-1} \cap R}) \psi^v(dx_t^v) \widetilde{\pi}_{t-1}(dx_{t-1}^{k_{t-1}})\end{aligned}}{\begin{aligned}&\int \prod_{\omega \in B_t} f_t^\omega(x_t^\omega \mid k_t^{\omega(\mathcal{R})}, x_{t-1}^{k_{t-1} \cap \{\omega\}}) g_t^v(y_t^v \mid k_t^{v(\mathcal{R})}, x_t^v) \widetilde{f}_t^v(x_t^v, x_t^{N_t(v)}) \\ &\quad \times p_{t+1}^{v(\mathcal{R})}(k_{t+1}^{v(\mathcal{R})} \mid k_t, x_t^{k_t \cap v(\mathcal{R})}) f_{t+1}^v(x_{t+1}^{k_{t+1} \cap \{v\}} \mid k_{t+1}^{v(\mathcal{R})}, x_t^v) \\ &\quad \times \prod_{R \in \mathcal{R}} p_t^R(k_t^R \mid k_{t-1}, x_{t-1}^{k_{t-1} \cap R}) \psi^v(dx_t^v) \widetilde{\pi}_{t-1}(dx_{t-1}^{k_{t-1}})\end{aligned}},$$

$$\overline{\rho}(A) = \frac{\begin{aligned}&\int \mathbb{1}_A(x_{t-1}^{k_{t-1}}, \overline{x}_t^v) \prod_{\omega \in B_t} f_t^\omega(\overline{x}_t^\omega \mid k_t^{\omega(\mathcal{R})}, x_{t-1}^{k_{t-1} \cap \{\omega\}}) \\ &\quad \times g_t^v(y_t^v \mid k_t^{v(\mathcal{R})}, \overline{x}_t^v) \widetilde{f}_t^v(\overline{x}_t^v, \overline{x}_t^{N_t(v)}) \\ &\quad \times p_{t+1}^{v(\mathcal{R})}(k_{t+1}^{v(\mathcal{R})} \mid k_t, \overline{x}_t^{k_t \cap v(\mathcal{R})}) f_{t+1}^v(x_{t+1}^{k_{t+1} \cap \{v\}} \mid k_{t+1}^{v(\mathcal{R})}, \overline{x}_t^v) \\ &\quad \times \prod_{R \in \mathcal{R}} p_t^R(k_t^R \mid k_{t-1}, x_{t-1}^{k_{t-1} \cap R}) \psi^v(d\overline{x}_t^v) \widetilde{\pi}_{t-1}(dx_{t-1}^{k_{t-1}})\end{aligned}}{\begin{aligned}&\int \prod_{\omega \in B_t} f_t^\omega(\overline{x}_t^\omega \mid k_t^{\omega(\mathcal{R})}, x_{t-1}^{k_{t-1} \cap \{\omega\}}) g_t^v(y_t^v \mid k_t^{v(\mathcal{R})}, \overline{x}_t^v) \widetilde{f}_t^v(\overline{x}_t^v, \overline{x}_t^{N_t(v)}) \\ &\quad \times p_{t+1}^{v(\mathcal{R})}(k_{t+1}^{v(\mathcal{R})} \mid k_t, \overline{x}_t^{k_t \cap v(\mathcal{R})}) f_{t+1}^v(x_{t+1}^{k_{t+1} \cap \{v\}} \mid k_{t+1}^{v(\mathcal{R})}, \overline{x}_t^v) \\ &\quad \times \prod_{R \in \mathcal{R}} p_t^R(k_t^R \mid k_{t-1}, x_{t-1}^{k_{t-1} \cap R}) \psi^v(d\overline{x}_t^v) \widetilde{\pi}_{t-1}(dx_{t-1}^{k_{t-1}})\end{aligned}}.$$

In accordance with the notations used in this paper, $f_{t+1}^v(x_{t+1}^{k_{t+1}\cap\{v\}} \mid k_{t+1}^{v(\mathcal{R})}, x_t^v)$ represents $f_{t+1}^v(x_{t+1}^v \mid k_{t+1}^{v(\mathcal{R})}, x_t^v)$ if $v \in k_{t+1}$, and vanishes otherwise. Note that for any $t \in [T]$ and any set $A^v \subseteq \mathbb{X}^v$, we have

$$(\widetilde{\mathsf{F}}_t \overline{\mu}_{t-1})_{\chi_t, \chi_{t+1}}^v(A^v) = \frac{\begin{array}{l} \int \mathbb{1}_A(x_t^v) \prod_{\omega \in B_t} f_t^\omega(x_t^\omega \mid k_t^{\omega(\mathcal{R})}, x_{t-1}^{k_{t-1}\cap\{\omega\}}) \\ \times g_t^v(y_t^v \mid k_t^{v(\mathcal{R})}, x_t^v) \widetilde{f}_t^v(x_t^v, x_t^{N_t(v)}) \\ \times p_{t+1}^{v(\mathcal{R})}(k_{t+1}^{v(\mathcal{R})} \mid k_t, x_t^{k_t \cap v(\mathcal{R})}) f_{t+1}^v(x_{t+1}^{k_{t+1}\cap\{v\}} \mid k_{t+1}^{v(\mathcal{R})}, x_t^v) \\ \times \prod_{R \in \mathcal{R}} p_t^R(k_t^R \mid k_{t-1}, x_{t-1}^{k_{t-1}\cap R}) \widetilde{\pi}_{t-1}(dx_{t-1}^{k_{t-1}}) \psi^v(dx_t^v) \end{array}}{\begin{array}{l} \int \prod_{\omega \in B_t} f_t^\omega(x_t^\omega \mid k_t^{\omega(\mathcal{R})}, x_{t-1}^{k_{t-1}\cap\{\omega\}}) \\ \times g_t^v(y_t^v \mid k_t^{v(\mathcal{R})}, x_t^v) \widetilde{f}_t^v(x_t^v, x_t^{N_t(v)}) \\ \times p_{t+1}^{v(\mathcal{R})}(k_{t+1}^{v(\mathcal{R})} \mid k_t, x_t^{k_t \cap v(\mathcal{R})}) f_{t+1}^v(x_{t+1}^{k_{t+1}\cap\{v\}} \mid k_{t+1}^{v(\mathcal{R})}, x_t^v) \\ \times \prod_{R \in \mathcal{R}} p_t^R(k_t^R \mid k_{t-1}, x_{t-1}^{k_{t-1}\cap R}) \widetilde{\pi}_{t-1}(dx_{t-1}^{k_{t-1}}) \psi^v(dx_t^v) \end{array}},$$

$$(\widetilde{\mathsf{F}}_t \overline{\mu}_{t-1})_{\overline{\chi}_t, \chi_{t+1}}^v(A^v) = \frac{\begin{array}{l} \int \mathbb{1}_A(\overline{x}_t^v) \prod_{\omega \in B_t} f_t^\omega(\overline{x}_t^\omega \mid k_t^{\omega(\mathcal{R})}, x_{t-1}^{k_{t-1}\cap\{\omega\}}) \\ \times g_t^v(y_t^v \mid k_t^{v(\mathcal{R})}, \overline{x}_t^v) \widetilde{f}_t^v(\overline{x}_t^v, \overline{x}_t^{N_t(v)}) \\ \times p_{t+1}^{v(\mathcal{R})}(k_{t+1}^{v(\mathcal{R})} \mid k_t, \overline{x}_t^{k_t \cap v(\mathcal{R})}) f_{t+1}^v(x_{t+1}^{k_{t+1}\cap\{v\}} \mid k_{t+1}^{v(\mathcal{R})}, \overline{x}_t^v) \\ \times \prod_{R \in \mathcal{R}} p_t^R(k_t^R \mid k_{t-1}, x_{t-1}^{k_{t-1}\cap R}) \widetilde{\pi}_{t-1}(dx_{t-1}^{k_{t-1}}) \psi^v(d\overline{x}_t^v) \end{array}}{\begin{array}{l} \int \prod_{\omega \in B_t} f_t^\omega(\overline{x}_t^\omega \mid k_t^{\omega(\mathcal{R})}, x_{t-1}^{k_{t-1}\cap\{\omega\}}) \\ \times g_t^v(y_t^v \mid k_t^{v(\mathcal{R})}, \overline{x}_t^v) \widetilde{f}_t^v(\overline{x}_t^v, \overline{x}_t^{N_t(v)}) \\ \times p_{t+1}^{v(\mathcal{R})}(k_{t+1}^{v(\mathcal{R})} \mid k_t, \overline{x}_t^{k_t \cap v(\mathcal{R})}) f_{t+1}^v(x_{t+1}^{k_{t+1}\cap\{v\}} \mid k_{t+1}^{v(\mathcal{R})}, \overline{x}_t^v) \\ \times \prod_{R \in \mathcal{R}} p_t^R(k_t^R \mid k_{t-1}, x_{t-1}^{k_{t-1}\cap R}) \widetilde{\pi}_{t-1}(dx_{t-1}^{k_{t-1}}) \psi^v(d\overline{x}_t^v) \end{array}}.$$

Then we have

$$\left\| (\widetilde{\mathsf{F}}_t \widetilde{\pi}_{t-1})_{\chi_t, \chi_{t+1}}^v - (\widetilde{\mathsf{F}}_t \widetilde{\pi}_{t-1})_{\overline{\chi}_t, \chi_{t+1}}^v \right\| = \|\rho - \overline{\rho}\|_{(t,v)}. \tag{S6}$$

In the following steps, we are going to use the Dobrushin comparison theorem (Theorem 3.3) to bound $\|\rho - \overline{\rho}\|_{(t,v)}$. We will bound $C_{ij}$ and $b_i$ with $i = (\tau, l)$ and $j = (\tau', l')$.

**Step 1.** In this step, we consider $\tau = t - 1$, which implies $l \in k_{t-1}$.

**Step 1.1.** When $l$ is not in $k_t$, we have

$$\rho_{(x_{t-1}^{k_{t-1}}, x_t^v)}^i(A) = \overline{\rho}_{(x_{t-1}^{k_{t-1}}, x_t^v)}^i(A) = \frac{\int \mathbb{1}_A(x_{t-1}^l) p_t^{l(\mathcal{R})}(k_t^{l(\mathcal{R})} \mid k_{t-1}, x_{t-1}^{l(\mathcal{R})}) \widetilde{\pi}_{t-1}^l(dx_{t-1}^l)}{\int p_t^{l(\mathcal{R})}(k_t^{l(\mathcal{R})} \mid k_{t-1}, x_{t-1}^{l(\mathcal{R})}) \widetilde{\pi}_{t-1}^l(dx_{t-1}^l)},$$

which reveals that $b_i = 0$. Furthermore, when $\tau' = t - 1$ which implies $l' \in k_{t-1}$. Noting that

$$\rho_{(\overline{x}_{t-1}^{k_{t-1}}, \overline{x}_t^v)}^i(A) = \frac{\int \mathbb{1}_A(\overline{x}_{t-1}^l) p_t^{l(\mathcal{R})}(k_t^{l(\mathcal{R})} \mid k_{t-1}, \overline{x}_{t-1}^{l(\mathcal{R})}) \widetilde{\pi}_{t-1}^l(d\overline{x}_{t-1}^l)}{\int p_t^{l(\mathcal{R})}(k_t^{l(\mathcal{R})} \mid k_{t-1}, \overline{x}_{t-1}^{l(\mathcal{R})}) \widetilde{\pi}_{t-1}^l(d\overline{x}_{t-1}^l)},$$

then we have

$$\rho_{(x_{t-1}^{k_{t-1}}, x_t^v)}^i(A) \geq \left(\frac{\kappa_d}{\kappa_u}\right) \frac{\int \mathbb{1}_A(x_{t-1}^l) \widetilde{\pi}_{t-1}^l(dx_{t-1}^l)}{\int \widetilde{\pi}_{t-1}^l(dx_{t-1}^l)} = \left(\frac{\kappa_d}{\kappa_u}\right) \widetilde{\pi}_{t-1}^l(A),$$

$$\rho^i_{(\overline{x}^{k_{t-1}}_{t-1}, \overline{x}^l_t)}(A) \geq \left(\frac{\kappa_d}{\kappa_u}\right) \frac{\mathbb{1}_A(\overline{x}^l_{t-1})\widetilde{\pi}^l_{t-1}(d\overline{x}^l_{t-1})}{\int \widetilde{\pi}^l_{t-1}(d\overline{x}^l_{t-1})} = \left(\frac{\kappa_d}{\kappa_u}\right)\widetilde{\pi}^l_{t-1}(A).$$

Hence, we have by Theorem S1.2 that $C_{ij} \leq 1 - \frac{\kappa_d}{\kappa_u}$ if $l' \in l(\mathcal{R})$ and $C_{ij} = 0$ otherwise. Next, when $\tau' = t$, for any $l' \in k_t$, we have $C_{ij} = 0$.

**Step 1.2.** When $l$ is also in $k_t$, we have

$$\rho^i_{(x^{k_{t-1}}_{t-1}, x^v_t)}(A) = \frac{\int \mathbb{1}_A(x^l_{t-1}) f^l_t(x^l_t \mid k^{l(\mathcal{R})}_t, x^l_{t-1}) p^{l(\mathcal{R})}_t(k^{l(\mathcal{R})}_t \mid k_{t-1}, x^{l(\mathcal{R})}_{t-1})\widetilde{\pi}^l_{t-1}(dx^l_{t-1})}{\int f^l_t(x^l_t \mid k^{l(\mathcal{R})}_t, x^l_{t-1}) p^{l(\mathcal{R})}_t(k^{l(\mathcal{R})}_t \mid k_{t-1}, x^{l(\mathcal{R})}_{t-1})\widetilde{\pi}^l_{t-1}(dx^l_{t-1})},$$

which is $\widetilde{\pi}^l_{\chi_{t-1},\chi_t}$ defined in equation (S2). In this case, if $\tau' = t - 1$, by the definition of $C^{\mu_{s-1}}_{ll'}$ in equation (S4), we have $C_{ij} \leq C^{\widetilde{\pi}_{t-1}}_{ll'}$. If $\tau' = t$ and $l' = v = l$, since

$$\rho^i_{(x^{k_{t-1}}_{t-1}, x^v_t)}(A) \geq \left(\frac{\epsilon_d}{\epsilon_u}\right) \frac{\int \mathbb{1}_A(x^l_{t-1}) p^{l(\mathcal{R})}_t(k^l_t \mid k_{t-1}, x^{l(\mathcal{R})}_{t-1})\widetilde{\pi}^l_{t-1}(dx^l_{t-1})}{\int p^{l(\mathcal{R})}_t(k^l_t \mid k_{t-1}, x^{l(\mathcal{R})}_{t-1})\widetilde{\pi}^l_{t-1}(dx^l_{t-1})},$$

we have $C_{ij} \leq 1 - \frac{\epsilon_d}{\epsilon_u}$ by Theorem S1.2, and $C_{ij} = 0$ if $\tau' = t$ and $l' \neq l$. Now, we need to calculate $b_i$. In this case, $\overline{\rho}^i_{(x^{k_{t-1}}_{t-1}, x^v_t)}$ is given by

$$\overline{\rho}^i_{(x^{k_{t-1}}_{t-1}, x^v_t)}(A) = \frac{\int \mathbb{1}_A(x^l_{t-1}) f^l_t(\overline{x}^l_t \mid k^{l(\mathcal{R})}_t, x^l_{t-1}) p^{l(\mathcal{R})}_t(k^{l(\mathcal{R})}_t \mid k_{t-1}, x^{l(\mathcal{R})}_{t-1})\widetilde{\pi}^l_{t-1}(dx^l_{t-1})}{\int f^l_t(\overline{x}^l_t \mid k^{l(\mathcal{R})}_t, x^l_{t-1}) p^{l(\mathcal{R})}_t(k^{l(\mathcal{R})}_t \mid k_{t-1}, x^{l(\mathcal{R})}_{t-1})\widetilde{\pi}^l_{t-1}(dx^l_{t-1})}.$$

Recalling that at the beginning of this proof we set $v \in B_t$, $v' \in k_t$, $v \neq v'$, $x^{k_t}_t, \overline{x}^{k_t}_t \in \mathcal{X}$ such that $x^{k_t\backslash\{v'\}}_t = \overline{x}^{k_t\backslash\{v'\}}_t$, then we have $b_i = 0$, since $\rho^i_{(x^{k_{t-1}}_{t-1}, x^v_t)} = \overline{\rho}^i_{(x^{k_{t-1}}_{t-1}, x^v_t)}$ if $v' \neq l$. Next, if $v' = l$, since

$$\rho^i_{(x^{k_{t-1}}_{t-1}, x^v_t)}(A), \overline{\rho}^i_{(x^{k_{t-1}}_{t-1}, x^v_t)}(A) \geq \left(\frac{\epsilon_d}{\epsilon_u}\right) \frac{\int \mathbb{1}_A(x^l_{t-1}) p^{l(\mathcal{R})}_t(k^l_t \mid k_{t-1}, x^{l(\mathcal{R})}_{t-1})\widetilde{\pi}^l_{t-1}(dx^l_{t-1})}{\int p^{l(\mathcal{R})}_t(k^l_t \mid k_{t-1}, x^{l(\mathcal{R})}_{t-1})\widetilde{\pi}^l_{t-1}(dx^l_{t-1})},$$

by Theorem S1.2, we have $b_i \leq 2\left(1 - \frac{\epsilon_d}{\epsilon_u}\right)$.

**Step 2.** In this step, we consider $\tau = t$, which implies $l = v$.

**Step 2.1.** When $v$ is not in $k_{t-1}$, we have

$$\rho^i_{(x^{k_{t-1}}_{t-1}, x^v_t)}(A) = \frac{\begin{aligned}&\int \mathbb{1}_A(x^v_t) f^v_t(x^v_t \mid k^{v(\mathcal{R})}_t) g^v_t(y^v_t \mid k^{v(\mathcal{R})}_t, x^v_t) \widetilde{f}^v_t(x^v_t, x^{N_t(v)}_t)\\ &\times p^{v(\mathcal{R})}_{t+1}(k^{v(\mathcal{R})}_{t+1} \mid k_t, x^{v(\mathcal{R})}_t) f^v_{t+1}(x^{k_{t+1}\cap\{v\}}_{t+1} \mid k^{v(\mathcal{R})}_{t+1}, x^{v(\mathcal{R})}_t) \psi^v(dx^v_t)\end{aligned}}{\begin{aligned}&\int f^v_t(x^v_t \mid k^{v(\mathcal{R})}_t) g^v_t(y^v_t \mid k^{v(\mathcal{R})}_t, x^v_t) \widetilde{f}^v_t(x^v_t, x^{N_t(v)}_t)\\ &\times p^{v(\mathcal{R})}_{t+1}(k^{v(\mathcal{R})}_{t+1} \mid k_t, x^{v(\mathcal{R})}_t) f^v_{t+1}(x^{k_{t+1}\cap\{v\}}_{t+1} \mid k^{v(\mathcal{R})}_{t+1}, x^v_t) \psi^v(dx^v_t)\end{aligned}}.$$

Hence, if $\tau' = t - 1$, for any $l' \in k_{t-1}$, we have $C_{ij} = 0$.

Next, when $v$ is in $k_{t-1}$, we have

$$\rho^i_{(x^{k_{t-1}}_{t-1}, x^v_t)}(A) = \frac{\begin{aligned}&\int \mathbb{1}_A(x^v_t) f^v_t(x^v_t \mid k^{v(\mathcal{R})}_t, x^v_{t-1}) g^v_t(y^v_t \mid k^{v(\mathcal{R})}_t, x^v_t) \widetilde{f}^v_t(x^v_t, x^{N_t(v)}_t)\\ &\times p^{v(\mathcal{R})}_{t+1}(k^{v(\mathcal{R})}_{t+1} \mid k_t, x^{v(\mathcal{R})}_t) f^v_{t+1}(x^{k_{t+1}\cap\{v\}}_{t+1} \mid k^{v(\mathcal{R})}_{t+1}, x^v_t) \psi^v(dx^v_t)\end{aligned}}{\begin{aligned}&\int f^v_t(x^v_t \mid k^{v(\mathcal{R})}_t, x^v_{t-1}) g^v_t(y^v_t \mid k^{v(\mathcal{R})}_t, x^v_t) \widetilde{f}^v_t(x^v_t, x^{N_t(v)}_t)\\ &\times p^{v(\mathcal{R})}_{t+1}(k^{v(\mathcal{R})}_{t+1} \mid k_t, x^{v(\mathcal{R})}_t) f^v_{t+1}(x^{k_{t+1}\cap\{v\}}_{t+1} \mid k^{v(\mathcal{R})}_{t+1}, x^v_t) \psi^v(dx^v_t)\end{aligned}}$$

$$\geq \left(\frac{\epsilon_d}{\epsilon_u}\right) \frac{\int \mathbb{1}_A(x_t^v) g_t^v(y_t^v \mid k_t^{v(\mathcal{R})}, x_t^v) \widetilde{f}_t^v(x_t^v, x_t^{N_t(v)})}{\int g_t^v(y_t^v \mid k_t^{v(\mathcal{R})}, x_t^v) \widetilde{f}_t^v(x_t^v, x_t^{N_t(v)})} \times \frac{p_{t+1}^{v(\mathcal{R})}(k_{t+1}^{v(\mathcal{R})} \mid k_t, x_t^{v(\mathcal{R})}) f_{t+1}^v(x_{t+1}^{k_{t+1} \cap \{v\}} \mid k_{t+1}^{v(\mathcal{R})}, x_t^v) \psi^v(dx_t^v)}{\times p_{t+1}^{v(\mathcal{R})}(k_{t+1}^{v(\mathcal{R})} \mid k_t, x_t^{v(\mathcal{R})}) f_{t+1}^v(x_{t+1}^{k_{t+1} \cap \{v\}} \mid k_{t+1}^{v(\mathcal{R})}, x_t^v) \psi^v(dx_t^v)}.$$

Then, when $\tau' = t-1$, we have $C_{ij} \leq 1 - \frac{\epsilon_d}{\epsilon_u}$ if $l' = v$ by Theorem S1.2, and $C_{ij} = 0$ otherwise.

**Step 2.2.** Now, we need to calculate $b_i$. Note that

$$\rho^i_{(x_{t-1}^{k_{t-1}}, x_t^v)}(A) = \frac{\int \mathbb{1}_A(x_t^v) f_t^v(x_t^v \mid k_t^{v(\mathcal{R})}, x_{t-1}^{k_{t-1} \cap \{v\}}) g_t^v(y_t^v \mid k_t^{v(\mathcal{R})}, x_t^v) \widetilde{f}_t^v(x_t^v, x_t^{N_t(v)})}{\int f_t^v(x_t^v \mid k_t^{v(\mathcal{R})}, x_{t-1}^{k_{t-1} \cap \{v\}}) g_t^v(y_t^v \mid k_t^{v(\mathcal{R})}, x_t^v) \widetilde{f}_t^v(x_t^v, x_t^{N_t(v)})} \times \frac{p_{t+1}^{v(\mathcal{R})}(k_{t+1}^{v(\mathcal{R})} \mid k_t, x_t^{k_t \cap v(\mathcal{R})}) f_{t+1}^v(x_{t+1}^{k_{t+1} \cap \{v\}} \mid k_{t+1}^{v(\mathcal{R})}, x_t^v) \psi^v(dx_t^v)}{\times p_{t+1}^{v(\mathcal{R})}(k_{t+1}^{v(\mathcal{R})} \mid k_t, x_t^{k_t \cap v(\mathcal{R})}) f_{t+1}^v(x_{t+1}^{k_{t+1} \cap \{v\}} \mid k_{t+1}^{v(\mathcal{R})}, x_t^v) \psi^v(dx_t^v)}$$

and

$$\overline{\rho}^i_{(x_{t-1}^{k_{t-1}}, x_t^v)}(A) = \frac{\int \mathbb{1}_A(\overline{x}_t^v) f_t^v(\overline{x}_t^v \mid k_t^{v(\mathcal{R})}, x_{t-1}^{k_{t-1} \cap \{v\}}) g_t^v(y_t^v \mid k_t^{v(\mathcal{R})}, \overline{x}_t^v) \widetilde{f}_t^v(\overline{x}_t^v, \overline{x}_t^{N_t(v)})}{\int f_t^v(\overline{x}_t^v \mid k_t^{v(\mathcal{R})}, x_{t-1}^{k_{t-1} \cap \{v\}}) g_t^v(y_t^v \mid k_t^{v(\mathcal{R})}, \overline{x}_t^v) \widetilde{f}_t^v(\overline{x}_t^v, \overline{x}_t^{N_t(v)})} \times \frac{p_{t+1}^{v(\mathcal{R})}(k_{t+1}^{v(\mathcal{R})} \mid k_t, \overline{x}_t^{k_t \cap v(\mathcal{R})}) f_{t+1}^v(x_{t+1}^{k_{t+1} \cap \{v\}} \mid k_{t+1}^{v(\mathcal{R})}, \overline{x}_t^v) \psi^v(d\overline{x}_t^v)}{\times p_{t+1}^{v(\mathcal{R})}(k_{t+1}^{v(\mathcal{R})} \mid k_t, \overline{x}_t^{k_t \cap v(\mathcal{R})}) f_{t+1}^v(x_{t+1}^{k_{t+1} \cap \{v\}} \mid k_{t+1}^{v(\mathcal{R})}, \overline{x}_t^v) \psi^v(d\overline{x}_t^v)}.$$

If $v' \in N_t(v) \cup v(\mathcal{R})$, since

$$\rho^i_{(x_{t-1}^{k_{t-1}}, x_t^v)}(A) \geq \frac{\epsilon_d'}{\epsilon_u'} \left(\frac{\kappa_d}{\kappa_u}\right) \frac{\int \mathbb{1}_A(x_t^v) f_t^v(x_t^v \mid k_t^{v(\mathcal{R})}, x_{t-1}^{k_{t-1} \cap \{v\}}) g_t^v(y_t^v \mid k_t^{v(\mathcal{R})}, x_t^v)}{\int f_t^v(x_t^v \mid k_t^{v(\mathcal{R})}, x_{t-1}^{k_{t-1} \cap \{v\}}) g_t^v(y_t^v \mid k_t^{v(\mathcal{R})}, x_t^v)} \times \frac{f_{t+1}^v(x_{t+1}^{k_{t+1} \cap \{v\}} \mid k_{t+1}^{v(\mathcal{R})}, x_t^v) \psi^v(dx_t^v)}{\times f_{t+1}^v(x_{t+1}^{k_{t+1} \cap \{v\}} \mid k_{t+1}^{v(\mathcal{R})}, x_t^v) \psi^v(dx_t^v)}$$

and

$$\overline{\rho}^i_{(x_{t-1}^{k_{t-1}}, x_t^v)}(A) \geq \frac{\epsilon_d'}{\epsilon_u'} \left(\frac{\kappa_d}{\kappa_u}\right) \frac{\int \mathbb{1}_A(\overline{x}_t^v) f_t^v(\overline{x}_t^v \mid k_t^{v(\mathcal{R})}, x_{t-1}^{k_{t-1} \cap \{v\}}) g_t^v(y_t^v \mid k_t^{v(\mathcal{R})}, \overline{x}_t^v)}{\int f_t^v(\overline{x}_t^v \mid k_t^{v(\mathcal{R})}, x_{t-1}^{k_{t-1} \cap \{v\}}) g_t^v(y_t^v \mid k_t^{v(\mathcal{R})}, \overline{x}_t^v)} \times \frac{f_{t+1}^v(x_{t+1}^{k_{t+1} \cap \{v\}} \mid k_{t+1}^{v(\mathcal{R})}, \overline{x}_t^v) \psi^v(d\overline{x}_t^v)}{\times f_{t+1}^v(x_{t+1}^{k_{t+1} \cap \{v\}} \mid k_{t+1}^{v(\mathcal{R})}, \overline{x}_t^v) \psi^v(d\overline{x}_t^v)},$$

by Theorem S1.2 and the assumption that $x_t^{k_t \setminus \{v'\}} = \overline{x}_t^{k_t \setminus \{v'\}}$, we have $b_i \leq 2\left(1 - \frac{\epsilon_d'}{\epsilon_u'} \frac{\kappa_d}{\kappa_u}\right)$, and $b_i = 0$ otherwise.

**Step 3.** In this step, we summary the results of $C_{ij}$ obtained in the previous two steps and aim to bound the following quantity:

$$\max_{(\tau,l) \in I} \sum_{(\tau',l') \in I} e^{\beta_T |\tau - \tau'|} e^{\beta_T d(l,l')} C_{(\tau,l)(\tau',l')}$$

$$= \max \left\{ \max_{(t-1,l) \in I} \sum_{(\tau',l') \in I} e^{\beta_T |t-1-\tau'|} e^{\beta_T d(l,l')} C_{(t-1,l)(\tau',l')}, \right.$$

$$\max_{(t,v)\in I} \sum_{(\tau',l')\in I} e^{\beta_T|t-\tau'|}e^{\beta_T d(v,l')}C_{(t,v)(\tau',l')}\Big\}. \tag{S7}$$

**Step 3.1.** We handled the first item in Step 1 and showed that

$$\max_{(t-1,l)\in I} \sum_{(\tau',l')\in I} e^{\beta_T|t-1-\tau'|}e^{\beta_T d(l,l')}C_{(t-1,l)(\tau',l')}$$

$$= \max\Bigg\{ \max_{(t-1,l)\in I} \sum_{(\tau',l')\in I} e^{\beta_T|t-1-\tau'|}e^{\beta_T d(l,l')}C_{(t-1,l)(\tau',l')}\mathbb{1}_{\{l\in k_{t-1},\, l\notin k_t\}}, \tag{S8}$$

$$\max_{(t-1,l)\in I} \sum_{(\tau',l')\in I} e^{\beta_T|t-1-\tau'|}e^{\beta_T d(l,l')}C_{(t-1,l)(\tau',l')}\mathbb{1}_{\{l\in k_{t-1},\, l\in k_t\}}\Bigg\}.$$

Specifically, in Step 1, we obtained that

$$\max_{(t-1,l)\in I} \sum_{(\tau',l')\in I} e^{\beta_T|t-1-\tau'|}e^{\beta_T d(l,l')}C_{(t-1,l)(\tau',l')}\mathbb{1}_{\{l\in k_{t-1},\, l\notin k_t\}}$$

$$= \max_{(t-1,l)\in I} \sum_{(t-1,l')\in I} e^{\beta_T|t-1-(t-1)|}e^{\beta_T d(l,l')}C_{(t-1,l)(t-1,l')}\mathbb{1}_{\{l\in k_{t-1},\, l\notin k_t,\, l'\in k_{t-1}\}}$$

$$+ \max_{(t-1,l)\in I} e^{\beta_T|t-1-t|}e^{\beta_T d(l,v)}C_{(t-1,l)(t,v)}\mathbb{1}_{\{l\in k_{t-1},\, l\notin k_t\}}$$

$$\leq \max_{(t-1,l)\in I} \sum_{(t-1,l')\in I} e^{\beta_T|t-1-(t-1)|}e^{\beta_T d(l,l')}\left(1 - \frac{\kappa_d}{\kappa_u}\right)\mathbb{1}_{\{l\in k_{t-1},\, l\notin k_t,\, l'\in k_{t-1},\, l'\in l(\mathcal{R})\}}$$

$$\leq e^{\beta_T r_T^{\mathcal{R}}}\left(1 - \frac{\kappa_d}{\kappa_u}\right)\Delta_T^{\mathcal{R}},$$

and

$$\max_{(t-1,l)\in I} \sum_{(\tau',l')\in I} e^{\beta_T|t-1-\tau'|}e^{\beta_T d(l,l')}C_{(t-1,l)(\tau',l')}\mathbb{1}_{\{l\in k_{t-1},\, l\in k_t\}}$$

$$= \max_{(t-1,l)\in I} \sum_{(t-1,l')\in I} e^{\beta_T|t-1-(t-1)|}e^{\beta_T d(l,l')}C_{(t-1,l)(t-1,l')}\mathbb{1}_{\{l\in k_{t-1},\, l\in k_t,\, l'\in k_{t-1}\}}$$

$$+ \max_{(t-1,l)\in I} e^{\beta_T|t-1-t|}e^{\beta_T d(l,v)}C_{(t-1,l)(t,v)}\mathbb{1}_{\{l\in k_{t-1},\, l\in k_t\}}$$

$$\leq \max_{(t-1,l)\in I} \sum_{(t-1,l')\in I} e^{\beta_T|t-1-(t-1)|}e^{\beta_T d(l,l')}C_{ll'}^{\tilde{\pi}_{t-1}}\mathbb{1}_{\{l\in k_{t-1},\, l\in k_t,\, l'\in k_{t-1}\}}$$

$$+ \max_{(t-1,l)\in I} e^{\beta_T|t-1-t|}e^{\beta_T d(l,v)}\left(1 - \frac{\epsilon_d}{\epsilon_u}\right)\mathbb{1}_{\{l\in k_{t-1},\, l\in k_t,\, l=v\}}$$

$$\leq \max_{l\in k_{t-1}} \sum_{l'\in k_{t-1}} e^{\beta_T d(l,l')}C_{ll'}^{\tilde{\pi}_{t-1}} + e^{\beta_T}\left(1 - \frac{\epsilon_d}{\epsilon_u}\right).$$

Then plugging the above two inequalities into equation (S8) yields that

$$\max_{(t-1,l)\in I} \sum_{(\tau',l')\in I} e^{\beta_T|t-1-\tau'|}e^{\beta_T d(l,l')}C_{(t-1,l)(\tau',l')} \tag{S9}$$

$$\leq \max\left\{ e^{\beta_T r_T^{\mathcal{R}}} \left(1 - \frac{\kappa_d}{\kappa_u}\right) \Delta_T^{\mathcal{R}}, \ \frac{1}{3} + e^{\beta_T} \left(1 - \frac{\epsilon_d}{\epsilon_u}\right) \right\},$$

where we used the induction assumption in equation (S5) that

$$\max_{l \in k_{t-1}} \sum_{l' \in k_{t-1}} e^{\beta_T d(l,l')} C_{ll'}^{\overline{\pi}_{t-1}} \leq \frac{1}{3}.$$

**Step 3.2.** We handled the second item in Step 2 and showed that

$$\max_{(t,v) \in I} \sum_{(\tau',l') \in I} e^{\beta_T |t - \tau'|} e^{\beta_T d(v,l')} C_{(t,v)(\tau',l')}$$

$$= \max\left\{ \max_{(t,v) \in I} \sum_{(t-1,l') \in I} e^{\beta_T |t-(t-1)|} e^{\beta_T d(v,l')} C_{(t,v)(t-1,l')} \mathbb{1}_{\{v \notin k_{t-1}, \ l' \in k_{t-1}\}}, \right.$$

$$\left. \max_{(t,v) \in I} \sum_{(t-1,l') \in I} e^{\beta_T |t-(t-1)|} e^{\beta_T d(v,l')} C_{(t,v)(t-1,l')} \mathbb{1}_{\{v \in k_{t-1}, \ l' \in k_{t-1}\}} \right\}$$

$$\leq \max\left\{ 0, \ \max_{(t,v) \in I} \sum_{(t-1,l') \in I} e^{\beta_T} e^{\beta_T d(v,l')} \left(1 - \frac{\epsilon_d}{\epsilon_u}\right) \mathbb{1}_{\{v \in k_{t-1}, \ l' \in k_{t-1}, \ l' = v\}} \right\}$$

$$= e^{\beta_T} \left(1 - \frac{\epsilon_d}{\epsilon_u}\right). \tag{S10}$$

Then for $\beta_T$ being the finite positive constant defined in equation (32) as

$$\beta_T = \frac{1}{r + r_T^{\mathcal{R}}} \log\left( \frac{1}{6\left(1 - \frac{\epsilon_d}{\epsilon_u} \frac{\epsilon_d'}{\epsilon_u'} \frac{\kappa_d}{\kappa_u}\right)(\Delta_T + \Delta_T^{\mathcal{R}})} \right),$$

since $r$, $\Delta_T$, $\Delta_T^{\mathcal{R}}$, and $r_T^{\mathcal{R}}$ are greater than and equal to one, we have

$$e^{\beta_T} \left(1 - \frac{\epsilon_d}{\epsilon_u}\right) \leq e^{\beta_T(r + r_T^{\mathcal{R}})} \left(1 - \frac{\epsilon_d}{\epsilon_u}\right) \leq e^{\beta_T(r + r_T^{\mathcal{R}})} \left[1 - \frac{\epsilon_d}{\epsilon_u} \frac{\epsilon_d'}{\epsilon_u'} \frac{\kappa_d}{\kappa_u}\right] (\Delta_T + \Delta_T^{\mathcal{R}}) = \frac{1}{6} \tag{S11}$$

and

$$e^{\beta_T r_T^{\mathcal{R}}} \left(1 - \frac{\kappa_d}{\kappa_u}\right) \Delta_T^{\mathcal{R}} \leq e^{\beta_T(r + r_T^{\mathcal{R}})} \left[1 - \frac{\epsilon_d}{\epsilon_u} \frac{\epsilon_d'}{\epsilon_u'} \frac{\kappa_d}{\kappa_u}\right] (\Delta_T + \Delta_T^{\mathcal{R}}) = \frac{1}{6}. \tag{S12}$$

Plugging equations (S9)-(S12) into equation (S7) yields that

$$\max_{(\tau,l) \in I} \sum_{(\tau',l') \in I} e^{\beta_T |\tau - \tau'|} e^{\beta_T d(l,l')} C_{(\tau,v)(\tau',v')} = \max\left\{ \frac{1}{6}, \ \frac{1}{3} + \frac{1}{6} \right\} = \frac{1}{2}.$$

By Theorem S1.3,

$$\max_{(\tau,l) \in I} \sum_{(\tau',l') \in I} e^{\beta_T |\tau - \tau'|} e^{\beta_T d(l,l')} D_{(\tau,l)(\tau',l')} \leq \frac{1}{1 - \frac{1}{2}} = 2. \tag{S13}$$

**Step 4.** We complete the proof in this step. We first summarize the results obtained in Steps 1 and 2 regarding $b_i$:

- When $i = (\tau, l) = (t-1, l)$,

$$b_i = b_{(t-1,l)} \mathbb{1}_{\{l \in k_{t-1},\, l \in k_t\}} + b_{(t-1,l)} \mathbb{1}_{\{l \in k_{t-1},\, l \notin k_t\}} \leq 2 \left(1 - \frac{\epsilon_d}{\epsilon_u}\right) \mathbb{1}_{\{l \in k_{t-1},\, l \in k_t,\, l=v'\}};$$

- When $i = (\tau, l) = (t, v)$,

$$b_i \leq 2 \left(1 - \frac{\epsilon_d'}{\epsilon_u'} \frac{\kappa_d}{\kappa_u}\right) \mathbb{1}_{\{v' \in N_t(v) \cup v(\mathcal{R})\}}.$$

Next, applying the Dobrushin comparison theorem (Theorem 3.3), we obtain that

$$\|\rho - \overline{\rho}\|_{(t,v)}$$
$$\leq 2 \left(1 - \frac{\epsilon_d}{\epsilon_u}\right) D_{(t,v)(t-1,v')} \mathbb{1}_{\{v' \in k_{t-1},\, v' \in k_t\}} + 2 \left(1 - \frac{\epsilon_d'}{\epsilon_u'} \frac{\kappa_d}{\kappa_u}\right) \mathbb{1}_{\{v' \in N_t(v) \cup v(\mathcal{R})\}} D_{(t,v)(t,v)}.$$

Therefore, by equation (S6) and the definition of $C_{vv'}^{\mu_{s-1}}$ in equation (S4), we have

$$C_{vv'}^{\widetilde{\pi}_t} = \frac{1}{2} \sup_{x_{t+1}^{k_{t+1}} \in \mathcal{X}} \sup_{\substack{x_t^{k_t}, \overline{x}_t^{k_t} \in \mathcal{X}: \\ x_t^{k_t \setminus \{v'\}} = \overline{x}_t^{k_t \setminus \{v'\}}}} \left\| (\widetilde{\mathsf{F}}_t \widetilde{\pi}_{t-1})_{\chi_t, \chi_{t+1}}^v - (\widetilde{\mathsf{F}}_t \widetilde{\pi}_{t-1})_{\overline{\chi}_t, \chi_{t+1}}^v \right\|$$
$$\leq \left(1 - \frac{\epsilon_d}{\epsilon_u}\right) D_{(t,v)(t-1,v')} \mathbb{1}_{\{v' \in k_{t-1},\, v' \in k_t\}} + \left(1 - \frac{\epsilon_d'}{\epsilon_u'} \frac{\kappa_d}{\kappa_u}\right) \mathbb{1}_{\{v' \in N_t(v) \cup v(\mathcal{R})\}} D_{(t,v)(t,v)},$$

which yields that

$$\max_{v \in k_t} \sum_{v' \in k_t} e^{\beta_T d(v,v')} C_{vv'}^{\widetilde{\pi}_t}$$
$$\leq \left(1 - \frac{\epsilon_d}{\epsilon_u}\right) \max_{v \in k_t} \sum_{v' \in k_{t-1} \cap k_t} e^{\beta_T d(v,v')} D_{(t,v)(t-1,v')}$$
$$\qquad\qquad + \left(1 - \frac{\epsilon_d'}{\epsilon_u'} \frac{\kappa_d}{\kappa_u}\right) \max_{v \in k_t} \sum_{v' \in N_t(v) \cup v(\mathcal{R})} e^{\beta_T d(v,v')} D_{(t,v)(t,v)}$$
$$\leq \left(1 - \frac{\epsilon_d}{\epsilon_u}\right) \max_{v \in k_t} \sum_{v' \in k_{t-1}} e^{\beta_T d(v,v')} D_{(t,v)(t-1,v')}$$
$$\qquad\qquad + \left(1 - \frac{\epsilon_d'}{\epsilon_u'} \frac{\kappa_d}{\kappa_u}\right) e^{\beta_T (r + r_T^{\mathcal{R}})} (\Delta_T + \Delta_T^{\mathcal{R}}) \max_{v \in k_t} D_{(t,v)(t,v)}$$
$$\leq \left(1 - \frac{\epsilon_d}{\epsilon_u} \frac{\epsilon_d'}{\epsilon_u'} \frac{\kappa_d}{\kappa_u}\right) e^{\beta_T (r + r_T^{\mathcal{R}})} (\Delta_T + \Delta_T^{\mathcal{R}}) \max_{v \in k_t} \sum_{(\tau',v') \in I} e^{\beta_T \{|t - \tau'| + d(v,v')\}} D_{(t,v)(\tau',v')}.$$

Then by equation (S13) and the definition of $\beta_T$ given in equation (32), we have

$$\max_{v \in k_t} \sum_{v' \in k_t} e^{\beta_T d(v,v')} C_{vv'}^{\widetilde{\pi}_t} \leq 2 \left(1 - \frac{\epsilon_d}{\epsilon_u} \frac{\epsilon_d'}{\epsilon_u'} \frac{\kappa_d}{\kappa_u}\right) e^{\beta_T (r + r_T^{\mathcal{R}})} (\Delta_T + \Delta_T^{\mathcal{R}}) = 2 \times \frac{1}{6} = \frac{1}{3},$$

which completes the proof. $\qquad\square$

Proposition S1.1 investigates the difference $\left\| (\widetilde{\mathsf{F}}_t \widetilde{\pi}_{t-1})_{\chi_t, \chi_{t+1}}^v - (\widetilde{\mathsf{F}}_t \widetilde{\pi}_{t-1})_{\overline{\chi}_t, \chi_{t+1}}^v \right\|$ under the scenario that $x_t^{k_t \setminus \{v'\}} = \overline{x}_t^{k_t \setminus \{v'\}}$, where the same opertaor $\widetilde{\mathsf{F}}_t$ is applied. In the next proposition, we investigate the one step difference caused by applying two different operators: $\mathsf{F}_t$ and $\widetilde{\mathsf{F}}_t$.

**Proposition S1.4.** *Under Assumption 3.1, for every $t \in [T]$, $B_t \in \mathcal{B}(k_t)$ and $J \subseteq B_t$, we have that*

$$\|\mathsf{F}_t \widetilde{\pi}_{t-1} - \widetilde{\mathsf{F}}_t \widetilde{\pi}_{t-1}\|_J \leq 4\left(1 - \frac{\epsilon_d}{\epsilon_u}\right) e^{-\beta_T d(J, \partial B_t)} \mathrm{card}(J).$$

*Proof.* For $t \in [T]$, define

$$I = (\{t-1\} \times k_{t-1}) \cup (\{t\} \times k_t) \quad \text{and} \quad \mathbb{S} = \mathcal{X}^2.$$

Fix $B_t \in \mathcal{B}(k_t)$ and define

$$\rho(A) = \frac{\begin{array}{c} \int \mathbb{1}_A(x_{t-1}^{k_{t-1}}, x_t^{k_t}) \prod_{\omega \in k_t} f_t^\omega(x_t^\omega \mid k_t^{\omega(\mathcal{R})}, x_{t-1}^{k_{t-1} \cap \{\omega\}}) g_t^\omega(y_t^\omega \mid k_t^{\omega(\mathcal{R})}, x_t^\omega) \widetilde{f}_t^\omega(x_t^\omega, x_t^{N_t(\omega)}) \\ \times \prod_{R \in \mathcal{R}} p_t^R(k_t^R \mid k_{t-1}, x_{t-1}^{k_{t-1} \cap R}) \widetilde{\pi}_{t-1}(dx_{t-1}^{k_{t-1}}) \psi(dx_t^{k_t}) \end{array}}{\begin{array}{c} \int \prod_{\omega \in k_t} f_t^\omega(x_t^\omega \mid k_t^{\omega(\mathcal{R})}, x_{t-1}^{k_{t-1} \cap \{\omega\}}) g_t^\omega(y_t^\omega \mid k_t^{\omega(\mathcal{R})}, x_t^\omega) \widetilde{f}_t^\omega(x_t^\omega, x_t^{N_t(\omega)}) \\ \times \prod_{R \in \mathcal{R}} p_t^R(k_t^R \mid k_{t-1}, x_{t-1}^{k_{t-1} \cap R}) \widetilde{\pi}_{t-1}(dx_{t-1}^{k_{t-1}}) \psi(dx_t^{k_t}) \end{array}},$$

and

$$\widetilde{\rho}(A) = \frac{\begin{array}{c} \int \mathbb{1}_A(x_{t-1}^{k_{t-1}}, x_t^{k_t}) \prod_{v \in B_t} f_t^v(x_t^v \mid k_t^{v(\mathcal{R})}, x_{t-1}^{k_{t-1} \cap \{v\}}) \\ \times \prod_{\omega \in k_t} g_t^\omega(y_t^\omega \mid k_t^{\omega(\mathcal{R})}, x_t^\omega) \widetilde{f}_t^\omega(x_t^\omega, x_t^{N_t(\omega)}) \\ \times \prod_{R \in \mathcal{R}} p_t^R(k_t^R \mid k_{t-1}, x_{t-1}^{k_{t-1} \cap R}) \widetilde{\pi}_{t-1}(dx_{t-1}^{k_{t-1}}) \psi(dx_t^{k_t}) \end{array}}{\begin{array}{c} \int \prod_{v \in B_t} f_t^v(x_t^v \mid k_t^{v(\mathcal{R})}, x_{t-1}^{k_{t-1} \cap \{v\}}) \\ \times \prod_{\omega \in k_t} g_t^\omega(y_t^\omega \mid k_t^{\omega(\mathcal{R})}, x_t^\omega) \widetilde{f}_t^\omega(x_t^\omega, x_t^{N_t(\omega)}) \\ \times \prod_{R \in \mathcal{R}} p_t^R(k_t^R \mid k_{t-1}, x_{t-1}^{k_{t-1} \cap R}) \widetilde{\pi}_{t-1}(dx_{t-1}^{k_{t-1}}) \psi(dx_t^{k_t}) \end{array}}.$$

Then for any $J \subseteq B_t$ and $B_t \in \mathcal{B}(k_t)$, we have

$$\|\mathsf{F}_t \widetilde{\pi}_{t-1} - \widetilde{\mathsf{F}}_t \widetilde{\pi}_{t-1}\|_J = \|\rho - \widetilde{\rho}\|_{\{t\} \times J}. \tag{S14}$$

In the following steps, we are going to use the Dobrushin comparison theorem (Theorem 3.3) to bound $\|\rho - \widetilde{\rho}\|_{\{t\} \times J}$. We will bound $C_{ij}$ and $b_i$ with $i = (\tau, l)$ and $j = (\tau', l')$.

**Step 1.** We first consider $\tau = t - 1$, which implies $l \in k_{t-1}$.

**Step 1.1.** When $l$ is not in $k_t$, we have

$$\rho^i_{(x_{t-1}^{k_{t-1}}, x_t^{k_t})}(A) = \widetilde{\rho}^i_{(x_{t-1}^{k_{t-1}}, x_t^{k_t})}(A) = \frac{\int \mathbb{1}_A(x_{t-1}^l) p_t^{l(\mathcal{R})}(k_t^{l(\mathcal{R})} \mid k_{t-1}, x_{t-1}^{l(\mathcal{R})}) \widetilde{\pi}_{t-1}^l(dx_{t-1}^l)}{\int p_t^{l(\mathcal{R})}(k_t^{l(\mathcal{R})} \mid k_{t-1}, x_{t-1}^{l(\mathcal{R})}) \widetilde{\pi}_{t-1}^l(dx_{t-1}^l)},$$

which reveals that $b_i = 0$. Furthermore, when $\tau' = t - 1$ which implies $l' \in k_{t-1}$, since

$$\rho^i_{(x_{t-1}^{k_{t-1}}, x_t^{k_t})}(A) \geq \left(\frac{\kappa_d}{\kappa_u}\right) \frac{\int \mathbb{1}_A(x_{t-1}^l) \widetilde{\pi}_{t-1}^l(dx_{t-1}^l)}{\int \widetilde{\pi}_{t-1}^l(dx_{t-1}^l)} = \left(\frac{\kappa_d}{\kappa_u}\right) \widetilde{\pi}_{t-1}^l(A)$$

$$\text{and} \quad \rho^i_{(\overline{x}_{t-1}^{k_{t-1}}, \overline{x}_t^{k_t})}(A) \geq \left(\frac{\kappa_d}{\kappa_u}\right) \frac{\int \mathbb{1}_A(\overline{x}_{t-1}^l) \widetilde{\pi}_{t-1}^l(d\overline{x}_{t-1}^l)}{\int \widetilde{\pi}_{t-1}^l(d\overline{x}_{t-1}^l)} = \left(\frac{\kappa_d}{\kappa_u}\right) \widetilde{\pi}_{t-1}^l(A),$$

we have $C_{ij} \leq 1 - \frac{\kappa_d}{\kappa_u}$ if $l' \in l(\mathcal{R})$ by Theorem S1.2 and $C_{ij} = 0$ otherwise. When $\tau' = t$, for any $l' \in k_t$, we have $C_{ij} = 0$.

**Step 1.2.** When $l$ is also in $k_t$, we have

$$\rho^i_{(x^{k_{t-1}}_{t-1}, x^{k_t}_t)}(A) = \frac{\int \mathbb{1}_A(x^l_{t-1})f^l_t(x^l_t \mid k^{l(\mathcal{R})}_t, x^l_{t-1})p^{l(\mathcal{R})}_t(k^{l(\mathcal{R})}_t \mid k_{t-1}, x^{l(\mathcal{R})}_{t-1})\widetilde{\pi}^l_{t-1}(dx^l_{t-1})}{\int f^l_t(x^l_t \mid k^{l(\mathcal{R})}_t, x^l_{t-1})p^{l(\mathcal{R})}_t(k^{l(\mathcal{R})}_t \mid k_{t-1}, x^{l(\mathcal{R})}_{t-1})\widetilde{\pi}^l_{t-1}(dx^l_{t-1})},$$

which is $\widetilde{\pi}^l_{\chi_{t-1},\chi_t}$ defined in equation (S2). In this case, if $\tau' = t-1$, by the definition of $C^{\mu_{s-1}}_{ll'}$ in equation (S4), we have $C_{ij} \leq C^{\widetilde{\pi}_{t-1}}_{ll'}$. If $\tau' = t$ and $l' = l$, since

$$\rho^i_{(x^{k_{t-1}}_{t-1}, x^{k_t}_t)}(A) \geq \left(\frac{\epsilon_d}{\epsilon_u}\right)\frac{\int \mathbb{1}_A(x^l_{t-1})p^{l(\mathcal{R})}_t(k^{l(\mathcal{R})}_t \mid k_{t-1}, x^{l(\mathcal{R})}_{t-1})\widetilde{\pi}^l_{t-1}(dx^l_{t-1})}{\int p^{l(\mathcal{R})}_t(k^{l(\mathcal{R})}_t \mid k_{t-1}, x^{l(\mathcal{R})}_{t-1})\widetilde{\pi}^l_{t-1}(dx^l_{t-1})},$$

we have $C_{ij} \leq 1 - \frac{\epsilon_d}{\epsilon_u}$ by Theorem S1.2, and $C_{ij} = 0$ otherwise.

To calculate $b_i$, note that if $l \in B_t$ we have

$$\widetilde{\rho}^i_{(x^{k_{t-1}}_{t-1}, x^{k_t}_t)}(A) = \rho^i_{(x^{k_{t-1}}_{t-1}, x^{k_t}_t)}(A)$$
$$= \frac{\int \mathbb{1}_A(x^l_{t-1})f^l_t(x^l_t \mid k^{l(\mathcal{R})}_t, x^l_{t-1})p^{l(\mathcal{R})}_t(k^{l(\mathcal{R})}_t \mid k_{t-1}, x^{l(\mathcal{R})}_{t-1})\widetilde{\pi}^l_{t-1}(dx^l_{t-1})}{\int f^l_t(x^l_t \mid k^{l(\mathcal{R})}_t, x^l_{t-1})p^{l(\mathcal{R})}_t(k^{l(\mathcal{R})}_t \mid k_{t-1}, x^{l(\mathcal{R})}_{t-1})\widetilde{\pi}^l_{t-1}(dx^l_{t-1})},$$

which yields that $b_i = 0$. If $l \notin B_t$ we have

$$\widetilde{\rho}^i_{(x^{k_{t-1}}_{t-1}, x^{k_t}_t)}(A) = \frac{\int \mathbb{1}_A(x^l_{t-1})p^{l(\mathcal{R})}_t(k^{l(\mathcal{R})}_t \mid k_{t-1}, x^{l(\mathcal{R})}_{t-1})\widetilde{\pi}^l_{t-1}(dx^l_{t-1})}{\int p^{l(\mathcal{R})}_t(k^{l(\mathcal{R})}_t \mid k_{t-1}, x^{l(\mathcal{R})}_{t-1})\widetilde{\pi}^l_{t-1}(dx^l_{t-1})},$$

which is different to $\rho^i_{(x^{k_{t-1}}_{t-1}, x^{k_t}_t)}(A)$. However, by Assumption 3.1 and the fact that $\frac{\epsilon_d}{\epsilon_u} \leq 1$, we have

$$\rho^i_{(x^{k_{t-1}}_{t-1}, x^{k_t}_t)}(A), \widetilde{\rho}^i_{(x^{k_{t-1}}_{t-1}, x^{k_t}_t)}(A) \geq \left(\frac{\epsilon_d}{\epsilon_u}\right)\frac{\int \mathbb{1}_A(x^l_{t-1})p^{l(\mathcal{R})}_t(k^{l(\mathcal{R})}_t \mid k_{t-1}, x^{l(\mathcal{R})}_{t-1})\widetilde{\pi}^l_{t-1}(dx^l_{t-1})}{\int p^{l(\mathcal{R})}_t(k^{l(\mathcal{R})}_t \mid k_{t-1}, x^{l(\mathcal{R})}_{t-1})\widetilde{\pi}^l_{t-1}(dx^l_{t-1})}.$$

Then, by Theorem S1.2, we have $b_i \leq 2\left(1 - \frac{\epsilon_d}{\epsilon_u}\right)$ if $l \notin B_t$.

**Step 2.** Now, we consider $\tau = t$, which implies $l \in k_t$.

**Step 2.1.** Firstly, when $l$ is not in $k_{t-1}$, we have

$$\rho^i_{(x^{k_{t-1}}_{t-1}, x^{k_t}_t)}(A) = \frac{\int \mathbb{1}_A(x^l_t)f^l_t(x^l_t \mid k^{l(\mathcal{R})}_t)g^l_t(y^l_t \mid k^{l(\mathcal{R})}_t, x^l_t)\widetilde{f}^l_t(x^l_t, x^{N_t(l)}_t)\psi^l(dx^l_t)}{\int f^l_t(x^l_t \mid k^{l(\mathcal{R})}_t)g^l_t(y^l_t \mid k^{l(\mathcal{R})}_t, x^l_t)\widetilde{f}^l_t(x^l_t, x^{N_t(l)}_t)\psi^l(dx^l_t)}.$$

Hence, if $\tau' = t-1$, for any $l' \in k_{t-1}$, we have $C_{ij} = 0$. But, if $\tau' = t$ which implies $l' \in k_t$, for any $l' \in N_t(l)$, given that

$$\rho^i_{(x^{k_{t-1}}_{t-1}, x^{k_t}_t)}(A) \geq \left(\frac{\epsilon'_d}{\epsilon'_u}\right)\frac{\int \mathbb{1}_A(x^l_t)f^l_t(x^l_t \mid k^{l(\mathcal{R})}_t)g^l_t(y^l_t \mid k^{l(\mathcal{R})}_t, x^l_t)\psi^l(dx^l_t)}{\int f^l_t(x^l_t \mid k^{l(\mathcal{R})}_t)g^l_t(y^l_t \mid k^{l(\mathcal{R})}_t, x^l_t)\psi^l(dx^l_t)},$$

we have $C_{ij} \leq 1 - \frac{\epsilon'_d}{\epsilon'_u}$ by Theorem S1.2, and $C_{ij} = 0$ otherwise. The value of $b_i$ depends on whether $l \in B_t$ or not. If $l \in B_t$, then we have

$$\rho^i_{(x^{k_{t-1}}_{t-1}, x^{k_t}_t)}(A) = \widetilde{\rho}^i_{(x^{k_{t-1}}_{t-1}, x^{k_t}_t)}(A)$$

which gives $b_i = 0$. If $l \notin B_t$,

$$\widetilde{\rho}^i_{(x^{k_{t-1}}_{t-1}, x^{k_t}_t)}(A) = \frac{\int \mathbb{1}_A(x^l_t) g^l_t(y^l_t \mid k^{l(\mathcal{R})}_t, x^l_t) \widetilde{f}^l_t(x^l_t, x^{N_t(l)}_t) \psi^l(dx^l_t)}{\int g^l_t(y^l_t \mid k^{l(\mathcal{R})}_t, x^l_t) \widetilde{f}^l_t(x^l_t, x^{N_t(l)}_t) \psi^l(dx^l_t)},$$

which is different to $\rho^i_{(x^{k_{t-1}}_{t-1}, x^{k_t}_t)}(A)$. However, by Assumption 3.1 and the fact that $\frac{\epsilon_d}{\epsilon_u} \leq 1$, we have

$$\rho^i_{(x^{k_{t-1}}_{t-1}, x^{k_t}_t)}(A), \; \widetilde{\rho}^i_{(x^{k_{t-1}}_{t-1}, x^{k_t}_t)}(A) \geq \left(\frac{\epsilon_d}{\epsilon_u}\right) \frac{\int \mathbb{1}_A(x^l_t) g^l_t(y^l_t \mid k^{l(\mathcal{R})}_t, x^l_t) \widetilde{f}^l_t(x^l_t, x^{N_t(l)}_t) \psi^l(dx^l_t)}{\int g^l_t(y^l_t \mid k^{l(\mathcal{R})}_t, x^l_t) \widetilde{f}^l_t(x^l_t, x^{N_t(l)}_t) \psi^l(dx^l_t)},$$

which yields $b_i \leq 2\left(1 - \frac{\epsilon_d}{\epsilon_u}\right)$ by Theorem S1.2, under the case that $l \notin B_t$.

**Step 2.2.** Next, we discuss the case that $l \in k_{t-1}$. Under this case,

$$\rho^i_{(x^{k_{t-1}}_{t-1}, x^{k_t}_t)}(A) = \frac{\int \mathbb{1}_A(x^l_t) f^l_t(x^l_t \mid k^{l(\mathcal{R})}_t, x^l_{t-1}) g^l_t(y^l_t \mid k^{l(\mathcal{R})}_t, x^l_t) \widetilde{f}^l_t(x^l_t, x^{N_t(l)}_t) \psi^l(dx^l_t)}{\int f^l_t(x^l_t \mid k^{l(\mathcal{R})}_t, x^l_{t-1}) g^l_t(y^l_t \mid k^{l(\mathcal{R})}_t, x^l_t) \widetilde{f}^l_t(x^l_t, x^{N_t(l)}_t) \psi^l(dx^l_t)}.$$

Hence, if $\tau' = t - 1$, when $l' \neq l$, we have $C_{ij} = 0$; when $l' = l$, even in the presence of interaction, we still have

$$\rho^i_{(x^{k_{t-1}}_{t-1}, x^{k_t}_t)}(A) \geq \left(\frac{\epsilon_d}{\epsilon_u}\right) \frac{\int \mathbb{1}_A(x^l_t) g^l_t(y^l_t \mid k^{l(\mathcal{R})}_t, x^l_t) \widetilde{f}^l_t(x^l_t, x^{N_t(l)}_t) \psi^l(dx^l_t)}{\int g^l_t(y^l_t \mid k^{l(\mathcal{R})}_t, x^l_t) \widetilde{f}^l_t(x^l_t, x^{N_t(l)}_t) \psi^l(dx^l_t)},$$

which yields that $C_{ij} \leq 1 - \frac{\epsilon_d}{\epsilon_u}$ by Theorem S1.2. If $\tau' = t$, when $l' \in N_t(l)$, we have

$$\rho^i_{(x^{k_{t-1}}_{t-1}, x^{k_t}_t)}(A) \geq \left(\frac{\epsilon'_d}{\epsilon'_u}\right) \frac{\int \mathbb{1}_A(x^l_t) f^l_t(x^l_t \mid k^{l(\mathcal{R})}_t, x^l_{t-1}) g^l_t(y^l_t \mid k^{l(\mathcal{R})}_t, x^l_t) \psi^l(dx^l_t)}{\int f^l_t(x^l_t \mid k^{l(\mathcal{R})}_t, x^l_{t-1}) g^l_t(y^l_t \mid k^{l(\mathcal{R})}_t, x^l_t) \psi^l(dx^l_t)},$$

which yields that $C_{ij} \leq 1 - \frac{\epsilon'_d}{\epsilon'_u}$ by Theorem S1.2; we have $C_{ij} = 0$ when $l' \notin N_t(l)$.

The value of $b_i$ depends on whether $l \in B_t$ or not. If $l \in B_t$, then we have

$$\rho^i_{(x^{k_{t-1}}_{t-1}, x^{k_t}_t)}(A) = \widetilde{\rho}^i_{(x^{k_{t-1}}_{t-1}, x^{k_t}_t)}(A)$$

which gives $b_i = 0$. If $l \notin B_t$,

$$\widetilde{\rho}^i_{(x^{k_{t-1}}_{t-1}, x^{k_t}_t)}(A) = \frac{\int \mathbb{1}_A(x^l_t) g^l_t(y^l_t \mid k^{l(\mathcal{R})}_t, x^l_t) \widetilde{f}^l_t(x^l_t, x^{N_t(l)}_t) \psi^l(dx^l_t)}{\int g^l_t(y^l_t \mid k^{l(\mathcal{R})}_t, x^l_t) \widetilde{f}^l_t(x^l_t, x^{N_t(l)}_t) \psi^l(dx^l_t)},$$

which is different to $\rho^i_{(x^{k_{t-1}}_{t-1}, x^{k_t}_t)}(A)$. However, by Assumption 3.1 and the fact that $\frac{\epsilon_d}{\epsilon_u} \leq 1$, we have

$$\rho^i_{(x^{k_{t-1}}_{t-1}, x^{k_t}_t)}(A), \; \widetilde{\rho}^i_{(x^{k_{t-1}}_{t-1}, x^{k_t}_t)}(A) \geq \left(\frac{\epsilon_d}{\epsilon_u}\right) \frac{\int \mathbb{1}_A(x^l_t) g^l_t(y^l_t \mid k^{l(\mathcal{R})}_t, x^l_t) \widetilde{f}^l_t(x^l_t, x^{N_t(l)}_t) \psi^l(dx^l_t)}{\int g^l_t(y^l_t \mid k^{l(\mathcal{R})}_t, x^l_t) \widetilde{f}^l_t(x^l_t, x^{N_t(l)}_t) \psi^l(dx^l_t)},$$

which yields $b_i \leq 2\left(1 - \frac{\epsilon_d}{\epsilon_u}\right)$ by Theorem S1.2, under the case that $l \notin B_t$.

**Step 3.** In this step, we summarize the results of $C_{ij}$ obtained in the previous two steps and aim to bound the following quantity:

$$\max_{(\tau,l)\in I} \sum_{(\tau',l')\in I} e^{\beta_T|\tau-\tau'|} e^{\beta_T d(l,l')} C_{(\tau,l)(\tau',l')}$$

$$= \max\Bigg\{ \max_{(t-1,l)\in I} \sum_{(\tau',l')\in I} e^{\beta_T|t-1-\tau'|} e^{\beta_T d(l,l')} C_{(t-1,l)(\tau',l')},$$

$$\max_{(t,l)\in I} \sum_{(\tau',l')\in I} e^{\beta_T|t-\tau'|} e^{\beta_T d(l,l')} C_{(t,l)(\tau',l')} \Bigg\}. \tag{S15}$$

**Step 3.1.** We handled the first item in Step 1 and showed that

$$\max_{(t-1,l)\in I} \sum_{(\tau',l')\in I} e^{\beta_T|t-1-\tau'|} e^{\beta_T d(l,l')} C_{(t-1,l)(\tau',l')}$$

$$= \max\Bigg\{ \max_{(t-1,l)\in I} \sum_{(\tau',l')\in I} e^{\beta_T|t-1-\tau'|} e^{\beta_T d(l,l')} C_{(t-1,l)(\tau',l')} \mathbb{1}_{\{l\in k_{t-1},\ l\notin k_t\}}, \tag{S16}$$

$$\max_{(t-1,l)\in I} \sum_{(\tau',l')\in I} e^{\beta_T|t-1-\tau'|} e^{\beta_T d(l,l')} C_{(t-1,l)(\tau',l')} \mathbb{1}_{\{l\in k_{t-1},\ l\in k_t\}} \Bigg\}.$$

Specifically, in Step 1, we obtained that

$$\max_{(t-1,l)\in I} \sum_{(\tau',l')\in I} e^{\beta_T|t-1-\tau'|} e^{\beta_T d(l,l')} C_{(t-1,l)(\tau',l')} \mathbb{1}_{\{l\in k_{t-1},\ l\notin k_t\}}$$

$$= \max_{(t-1,l)\in I} \sum_{(t-1,l')\in I} e^{\beta_T|t-1-(t-1)|} e^{\beta_T d(l,l')} C_{(t-1,l)(t-1,l')} \mathbb{1}_{\{l\in k_{t-1},\ l\notin k_t,\ l'\in k_{t-1}\}}$$

$$+ \max_{(t-1,l)\in I} \sum_{(t,l')\in I} e^{\beta_T|t-1-t|} e^{\beta_T d(l,l')} C_{(t-1,l)(t,l')} \mathbb{1}_{\{l\in k_{t-1},\ l\notin k_t,\ l'\in k_t\}}$$

$$\leq \max_{(t-1,l)\in I} \sum_{(t-1,l')\in I} e^{\beta_T|t-1-(t-1)|} e^{\beta_T d(l,l')} \left(1-\frac{\kappa_d}{\kappa_u}\right) \mathbb{1}_{\{l\in k_{t-1},\ l\notin k_t,\ l'\in k_{t-1},\ l'\in l(\mathcal{R})\}}$$

$$\leq e^{\beta_T r_T^{\mathcal{R}}} \left(1-\frac{\kappa_d}{\kappa_u}\right) \Delta_T^{\mathcal{R}},$$

and

$$\max_{(t-1,l)\in I} \sum_{(\tau',l')\in I} e^{\beta_T|t-1-\tau'|} e^{\beta_T d(l,l')} C_{(t-1,l)(\tau',l')} \mathbb{1}_{\{l\in k_{t-1},\ l\in k_t\}}$$

$$= \max_{(t-1,l)\in I} \sum_{(t-1,l')\in I} e^{\beta_T|t-1-(t-1)|} e^{\beta_T d(l,l')} C_{(t-1,l)(t-1,l')} \mathbb{1}_{\{l\in k_{t-1},\ l\in k_t,\ l'\in k_{t-1}\}}$$

$$+ \max_{(t-1,l)\in I} \sum_{(t,l')\in I} e^{\beta_T|t-1-t|} e^{\beta_T d(l,l')} C_{(t-1,l)(t,l')} \mathbb{1}_{\{l\in k_{t-1},\ l\in k_t,\ l'\in k_t\}}$$

$$\leq \max_{(t-1,l)\in I} \sum_{(t-1,l')\in I} e^{\beta_T|t-1-(t-1)|} e^{\beta_T d(l,l')} C_{ll'}^{\widetilde{\pi}_{t-1}} \mathbb{1}_{\{l\in k_{t-1},\ l\in k_t,\ l'\in k_{t-1}\}}$$

$$+ \max_{(t-1,l)\in I} e^{\beta_T|t-1-t|} e^{\beta_T d(l,l')} \left(1 - \frac{\epsilon_d}{\epsilon_u}\right) \mathbb{1}_{\{l\in k_{t-1},\, l\in k_t,\, l'\in k_t,\, l'=l\}}$$

$$\leq \max_{l\in k_{t-1}} \sum_{l'\in k_{t-1}} e^{\beta_T d(l,l')} C_{ll'}^{\widetilde{\pi}_{t-1}} + e^{\beta_T}\left(1 - \frac{\epsilon_d}{\epsilon_u}\right).$$

Then plugging the above two inequalities into equation (S16),

$$\max_{(t-1,l)\in I} \sum_{(\tau',l')\in I} e^{\beta_T|t-1-\tau'|} e^{\beta_T d(l,l')} C_{(t-1,l)(\tau',l')} \tag{S17}$$

$$\leq \max\left\{ e^{\beta_T r_T^{\mathcal{R}}}\left(1 - \frac{\kappa_d}{\kappa_u}\right)\Delta_T^{\mathcal{R}},\; \frac{1}{3} + e^{\beta_T}\left(1 - \frac{\epsilon_d}{\epsilon_u}\right)\right\},$$

where we used Proposition S1.1.

**Step 3.2.** We handled the second item in Step 2 and showed that

$$\max_{(t,l)\in I} \sum_{(\tau',l')\in I} e^{\beta_T|t-\tau'|} e^{\beta_T d(l,l')} C_{(t,l)(\tau',l')}$$

$$= \max\left\{ \max_{(t,l)\in I} \sum_{(\tau',l')\in I} e^{\beta_T|t-\tau'|} e^{\beta_T d(l,l')} C_{(t,l)(\tau',l')} \mathbb{1}_{\{l\in k_t,\, l\notin k_{t-1}\}},\right. \tag{S18}$$

$$\left. \max_{(t,l)\in I} \sum_{(\tau',l')\in I} e^{\beta_T|t-\tau'|} e^{\beta_T d(l,l')} C_{(t,l)(\tau',l')} \mathbb{1}_{\{l\in k_t,\, l\in k_{t-1}\}}\right\}.$$

Specifically, in Step 2, we obtained that

$$\max_{(t,l)\in I} \sum_{(\tau',l')\in I} e^{\beta_T|t-\tau'|} e^{\beta_T d(l,l')} C_{(t,l)(\tau',l')} \mathbb{1}_{\{l\in k_t,\, l\notin k_{t-1}\}}$$

$$= \max_{(t,l)\in I} \sum_{(t-1,l')\in I} e^{\beta_T|t-(t-1)|} e^{\beta_T d(l,l')} C_{(t,l)(t-1,l')} \mathbb{1}_{\{l\in k_t,\, l\notin k_{t-1},\, l'\in k_{t-1}\}}$$

$$+ \max_{(t,l)\in I} \sum_{(t,l')\in I} e^{\beta_T|t-t|} e^{\beta_T d(l,l')} C_{(t,l)(t,l')} \mathbb{1}_{\{l\in k_t,\, l\notin k_{t-1},\, l'\in k_t\}}$$

$$\leq \max_{(t,l)\in I} \sum_{(t,l')\in I} e^{\beta_T|t-t|} e^{\beta_T d(l,l')} \left(1 - \frac{\epsilon_d'}{\epsilon_u'}\right) \mathbb{1}_{\{l\in k_t,\, l\notin k_t,\, l'\in N_t(l)\}}$$

$$\leq e^{\beta_T r}\left(1 - \frac{\epsilon_d'}{\epsilon_u'}\right)\Delta_T,$$

and

$$\max_{(t,l)\in I} \sum_{(\tau',l')\in I} e^{\beta_T|t-\tau'|} e^{\beta_T d(l,l')} C_{(t,l)(\tau',l')} \mathbb{1}_{\{l\in k_t,\, l\in k_{t-1}\}}$$

$$= \max_{(t,l)\in I} \sum_{(t-1,l')\in I} e^{\beta_T|t-(t-1)|} e^{\beta_T d(l,l')} C_{(t,l)(t-1,l')} \mathbb{1}_{\{l\in k_t,\, l\in k_{t-1},\, l'\in k_{t-1}\}}$$

$$+ \max_{(t,l)\in I} \sum_{(t,l')\in I} e^{\beta_T|t-t|} e^{\beta_T d(l,l')} C_{(t,l)(t,l')} \mathbb{1}_{\{l\in k_t,\, l\in k_{t-1},\, l'\in k_t\}}$$

$$\leq \max_{(t,l)\in I} \sum_{(t-1,l')\in I} e^{\beta_T |t-(t-1)|} e^{\beta_T d(l,l')} \left(1 - \frac{\epsilon_d}{\epsilon_u}\right) \mathbb{1}_{\{l\in k_t,\ l\in k_{t-1},\ l'\in k_{t-1},\ l'=l\}}$$

$$+ \max_{(t,l)\in I} \sum_{(t,l')\in I} e^{\beta_T |t-t|} e^{\beta_T d(l,l')} \left(1 - \frac{\epsilon'_d}{\epsilon'_u}\right) \mathbb{1}_{\{l\in k_{t-1},\ l\notin k_t,\ l'\in N_t(l)\}}$$

$$\leq e^{\beta_T} \left(1 - \frac{\epsilon_d}{\epsilon_u}\right) + e^{\beta_T r} \left(1 - \frac{\epsilon'_d}{\epsilon'_u}\right)\Delta_T.$$

Then plugging the above two inequalities into equation (S18),

$$\max_{(t,l)\in I} \sum_{(\tau',l')\in I} e^{\beta_T|t-\tau'|} e^{\beta_T d(l,l')} C_{(t,l)(\tau',l')} \tag{S19}$$

$$= \max\left\{ e^{\beta_T r} \left(1 - \frac{\epsilon'_d}{\epsilon'_u}\right)\Delta_T,\ e^{\beta_T}\left(1 - \frac{\epsilon_d}{\epsilon_u}\right) + e^{\beta_T r}\left(1 - \frac{\epsilon'_d}{\epsilon'_u}\right)\Delta_T \right\}.$$

By the definition of $\beta_T$ given in equation (32), we have

$$e^{\beta_T r r_T^{\mathcal{R}}} \left(1 - \frac{\kappa_d}{\kappa_u}\right)\Delta_T^{\mathcal{R}} \leq \frac{1}{6}, \quad e^{\beta_T}\left(1 - \frac{\epsilon_d}{\epsilon_u}\right) \leq \frac{1}{6} \quad \text{and} \quad e^{\beta_T r}\left(1 - \frac{\epsilon'_d}{\epsilon'_u}\right)\Delta_T \leq \frac{1}{6}.$$

Then plugging equation (S17) and (S19) into equation (S15), we obtain

$$\max_{(\tau,l)\in I} \sum_{(\tau',l')\in I} e^{\beta_T|\tau-\tau'|} e^{\beta_T d(l,l')} C_{(\tau,l)(\tau',l')} = \max\left\{\frac{1}{6}+\frac{1}{6},\ \frac{1}{3}+\frac{1}{6}\right\} = \frac{1}{2}.$$

By Theorem S1.3,

$$\max_{(\tau,l)\in I} \sum_{(\tau',l')\in I} e^{\beta_T|\tau-\tau'|} e^{\beta_T d(l,l')} D_{(\tau,l)(\tau',l')} \leq \frac{1}{1-\frac{1}{2}} = 2. \tag{S20}$$

**Step 4.** We complete the proof in this step. We first summarize the results obtained in Steps 1 and 2 regarding $b_i$:

- When $i = (\tau,l) = (t-1,l)$,

$$b_i = b_{(t-1,l)} \mathbb{1}_{\{l\in k_{t-1},\ l\in k_t\}} + b_{(t-1,l)} \mathbb{1}_{\{l\in k_{t-1},\ l\notin k_t\}}$$

$$= b_{(t-1,l)} \mathbb{1}_{\{l\in k_{t-1},\ l\in k_t\}} + 0$$

$$= b_{(t-1,l)} \mathbb{1}_{\{l\in k_{t-1},\ l\in k_t,\ l\in B_t\}} + b_{(t-1,l)} \mathbb{1}_{\{l\in k_{t-1},\ l\in k_t,\ l\notin B_t\}}$$

$$\leq 2\left(1 - \frac{\epsilon_d}{\epsilon_u}\right) \mathbb{1}_{\{l\in k_{t-1},\ l\in k_t,\ l\notin B_t\}};$$

- When $i = (\tau,l) = (t,l)$,

$$b_i = b_{(t,l)} \mathbb{1}_{\{l\in k_t,\ l\in k_{t-1}\}} + b_{(t,l)} \mathbb{1}_{\{l\in k_t,\ l\notin k_{t-1}\}}$$

$$= b_{(t,l)} \mathbb{1}_{\{l\in k_t,\ l\in k_{t-1}\ l\in B_t\}} + b_{(t,l)} \mathbb{1}_{\{l\in k_t,\ l\in k_{t-1}\ l\notin B_t\}}$$

$$+ b_{(t,l)} \mathbb{1}_{\{l\in k_t,\ l\notin k_{t-1}\ l\in B_t\}} + b_{(t,l)} \mathbb{1}_{\{l\in k_t,\ l\notin k_{t-1}\ l\notin B_t\}}$$

$$= 0 + b_{(t,l)} \mathbb{1}_{\{l\in k_t,\ l\in k_{t-1}\ l\notin B_t\}} + 0 + b_{(t,l)} \mathbb{1}_{\{l\in k_t,\ l\notin k_{t-1}\ l\notin B_t\}}$$

$$\leq 2\left(1-\frac{\epsilon_d}{\epsilon_u}\right)\mathbb{1}_{\{l\in k_t,\; l\in k_{t-1}\; l\notin B_t\}}+2\left(1-\frac{\epsilon_d}{\epsilon_u}\right)\mathbb{1}_{\{l\in k_t,\; l\notin k_{t-1}\; l\notin B_t\}}$$

$$=2\left(1-\frac{\epsilon_d}{\epsilon_u}\right)\mathbb{1}_{\{l\in k_t,\; l\notin B_t\}}.$$

Therefore,

$$\|\rho-\widetilde{\rho}\|_{\{t\}\times J}\leq 2\left(1-\frac{\epsilon_d}{\epsilon_u}\right)\sum_{l\in J}\left\{\sum_{l'\in k_t\backslash B_t}D_{(t,l)(t-1,l')}+\sum_{l'\in k_t\backslash B_t}D_{(t,l)(t,l')}\right\}$$

$$\leq 2\left(1-\frac{\epsilon_d}{\epsilon_u}\right)\sum_{l\in J}e^{-\beta_T d(l,\partial B_t)}\left\{\sum_{l'\in k_t\backslash B_t}e^{\beta_T}e^{\beta_T d(l,l')}D_{(t,l)(t-1,l')}\right.$$

$$\left.+\sum_{l'\in k_t\backslash B_t}e^{\beta_T d(l,l')}D_{(t,l)(t,l')}\right\}$$

$$\leq 4\left(1-\frac{\epsilon_d}{\epsilon_u}\right)\sum_{l\in J}\left(\max_{l\in J}e^{-\beta_T d(l,\partial B_t)}\right)$$

$$\leq 4\left(1-\frac{\epsilon_d}{\epsilon_u}\right)e^{-\beta_T d(J,\partial B_t)}\mathrm{card}(J).$$

Here, we derived the first inequality using the Dobrushin comparison theorem (Theorem 3.3). The second inequality was obtained by applying the definition of $d(l,\partial B_t)$ and the fact that it is smaller than $d(l,l')$ for any $l\in J\subseteq B_t$ and $l'\in k_t\backslash B_t$. We established the third inequality using equation (S20). The fourth inequality was derived from the definition of $d(J,\partial B_t)$, which is the smallest of $d(l,\partial B_t)$ for any $l\in J\subseteq B_t$.

At last, by equation (S14), we complete the proof. $\square$

In Proposition S1.4, we investigated the one-step error $\|\mathsf{F}_t\widetilde{\pi}_{t-1}-\widetilde{\mathsf{F}}_t\widetilde{\pi}_{t-1}\|_J$ generated by two operators. In the following proposition, we will show that error will decay exponentially not only in the time dimension but also in the spatial dimension.

**Proposition S1.5.** *Under Assumption 3.1, we have that for every $J\subseteq B_t$, $B_t\in\mathcal{B}(k_t)$ and $s\in[t-1]$,*

$$\left\|\mathsf{F}_t\cdots\mathsf{F}_{s+1}\mathsf{F}_s\widetilde{\pi}_{s-1}-\mathsf{F}_t\cdots\mathsf{F}_{s+1}\widetilde{\mathsf{F}}_s\widetilde{\pi}_{s-1}\right\|_J$$

$$\leq 2e^{-\beta_T(t-s)}\sum_{v\in J}\max_{v'\in k_s}e^{-\beta_T d(v,v')}\sup_{x_s^{k_s},x_{s+1}^{k_{s+1}}\in\mathcal{X}}\left\|(\mathsf{F}_s\widetilde{\pi}_{s-1})^{v'}_{\chi_s,\chi_{s+1}}-(\widetilde{\mathsf{F}}_s\widetilde{\pi}_{s-1})^{v'}_{\chi_s,\chi_{s+1}}\right\|,$$

*where, according to the definition of $\mu^v_{\chi_{s-1},\chi_s}$ given in equation (S2), for $B_s\in\mathcal{B}(k_s)$ and $v'\in B_s$,*

$$(\mathsf{F}_s\widetilde{\pi}_{s-1})^{v'}_{\chi_s,\chi_{s+1}}(A)$$

$$=\frac{\begin{aligned}&\int\mathbb{1}_A(x_s^{v'})\prod_{\omega\in k_s}f_s^{\omega}(x_s^{\omega}\mid k_s^{\omega(\mathcal{R})},x_{s-1}^{k_{s-1}\cap\{\omega\}})g_s^{v'}(Y_s^{v'}\mid k_s^{v'(\mathcal{R})},x_s^{v'})\widetilde{f}_s^{v'}(x_s^{v'},x_s^{N_s(v')})\\&\times p_{s+1}^{v'(\mathcal{R})}(k_{s+1}^{v'(\mathcal{R})}\mid k_s,x_s^{v'(\mathcal{R})})f_{s+1}^{v'}(x_{s+1}^{k_{s+1}\cap\{v'\}}\mid k_s^{v'(\mathcal{R})},x_s^{v'})\\&\times\prod_{R\in\mathcal{R}}p_s^R(k_s^R\mid k_{s-1},x_{s-1}^{k_{s-1}\cap R})\widetilde{\pi}_{s-1}(dx_{s-1}^{k_{s-1}})\psi^{v'}(dx_s^{v'})\end{aligned}}{\begin{aligned}&\int\prod_{\omega\in k_s}f_s^{\omega}(x_s^{\omega}\mid k_s^{\omega(\mathcal{R})},x_{s-1}^{k_{s-1}\cap\{\omega\}})g_s^{v'}(Y_s^{v'}\mid k_s^{v'(\mathcal{R})},x_s^{v'})\widetilde{f}_s^{v'}(x_s^{v'},x_s^{N_s(v')})\\&\times p_{s+1}^{v'(\mathcal{R})}(k_{s+1}^{v'(\mathcal{R})}\mid k_s,x_s^{v'(\mathcal{R})})f_{s+1}^{v'}(x_{s+1}^{k_{s+1}\cap\{v'\}}\mid k_s^{v'(\mathcal{R})},x_s^{v'})\\&\times\prod_{R\in\mathcal{R}}p_s^R(k_s^R\mid k_{s-1},x_{s-1}^{k_{s-1}\cap R})\widetilde{\pi}_{s-1}(dx_{s-1}^{k_{s-1}})\psi^{v'}(dx_s^{v'})\end{aligned}}\tag{S21}$$

*and*

$$
\begin{aligned}
&(\widetilde{\mathsf{F}}_s \widetilde{\pi}_{s-1})^{v'}_{\mathcal{X}_s, \mathcal{X}_{s+1}}(A) \\
&= \frac{\begin{aligned}&\int \mathbb{1}_A(x^{v'}_s) \prod_{\omega \in B_s} f^\omega_s(x^\omega_s \mid k^{\omega(\mathcal{R})}_s, x^{k_{s-1} \cap \{\omega\}}_{s-1}) g^{v'}_s(Y^{v'}_s \mid k^{v'(\mathcal{R})}_s, x^{v'}_s) \widetilde{f}^{v'}_s(x^{v'}_s, x^{N_s(v')}_s) \\ &\quad \times p^{v'(\mathcal{R})}_{s+1}(k^{v'(\mathcal{R})}_{s+1} \mid k_s, x^{v'(\mathcal{R})}_s) f^{v'}_{s+1}(x^{k_{s+1} \cap \{v'\}}_{s+1} \mid k^{v'(\mathcal{R})}_{s+1}, x^{v'}_s) \\ &\quad \times \prod_{R \in \mathcal{R}} p^R_s(k^R_s \mid k_{s-1}, x^{k_{s-1} \cap R}_{s-1}) \widetilde{\pi}_{s-1}(dx^{k_{s-1}}_{s-1}) \psi^{v'}(dx^{v'}_s)\end{aligned}}{\begin{aligned}&\int \prod_{\omega \in B_s} f^\omega_s(x^\omega_s \mid k^{\omega(\mathcal{R})}_s, x^{k_{s-1} \cap \{\omega\}}_{s-1}) g^{v'}_s(Y^{v'}_s \mid k^{v'(\mathcal{R})}_s, x^{v'}_s) \widetilde{f}^{v'}_s(x^{v'}_s, x^{N_s(v')}_s) \\ &\quad \times p^{v'(\mathcal{R})}_{s+1}(k^{v'(\mathcal{R})}_{s+1} \mid k_s, x^{v'(\mathcal{R})}_s) f^{v'}_{s+1}(x^{k_{s+1} \cap \{v'\}}_{s+1} \mid k^{v'(\mathcal{R})}_{s+1}, x^{v'}_s) \\ &\quad \times \prod_{R \in \mathcal{R}} p^R_s(k^R_s \mid k_{s-1}, x^{k_{s-1} \cap R}_{s-1}) \widetilde{\pi}_{s-1}(dx^{k_{s-1}}_{s-1}) \psi^{v'}(dx^{v'}_s)\end{aligned}}. \quad \text{(S22)}
\end{aligned}
$$

The proof will need the following result and we provide it here for readers' convenience.

**Theorem S1.6** (Lemma 4.2 of Rebeschini and Van Handel [2015]). *Let $\rho$ and $\rho'$ be probability measures and let $\Lambda$ be a bounded and strictly positive measurable function. Define*

$$
\rho_\Lambda(A) := \frac{\int \mathbb{1}_A(x) \Lambda(x) \rho(x)}{\int \Lambda(x) \rho(x)} \quad \text{and} \quad \rho'_\Lambda(A) := \frac{\int \mathbb{1}_A(x) \Lambda(x) \rho'(x)}{\int \Lambda(x) \rho'(x)}.
$$

*Then*

$$
\|\rho_\Lambda - \rho'_\Lambda\| \le 2 \frac{\sup_x \Lambda(x)}{\inf_x \Lambda(x)} \|\rho - \rho'\| \quad \text{and} \quad \|\|\rho_\Lambda - \rho'_\Lambda\|\| \le 2 \frac{\sup_x \Lambda(x)}{\inf_x \Lambda(x)} \|\|\rho - \rho'\|\|.
$$

*Proof of Proposition S1.5.* For $s \in [t-1]$ and $t \in [T]$, define

$$
I = (\{s\} \times k_s) \cup \ldots \cup (\{t\} \times k_t) \quad \text{and} \quad \mathbb{S} = \mathcal{X}^{t-s+1},
$$

and the probability measures on $\mathbb{S}$ as follows:

$$
\rho = \mathbb{P}^{\widetilde{\mathsf{F}}_s \widetilde{\pi}_{s-1}} \left( X^{k_s}_s, X^{k_{s+1}}_{s+1}, \ldots, X^{k_t}_t \in \cdot \mid Y_{s+1}, \ldots, Y_t \right),
$$

$$
\widetilde{\rho} = \mathbb{P}^{\mathsf{F}_s \widetilde{\pi}_{s-1}} \left( X^{k_s}_s, X^{k_{s+1}}_{s+1}, \ldots, X^{k_t}_t \in \cdot \mid Y_{s+1}, \ldots, Y_t \right).
$$

Then we have

$$
\left\| \mathsf{F}_t \cdots \mathsf{F}_{s+1} \widetilde{\mathsf{F}}_s \widetilde{\pi}_{s-1} - \mathsf{F}_t \cdots \mathsf{F}_{s+1} \mathsf{F}_s \widetilde{\pi}_{s-1} \right\|_J = \|\rho - \widetilde{\rho}\|_{\{t\} \times J}.
$$

In the following steps, we are going to use the Dobrushin comparison theorem (Theorem 3.3) to bound $\|\rho - \widetilde{\rho}\|_{\{t\} \times J}$. We will bound $C_{ij}$ and $b_i$ with $i = (\tau, v)$ and $j = (\tau', v')$.

**Step 1.** Consider $\tau = s$ which implies $v \in k_s$. If $v \notin k_{s+1}$, we have

$$
\rho^i_{(x^{k_s}_s, \ldots, x^{k_t}_t)}(A) = \frac{\int \mathbb{1}_A(x^v_s) p^{v(\mathcal{R})}_{s+1}(k^{v(\mathcal{R})}_{s+1} \mid k_s, x^{v(\mathcal{R})}_s) \widetilde{\pi}^v_{\mathcal{X}_s}(dx^v_s)}{\int p^{v(\mathcal{R})}_{s+1}(k^{v(\mathcal{R})}_{s+1} \mid k_s, x^{v(\mathcal{R})}_s) \widetilde{\pi}^v_{\mathcal{X}_s}(dx^v_s)}.
$$

Then when $\tau' = s$, if $v' \in v(\mathcal{R})$ we have by Theorem S1.2 that $C_{ij} \le \left(1 - \frac{\kappa_d}{\kappa_u}\right)$, since

$$
\rho^i_{(x^{k_s}_s, \ldots, x^{k_t}_t)}(A), \ \rho^i_{(\overline{x}^{k_s}_s, \ldots, \overline{x}^{k_t}_t)}(A) \ge \frac{\kappa_d}{\kappa_u} \widetilde{\pi}^v_{\mathcal{X}_s}(A);
$$

if $v' \notin v(\mathcal{R})$ we have $C_{ij} = 0$. When $\tau' \in \{s+1, \ldots, t\}$, we have $C_{ij} = 0$.

Next, if $v \in k_{s+1}$, we have

$$\rho^i_{(x_s^{k_s}, \dots, x_t^{k_t})}(A) = \frac{\int \mathbb{1}_A(x_s^v) f_{s+1}^v(x_{s+1}^{k_{s+1} \cap \{v\}} \mid k_{s+1}^{v(\mathcal{R})}, x_s^v) p_{s+1}^{v(\mathcal{R})}(k_{s+1}^{v(\mathcal{R})} \mid k_s, x_s^{v(\mathcal{R})}) \widetilde{\pi}_{\chi_s}^v(dx_s^v)}{\int f_{s+1}^v(x_{s+1}^{k_{s+1} \cap \{v\}} \mid k_{s+1}^{v(\mathcal{R})}, x_s^v) p_{s+1}^{v(\mathcal{R})}(k_{s+1}^{v(\mathcal{R})} \mid k_s, x_s^{v(\mathcal{R})}) \widetilde{\pi}_{\chi_s}^v(dx_s^v)}.$$

We have $\rho^i_{(x_s^{k_s}, \dots, x_t^{k_t})}(A) = \widetilde{\pi}^v_{\chi_s, \chi_{s+1}}(A)$ according to the definition of $\mu^v_{\chi_{s-1}, \chi_s}$ given in equation (S2). Therefore, when $\tau' = s$ which implies $v' \in k_s$, by the definition of $C^{\widetilde{\pi}_s}_{vv'}$ in equation (S4), we know that $C_{ij} \leq C^{\widetilde{\pi}_{s'}}_{vv'}$. When $\tau' = s+1$ which implies $v' \in k_{s+1}$, if $v = v'$ we have by Theorem S1.2 that $C_{ij} \leq \left(1 - \frac{\epsilon_d}{\epsilon_u}\right)$, since

$$\rho^i_{(x_s^{k_s}, \dots, x_t^{k_t})}(A) \geq \left(\frac{\epsilon_d}{\epsilon_u}\right) \frac{\int \mathbb{1}_A(x_s^v) p_{s+1}^{v(\mathcal{R})}(k_{s+1}^{v(\mathcal{R})} \mid k_s, x_s^{v(\mathcal{R})}) \widetilde{\pi}_{\chi_s}^v(dx_s^v)}{\int p_{s+1}^{v(\mathcal{R})}(k_{s+1}^{v(\mathcal{R})} \mid k_s, x_s^{v(\mathcal{R})}) \widetilde{\pi}_{\chi_s}^v(dx_s^v)};$$

$C_{ij} = 0$ otherwise.

**Step 2.** Consider $\tau \in \{s+1, \dots, t-1\}$, which implies $v \in k_\tau$.

**Step 2.1.** When $v \notin k_{\tau-1}$ and $v \notin k_{\tau+1}$, we have

$$\rho^i_{(x_s^{k_s}, \dots, x_t^{k_t})}(A) = \frac{\int \mathbb{1}_A(x_\tau^v) f_\tau^v(x_\tau^v \mid k_\tau^{v(\mathcal{R})}) g_\tau^v(y_\tau^v \mid k_\tau^{v(\mathcal{R})}, x_\tau^v) \widetilde{f}_\tau^v(x_\tau^v, x_\tau^{N_\tau(v)})}{\int f_\tau^v(x_\tau^v \mid k_\tau^{v(\mathcal{R})}) g_\tau^v(y_\tau^v \mid k_\tau^{v(\mathcal{R})}, x_\tau^v) \widetilde{f}_\tau^v(x_\tau^v, x_\tau^{N_\tau(v)})} \begin{array}{c} \times p_{\tau+1}^{v(\mathcal{R})}(k_{\tau+1}^{v(\mathcal{R})} \mid k_\tau, x_\tau^{v(\mathcal{R})}) \psi^v(dx_\tau^v) \\ \times p_{\tau+1}^{v(\mathcal{R})}(k_{\tau+1}^{v(\mathcal{R})} \mid k_\tau, x_\tau^{v(\mathcal{R})}) \psi^v(dx_\tau^v) \end{array}.$$

Then when $\tau' = \tau - 1$, we have $C_{ij} = 0$. When $\tau' = \tau$, if $v' \in N_\tau(v) \cup v(\mathcal{R})$ we have by Theorem S1.2 that $C_{ij} \leq \left(1 - \frac{\epsilon'_d}{\epsilon'_u} \frac{\kappa_d}{\kappa_u}\right)$, since

$$\rho^i_{(x_s^{k_s}, \dots, x_t^{k_t})}(A) \geq \left(\frac{\epsilon'_d}{\epsilon'_u} \frac{\kappa_d}{\kappa_u}\right) \frac{\int \mathbb{1}_A(x_\tau^v) f_\tau^v(x_\tau^v \mid k_\tau^{v(\mathcal{R})}) g_\tau^v(y_\tau^v \mid k_\tau^{v(\mathcal{R})}, x_\tau^v) \psi^v(dx_\tau^v)}{\int f_\tau^v(x_\tau^v \mid k_\tau^{v(\mathcal{R})}) g_\tau^v(y_\tau^v \mid k_\tau^{v(\mathcal{R})}, x_\tau^v) \psi^v(dx_\tau^v)};$$

if $v' \notin N_\tau(v) \cup v(\mathcal{R})$ we have $C_{ij} = 0$. When $\tau' \neq \tau$ and $\tau' \neq \tau - 1$, we have $C_{ij} = 0$.

**Step 2.2.** When $v \in k_{\tau-1}$ and $v \notin k_{\tau+1}$, we have

$$\rho^i_{(x_s^{k_s}, \dots, x_t^{k_t})}(A) = \frac{\int \mathbb{1}_A(x_\tau^v) f_\tau^v(x_\tau^v \mid k_\tau^{v(\mathcal{R})}, x_{\tau-1}^v) g_\tau^v(y_\tau^v \mid k_\tau^{v(\mathcal{R})}, x_\tau^v) \widetilde{f}_\tau^v(x_\tau^v, x_\tau^{N_\tau(v)})}{\int \int f_\tau^v(x_\tau^v \mid k_\tau^{v(\mathcal{R})}, x_{\tau-1}^v) g_\tau^v(y_\tau^v \mid k_\tau^{v(\mathcal{R})}, x_\tau^v) \widetilde{f}_\tau^v(x_\tau^v, x_\tau^{N_\tau(v)})} \begin{array}{c} \times p_{\tau+1}^{v(\mathcal{R})}(k_{\tau+1}^{v(\mathcal{R})} \mid k_\tau, x_\tau^{v(\mathcal{R})}) \psi^v(dx_\tau^v) \\ \times p_{\tau+1}^{v(\mathcal{R})}(k_{\tau+1}^{v(\mathcal{R})} \mid k_\tau, x_\tau^{v(\mathcal{R})}) \psi^v(dx_\tau^v) \end{array}.$$

Then when $\tau' = \tau - 1$ which implies $v' \in k_{\tau-1}$, if $v' = v$ we have by Theorem S1.2 that $C_{ij} \leq \left(1 - \frac{\epsilon_d}{\epsilon_u}\right)$, since

$$\rho^i_{(x_s^{k_s}, \dots, x_t^{k_t})}(A)$$
$$\geq \left(\frac{\epsilon_d}{\epsilon_u}\right) \frac{\int \mathbb{1}_A(x_\tau^v) g_\tau^v(y_\tau^v \mid k_\tau^{v(\mathcal{R})}, x_\tau^v) \widetilde{f}_\tau^v(x_\tau^v, x_\tau^{N_\tau(v)}) p_{\tau+1}^{v(\mathcal{R})}(k_{\tau+1}^{v(\mathcal{R})} \mid k_\tau, x_\tau^{v(\mathcal{R})}) \psi^v(dx_\tau^v)}{\int g_\tau^v(y_\tau^v \mid k_\tau^{v(\mathcal{R})}, x_\tau^v) \widetilde{f}_\tau^v(x_\tau^v, x_\tau^{N_\tau(v)}) p_{\tau+1}^{v(\mathcal{R})}(k_{\tau+1}^{v(\mathcal{R})} \mid k_\tau, x_\tau^{v(\mathcal{R})}) \psi^v(dx_\tau^v)};$$

if $v' \neq v$ we have $C_{ij} = 0$. When $\tau' = \tau$, if $v' \in N_\tau(v) \cup v(\mathcal{R})$ we have by Theorem S1.2 that $C_{ij} \leq \left(1 - \frac{\epsilon'_d}{\epsilon'_u} \frac{\kappa_d}{\kappa_u}\right)$, since

$$\rho^i_{(x_s^{k_s}, \ldots, x_t^{k_t})}(A) \geq \left(\frac{\epsilon'_d}{\epsilon'_u} \frac{\kappa_d}{\kappa_u}\right) \frac{\int \mathbb{1}_A(x_\tau^v) f_\tau^v(x_\tau^v \mid k_\tau^{v(\mathcal{R})}, x_{\tau-1}^v) g_\tau^v(y_\tau^v \mid k_\tau^{v(\mathcal{R})}, x_\tau^v) \psi^v(dx_\tau^v)}{\int f_\tau^v(x_\tau^v \mid k_\tau^{v(\mathcal{R})}, x_{\tau-1}^v) g_\tau^v(y_\tau^v \mid k_\tau^{v(\mathcal{R})}, x_\tau^v) \psi^v(dx_\tau^v)};$$

if $v' \notin N_\tau(v) \cup v(\mathcal{R})$ we have $C_{ij} = 0$. When $\tau' \neq \tau$ and $\tau' \neq \tau - 1$, we have $C_{ij} = 0$.

**Step 2.3.** When $v \notin k_{\tau-1}$ and $v \in k_{\tau+1}$, we have

$$\rho^i_{(x_s^{k_s}, \ldots, x_t^{k_t})}(A) = \frac{\int \mathbb{1}_A(x_\tau^v) f_\tau^v(x_\tau^v \mid k_\tau^{v(\mathcal{R})}) g_\tau^v(y_\tau^v \mid k_\tau^{v(\mathcal{R})}, x_\tau^v) \widetilde{f}_\tau^v(x_\tau^v, x_\tau^{N_\tau(v)})}{\times f_{\tau+1}^v(x_{\tau+1}^v \mid k_{\tau+1}^{v(\mathcal{R})}, x_\tau^v) p_{\tau+1}^{v(\mathcal{R})}(k_{\tau+1}^{v(\mathcal{R})} \mid k_\tau, x_\tau^{v(\mathcal{R})}) \psi^v(dx_\tau^v)}{\int f_\tau^v(x_\tau^v \mid k_\tau^{v(\mathcal{R})}) g_\tau^v(y_\tau^v \mid k_\tau^{v(\mathcal{R})}, x_\tau^v) \widetilde{f}_\tau^v(x_\tau^v, x_\tau^{N_\tau(v)})} {\times f_{\tau+1}^v(x_{\tau+1}^v \mid k_{\tau+1}^{v(\mathcal{R})}, x_\tau^v) p_{\tau+1}^{v(\mathcal{R})}(k_{\tau+1}^{v(\mathcal{R})} \mid k_\tau, x_\tau^{v(\mathcal{R})}) \psi^v(dx_\tau^v)}.$$

Then when $\tau' = \tau - 1$ which implies $v' \in k_{\tau-1}$, we have $C_{ij} = 0$. When $\tau' = \tau$ which implies $v' \in k_\tau$, if $v' \in N_\tau(v) \cup v(\mathcal{R})$ we have by Theorem S1.2 that $C_{ij} \leq \left(1 - \frac{\epsilon'_d}{\epsilon'_u} \frac{\kappa_d}{\kappa_u}\right)$, since

$$\rho^i_{(x_s^{k_s}, \ldots, x_t^{k_t})}(A) \geq \left(\frac{\epsilon'_d}{\epsilon'_u} \frac{\kappa_d}{\kappa_u}\right) \frac{\int \mathbb{1}_A(x_\tau^v) f_\tau^v(x_\tau^v \mid k_\tau^{v(\mathcal{R})}) g_\tau^v(y_\tau^v \mid k_\tau^{v(\mathcal{R})}, x_\tau^v)}{\times f_{\tau+1}^v(x_{\tau+1}^v \mid k_{\tau+1}^{v(\mathcal{R})}, x_\tau^v) \psi^v(dx_\tau^v)}{\int f_\tau^v(x_\tau^v \mid k_\tau^{v(\mathcal{R})}) g_\tau^v(y_\tau^v \mid k_\tau^{v(\mathcal{R})}, x_\tau^v)} {\times f_{\tau+1}^v(x_{\tau+1}^v \mid k_{\tau+1}^{v(\mathcal{R})}, x_\tau^v) \psi^v(dx_\tau^v)};$$

if $v' \notin N_\tau(v) \cup v(\mathcal{R})$ we have $C_{ij} = 0$. When $\tau' = \tau + 1$ which implies $v' \in k_{\tau+1}$, if $v' = v$ we have by Theorem S1.2 that $C_{ij} \leq \left(1 - \frac{\epsilon_d}{\epsilon_u}\right)$, since

$$\rho^i_{(x_s^{k_s}, \ldots, x_t^{k_t})}(A) \geq \left(\frac{\epsilon_d}{\epsilon_u}\right) \frac{\int \mathbb{1}_A(x_\tau^v) f_\tau^v(x_\tau^v \mid k_\tau^{v(\mathcal{R})}) g_\tau^v(y_\tau^v \mid k_\tau^{v(\mathcal{R})}, x_\tau^v) \widetilde{f}_\tau^v(x_\tau^v, x_\tau^{N_\tau(v)})}{\times p_{\tau+1}^{v(\mathcal{R})}(k_{\tau+1}^{v(\mathcal{R})} \mid k_\tau, x_\tau^{v(\mathcal{R})}) \psi^v(dx_\tau^v)}{\int f_\tau^v(x_\tau^v \mid k_\tau^{v(\mathcal{R})}) g_\tau^v(y_\tau^v \mid k_\tau^{v(\mathcal{R})}, x_\tau^v) \widetilde{f}_\tau^v(x_\tau^v, x_\tau^{N_\tau(v)})} {\times p_{\tau+1}^{v(\mathcal{R})}(k_{\tau+1}^{v(\mathcal{R})} \mid k_\tau, x_\tau^{v(\mathcal{R})}) \psi^v(dx_\tau^v)};$$

if $v' \neq v$ we have $C_{ij} = 0$.

**Step 2.4.** When $v \in k_{\tau-1}$ and $v \in k_{\tau+1}$, we have

$$\rho^i_{(x_s^{k_s}, \ldots, x_t^{k_t})}(A) = \frac{\int \mathbb{1}_A(x_\tau^v) f_\tau^v(x_\tau^v \mid k_\tau^{v(\mathcal{R})}, x_{\tau-1}^v) g_\tau^v(y_\tau^v \mid k_\tau^{v(\mathcal{R})}, x_\tau^v) \widetilde{f}_\tau^v(x_\tau^v, x_\tau^{N_\tau(v)})}{\times f_{\tau+1}^v(x_{\tau+1}^v \mid k_{\tau+1}^{v(\mathcal{R})}, x_\tau^v) p_{\tau+1}^{v(\mathcal{R})}(k_{\tau+1}^{v(\mathcal{R})} \mid k_\tau, x_\tau^{v(\mathcal{R})}) \psi^v(dx_\tau^v)}{\int f_\tau^v(x_\tau^v \mid k_\tau^{v(\mathcal{R})}, x_{\tau-1}^v) g_\tau^v(y_\tau^v \mid k_\tau^{v(\mathcal{R})}, x_\tau^v) \widetilde{f}_\tau^v(x_\tau^v, x_\tau^{N_\tau(v)})} {\times f_{\tau+1}^v(x_{\tau+1}^v \mid k_{\tau+1}^{v(\mathcal{R})}, x_\tau^v) p_{\tau+1}^{v(\mathcal{R})}(k_{\tau+1}^{v(\mathcal{R})} \mid k_\tau, x_\tau^{v(\mathcal{R})}) \psi^v(dx_\tau^v)}.$$

Then if $\tau' = \tau - 1$ which implies $v' \in k_{\tau-1}$, when $v' = v$ we have by Theorem S1.2 that $C_{ij} \leq \left(1 - \frac{\epsilon_d}{\epsilon_u}\right)$, since

$$\rho^i_{(x_s^{k_s}, \ldots, x_t^{k_t})}(A) \geq \left(\frac{\epsilon_d}{\epsilon_u}\right) \frac{\int \mathbb{1}_A(x_\tau^v) g_\tau^v(y_\tau^v \mid k_\tau^{v(\mathcal{R})}, x_\tau^v) \widetilde{f}_\tau^v(x_\tau^v, x_\tau^{N_\tau(v)})}{\times f_{\tau+1}^v(x_{\tau+1}^v \mid k_{\tau+1}^{v(\mathcal{R})}, x_\tau^v) p_{\tau+1}^{v(\mathcal{R})}(k_{\tau+1}^{v(\mathcal{R})} \mid k_\tau, x_\tau^{v(\mathcal{R})}) \psi^v(dx_\tau^v)}{\int g_\tau^v(y_\tau^v \mid k_\tau^{v(\mathcal{R})}, x_\tau^v) \widetilde{f}_\tau^v(x_\tau^v, x_\tau^{N_\tau(v)})} {\times f_{\tau+1}^v(x_{\tau+1}^v \mid k_{\tau+1}^{v(\mathcal{R})}, x_\tau^v) p_{\tau+1}^{v(\mathcal{R})}(k_{\tau+1}^{v(\mathcal{R})} \mid k_\tau, x_\tau^{v(\mathcal{R})}) \psi^v(dx_\tau^v)};$$

$C_{ij} = 0$ otherwise. If $\tau' = \tau$ which implies $v' \in k_\tau$, when $v' \in N_\tau(v) \cup v(\mathcal{R})$ we have by Theorem S1.2 that $C_{ij} \leq \left(1 - \frac{\epsilon'_d}{\epsilon'_u} \frac{\kappa_d}{\kappa_u}\right)$, since

$$\rho^i_{(x_s^{k_s}, \ldots, x_t^{k_t})}(A) \geq \left(\frac{\epsilon'_d}{\epsilon'_u} \frac{\kappa_d}{\kappa_u}\right) \frac{\begin{aligned}\int \mathbb{1}_A(x_\tau^v) f_\tau^v(x_\tau^v \mid k_\tau^{v(\mathcal{R})}, x_{\tau-1}^v) g_\tau^v(y_\tau^v \mid k_\tau^{v(\mathcal{R})}, x_\tau^v) \\ \times f_{\tau+1}^v(x_{\tau+1}^v \mid k_{\tau+1}^{v(\mathcal{R})}, x_\tau^v) \psi^v(dx_\tau^v)\end{aligned}}{\begin{aligned}\int f_\tau^v(x_\tau^v \mid k_\tau^{v(\mathcal{R})}, x_{\tau-1}^v) g_\tau^v(y_\tau^v \mid k_\tau^{v(\mathcal{R})}, x_\tau^v) \\ \times f_{\tau+1}^v(x_{\tau+1}^v \mid k_{\tau+1}^{v(\mathcal{R})}, x_\tau^v) \psi^v(dx_\tau^v)\end{aligned}};$$

$C_{ij} = 0$ otherwise. When $\tau' = \tau + 1$ which implies $v' \in k_{\tau+1}$, if $v' = v$ we have by Theorem S1.2 that $C_{ij} \leq \left(1 - \frac{\epsilon_d}{\epsilon_u}\right)$, since

$$\rho^i_{(x_s^{k_s}, \ldots, x_t^{k_t})}(A) \geq \left(\frac{\epsilon_d}{\epsilon_u}\right) \frac{\begin{aligned}\int \mathbb{1}_A(x_\tau^v) f_\tau^v(x_\tau^v \mid k_\tau^{v(\mathcal{R})}, x_{\tau-1}^v) g_\tau^v(y_\tau^v \mid k_\tau^{v(\mathcal{R})}, x_\tau^v) \widetilde{f}_\tau^v(x_\tau^v, x_\tau^{N_\tau(v)}) \\ \times p_{\tau+1}^{v(\mathcal{R})}(k_{\tau+1}^{v(\mathcal{R})} \mid k_\tau, x_\tau^{v(\mathcal{R})}) \psi^v(dx_\tau^v)\end{aligned}}{\begin{aligned}\int f_\tau^v(x_\tau^v \mid k_\tau^{v(\mathcal{R})}, x_{\tau-1}^v) g_\tau^v(y_\tau^v \mid k_\tau^{v(\mathcal{R})}, x_\tau^v) \widetilde{f}_\tau^v(x_\tau^v, x_\tau^{N_\tau(v)}) \\ \times p_{\tau+1}^{v(\mathcal{R})}(k_{\tau+1}^{v(\mathcal{R})} \mid k_\tau, x_\tau^{v(\mathcal{R})}) \psi^v(dx_\tau^v)\end{aligned}};$$

if $v' \neq v$ we have $C_{ij} = 0$.

**Step 3.** When $\tau = t$, which implies $v \in k_t$.

**Step 3.1.** When $v \in k_{t-1}$ we have

$$\rho^i_{(x_s^{k_s}, \ldots, x_t^{k_t})}(A) = \frac{\int \mathbb{1}_A(x_t^v) f_t^v(x_t^v \mid k_t^{v(\mathcal{R})}, x_{t-1}^v) g_t^v(y_t^v \mid k_t^{v(\mathcal{R})}, x_t^v) \widetilde{f}_t^v(x_t^v, x_t^{N_t(v)}) \psi^v(dx_t^v)}{\int f_t^v(x_t^v \mid k_t^{v(\mathcal{R})}, x_{t-1}^v) g_t^v(y_t^v \mid k_t^{v(\mathcal{R})}, x_t^v) \widetilde{f}_t^v(x_t^v, x_t^{N_t(v)}) \psi^v(dx_t^v)}.$$

Then if $\tau' = t-1$ which implies $v' \in k_{t-1}$, when $v' = v$ we have by Theorem S1.2 that $C_{ij} \leq \left(1 - \frac{\epsilon_d}{\epsilon_u}\right)$, since

$$\rho^i_{(x_s^{k_s}, \ldots, x_t^{k_t})}(A) \geq \left(\frac{\epsilon_d}{\epsilon_u}\right) \frac{\int \mathbb{1}_A(x_t^v) g_t^v(y_t^v \mid k_t^{v(\mathcal{R})}, x_t^v) \widetilde{f}_t^v(x_t^v, x_t^{N_t(v)}) \psi^v(dx_t^v)}{\int g_t^v(y_t^v \mid k_t^{v(\mathcal{R})}, x_t^v) \widetilde{f}_t^v(x_t^v, x_t^{N_t(v)}) \psi^v(dx_t^v)};$$

$C_{ij} = 0$ otherwise. If $\tau' = t$ which implies $v' \in k_t$, when $v' \in N_t(v)$ we have by Theorem S1.2 that $C_{ij} \leq \left(1 - \frac{\epsilon'_d}{\epsilon'_u}\right)$ by Theorem S1.2, since

$$\rho^i_{(x_s^{k_s}, \ldots, x_t^{k_t})}(A) \geq \left(\frac{\epsilon'_d}{\epsilon'_u}\right) \frac{\int \mathbb{1}_A(x_t^v) f_t^v(x_t^v \mid k_t^{v(\mathcal{R})}, x_{t-1}^v) g_t^v(y_t^v \mid k_t^{v(\mathcal{R})}, x_t^v) \psi^v(dx_t^v)}{\int f_t^v(x_t^v \mid k_t^{v(\mathcal{R})}, x_{t-1}^v) g_t^v(y_t^v \mid k_t^{v(\mathcal{R})}, x_t^v) \psi^v(dx_t^v)};$$

$C_{ij} = 0$ otherwise.

**Step 3.2.** When $v \notin k_{t-1}$ we have

$$\rho^i_{(x_s^{k_s}, \ldots, x_t^{k_t})}(A) = \frac{\int \mathbb{1}_A(x_t^v) f_t^v(x_t^v \mid k_t^{v(\mathcal{R})}) g_t^v(y_t^v \mid k_t^{v(\mathcal{R})}, x_t^v) \widetilde{f}_t^v(x_t^v, x_t^{N_t(v)}) \psi^v(dx_t^v)}{\int f_t^v(x_t^v \mid k_t^{v(\mathcal{R})}) g_t^v(y_t^v \mid k_t^{v(\mathcal{R})}, x_t^v) \widetilde{f}_t^v(x_t^v, x_t^{N_t(v)}) \psi^v(dx_t^v)}.$$

Then if $\tau' = t - 1$ we have $C_{ij} = 0$. If $\tau' = t$ which implies $v' \in k_t$, when $v' \in N_t(v)$ we have by Theorem S1.2 that $C_{ij} \leq \left(1 - \frac{\epsilon'_d}{\epsilon'_u}\right)$, since

$$\rho^i_{(x_s^{k_s}, \ldots, x_t^{k_t})}(A) \geq \left(\frac{\epsilon'_d}{\epsilon'_u}\right) \frac{\int \mathbb{1}_A(x_t^v) f_t^v(x_t^v \mid k_t^{v(\mathcal{R})}) g_t^v(y_t^v \mid k_t^{v(\mathcal{R})}, x_t^v) \psi^v(dx_t^v)}{\int f_t^v(x_t^v \mid k_t^{v(\mathcal{R})}) g_t^v(y_t^v \mid k_t^{v(\mathcal{R})}, x_t^v) \psi^v(dx_t^v)};$$

$C_{ij} = 0$ otherwise.

**Step 4.** In this step, we summary the results of $C_{ij}$ obtained in the previous three steps and aim to bound the following quantity:

$$\max_{(\tau, v) \in I} \sum_{(\tau', v') \in I} e^{\beta_T |\tau - \tau'|} e^{\beta_T d(v, v')} C_{(\tau, v)(\tau', v')}$$

$$= \max \Bigg\{ \max_{(s, v) \in I} \sum_{(\tau', v') \in I} e^{\beta_T |s - \tau'|} e^{\beta_T d(v, v')} C_{(s, v)(\tau', v')},$$

$$\max_{\substack{(\tau, v) \in I \\ \tau \in \{s+1, \ldots, t-1\}}} \sum_{(\tau', v') \in I} e^{\beta_T |\tau - \tau'|} e^{\beta_T d(v, v')} C_{(\tau, v)(\tau', v')} \tag{S23}$$

$$\max_{(t, v) \in I} \sum_{(\tau', v') \in I} e^{\beta_T |t - \tau'|} e^{\beta_T d(v, v')} C_{(t, v)(\tau', v')} \Bigg\},$$

where each of the above three items will be analyzed in the subsequent substeps.

**Step 4.1.** We handled the first item of equation (S23) in Step 1 and showed that

$$\max_{(s, v) \in I} \sum_{(\tau', v') \in I} e^{\beta_T |s - \tau'|} e^{\beta_T d(v, v')} C_{(s, v)(\tau', v')}$$

$$= \max \Bigg\{ \max_{(s, v) \in I} \sum_{(\tau', v') \in I} e^{\beta_T |s - \tau'|} e^{\beta_T d(v, v')} C_{(s, v)(\tau', v')} \mathbb{1}_{\{v \in k_s, \, v \notin k_{s+1}\}},$$

$$\max_{(s, v) \in I} \sum_{(\tau', v') \in I} e^{\beta_T |s - \tau'|} e^{\beta_T d(v, v')} C_{(s, v)(\tau', v')} \mathbb{1}_{\{v \in k_s, \, v \in k_{s+1}\}} \Bigg\}.$$

Specifically, in Step 1, we obtained that

$$\max_{(s, v) \in I} \sum_{(\tau', v') \in I} e^{\beta_T |s - \tau'|} e^{\beta_T d(v, v')} C_{(s, v)(\tau', v')} \mathbb{1}_{\{v \in k_s, \, v \notin k_{s+1}\}}$$

$$= \max_{(s, v) \in I} \sum_{(s, v') \in I} e^{\beta_T |s - s|} e^{\beta_T d(v, v')} C_{(s, v)(s, v')} \mathbb{1}_{\{v \in k_s, \, v \notin k_{s+1}, \, v' \in k_s\}}$$

$$+ \max_{(s, v) \in I} \sum_{\substack{(\tau', v') \in I \\ \tau' \in \{s+1, \ldots, t\}}} e^{\beta_T |s - \tau'|} e^{\beta_T d(v, v')} C_{(s, v)(\tau', v')} \mathbb{1}_{\{v \in k_s, \, v \notin k_{s+1}, \, v' \in k_{\tau'}\}}$$

$$= \max_{(s, v) \in I} \sum_{(s, v') \in I} e^{\beta_T |s - s|} e^{\beta_T d(v, v')} C_{(s, v)(s, v')} \mathbb{1}_{\{v \in k_s, \, v \notin k_{s+1}, \, v' \in k_s, \, v' \in v(\mathcal{R})\}}$$

$$+ \max_{(s, v) \in I} \sum_{(s, v') \in I} e^{\beta_T |s - s|} e^{\beta_T d(v, v')} C_{(s, v)(s, v')} \mathbb{1}_{\{v \in k_s, \, v \notin k_{s+1}, \, v' \in k_s, \, v' \notin v(\mathcal{R})\}} + 0$$

$$= \max_{(s,v) \in I} \sum_{(s,v') \in I} e^{\beta_T |s-s|} e^{\beta_T d(v,v')} C_{(s,v)(s,v')} \mathbb{1}_{\{v \in k_s,\ v \notin k_{s+1},\ v' \in k_s,\ v' \in v(\mathcal{R})\}} + 0 + 0$$

$$\leq \max_{(s,v) \in I} \sum_{(s,v') \in I} e^{\beta_T d(v,v')} \left(1 - \frac{\kappa_d}{\kappa_u}\right) \mathbb{1}_{\{v \in k_s,\ v \notin k_{s+1},\ v' \in k_s,\ v' \in v(\mathcal{R})\}}$$

$$\leq e^{\beta_T r_T^{\mathcal{R}}} \left(1 - \frac{\kappa_d}{\kappa_u}\right) \Delta_T^{\mathcal{R}},$$

and

$$\max_{(s,v) \in I} \sum_{(\tau',v') \in I} e^{\beta_T |s-\tau'|} e^{\beta_T d(v,v')} C_{(s,v)(\tau',v')} \mathbb{1}_{\{v \in k_s,\ v \in k_{s+1}\}}$$

$$= \max_{(s,v) \in I} \sum_{(s,v') \in I} e^{\beta_T |s-s|} e^{\beta_T d(v,v')} C_{(s,v)(s,v')} \mathbb{1}_{\{v \in k_s,\ v \in k_{s+1},\ v' \in k_s\}}$$

$$\quad + \max_{(s,v) \in I} \sum_{(s+1,v') \in I} e^{\beta_T |s-(s+1)|} e^{\beta_T d(v,v')} C_{(s,v)(s+1,v')} \mathbb{1}_{\{v \in k_s,\ v \in k_{s+1},\ v' \in k_{s+1}\}}$$

$$\quad + \max_{(s,v) \in I} \sum_{\substack{(\tau',v') \in I \\ \tau' \in \{s+2,\dots,t\}}} e^{\beta_T |s-\tau'|} e^{\beta_T d(v,v')} C_{(s,v)(\tau',v')} \mathbb{1}_{\{v \in k_s,\ v \in k_{s+1},\ v' \in k_{\tau'}\}}$$

$$= \max_{(s,v) \in I} \sum_{(s,v') \in I} e^{\beta_T |s-s|} e^{\beta_T d(v,v')} C_{(s,v)(s,v')} \mathbb{1}_{\{v \in k_s,\ v \in k_{s+1},\ v' \in k_s\}}$$

$$\quad + \max_{(s,v) \in I} \sum_{(s+1,v') \in I} e^{\beta_T |s-(s+1)|} e^{\beta_T d(v,v')} C_{(s,v)(s+1,v')} \mathbb{1}_{\{v \in k_s,\ v \in k_{s+1},\ v' \in k_{s+1}\}} + 0$$

$$\leq \max_{(s,v) \in I} \sum_{(s,v') \in I} e^{\beta_T d(v,v')} C_{vv'}^{\widetilde{\pi}_s} \mathbb{1}_{\{v \in k_s,\ v' \in k_s\}}$$

$$\quad + \max_{(s,v) \in I} \sum_{(s+1,v') \in I} e^{\beta_T} e^{\beta_T d(v,v')} \left(1 - \frac{\epsilon_d}{\epsilon_u}\right) \mathbb{1}_{\{v \in k_s,\ v \in k_{s+1},\ v' \in k_{s+1},\ v'=v\}} + 0$$

$$\leq \max_{v \in k_s} \sum_{v' \in k_s} e^{\beta_T d(v,v')} C_{vv'}^{\widetilde{\pi}_s} + e^{\beta_T} \left(1 - \frac{\epsilon_d}{\epsilon_u}\right).$$

Hence, we have

$$\max_{(s,v) \in I} \sum_{(\tau',v') \in I} e^{\beta_T |s-\tau'|} e^{\beta_T d(v,v')} C_{(s,v)(\tau',v')}$$

$$\leq \max \left\{ e^{\beta_T r_T^{\mathcal{R}}} \left(1 - \frac{\kappa_d}{\kappa_u}\right) \Delta_T^{\mathcal{R}},\ \max_{v \in k_s} \sum_{v' \in k_s} e^{\beta_T d(v,v')} C_{vv'}^{\widetilde{\pi}_s} + e^{\beta_T} \left(1 - \frac{\epsilon_d}{\epsilon_u}\right) \right\}.$$

By the definition of $\beta_T$ given in equation (32), we have

$$e^{\beta_T r_T^{\mathcal{R}}} \left(1 - \frac{\kappa_d}{\kappa_u}\right) \Delta_T^{\mathcal{R}} \leq \frac{1}{6} \quad \text{and} \quad e^{\beta_T} \left(1 - \frac{\epsilon_d}{\epsilon_u}\right) \leq \frac{1}{6}.$$

Thus by Proposition S1.1, we have

$$\max_{(s,v) \in I} \sum_{(\tau',v') \in I} e^{\beta_T |s-\tau'|} e^{\beta_T d(v,v')} C_{(s,v)(\tau',v')} \leq \frac{1}{3} + \frac{1}{6} = \frac{1}{2}. \tag{S24}$$

**Step 4.2.** We handled the second item of equation [S23] in Step 2 and showed that

$$\max_{\tau \in \{s+1,\dots,t-1\}} \sum_{(\tau',v') \in I} e^{\beta_T |\tau - \tau'|} e^{\beta_T d(v,v')} C_{(\tau,v)(\tau',v')}$$

$$= \max \left\{ \max_{\substack{(\tau,v) \in I \\ \tau \in \{s+1,\dots,t-1\}}} \sum_{(\tau',v') \in I} e^{\beta_T |\tau - \tau'|} e^{\beta_T d(v,v')} C_{(\tau,v)(\tau',v')} \mathbb{1}_{\{v \in k_\tau,\ v' \in k_{\tau'},\ v \notin k_{\tau-1},\ v \notin k_{\tau+1}\}}, \right.$$

$$\max_{\substack{(\tau,v) \in I \\ \tau \in \{s+1,\dots,t-1\}}} \sum_{(\tau',v') \in I} e^{\beta_T |\tau - \tau'|} e^{\beta_T d(v,v')} C_{(\tau,v)(\tau',v')} \mathbb{1}_{\{v \in k_\tau,\ v' \in k_{\tau'},\ v \in k_{\tau-1},\ v \notin k_{\tau+1}\}},$$

$$\max_{\substack{(\tau,v) \in I \\ \tau \in \{s+1,\dots,t-1\}}} \sum_{(\tau',v') \in I} e^{\beta_T |\tau - \tau'|} e^{\beta_T d(v,v')} C_{(\tau,v)(\tau',v')} \mathbb{1}_{\{v \in k_\tau,\ v' \in k_{\tau'},\ v \notin k_{\tau-1},\ v \in k_{\tau+1}\}},$$

$$\left. \max_{\substack{(\tau,v) \in I \\ \tau \in \{s+1,\dots,t-1\}}} \sum_{(\tau',v') \in I} e^{\beta_T |\tau - \tau'|} e^{\beta_T d(v,v')} C_{(\tau,v)(\tau',v')} \mathbb{1}_{\{v \in k_\tau,\ v' \in k_{\tau'},\ v \in k_{\tau-1},\ v \in k_{\tau+1}\}} \right\}.$$

Specifically, in Step 2, we obtained that

$$\max_{\substack{(\tau,v) \in I \\ \tau \in \{s+1,\dots,t-1\}}} \sum_{(\tau',v') \in I} e^{\beta_T |\tau - \tau'|} e^{\beta_T d(v,v')} C_{(\tau,v)(\tau',v')} \mathbb{1}_{\{v \in k_\tau,\ v' \in k_{\tau'},\ v \notin k_{\tau-1},\ v \notin k_{\tau+1}\}}$$

$$= \max_{\substack{(\tau,v) \in I \\ \tau \in \{s+1,\dots,t-1\}}} \sum_{(\tau-1,v') \in I} e^{\beta_T |\tau - (\tau-1)|} e^{\beta_T d(v,v')} C_{(\tau,v)(\tau-1,v')} \mathbb{1}_{\{v \in k_\tau,\ v' \in k_{\tau-1},\ v \notin k_{\tau-1},\ v \notin k_{\tau+1}\}}$$

$$+ \max_{\substack{(\tau,v) \in I \\ \tau \in \{s+1,\dots,t-1\}}} \sum_{(\tau,v') \in I} e^{\beta_T |\tau - \tau|} e^{\beta_T d(v,v')} C_{(\tau,v)(\tau,v')} \mathbb{1}_{\{v \in k_\tau,\ v' \in k_\tau,\ v \notin k_{\tau-1},\ v \notin k_{\tau+1}\}}$$

$$+ \max_{\substack{(\tau,v) \in I \\ \tau \in \{s+1,\dots,t-1\}}} \sum_{\substack{(\tau',v') \in I \\ \tau' \neq \tau-1,\tau}} e^{\beta_T |\tau - \tau'|} e^{\beta_T d(v,v')} C_{(\tau,v)(\tau',v')} \mathbb{1}_{\{v \in k_\tau,\ v' \in k_{\tau'},\ v \notin k_{\tau-1},\ v \notin k_{\tau+1}\}}$$

$$= 0 + \max_{\substack{(\tau,v) \in I \\ \tau \in \{s+1,\dots,t-1\}}} \sum_{(\tau,v') \in I} e^{\beta_T |\tau - \tau|} e^{\beta_T d(v,v')} C_{(\tau,v)(\tau,v')} \mathbb{1}_{\{v \in k_\tau,\ v' \in k_\tau,\ v \notin k_{\tau-1},\ v \notin k_{\tau+1}\}} + 0$$

$$\leq \max_{\substack{(\tau,v) \in I \\ \tau \in \{s+1,\dots,t-1\}}} \sum_{(\tau,v') \in I} e^{\beta_T d(v,v')} \left(1 - \frac{\epsilon_d'}{\epsilon_u'} \frac{\kappa_d}{\kappa_u}\right) \mathbb{1}_{\{v \in k_\tau,\ v' \in k_\tau,\ v \notin k_{\tau-1},\ v \notin k_{\tau+1}\ v' \in N_\tau(v) \cup v(\mathcal{R})\}}$$

$$\leq e^{\beta_T (r + r_T^{\mathcal{R}})} \left(1 - \frac{\epsilon_d'}{\epsilon_u'} \frac{\kappa_d}{\kappa_u}\right) (\Delta_T + \Delta_T^{\mathcal{R}})$$

and

$$\max_{\substack{(\tau,v) \in I \\ \tau \in \{s+1,\dots,t-1\}}} \sum_{(\tau',v') \in I} e^{\beta_T |\tau - \tau'|} e^{\beta_T d(v,v')} C_{(\tau,v)(\tau',v')} \mathbb{1}_{\{v \in k_\tau,\ v' \in k_{\tau'},\ v \in k_{\tau-1},\ v \notin k_{\tau+1}\}}$$

$$= \max_{\substack{(\tau,v) \in I \\ \tau \in \{s+1,\dots,t-1\}}} \sum_{(\tau-1,v') \in I} e^{\beta_T |\tau - (\tau-1)|} e^{\beta_T d(v,v')} C_{(\tau,v)(\tau-1,v')} \mathbb{1}_{\{v \in k_\tau,\ v \in k_{\tau-1},\ v \notin k_{\tau+1},\ v' \in k_{\tau-1}\}}$$

$$+ \max_{\substack{(\tau,v) \in I \\ \tau \in \{s+1,\dots,t-1\}}} \sum_{(\tau,v') \in I} e^{\beta_T |\tau - \tau|} e^{\beta_T d(v,v')} C_{(\tau,v)(\tau,v')} \mathbb{1}_{\{v \in k_\tau,\ v \in k_{\tau-1},\ v \notin k_{\tau+1},\ v' \in k_\tau\}}$$

$$+ \max_{\substack{(\tau,v) \in I \\ \tau \in \{s+1,\dots,t-1\}}} \sum_{\substack{(\tau',v') \in I \\ \tau' \neq \tau-1,\tau}} e^{\beta_T |\tau - \tau'|} e^{\beta_T d(v,v')} C_{(\tau,v)(\tau',v')} \mathbb{1}_{\{v \in k_\tau,\ v \in k_{\tau-1},\ v \notin k_{\tau+1},\ v' \in k_{\tau'}\}}$$

$$= \max_{\substack{(\tau,v) \in I \\ \tau \in \{s+1,\dots,t-1\}}} \sum_{(\tau-1,v') \in I} e^{\beta_T |\tau - (\tau-1)|} e^{\beta_T d(v,v')} C_{(\tau,v)(\tau-1,v')} \mathbb{1}_{\{v \in k_\tau,\ v \in k_{\tau-1},\ v \notin k_{\tau+1},\ v' \in k_{\tau-1}\}}$$

$$+ \max_{\tau \in \{s+1,\ldots,t-1\}} \sum_{(\tau',v') \in I} e^{\beta_T |\tau-\tau|} e^{\beta_T d(v,v')} C_{(\tau,v)(\tau,v')} \mathbb{1}_{\{v \in k_\tau,\, v \in k_{\tau-1},\, v \notin k_{\tau+1},\, v' \in k_\tau\}} + 0$$

$$\leq \max_{\tau \in \{s+1,\ldots,t-1\}} \sum_{(\tau-1,v') \in I} e^{\beta_T} e^{\beta_T d(v,v')} \left(1 - \frac{\epsilon_d}{\epsilon_u}\right) \mathbb{1}_{\{v \in k_\tau,\, v \in k_{\tau-1},\, v \notin k_{\tau+1},\, v' \in k_{\tau-1},\, v'=v\}}$$

$$+ \max_{\tau \in \{s+1,\ldots,t-1\}} \sum_{(\tau,v') \in I} e^{\beta_T d(v,v')} \left(1 - \frac{\epsilon_d'}{\epsilon_u'} \frac{\kappa_d}{\kappa_u}\right) \mathbb{1}_{\{v \in k_\tau,\, v \in k_{\tau-1},\, v \notin k_{\tau+1},\, v' \in k_\tau,\, v' \in N_\tau(v) \cup v(\mathcal{R})\}}$$

$$\leq e^{\beta_T} \left(1 - \frac{\epsilon_d}{\epsilon_u}\right) + e^{\beta_T(r+r_T^{\mathcal{R}})} \left(1 - \frac{\epsilon_d'}{\epsilon_u'} \frac{\kappa_d}{\kappa_u}\right) (\Delta_T + \Delta_T^{\mathcal{R}})$$

and

$$\max_{\tau \in \{s+1,\ldots,t-1\}} \sum_{(\tau',v') \in I} e^{\beta_T |\tau-\tau'|} e^{\beta_T d(v,v')} C_{(\tau,v)(\tau',v')} \mathbb{1}_{\{v \in k_\tau,\, v' \in k_{\tau'},\, v \notin k_{\tau-1},\, v \in k_{\tau+1}\}}$$

$$= \max_{\tau \in \{s+1,\ldots,t-1\}} \sum_{(\tau-1,v') \in I} e^{\beta_T |\tau-(\tau-1)|} e^{\beta_T d(v,v')} C_{(\tau,v)(\tau-1,v')} \mathbb{1}_{\{v \in k_\tau,\, v \notin k_{\tau-1},\, v \in k_{\tau+1},\, v' \in k_{\tau-1}\}}$$

$$+ \max_{\tau \in \{s+1,\ldots,t-1\}} \sum_{(\tau,v') \in I} e^{\beta_T |\tau-\tau|} e^{\beta_T d(v,v')} C_{(\tau,v)(\tau,v')} \mathbb{1}_{\{v \in k_\tau,\, v \notin k_{\tau-1},\, v \in k_{\tau+1},\, v' \in k_\tau\}}$$

$$+ \max_{\tau \in \{s+1,\ldots,t-1\}} \sum_{(\tau+1,v') \in I} e^{\beta_T |\tau-(\tau+1)|} e^{\beta_T d(v,v')} C_{(\tau,v)(\tau+1,v')} \mathbb{1}_{\{v \in k_\tau,\, v \notin k_{\tau-1},\, v \in k_{\tau+1},\, v' \in k_{\tau+1}\}}$$

$$+ \max_{\tau \in \{s+1,\ldots,t-1\}} \sum_{\substack{(\tau',v') \in I \\ \tau' \neq \tau-1,\tau,\tau+1}} e^{\beta_T |\tau-\tau'|} e^{\beta_T d(v,v')} C_{(\tau,v)(\tau',v')} \mathbb{1}_{\{v \in k_\tau,\, v \notin k_{\tau-1},\, v \in k_{\tau+1},\, v' \in k_{\tau'}\}}$$

$$= 0 + \max_{\tau \in \{s+1,\ldots,t-1\}} \sum_{(\tau,v') \in I} e^{\beta_T |\tau-\tau|} e^{\beta_T d(v,v')} C_{(\tau,v)(\tau,v')} \mathbb{1}_{\{v \in k_\tau,\, v \notin k_{\tau-1},\, v \in k_{\tau+1},\, v' \in k_\tau\}}$$

$$+ \max_{\tau \in \{s+1,\ldots,t-1\}} \sum_{(\tau+1,v') \in I} e^{\beta_T |\tau-(\tau+1)|} e^{\beta_T d(v,v')} C_{(\tau,v)(\tau+1,v')} \mathbb{1}_{\{v \in k_\tau,\, v \notin k_{\tau-1},\, v \in k_{\tau+1},\, v' \in k_{\tau+1}\}} + 0$$

$$\leq \max_{\tau \in \{s+1,\ldots,t-1\}} \sum_{(\tau,v') \in I} e^{\beta_T d(v,v')} \left(1 - \frac{\epsilon_d'}{\epsilon_u'} \frac{\kappa_d}{\kappa_u}\right) \mathbb{1}_{\{v \in k_\tau,\, v \notin k_{\tau-1},\, v \in k_{\tau+1},\, v' \in k_\tau,\, v' \in N_\tau(v) \cup v(\mathcal{R})\}}$$

$$+ \max_{\tau \in \{s+1,\ldots,t-1\}} \sum_{(\tau+1,v') \in I} e^{\beta_T} e^{\beta_T d(v,v')} \left(1 - \frac{\epsilon_d}{\epsilon_u}\right) \mathbb{1}_{\{v \in k_\tau,\, v \notin k_{\tau-1},\, v \in k_{\tau+1},\, v' \in k_{\tau+1},\, v'=v\}}$$

$$\leq e^{\beta_T(r+r_T^{\mathcal{R}})} \left(1 - \frac{\epsilon_d'}{\epsilon_u'} \frac{\kappa_d}{\kappa_u}\right) (\Delta_T + \Delta_T^{\mathcal{R}}) + e^{\beta_T} \left(1 - \frac{\epsilon_d}{\epsilon_u}\right)$$

and

$$\max_{\tau \in \{s+1,\ldots,t-1\}} \sum_{(\tau',v') \in I} e^{\beta_T |\tau-\tau'|} e^{\beta_T d(v,v')} C_{(\tau,v)(\tau',v')} \mathbb{1}_{\{v \in k_\tau,\, v' \in k_{\tau'},\, v \in k_{\tau-1},\, v \in k_{\tau+1}\}}$$

$$= \max_{\tau \in \{s+1,\ldots,t-1\}} \sum_{(\tau-1,v') \in I} e^{\beta_T |\tau-(\tau-1)|} e^{\beta_T d(v,v')} C_{(\tau,v)(\tau-1,v')} \mathbb{1}_{\{v \in k_\tau,\, v \in k_{\tau-1},\, v \in k_{\tau+1},\, v' \in k_{\tau-1}\}}$$

$$+ \max_{\tau \in \{s+1,\ldots,t-1\}} \sum_{(\tau,v') \in I} e^{\beta_T |\tau-\tau|} e^{\beta_T d(v,v')} C_{(\tau,v)(\tau,v')} \mathbb{1}_{\{v \in k_\tau,\, v \in k_{\tau-1},\, v \in k_{\tau+1},\, v' \in k_\tau\}}$$

$$+ \max_{\tau \in \{s+1,\ldots,t-1\}} \sum_{(\tau+1,v') \in I} e^{\beta_T |\tau-(\tau+1)|} e^{\beta_T d(v,v')} C_{(\tau,v)(\tau+1,v')} \mathbb{1}_{\{v \in k_\tau,\, v \in k_{\tau-1},\, v \in k_{\tau+1},\, v' \in k_{\tau+1}\}}$$

$$+ \max_{\substack{(\tau,v)\in I \\ \tau\in\{s+1,\dots,t-1\}}} \sum_{\substack{(\tau',v')\in I \\ \tau'\neq\tau-1,\tau,\tau+1}} e^{\beta_T|\tau-\tau'|} e^{\beta_T d(v,v')} C_{(\tau,v)(\tau',v')} \mathbb{1}_{\{v\in k_\tau,\, v\in k_{\tau-1},\, v\in k_{\tau+1},\, v'\in k_{\tau'}\}}$$

$$\leq \max_{\substack{(\tau,v)\in I \\ \tau\in\{s+1,\dots,t-1\}}} \sum_{(\tau-1,v')\in I} e^{\beta_T} e^{\beta_T d(v,v')} \left(1-\frac{\epsilon_d}{\epsilon_u}\right) \mathbb{1}_{\{v\in k_\tau,\, v\in k_{\tau-1},\, v\in k_{\tau+1},\, v'\in k_{\tau-1},\, v'=v\}}$$

$$+ \max_{\substack{(\tau,v)\in I \\ \tau\in\{s+1,\dots,t-1\}}} \sum_{(\tau,v')\in I} e^{\beta_T d(v,v')} \left(1-\frac{\epsilon'_d}{\epsilon'_u}\frac{\kappa_d}{\kappa_u}\right) \mathbb{1}_{\{v\in k_\tau,\, v\in k_{\tau-1},\, v\in k_{\tau+1},\, v'\in k_\tau,\, v'\in N_\tau(v)\cup v(\mathcal{R})\}}$$

$$+ \max_{\substack{(\tau,v)\in I \\ \tau\in\{s+1,\dots,t-1\}}} \sum_{(\tau+1,v')\in I} e^{\beta_T} e^{\beta_T d(v,v')} \left(1-\frac{\epsilon_d}{\epsilon_u}\right) \mathbb{1}_{\{v\in k_\tau,\, v\in k_{\tau-1},\, v\in k_{\tau+1},\, v'\in k_{\tau+1},\, v'=v\}}$$

$$\leq e^{\beta_T(r+r_T^{\mathcal{R}})} \left(1-\frac{\epsilon'_d}{\epsilon'_u}\frac{\kappa_d}{\kappa_u}\right)(\Delta_T+\Delta_T^{\mathcal{R}}) + 2e^{\beta_T}\left(1-\frac{\epsilon_d}{\epsilon_u}\right).$$

Then we have

$$\max_{\substack{(\tau,v)\in I \\ \tau\in\{s+1,\dots,t-1\}}} \sum_{(\tau',v')\in I} e^{\beta_T|\tau-\tau'|} e^{\beta_T d(v,v')} C_{(\tau,v)(\tau',v')}$$

$$= \max\left\{ e^{\beta_T(r+r_T^{\mathcal{R}})} \left(1-\frac{\epsilon'_d}{\epsilon'_u}\frac{\kappa_d}{\kappa_u}\right)(\Delta_T+\Delta_T^{\mathcal{R}}), \right.$$

$$e^{\beta_T}\left(1-\frac{\epsilon_d}{\epsilon_u}\right) + e^{\beta_T(r+r_T^{\mathcal{R}})} \left(1-\frac{\epsilon'_d}{\epsilon'_u}\frac{\kappa_d}{\kappa_u}\right)(\Delta_T+\Delta_T^{\mathcal{R}}),$$

$$e^{\beta_T(r+r_T^{\mathcal{R}})} \left(1-\frac{\epsilon'_d}{\epsilon'_u}\frac{\kappa_d}{\kappa_u}\right)(\Delta_T+\Delta_T^{\mathcal{R}}) + e^{\beta_T}\left(1-\frac{\epsilon_d}{\epsilon_u}\right),$$

$$\left. e^{\beta_T(r+r_T^{\mathcal{R}})} \left(1-\frac{\epsilon'_d}{\epsilon'_u}\frac{\kappa_d}{\kappa_u}\right)(\Delta_T+\Delta_T^{\mathcal{R}}) + 2e^{\beta_T}\left(1-\frac{\epsilon_d}{\epsilon_u}\right) \right\}$$

$$= e^{\beta_T(r+r_T^{\mathcal{R}})} \left(1-\frac{\epsilon'_d}{\epsilon'_u}\frac{\kappa_d}{\kappa_u}\right)(\Delta_T+\Delta_T^{\mathcal{R}}) + 2e^{\beta_T}\left(1-\frac{\epsilon_d}{\epsilon_u}\right).$$

By the definition of $\beta_T$ given in equation (32), we have

$$e^{\beta_T(r+r_T^{\mathcal{R}})} \left(1-\frac{\epsilon'_d}{\epsilon'_u}\frac{\kappa_d}{\kappa_u}\right)(\Delta_T+\Delta_T^{\mathcal{R}}) \leq \frac{1}{6} \quad \text{and} \quad e^{\beta_T}\left(1-\frac{\epsilon_d}{\epsilon_u}\right) \leq \frac{1}{6}.$$

Thus we have

$$\max_{\substack{(\tau,v)\in I \\ \tau\in\{s+1,\dots,t-1\}}} \sum_{(\tau',v')\in I} e^{\beta_T|\tau-\tau'|} e^{\beta_T d(v,v')} C_{(\tau,v)(\tau',v')} \leq \frac{1}{2}. \tag{S25}$$

**Step 4.3.** We handled the third item of equation (S23) in Step 3 and showed that

$$\max_{(t,v)\in I} \sum_{(\tau',v')\in I} e^{\beta_T|t-\tau'|} e^{\beta_T d(v,v')} C_{(t,v)(\tau',v')}$$

$$= \max\left\{ \max_{(t,v)\in I} \sum_{(\tau',v')\in I} e^{\beta_T|t-\tau'|} e^{\beta_T d(v,v')} C_{(t,v)(\tau',v')} \mathbb{1}_{\{v\in k_t,\, v\notin k_{t-1},\, v'\in k_{\tau'}\}}, \right.$$

$$\max_{(t,v)\in I} \sum_{(\tau',v')\in I} e^{\beta_T|t-\tau'|} e^{\beta_T d(v,v')} C_{(t,v)(\tau',v')} \mathbb{1}_{\{v\in k_t,\ v\in k_{t-1},\ v'\in k_{\tau'}\}} \Bigg\}.$$

Specifically, in Step 3, we obtained that

$$\max_{(t,v)\in I} \sum_{(\tau',v')\in I} e^{\beta_T|t-\tau'|} e^{\beta_T d(v,v')} C_{(t,v)(\tau',v')} \mathbb{1}_{\{v\in k_t,\ v\notin k_{t-1},\ v'\in k_{\tau'}\}}$$

$$= \max_{(t,v)\in I} \sum_{(t-1,v')\in I} e^{\beta_T|t-(t-1)|} e^{\beta_T d(v,v')} C_{(t,v)(t-1,v')} \mathbb{1}_{\{v\in k_t,\ v\notin k_{t-1},\ v'\in k_{t-1}\}}$$

$$\quad + \max_{(t,v)\in I} \sum_{(t,v')\in I} e^{\beta_T|t-t|} e^{\beta_T d(v,v')} C_{(t,v)(t,v')} \mathbb{1}_{\{v\in k_t,\ v\in k_{t-1},\ v'\in k_t\}}$$

$$= 0 + \max_{(t,v)\in I} \sum_{(t,v')\in I} e^{\beta_T d(v,v')} C_{(t,v)(t,v')} \mathbb{1}_{\{v\in k_t,\ v\in k_{t-1},\ v'\in k_t\}}$$

$$\leq \max_{(t,v)\in I} \sum_{(t,v')\in I} e^{\beta_T d(v,v')} \left(1 - \frac{\epsilon_d'}{\epsilon_u'}\right) \mathbb{1}_{\{v\in k_t,\ v\in k_{t-1},\ v'\in k_t,\ v'\in N_t(v)\}}$$

$$\leq e^{\beta_T r} \left(1 - \frac{\epsilon_d'}{\epsilon_u'}\right) \Delta_T$$

and

$$\max_{(t,v)\in I} \sum_{(\tau',v')\in I} e^{\beta_T|t-\tau'|} e^{\beta_T d(v,v')} C_{(t,v)(\tau',v')} \mathbb{1}_{\{v\in k_t,\ v\in k_{t-1},\ v'\in k_{\tau'}\}}$$

$$= \max_{(t,v)\in I} \sum_{(t-1,v')\in I} e^{\beta_T|t-(t-1)|} e^{\beta_T d(v,v')} C_{(t,v)(t-1,v')} \mathbb{1}_{\{v\in k_t,\ v\in k_{t-1},\ v'\in k_{t-1}\}}$$

$$\quad + \max_{(t,v)\in I} \sum_{(t,v')\in I} e^{\beta_T|t-t|} e^{\beta_T d(v,v')} C_{(t,v)(t,v')} \mathbb{1}_{\{v\in k_t,\ v\in k_{t-1},\ v'\in k_t\}}$$

$$\leq \max_{(t,v)\in I} \sum_{(t-1,v')\in I} e^{\beta_T} e^{\beta_T d(v,v')} \left(1 - \frac{\epsilon_d}{\epsilon_u}\right) \mathbb{1}_{\{v\in k_t,\ v\in k_{t-1},\ v'\in k_{t-1},\ v'=v\}}$$

$$\quad + \max_{(t,v)\in I} \sum_{(t,v')\in I} e^{\beta_T d(v,v')} \left(1 - \frac{\epsilon_d'}{\epsilon_u'}\right) \mathbb{1}_{\{v\in k_t,\ v\in k_{t-1},\ v'\in k_t,\ v'\in N_t(v)\}}$$

$$\leq e^{\beta_T} \left(1 - \frac{\epsilon_d}{\epsilon_u}\right) + e^{\beta_T r} \left(1 - \frac{\epsilon_d'}{\epsilon_u'}\right) \Delta_T.$$

Then we have

$$\max_{(t,v)\in I} \sum_{(\tau',v')\in I} e^{\beta_T|t-\tau'|} e^{\beta_T d(v,v')} C_{(t,v)(\tau',v')}$$

$$= \max\left\{ e^{\beta_T r} \left(1 - \frac{\epsilon_d'}{\epsilon_u'}\right) \Delta_T,\ \ e^{\beta_T} \left(1 - \frac{\epsilon_d}{\epsilon_u}\right) + e^{\beta_T r} \left(1 - \frac{\epsilon_d'}{\epsilon_u'}\right) \Delta_T \right\}.$$

By the definition of $\beta_T$ given in equation (32), we have

$$e^{\beta_T r} \left(1 - \frac{\epsilon_d'}{\epsilon_u'}\right) \Delta_T \leq \frac{1}{6} \quad \text{and} \quad e^{\beta_T} \left(1 - \frac{\epsilon_d}{\epsilon_u}\right) \leq \frac{1}{6}.$$

Thus by Proposition S1.1, we have

$$\max_{(t,v)\in I}\sum_{(\tau',v')\in I}e^{\beta_T|t-\tau'|}e^{\beta_T d(v,v')}C_{(t,v)(\tau',v')}\leq \frac{1}{3}. \tag{S26}$$

**Step 5.** Plugging equations (S24), (S25) and (S26) into (S23), we have

$$\max_{(\tau,v)\in I}\sum_{(\tau',v')\in I}e^{\beta_T|\tau-\tau'|}e^{\beta_T d(v,v')}C_{(\tau,v)(\tau',v')}\leq \max\left\{\frac{1}{2},\frac{1}{2},\frac{1}{3}\right\}=\frac{1}{2}.$$

By Theorem S1.3,

$$\max_{(\tau,v)\in I}\sum_{(\tau',v')\in I}e^{\beta_T|\tau-\tau'|}e^{\beta_T d(v,v')}D_{(t,v)(\tau',v')}\leq \frac{1}{1-\frac{1}{2}}=2. \tag{S27}$$

Note that, $\rho^i_{(x_s^{k_s},\dots,x_t^{k_t})}=\widetilde{\rho}^i_{(x_s^{k_s},\dots,x_t^{k_t})}$ when $i=(\tau,v)$ with $\tau>s$. For $\tau=s$, we have

$$\widetilde{\rho}^i_{(x_s^{k_s},\dots,x_t^{k_t})}=(\mathsf{F}_s\widetilde{\pi}_{s-1})^v_{\chi_s,\chi_{s+1}} \quad\text{and}\quad \rho^i_{(x_s^{k_s},\dots,x_t^{k_t})}=(\widetilde{\mathsf{F}}_s\widetilde{\pi}_{s-1})^v_{\chi_s,\chi_{s+1}}.$$

By Theorem S1.6, we have

$$
\begin{aligned}
&\|\rho-\widetilde{\rho}\|_{\{t\}\times J}\\
&\leq \sum_{v\in J}\sum_{v'\in k_s}D_{(t,v)(s,v')}\sup_{x_s^{k_s},x_{s+1}^{k_{s+1}}\in\mathcal{X}}\|(\widetilde{\mathsf{F}}_s\widetilde{\pi}_{s-1})^{v'}_{\chi_s,\chi_{s+1}}-(\mathsf{F}_s\widetilde{\pi}_{s-1})^{v'}_{\chi_s,\chi_{s+1}}\|\\
&= \sum_{v\in J}e^{-\beta_T(t-s)}\sum_{v'\in k_s}e^{\beta_T((t-s)+d(v,v'))}D_{(t,v)(s,v')}e^{-\beta_T d(v,v')}\\
&\qquad\qquad\times\sup_{x_s^{k_s},x_{s+1}^{k_{s+1}}\in\mathcal{X}}\left\|(\widetilde{\mathsf{F}}_s\widetilde{\pi}_{s-1})^{v'}_{\chi_s,\chi_{s+1}}-(\mathsf{F}_s\widetilde{\pi}_{s-1})^{v'}_{\chi_s,\chi_{s+1}}\right\|\\
&\leq \sum_{v\in J}e^{-\beta_T(t-s)}\left(\sum_{v'\in k_s}e^{\beta_T((t-s)+d(v,v'))}D_{(t,v)(s,v')}\right)\times\max_{v'\in k_s}e^{-\beta_T d(v,v')}\\
&\qquad\qquad\times\sup_{x_s^{k_s},x_{s+1}^{k_{s+1}}\in\mathcal{X}}\left\|(\widetilde{\mathsf{F}}_s\widetilde{\pi}_{s-1})^{v'}_{\chi_s,\chi_{s+1}}-(\mathsf{F}_s\widetilde{\pi}_{s-1})^{v'}_{\chi_s,\chi_{s+1}}\right\|\\
&\leq 2e^{-\beta_T(t-s)}\sum_{v\in J}\max_{v'\in k_s}e^{-\beta_T d(v,v')}\sup_{x_s^{k_s},x_{s+1}^{k_{s+1}}\in\mathcal{X}}\left\|(\widetilde{\mathsf{F}}_s\widetilde{\pi}_{s-1})^{v'}_{\chi_s,\chi_{s+1}}-(\mathsf{F}_s\widetilde{\pi}_{s-1})^{v'}_{\chi_s,\chi_{s+1}}\right\|,
\end{aligned}
$$

where we used equation (S27) in the last inequality. □

Different to Proposition S1.1 which examines the difference $\left\|\widetilde{\pi}^v_{\chi_t,\chi_{t+1}}-\widetilde{\pi}^v_{\overline{\chi}_t,\chi_{t+1}}\right\|$, and Proposition S1.4 which explores the discrepancy $\|\mathsf{F}_t\widetilde{\pi}_{t-1}-\widetilde{\mathsf{F}}_t\widetilde{\pi}_{t-1}\|_J$, the subsequent proposition endeavors to bound the difference $\left\|(\widetilde{\mathsf{F}}_s\widetilde{\pi}_{s-1})^{v'}_{\chi_s,\chi_{s+1}}-(\mathsf{F}_s\widetilde{\pi}_{s-1})^{v'}_{\chi_s,\chi_{s+1}}\right\|$ uniformly over $x_s^{k_s},x_{s+1}^{k_{s+1}}\in\mathcal{X}$, where $(\widetilde{\mathsf{F}}_s\widetilde{\pi}_{s-1})^{v'}_{\chi_s,\chi_{s+1}}$ is defined in equation (S22) and $(\mathsf{F}_s\widetilde{\pi}_{s-1})^{v'}_{\chi_s,\chi_{s+1}}$ is defined in equation (S21).

**Proposition S1.7.** *Under Assumption 3.1, we have that for every $s\in[T]$, $B_s\in\mathcal{B}(k_s)$, and $v'\in B_s$,*

$$\sup_{x_s^{k_s},x_{s+1}^{k_{s+1}}\in\mathcal{X}}\left\|(\widetilde{\mathsf{F}}_s\widetilde{\pi}_{s-1})^{v'}_{\chi_s,\chi_{s+1}}-(\mathsf{F}_s\widetilde{\pi}_{s-1})^{v'}_{\chi_s,\chi_{s+1}}\right\|\leq 4e^{-\beta_T}\left(1-\frac{\epsilon_d}{\epsilon_u}\right)e^{-\beta_T d(v',\partial B_s)}.$$

*Proof.* For $s \in [T]$, $B_s \in \mathcal{B}(k_s)$, and $v' \in B_s$, define

$$I = (\{s-1\} \times k_{s-1}) \cup (s, v') \quad \text{and} \quad \mathbb{S} = \mathcal{X} \times \mathbb{X}^{v'}.$$

Define the probability measures on $\mathbb{S}$ as follows:

$$\rho(A) = \frac{\begin{aligned}\int \mathbb{1}_A(x_{s-1}^{k_{s-1}}, x_s^{v'}) \prod_{\omega \in k_s} f_s^\omega(x_s^\omega \mid k_s^{\omega(\mathcal{R})}, x_{s-1}^{k_{s-1} \cap \{\omega\}}) \\ \times g_s^{v'}(Y_s^{v'} \mid k_s^{v'(\mathcal{R})}, x_s^{v'}) \widetilde{f}_s^{v'}(x_s^{v'}, x_s^{N_s(v')}) \\ \times p_{s+1}^{v'(\mathcal{R})}(k_{s+1}^{v'(\mathcal{R})} \mid k_s, x_s^{v'(\mathcal{R})}) f_{s+1}^{v'}(x_{s+1}^{k_{s+1} \cap \{v'\}} \mid k_{s+1}^{v'(\mathcal{R})}, x_s^{v'}) \\ \times \prod_{R \in \mathcal{R}} p_s^R(k_s^R \mid k_{s-1}, x_{s-1}^{k_{s-1} \cap R}) \widetilde{\pi}_{s-1}(dx_{s-1}^{k_{s-1}}) \psi^{v'}(dx_s^{v'})\end{aligned}}{\begin{aligned}\int \prod_{\omega \in k_s} f_s^\omega(x_s^\omega \mid k_s^{\omega(\mathcal{R})}, x_{s-1}^{k_{s-1} \cap \{\omega\}}) \\ \times g_s^{v'}(Y_s^{v'} \mid k_s^{v'(\mathcal{R})}, x_s^{v'}) \widetilde{f}_s^{v'}(x_s^{v'}, x_s^{N_s(v')}) \\ \times p_{s+1}^{v'(\mathcal{R})}(k_{s+1}^{v'(\mathcal{R})} \mid k_s, x_s^{v'(\mathcal{R})}) f_{s+1}^{v'}(x_{s+1}^{k_{s+1} \cap \{v'\}} \mid k_{s+1}^{v'(\mathcal{R})}, x_s^{v'}) \\ \times \prod_{R \in \mathcal{R}} p_s^R(k_s^R \mid k_{s-1}, x_{s-1}^{k_{s-1} \cap R}) \widetilde{\pi}_{s-1}(dx_{s-1}^{k_{s-1}}) \psi^{v'}(dx_s^{v'})\end{aligned}},$$

$$\widetilde{\rho}(A) = \frac{\begin{aligned}\int \mathbb{1}_A(x_{s-1}^{k_{s-1}}, x_s^{v'}) \prod_{\omega \in B_s} f_s^\omega(x_s^\omega \mid k_s^{\omega(\mathcal{R})}, x_{s-1}^{k_{s-1} \cap \{\omega\}}) \\ \times g_s^{v'}(Y_s^{v'} \mid k_s^{v'(\mathcal{R})}, x_s^{v'}) \widetilde{f}_s^{v'}(x_s^{v'}, x_s^{N_s(v')}) \\ \times p_{s+1}^{v'(\mathcal{R})}(k_{s+1}^{v'(\mathcal{R})} \mid k_s, x_s^{v'(\mathcal{R})}) f_{s+1}^{v'}(x_{s+1}^{k_{s+1} \cap \{v'\}} \mid k_{s+1}^{v'(\mathcal{R})}, x_s^{v'}) \\ \times \prod_{R \in \mathcal{R}} p_s^R(k_s^R \mid k_{s-1}, x_{s-1}^{k_{s-1} \cap R}) \widetilde{\pi}_{s-1}(dx_{s-1}^{k_{s-1}}) \psi^{v'}(dx_s^{v'})\end{aligned}}{\begin{aligned}\int \prod_{\omega \in B_s} f_s^\omega(x_s^\omega \mid k_s^{\omega(\mathcal{R})}, x_{s-1}^{k_{s-1} \cap \{\omega\}}) \\ \times g_s^{v'}(Y_s^{v'} \mid k_s^{v'(\mathcal{R})}, x_s^{v'}) \widetilde{f}_s^{v'}(x_s^{v'}, x_s^{N_s(v')}) \\ \times p_{s+1}^{v'(\mathcal{R})}(k_{s+1}^{v'(\mathcal{R})} \mid k_s, x_s^{v'(\mathcal{R})}) f_{s+1}^{v'}(x_{s+1}^{k_{s+1} \cap \{v'\}} \mid k_{s+1}^{v'(\mathcal{R})}, x_s^{v'}) \\ \times \prod_{R \in \mathcal{R}} p_s^R(k_s^R \mid k_{s-1}, x_{s-1}^{k_{s-1} \cap R}) \widetilde{\pi}_{s-1}(dx_{s-1}^{k_{s-1}}) \psi^{v'}(dx_s^{v'})\end{aligned}}.$$

Observing $(\mathsf{F}_s \widetilde{\pi}_{s-1})_{\chi_s, \chi_{s+1}}^{v'}$ in equation (S21) and $(\widetilde{\mathsf{F}}_s \widetilde{\pi}_{s-1})_{\chi_s, \chi_{s+1}}^{v'}$ in equation (S22), we can see that

$$\|(\mathsf{F}_s \widetilde{\pi}_{s-1})_{\chi_s, \chi_{s+1}}^{v'} - (\widetilde{\mathsf{F}}_s \widetilde{\pi}_{s-1})_{\chi_s, \chi_{s+1}}^{v'}\| = \|\rho - \widetilde{\rho}\|_{(s,v')}. \tag{S28}$$

In the following steps, we are going to use Dobrushin comparison theorem (Theorem 3.3) to bound $\|\rho - \widetilde{\rho}\|_{(s,v')}$. We will bound $C_{ij}$ and $b_i$ with $i = (\tau, l)$ and $j = (\tau', l')$.

**Step 1.** Consider $\tau = s - 1$ which implies $l \in k_{s-1}$.

**Step 1.1.** When $l \in k_s$, we have

$$\rho_{(x_{s-1}^{k_{s-1}}, x_s^{v'})}^i(A) = \frac{\int \mathbb{1}_A(x_{s-1}^l) f_s^l(x_s^l \mid k_s^{l(\mathcal{R})}, x_{s-1}^l) p_s^{l(\mathcal{R})}(k_s^{l(\mathcal{R})} \mid k_{s-1}, x_{s-1}^{l(\mathcal{R})}) \widetilde{\pi}_{\chi_{s-1}}^l(dx_{s-1}^l)}{\int f_s^l(x_s^l \mid k_s^{l(\mathcal{R})}, x_{s-1}^l) p_s^{l(\mathcal{R})}(k_s^{l(\mathcal{R})} \mid k_{s-1}, x_{s-1}^{l(\mathcal{R})}) \widetilde{\pi}_{\chi_{s-1}}^l(dx_{s-1}^l)}$$

where $\widetilde{\pi}_{\chi_{s-1}}^l$ is defined according to the definition of $\mu_{\chi_{s-1}}^v$ given in equation (S1). Therefore, by the definition of $\mu_{\chi_{s-1}, \chi_s}^v$ given in equation (S2), we have $\rho_{(x_{s-1}^{k_{s-1}}, x_s^{v'})}^i = \widetilde{\pi}_{\chi_{s-1}, \chi_s}^l$. If $\tau' = s - 1$ which implies $l' \in k_{s-1}$, by the definition of $C_{vv'}^{\mu_{s-1}}$ in equation (S4), we know that $C_{ij} \le C_{ll'}^{\widetilde{\pi}_{s-1}}$. If $\tau' = s$

which implies $l' \in k_s$, when $l' = l$ we have by Theorem S1.2 that $C_{ij} \leq \left(1 - \frac{\epsilon_d}{\epsilon_u}\right)$, since

$$\rho^i_{(x^{k_{s-1}}_{s-1}, x^{v'}_s)}(A) \geq \left(\frac{\epsilon_d}{\epsilon_u}\right) \frac{\int \mathbb{1}_A(x^l_{s-1}) p^{l(\mathcal{R})}_s(k^{l(\mathcal{R})}_s \mid k_{s-1}, x^{l(\mathcal{R})}_{s-1}) \widetilde{\pi}^l_{\chi_{s-1}}(dx^l_{s-1})}{\int p^{l(\mathcal{R})}_s(k^{l(\mathcal{R})}_s \mid k_{s-1}, x^{l(\mathcal{R})}_{s-1}) \widetilde{\pi}^l_{\chi_{s-1}}(dx^l_{s-1})};$$

$C_{ij} = 0$ otherwise.

**Step 1.2.** When $l \notin k_s$, we have

$$\rho^i_{(x^{k_{s-1}}_{s-1}, x^{v'}_s)}(A) = \frac{\int \mathbb{1}_A(x^l_{s-1}) p^{l(\mathcal{R})}_s(k^{l(\mathcal{R})}_s \mid k_{s-1}, x^{l(\mathcal{R})}_{s-1}) \widetilde{\pi}^l_{\chi_{s-1}}(dx^l_{s-1})}{\int p^{l(\mathcal{R})}_s(k^{l(\mathcal{R})}_s \mid k_{s-1}, x^{l(\mathcal{R})}_{s-1}) \widetilde{\pi}^l_{\chi_{s-1}}(dx^l_{s-1})}.$$

If $\tau' = s - 1$ which implies $l' \in k_{s-1}$, when $l' \in l(\mathcal{R})$ we have $C_{ij} \leq \left(1 - \frac{\kappa_d}{\kappa_u}\right)$, since

$$\rho^i_{(x^{k_{s-1}}_{s-1}, x^{v'}_s)}(A) \geq \left(\frac{\kappa_d}{\kappa_u}\right) \widetilde{\pi}^l_{\chi_{s-1}}(A)$$

and by Theorem S1.2, and $C_{ij} = 0$ otherwise. If $\tau' = s$ which implies $l' \in k_s$, we have $C_{ij} = 0$.

**Step 1.3.** Next, we calculate $b_{(s-1,l)}$. Note that, when $l \notin k_s$ we have $\rho^i_{(x^{k_{s-1}}_{s-1}, x^{v'}_s)}(A) = \widetilde{\rho}^i_{(x^{k_{s-1}}_{s-1}, x^{v'}_s)}(A)$ and hence $b_{(s-1,l)} = 0$. Also, when $l \in B_s \subseteq k_s$, we have $\rho^i_{(x^{k_{s-1}}_{s-1}, x^{v'}_s)}(A) = \widetilde{\rho}^i_{(x^{k_{s-1}}_{s-1}, x^{v'}_s)}(A)$ and hence $b_{(s-1,l)} = 0$. What makes a difference is the situation that when $l \in k_s$ but $l \notin B_s$, where

$$\widetilde{\rho}^i_{(x^{k_{s-1}}_{s-1}, x^{v'}_s)}(A) = \frac{\int \mathbb{1}_A(x^l_{s-1}) p^{l(\mathcal{R})}_s(k^{l(\mathcal{R})}_s \mid k_{s-1}, x^{l(\mathcal{R})}_{s-1}) \widetilde{\pi}^l_{\chi_{s-1}}(dx^l_{s-1})}{\int p^{l(\mathcal{R})}_s(k^{l(\mathcal{R})}_s \mid k_{s-1}, x^{l(\mathcal{R})}_{s-1}) \widetilde{\pi}^l_{\chi_{s-1}}(dx^l_{s-1})}.$$

However, since

$$\rho^i_{(x^{k_{s-1}}_{s-1}, x^{v'}_s)}(A), \ \widetilde{\rho}^i_{(x^{k_{s-1}}_{s-1}, x^{v'}_s)}(A) \geq \left(\frac{\epsilon_d}{\epsilon_u}\right) \frac{\int \mathbb{1}_A(x^l_{s-1}) p^{l(\mathcal{R})}_s(k^{l(\mathcal{R})}_s \mid k_{s-1}, x^{l(\mathcal{R})}_{s-1}) \widetilde{\pi}^l_{\chi_{s-1}}(dx^l_{s-1})}{\int p^{l(\mathcal{R})}_s(k^{l(\mathcal{R})}_s \mid k_{s-1}, x^{l(\mathcal{R})}_{s-1}) \widetilde{\pi}^l_{\chi_{s-1}}(dx^l_{s-1})},$$

by Theorem S1.2 we have $b_{(s-1,l)} \leq 2\left(1 - \frac{\epsilon_d}{\epsilon_u}\right)$.

**Step 2.** Consider $\tau = s$ which implies $l = v' \in k_s$.

**Step 2.1.** When $v' \in k_{s-1}$, we have

$$\rho^i_{(x^{k_{s-1}}_{s-1}, x^{v'}_s)}(A)$$
$$= \frac{\int \mathbb{1}_A(x^{v'}_s) f^{v'}_s(x^{v'}_s \mid k^{v'(\mathcal{R})}_s, x^{v'}_{s-1}) g^{v'}_s(Y^{v'}_s \mid k^{v'(\mathcal{R})}_s, x^{v'}_s) \widetilde{f}^{v'}_s(x^{v'}_s, x^{N_s(v')}_s)}{\int f^{v'}_s(x^{v'}_s \mid k^{v'(\mathcal{R})}_s, x^{v'}_{s-1}) g^{v'}_s(Y^{v'}_s \mid k^{v'(\mathcal{R})}_s, x^{v'}_s) \widetilde{f}^{v'}_s(x^{v'}_s, x^{N_s(v')}_s)} \\ \frac{\times p^{v'(\mathcal{R})}_{s+1}(k^{v'(\mathcal{R})}_{s+1} \mid k_s, x^{v'(\mathcal{R})}_s) f^{v'}_{s+1}(x^{k_{s+1} \cap \{v'\}}_{s+1} \mid k^{v'(\mathcal{R})}_{s+1}, x^{v'}_s) \psi^{v'}(dx^{v'}_s)}{\times p^{v'(\mathcal{R})}_{s+1}(k^{v'(\mathcal{R})}_{s+1} \mid k_s, x^{v'(\mathcal{R})}_s) f^{v'}_{s+1}(x^{k_{s+1} \cap \{v'\}}_{s+1} \mid k^{v'(\mathcal{R})}_{s+1}, x^{v'}_s) \psi^{v'}(dx^{v'}_s)}.$$

If $\tau' = s-1$ which implies $l' \in k_{s-1}$, when $l' = v'$ we have $C_{ij} \leq \left(1 - \frac{\epsilon_d}{\epsilon_u}\right)$, since

$$
\begin{aligned}
&\rho^i_{(x^{k_{s-1}}_{s-1}, x^{v'}_s)}(A) \\
&\geq \left(\frac{\epsilon_d}{\epsilon_u}\right) \frac{\begin{array}{l}\int \mathbb{1}_A(x^{v'}_s) g^{v'}_s(Y^{v'}_s \mid k^{v'(\mathcal{R})}_s, x^{v'}_s) \widetilde{f}^{v'}_s(x^{v'}_s, x^{N_s(v')}_s) \\ \times p^{v'(\mathcal{R})}_{s+1}(k^{v'(\mathcal{R})}_{s+1} \mid k_s, x^{v'(\mathcal{R})}_s) f^{v'}_{s+1}(x^{k_{s+1} \cap \{v'\}}_{s+1} \mid k^{v'(\mathcal{R})}_{s+1}, x^{v'}_s) \psi^{v'}(dx^{v'}_s)\end{array}}{\begin{array}{l}\int g^{v'}_s(Y^{v'}_s \mid k^{v'(\mathcal{R})}_s, x^{v'}_s) \widetilde{f}^{v'}_s(x^{v'}_s, x^{N_s(v')}_s) \\ \times p^{v'(\mathcal{R})}_{s+1}(k^{v'(\mathcal{R})}_{s+1} \mid k_s, x^{v'(\mathcal{R})}_s) f^{v'}_{s+1}(x^{k_{s+1} \cap \{v'\}}_{s+1} \mid k^{v'(\mathcal{R})}_{s+1}, x^{v'}_s) \psi^{v'}(dx^{v'}_s)\end{array}}
\end{aligned}
$$

and by Theorem S1.2, and $C_{ij} = 0$ otherwise. If $\tau' = s$ which implies $l' = v'$, we have $C_{ij} = 0$.

**Step 2.2.** When $v' \notin k_{s-1}$, we have

$$
\rho^i_{(x^{k_{s-1}}_{s-1}, x^{v'}_s)}(A) = \frac{\begin{array}{l}\int \mathbb{1}_A(x^{v'}_s) f^{v'}_s(x^{v'}_s \mid k^{v'(\mathcal{R})}_s) g^{v'}_s(Y^{v'}_s \mid k^{v'(\mathcal{R})}_s, x^{v'}_s) \widetilde{f}^{v'}_s(x^{v'}_s, x^{N_s(v')}_s) \\ \times p^{v'(\mathcal{R})}_{s+1}(k^{v'(\mathcal{R})}_{s+1} \mid k_s, x^{v'(\mathcal{R})}_s) f^{v'}_{s+1}(x^{k_{s+1} \cap \{v'\}}_{s+1} \mid k^{v'(\mathcal{R})}_{s+1}, x^{v'}_s) \psi^{v'}(dx^{v'}_s)\end{array}}{\begin{array}{l}\int f^{v'}_s(x^{v'}_s \mid k^{v'(\mathcal{R})}_s) g^{v'}_s(Y^{v'}_s \mid k^{v'(\mathcal{R})}_s, x^{v'}_s) \widetilde{f}^{v'}_s(x^{v'}_s, x^{N_s(v')}_s) \\ \times p^{v'(\mathcal{R})}_{s+1}(k^{v'(\mathcal{R})}_{s+1} \mid k_s, x^{v'(\mathcal{R})}_s) f^{v'}_{s+1}(x^{k_{s+1} \cap \{v'\}}_{s+1} \mid k^{v'(\mathcal{R})}_{s+1}, x^{v'}_s) \psi^{v'}(dx^{v'}_s)\end{array}}.
$$

If $\tau' = s-1$ which implies $l' \in k_{s-1}$, we have $C_{ij} = 0$. If $\tau' = s$ which implies $l' = v'$, we have $C_{ij} = 0$.

**Step 2.3.** Next, we calculate $b_{(s,v')}$. Note that $v' \in B_s$ is assumed in this proposition, we have $\rho^i_{(x^{k_{s-1}}_{s-1}, x^{v'}_s)}(A) = \widetilde{\rho}^i_{(x^{k_{s-1}}_{s-1}, x^{v'}_s)}(A)$ and hence $b_{(s,v')} = 0$.

**Step 3.** In this step, we summary the results of $C_{ij}$ obtained in the previous two steps and aim to bound the following quantity:

$$
\begin{aligned}
&\max_{(\tau,l) \in I} \sum_{(\tau',l') \in I} e^{\beta_T |\tau - \tau'|} e^{\beta_T d(l,l')} C_{(\tau,l)(\tau',l')} \\
&= \max \left\{ \max_{(s-1,l) \in I} \sum_{(\tau',l') \in I} e^{\beta_T |s-1-\tau'|} e^{\beta_T d(l,l')} C_{(s-1,l)(\tau',l')}, \right. \\
&\qquad\qquad\qquad \left. \max_{(s,v') \in I} \sum_{(\tau',l') \in I} e^{\beta_T |s-\tau'|} e^{\beta_T d(v',l')} C_{(s,v')(\tau',l')} \right\}. \qquad \text{(S29)}
\end{aligned}
$$

**Step 3.1.** We handled the first item of equation (S29) in Step 1 and showed that

$$
\begin{aligned}
&\max_{(s-1,l) \in I} \sum_{(\tau',l') \in I} e^{\beta_T |s-1-\tau'|} e^{\beta_T d(l,l')} C_{(s-1,l)(\tau',l')} \\
&= \max \left\{ \max_{(s-1,l) \in I} \sum_{(\tau',l') \in I} e^{\beta_T |s-1-\tau'|} e^{\beta_T d(l,l')} C_{(s-1,l)(\tau',l')} \mathbb{1}_{\{l \in k_{s-1}, \ l \in k_s\}}, \right. \\
&\qquad\qquad\qquad \left. \max_{(s-1,l) \in I} \sum_{(\tau',l') \in I} e^{\beta_T |s-1-\tau'|} e^{\beta_T d(l,l')} C_{(s-1,l)(\tau',l')} \mathbb{1}_{\{l \in k_{s-1}, \ l \notin k_s\}} \right\}.
\end{aligned}
$$

Specifically, in Step 1, we obtained that

$$\max_{(s-1,l)\in I} \sum_{(\tau',l')\in I} e^{\beta_T|s-1-\tau'|} e^{\beta_T d(l,l')} C_{(s-1,l)(\tau',l')} \mathbb{1}_{\{l\in k_{s-1},\, l\in k_s\}}$$

$$= \max_{(s-1,l)\in I} \sum_{(s-1,l')\in I} e^{\beta_T|s-1-(s-1)|} e^{\beta_T d(l,l')} C_{(s-1,l)(s-1,l')} \mathbb{1}_{\{l\in k_{s-1},\, l\in k_s,\, l'\in k_{s-1}\}}$$

$$\quad + \max_{(s-1,l)\in I} \sum_{(s,l')\in I} e^{\beta_T|s-1-s|} e^{\beta_T d(l,l')} C_{(s-1,l)(s,l')} \mathbb{1}_{\{l\in k_{s-1},\, l\in k_s,\, l'\in k_s\}}$$

$$\leq \max_{(s-1,l)\in I} \sum_{(s-1,l')\in I} e^{\beta_T d(l,l')} C_{ll'}^{\widetilde{\pi}_{s-1}} \mathbb{1}_{\{l\in k_{s-1},\, l\in k_s,\, l'\in k_{s-1}\}}$$

$$\quad + \max_{(s-1,l)\in I} \sum_{(s,l')\in I} e^{\beta_T} e^{\beta_T d(l,l')} \left(1 - \frac{\epsilon_d}{\epsilon_u}\right) \mathbb{1}_{\{l\in k_{s-1},\, l\in k_s,\, l'\in k_s,\, l'=l\}}$$

$$\leq \max_{l\in k_{s-1}} \sum_{l'\in k_{s-1}} e^{\beta_T d(l,l')} C_{ll'}^{\widetilde{\pi}_{s-1}} + e^{\beta_T} \left(1 - \frac{\epsilon_d}{\epsilon_u}\right),$$

and

$$\max_{(s-1,l)\in I} \sum_{(\tau',l')\in I} e^{\beta_T|s-1-\tau'|} e^{\beta_T d(l,l')} C_{(s-1,l)(\tau',l')} \mathbb{1}_{\{l\in k_{s-1},\, l\notin k_s\}}$$

$$= \max_{(s-1,l)\in I} \sum_{(s-1,l')\in I} e^{\beta_T|s-1-(s-1)|} e^{\beta_T d(l,l')} C_{(s-1,l)(s-1,l')} \mathbb{1}_{\{l\in k_{s-1},\, l\notin k_s,\, l'\in k_{s-1}\}}$$

$$\quad + \max_{(s-1,l)\in I} \sum_{(s,l')\in I} e^{\beta_T|s-1-s|} e^{\beta_T d(l,l')} C_{(s-1,l)(s,l')} \mathbb{1}_{\{l\in k_{s-1},\, l\notin k_s,\, l'\in k_s\}}$$

$$\leq \max_{(s-1,l)\in I} \sum_{(s-1,l')\in I} e^{\beta_T d(l,l')} \left(1 - \frac{\kappa_d}{\kappa_u}\right) \mathbb{1}_{\{l\in k_{s-1},\, l\notin k_s,\, l'\in k_{s-1},\, l'\in l(\mathcal{R})\}} + 0$$

$$\leq e^{\beta_T r_T^{\mathcal{R}}} \left(1 - \frac{\kappa_d}{\kappa_u}\right) \Delta_T^{\mathcal{R}}.$$

By the definition of $\beta_T$ given in equation (32), we have

$$e^{\beta_T r_T^{\mathcal{R}}} \left(1 - \frac{\kappa_d}{\kappa_u}\right) \Delta_T^{\mathcal{R}} \leq \frac{1}{6} \quad \text{and} \quad e^{\beta_T} \left(1 - \frac{\epsilon_d}{\epsilon_u}\right) \leq \frac{1}{6}.$$

Thus by Proposition S1.1, we have

$$\max_{(s-1,l)\in I} \sum_{(\tau',l')\in I} e^{\beta_T|s-1-\tau'|} e^{\beta_T d(l,l')} C_{(s-1,l)(\tau',l')} \leq \max\left\{\frac{1}{2}, \frac{1}{6}\right\} = \frac{1}{2}.$$

**Step 3.2.** We handled the second item of equation (S29) in Step 2 and showed that

$$\max_{(s,v')\in I} \sum_{(\tau',l')\in I} e^{\beta_T|s-\tau'|} e^{\beta_T d(v',l')} C_{(s,v')(\tau',l')}$$

$$= \max\left\{ \max_{(s,v')\in I} \sum_{(s-1,l')\in I} e^{\beta_T|s-(s-1)|} e^{\beta_T d(v',l')} C_{(s,v')(s-1,l')} \mathbb{1}_{\{v'\in k_s,\, v\in k_{s-1},\, l'\in k_{s-1}\}}, \right.$$

$$\max_{(s,v')\in I} \sum_{(s-1,l')\in I} e^{\beta_T|s-(s-1)|} e^{\beta_T d(v',l')} C_{(s,v')(s-1,l')} \mathbb{1}_{\{v'\in k_s,\ v'\notin k_{s-1},\ l'\in k_{s-1}\}} \Bigg\}$$

$$= \max\Bigg\{ \max_{(s,v')\in I} \sum_{(s-1,l')\in I} e^{\beta_T|s-(s-1)|} e^{\beta_T d(v',l')} C_{(s,v')(s-1,l')} \mathbb{1}_{\{v'\in k_s,\ v'\in k_{s-1},\ l'\in k_{s-1}\}}, 0 \Bigg\}$$

$$\leq \max_{(s,v')\in I} \sum_{(s-1,l')\in I} e^{\beta_T|s-(s-1)|} e^{\beta_T d(v',l')} \left(1 - \frac{\epsilon_d}{\epsilon_u}\right) \mathbb{1}_{\{v'\in k_s,\ v'\in k_{s-1},\ l'\in k_{s-1},\ l'=v'\}}$$

$$= e^{\beta_T} \left(1 - \frac{\epsilon_d}{\epsilon_u}\right)$$

$$\leq \frac{1}{6}.$$

Plugging the above two results into equation (S29), we obtain

$$\max_{(\tau,l)\in I} \sum_{(\tau',l')\in I} e^{\beta_T|\tau-\tau'|} e^{\beta_T d(l,l')} C_{(\tau,l)(\tau',l')} \leq \frac{1}{2}.$$

**Step 4.** Now, we are ready to finish the proof. By Theorem S1.3,

$$\max_{(s,v)\in I} \sum_{(\tau',v')\in I} e^{\beta_T|s-\tau'|} e^{\beta_T d(v,v')} D_{(s,v)(\tau',v')} \leq \frac{1}{1-\frac{1}{2}} = 2.$$

Recalling that in Step 1 we obtained that $b_{(s-1,l)} \leq 2\left(1 - \frac{\epsilon_d}{\epsilon_u}\right)$ when $l \in k_s \backslash B_s$ and in Step 2 we obtained that $b_{(s,v')} = 0$, we have by Theorem 3.3 that

$$\|\rho - \widetilde{\rho}\|_{(s,v')} \leq 2\left(1 - \frac{\epsilon_d}{\epsilon_u}\right) \sum_{l\in k_s\backslash B_s} D_{(s,v')(s-1,l)} \leq 4e^{-\beta_T}\left(1 - \frac{\epsilon_d}{\epsilon_u}\right) e^{-\beta_T d(v',\partial B_s)},$$

which completes the proof by equation (S28). □

In Proposition S1.5, we examined the difference $\left\| \mathsf{F}_t \cdots \mathsf{F}_{s+1} \mathsf{F}_s \widetilde{\pi}_{s-1} - \mathsf{F}_t \cdots \mathsf{F}_{s+1} \widetilde{\mathsf{F}}_s \widetilde{\pi}_{s-1} \right\|_J$ where $s \in [t-1]$, for every $J \subseteq B_t$ and $B_t \in \mathcal{B}(k_t)$. Next, it is necessary to investigate the long time behavior applying the $\widetilde{\mathsf{F}}_s$ operator. Recall that the influence path $\mathcal{P}^{B_t}$ for cluster $B_t$ is defined as

$$\mathcal{P}^{B_t} = \Big\{ [B_u \cdots B_t] : B_l \in \mathcal{B}(k_l),\ B_l \cap B_{l+1} \neq \emptyset,\ 0 \leq u \leq l < t \Big\},$$

where $[B_u \cdots B_t]$ is the cluster $B_u$ at time $u$ that has influence on the cluster $B_t$ at time $t$ after $t-u$ time steps; the vertex set of $\mathcal{P}^{B_t}$ is defined as

$$\mathcal{V}^{B_t} = \Big\{ [B_u \cdots B_t]v : [B_u \cdots B_t] \in \mathcal{P}^{B_t},\ v \in B_u \Big\};$$

the set of vertices at time 0 that could affact $B_t$ is defined as

$$\mathcal{V}_0^{B_t} = \Big\{ [B_0 \cdots B_t] : B_l \in \mathcal{B}(k_l),\ B_l \cap B_{l+1} \neq \emptyset,\ 0 \leq l < t \Big\}. \tag{S30}$$

Now, we define the location $v(i)$ of $i \in \mathcal{V}^{B_t}$ as

$$v([B_u \cdots B_t]v) = v;$$

define the depth $d(i)$ of $i \in \mathcal{V}^{B_t}$ as

$$d([B_u \cdots B_t]v) = u;$$

then the set of non-leaf vertices of the tree can be written as

$$\mathcal{V}_+^{B_t} = \{i \in \mathcal{V}^{B_t} : 0 < d(i) \leq t\}.$$

We first conduct preliminary analysis in the following lemma, which will only be used in Proposition S1.9.

**Lemma S1.8.** *Under Assumption 3.1, one has that for every $B_t \in \mathcal{B}(k_t)$ and $J \subseteq B_t$,*

$$\left\| \widetilde{\mathsf{F}}_t \cdots \widetilde{\mathsf{F}}_1 \delta_x - \widetilde{\mathsf{F}}_t \cdots \widetilde{\mathsf{F}}_1 \delta_{\overline{x}} \right\|_J \leq 4 e^{-\beta_\tau t} \mathrm{card}(J).$$

*Proof.* Define $\mathbb{S} = \prod_{i \in \mathcal{V}^{B_t}} \mathbb{X}^i$, where $\mathbb{X}^{[\tau]v} = \mathbb{X}^v$ for $[\tau]v \in \mathcal{V}^{B_t}$. Define two probability measures on $\mathbb{S}$ as follows:

$$\rho(A) = \cfrac{\begin{aligned}\int \mathbb{1}_A(x) \prod_{i \in \mathcal{V}_+^{B_t}} f_{d(i)}^i(x_{d(i)}^i \mid k_{d(i)}^{i(\mathcal{R})}, x_{d(i)-1}^{k_{d(i)-1} \cap \{i\}}) \\ \times g_{d(i)}^i(y_{d(i)}^i \mid k_{d(i)}^{i(\mathcal{R})}, x_{d(i)}^i) \widetilde{f}_{d(i)}^i(x_{d(i)}^i, x_{d(i)}^{N_{d(i)}(i)}) \\ \times \prod_{R \in \mathcal{R}} p_{d(i)}^R(k_{d(i)}^R \mid k_{d(i)-1}, x_{d(i)-1}^{k_{d(i)-1} \cap R}) \psi^{v(i)}(dx_{d(i)}^i) \prod_{\tau \in \mathcal{V}_0^{B_t}} \delta^\tau(dx_0^\tau)\end{aligned}}{\begin{aligned}\int \prod_{i \in \mathcal{V}_+^{B_t}} f_{d(i)}^i(x_{d(i)}^i \mid k_{d(i)}^{i(\mathcal{R})}, x_{d(i)-1}^{k_{d(i)-1} \cap \{i\}}) \\ \times g_{d(i)}^i(y_{d(i)}^i \mid k_{d(i)}^{i(\mathcal{R})}, x_{d(i)}^i) \widetilde{f}_{d(i)}^i(x_{d(i)}^i, x_{d(i)}^{N_{d(i)}(i)}) \\ \times \prod_{R \in \mathcal{R}} p_{d(i)}^R(k_{d(i)}^R \mid k_{d(i)-1}, x_{d(i)-1}^{k_{d(i)-1} \cap R}) \psi^{v(i)}(dx_{d(i)}^i) \prod_{\tau \in \mathcal{V}_0^{B_t}} \delta^\tau(dx_0^\tau)\end{aligned}},$$

$$\overline{\rho}(A) = \cfrac{\begin{aligned}\int \mathbb{1}_A(x) \prod_{i \in \mathcal{V}_+^{B_t}} f_{d(i)}^i(x_{d(i)}^i \mid k_{d(i)}^{i(\mathcal{R})}, x_{d(i)-1}^{k_{d(i)-1} \cap \{i\}}) \\ \times g_{d(i)}^i(y_{d(i)}^i \mid k_{d(i)}^{i(\mathcal{R})}, x_{d(i)}^i) \widetilde{f}_{d(i)}^i(x_{d(i)}^i, x_{d(i)}^{N_{d(i)}(i)}) \\ \times \prod_{R \in \mathcal{R}} p_{d(i)}^R(k_{d(i)}^R \mid k_{d(i)-1}, x_{d(i)-1}^{k_{d(i)-1} \cap R}) \psi^{v(i)}(dx_{d(i)}^i) \prod_{\tau \in \mathcal{V}_0^{B_t}} \delta^\tau(d\overline{x}_0^\tau)\end{aligned}}{\begin{aligned}\int \prod_{i \in \mathcal{V}_+^{B_t}} f_{d(i)}^i(x_{d(i)}^i \mid k_{d(i)}^{i(\mathcal{R})}, x_{d(i)-1}^{k_{d(i)-1} \cap \{i\}}) \\ \times g_{d(i)}^i(y_{d(i)}^i \mid k_{d(i)}^{i(\mathcal{R})}, x_{d(i)}^i) \widetilde{f}_{d(i)}^i(x_{d(i)}^i, x_{d(i)}^{N_{d(i)}(i)}) \\ \times \prod_{R \in \mathcal{R}} p_{d(i)}^R(k_{d(i)}^R \mid k_{d(i)-1}, x_{d(i)-1}^{k_{d(i)-1} \cap R}) \psi^{v(i)}(dx_{d(i)}^i) \prod_{\tau \in \mathcal{V}_0^{B_t}} \delta^\tau(d\overline{x}_0^\tau)\end{aligned}}.$$

where $\rho^{[B_u \cdots B_t]} = \rho^{B_u}$ for any $[B_u \cdots B_t] \in \mathcal{P}^{B_t}$ and any measure $\rho$. Therefore, for every $B_t \in \mathcal{B}(k_t)$ and $J \subseteq B_t$, we have

$$\left\| \widetilde{\mathsf{F}}_t \cdots \widetilde{\mathsf{F}}_1 \delta_x - \widetilde{\mathsf{F}}_t \cdots \widetilde{\mathsf{F}}_1 \delta_{\overline{x}} \right\|_J = \| \rho - \overline{\rho} \|_{[B_t]J}. \tag{S31}$$

In the following steps, we are going to use Dobrushin comparison theorem (Theorem 3.3) to bound $\| \rho - \overline{\rho} \|_{[B_t]J}$. We will bound $C_{ij}$ and $b_i$ with

$$i = [B_u \cdots B_t]v \quad \text{and} \quad j = [B_{u'} \cdots B_t]v'$$

where $0 \leq u, u' \leq t$. In the trivial case that $u = 0$, we have $\rho_x^i = \delta_{x_0^v}$ and $\overline{\rho}_x^i = \delta_{\overline{x}_0^v}$, which yields that $C_{ij} = 0$ and $b_i \leq 2$.

**Step 1.** Consider $u \in [t-1]$ which implies $v \in k_u$. We have $\rho_x^i = \overline{\rho}_x^i$ and hence $b_i = 0$. Next, we take care of $C_{ij}$.

**Step 1.1.** When $v \notin k_{u-1}$ and $v \notin k_{u+1}$, we have

$$\rho_x^i(A) = \frac{\begin{matrix}\int \mathbb{1}_A(x_u^v) f_u^v(x_u^v \mid k_u^{v(\mathcal{R})}) g_u^v(y_u^v \mid k_u^{v(\mathcal{R})}, x_u^v) \widetilde{f}_u^v(x_u^v, x_u^{N_u(v)}) \\ \times p_{u+1}^{v(\mathcal{R})}(k_{u+1}^v \mid k_u, x_u^{v(\mathcal{R})}) \psi^v(dx_u^v)\end{matrix}}{\begin{matrix}\int f_u^v(x_u^v \mid k_u^{v(\mathcal{R})}) g_u^v(y_u^v \mid k_u^{v(\mathcal{R})}, x_u^v) \widetilde{f}_u^v(x_u^v, x_u^{N_u(v)}) \\ \times p_{u+1}^{v(\mathcal{R})}(k_{u+1}^v \mid k_u, x_u^{v(\mathcal{R})}) \psi^v(dx_u^v)\end{matrix}}.$$

Then when $u' = u$, if $v' \in N_u(v) \cup v(\mathcal{R})$, we have by Theorem S1.2 that $C_{ij} \leq \left(1 - \frac{\epsilon_d'}{\epsilon_u'}\frac{\kappa_d}{\kappa_u}\right)$, since

$$\rho_x^i(A) \geq \left(\frac{\epsilon_d'}{\epsilon_u'}\frac{\kappa_d}{\kappa_u}\right) \frac{\int \mathbb{1}_A(x_u^v) f_u^v(x_u^v \mid k_u^{v(\mathcal{R})}) g_u^v(y_u^v \mid k_u^{v(\mathcal{R})}, x_u^v) \psi^v(dx_u^v)}{\int f_u^v(x_u^v \mid k_u^{v(\mathcal{R})}) g_u^v(y_u^v \mid k_u^{v(\mathcal{R})}, x_u^v) \psi^v(dx_u^v)}$$

if $v' \notin N_u(v) \cup v(\mathcal{R})$ we have $C_{ij} = 0$. When $u' \neq u$, we have $C_{ij} = 0$.

**Step 1.2.** When $v \in k_{u-1}$ and $v \notin k_{u+1}$, we have

$$\rho_x^i(A) = \frac{\begin{matrix}\int \mathbb{1}_A(x_u^v) f_u^v(x_u^v \mid k_u^{v(\mathcal{R})}, x_{u-1}^v) g_u^v(y_u^v \mid k_u^{v(\mathcal{R})}, x_u^v) \widetilde{f}_u^v(x_u^v, x_u^{N_u(v)}) \\ \times p_{u+1}^{v(\mathcal{R})}(k_{u+1}^v \mid k_u, x_u^{v(\mathcal{R})}) \psi^v(dx_u^v)\end{matrix}}{\begin{matrix}\int f_u^v(x_u^v \mid k_u^{v(\mathcal{R})}, x_{u-1}^v) g_u^v(y_u^v \mid k_u^{v(\mathcal{R})}, x_u^v) \widetilde{f}_u^v(x_u^v, x_u^{N_u(v)}) \\ \times p_{u+1}^{v(\mathcal{R})}(k_{u+1}^v \mid k_u, x_u^{v(\mathcal{R})}) \psi^v(dx_u^v)\end{matrix}}.$$

Then when $u' = u-1$ which implies $v' \in k_{u-1}$, if $v' = v$ we have by Theorem S1.2 that $C_{ij} \leq \left(1 - \frac{\epsilon_d}{\epsilon_u}\right)$, since

$$\rho_x^i(A) \geq \left(\frac{\epsilon_d}{\epsilon_u}\right) \frac{\int \mathbb{1}_A(x_u^v) g_u^v(y_u^v \mid k_u^{v(\mathcal{R})}, x_u^v) \widetilde{f}_u^v(x_u^v, x_u^{N_u(v)}) p_{u+1}^v(k_{u+1}^{v(\mathcal{R})} \mid k_u, x_u^{v(\mathcal{R})}) \psi^v(dx_u^v)}{\int g_u^v(y_u^v \mid k_u^{v(\mathcal{R})}, x_u^v) \widetilde{f}_u^v(x_u^v, x_u^{N_u(v)}) p_{u+1}^v(k_{u+1}^{v(\mathcal{R})} \mid k_u, x_u^{v(\mathcal{R})}) \psi^v(dx_u^v)};$$

if $v' \neq v$ we have $C_{ij} = 0$. When $u' = u$, if $v' \in N_u(v) \cup v(\mathcal{R})$, we have by Theorem S1.2 that $C_{ij} \leq \left(1 - \frac{\epsilon_d'}{\epsilon_u'}\frac{\kappa_d}{\kappa_u}\right)$, since

$$\rho_x^i(A) \geq \left(\frac{\epsilon_d'}{\epsilon_u'}\frac{\kappa_d}{\kappa_u}\right) \frac{\int \mathbb{1}_A(x_u^v) f_u^v(x_u^v \mid k_u^{v(\mathcal{R})}, x_{u-1}^v) g_u^v(y_u^v \mid k_u^{v(\mathcal{R})}, x_u^v) \psi^v(dx_u^v)}{\int f_u^v(x_u^v \mid k_u^{v(\mathcal{R})}, x_{u-1}^v) g_u^v(y_u^v \mid k_u^{v(\mathcal{R})}, x_u^v) \psi^v(dx_u^v)};$$

if $v' \notin N_u(v) \cup v(\mathcal{R})$ we have $C_{ij} = 0$. When $u' \notin \{u, u-1\}$, we have $C_{ij} = 0$.

**Step 1.3.** When $v \notin k_{u-1}$ and $v \in k_{u+1}$, we have

$$\rho_x^i(A) = \frac{\begin{matrix}\int \mathbb{1}_A(x_u^v) f_u^v(x_u^v \mid k_u^{v(\mathcal{R})}) g_u^v(y_u^v \mid k_u^{v(\mathcal{R})}, x_u^v) \widetilde{f}_u^v(x_u^v, x_u^{N_u(v)}) \\ \times f_{u+1}^v(x_{u+1}^v \mid k_{u+1}^{v(\mathcal{R})}, x_u^v) p_{u+1}^{v(\mathcal{R})}(k_{u+1}^{v(\mathcal{R})} \mid k_u, x_u^{v(\mathcal{R})}) \psi^v(dx_u^v)\end{matrix}}{\begin{matrix}\int f_u^v(x_u^v \mid k_u^{v(\mathcal{R})}) g_u^v(y_u^v \mid k_u^{v(\mathcal{R})}, x_u^v) \widetilde{f}_u^v(x_u^v, x_u^{N_u(v)}) \\ \times f_{u+1}^v(x_{u+1}^v \mid k_{u+1}^{v(\mathcal{R})}, x_u^v) p_{u+1}^{v(\mathcal{R})}(k_{u+1}^{v(\mathcal{R})} \mid k_u, x_u^{v(\mathcal{R})}) \psi^v(dx_u^v)\end{matrix}}.$$

Then $u' = u$ which implies $v' \in k_u$, if $v' \in N_u(v) \cup v(\mathcal{R})$, we have by Theorem S1.2 that $C_{ij} \leq \left(1 - \frac{\epsilon'_d}{\epsilon'_u} \frac{\kappa_d}{\kappa_u}\right)$, since

$$\rho^i_x(A) \geq \left(\frac{\epsilon'_d}{\epsilon'_u} \frac{\kappa_d}{\kappa_u}\right) \frac{\int \mathbb{1}_A(x^v_u) f^v_u(x^v_u \mid k^{v(\mathcal{R})}_u) g^v_u(y^v_u \mid k^{v(\mathcal{R})}_u, x^v_u) f^v_{u+1}(x^v_{u+1} \mid k^{v(\mathcal{R})}_{u+1}, x^v_u) \psi^v(dx^v_u)}{\int f^v_u(x^v_u \mid k^{v(\mathcal{R})}_u) g^v_u(y^v_u \mid k^{v(\mathcal{R})}_u, x^v_u) f^v_{u+1}(x^v_{u+1} \mid k^{v(\mathcal{R})}_{u+1}, x^v_u) \psi^v(dx^v_u)};$$

if $v' \notin N_u(v) \cup v(\mathcal{R})$ we have $C_{ij} = 0$. When $u' = u + 1$ which implies $v' \in k_{u+1}$, if $v' = v$ we have by Theorem S1.2 that $C_{ij} \leq \left(1 - \frac{\epsilon_d}{\epsilon_u}\right)$, since

$$\rho^i_x(A) \geq \left(\frac{\epsilon_d}{\epsilon_u}\right) \frac{\begin{array}{c}\int \mathbb{1}_A(x^v_u) f^v_u(x^v_u \mid k^{v(\mathcal{R})}_u) g^v_u(y^v_u \mid k^{v(\mathcal{R})}_u, x^v_u) \widetilde{f^v_u}(x^v_u, x^{N_u(v)}_u) \\ \times p^{v(\mathcal{R})}_{u+1}(k^{v(\mathcal{R})}_{u+1} \mid k_u, x^{v(\mathcal{R})}_u) \psi^v(dx^v_u)\end{array}}{\begin{array}{c}\int f^v_u(x^v_u \mid k^{v(\mathcal{R})}_u) g^v_u(y^v_u \mid k^{v(\mathcal{R})}_u, x^v_u) \widetilde{f^v_u}(x^v_u, x^{N_u(v)}_u) \\ \times p^{v(\mathcal{R})}_{u+1}(k^{v(\mathcal{R})}_{u+1} \mid k_u, x^{v(\mathcal{R})}_u) \psi^v(dx^v_u)\end{array}};$$

if $v' \neq v$ we have $C_{ij} = 0$. When $u' \notin \{u, u+1\}$, we have $C_{ij} = 0$.

**Step 1.4.** When $v \in k_{u-1}$ and $v \in k_{u+1}$, we have

$$\rho^i_x(A) = \frac{\begin{array}{c}\int \mathbb{1}_A(x^v_u) f^v_u(x^v_u \mid k^{v(\mathcal{R})}_u, x^v_{u-1}) g^v_u(y^v_u \mid k^{v(\mathcal{R})}_u, x^v_u) \widetilde{f^v_u}(x^v_u, x^{N_u(v)}_u) \\ \times f^v_{u+1}(x^v_{u+1} \mid k^{v(\mathcal{R})}_{u+1}, x^v_u) p^{v(\mathcal{R})}_{u+1}(k^{v(\mathcal{R})}_{u+1} \mid k_u, x^{v(\mathcal{R})}_u) \psi^v(dx^v_u)\end{array}}{\begin{array}{c}\int f^v_u(x^v_u \mid k^{v(\mathcal{R})}_u, x^v_{u-1}) g^v_u(y^v_u \mid k^{v(\mathcal{R})}_u, x^v_u) \widetilde{f^v_u}(x^v_u, x^{N_u(v)}_u) \\ \times f^v_{u+1}(x^v_{u+1} \mid k^{v(\mathcal{R})}_{u+1}, x^v_u) p^{v(\mathcal{R})}_{u+1}(k^{v(\mathcal{R})}_{u+1} \mid k_u, x^{v(\mathcal{R})}_u) \psi^v(dx^v_u)\end{array}}.$$

Then if $u' = u-1$ which implies $v' \in k_{u-1}$, when $v' = v$ we have by Theorem S1.2 that $C_{ij} \leq \left(1 - \frac{\epsilon_d}{\epsilon_u}\right)$, since

$$\rho^i_x(A) \geq \left(\frac{\epsilon_d}{\epsilon_u}\right) \frac{\begin{array}{c}\int \mathbb{1}_A(x^v_u) g^v_u(y^v_u \mid k^{v(\mathcal{R})}_u, x^v_u) \widetilde{f^v_u}(x^v_u, x^{N_u(v)}_u) \\ \times f^v_{u+1}(x^v_{u+1} \mid k^{v(\mathcal{R})}_{u+1}, x^v_u) p^{v(\mathcal{R})}_{u+1}(k^{v(\mathcal{R})}_{u+1} \mid k_u, x^{v(\mathcal{R})}_u) \psi^v(dx^v_u)\end{array}}{\begin{array}{c}\int g^v_u(y^v_u \mid k^{v(\mathcal{R})}_u, x^v_u) \widetilde{f^v_u}(x^v_u, x^{N_u(v)}_u) \\ \times f^v_{u+1}(x^v_{u+1} \mid k^{v(\mathcal{R})}_{u+1}, x^v_u) p^{v(\mathcal{R})}_{u+1}(k^{v(\mathcal{R})}_{u+1} \mid k_u, x^{v(\mathcal{R})}_u) \psi^v(dx^v_u)\end{array}};$$

$C_{ij} = 0$ otherwise. If $u' = u$ which implies $v' \in k_u$, when $v' \in N_u(v) \cup v(\mathcal{R})$ we have by Theorem S1.2 that $C_{ij} \leq \left(1 - \frac{\epsilon'_d}{\epsilon'_u} \frac{\kappa_d}{\kappa_u}\right)$, since

$$\rho^i_x(A) \geq \left(\frac{\epsilon'_d}{\epsilon'_u} \frac{\kappa_d}{\kappa_u}\right) \frac{\begin{array}{c}\int \mathbb{1}_A(x^v_u) f^v_u(x^v_u \mid k^{v(\mathcal{R})}_u, x^v_{u-1}) g^v_u(y^v_u \mid k^{v(\mathcal{R})}_u, x^v_u) \\ \times f^v_{u+1}(x^v_{u+1} \mid k^{v(\mathcal{R})}_{u+1}, x^v_u) \psi^v(dx^v_u)\end{array}}{\begin{array}{c}\int f^v_u(x^v_u \mid k^{v(\mathcal{R})}_u, x^v_{u-1}) g^v_u(y^v_u \mid k^{v(\mathcal{R})}_u, x^v_u) \\ \times f^v_{u+1}(x^v_{u+1} \mid k^{v(\mathcal{R})}_{u+1}, x^v_u) \psi^v(dx^v_u)\end{array}};$$

$C_{ij} = 0$ otherwise. When $u' = u+1$ which implies $v' \in k_{u+1}$, if $v' = v$ we have by Theorem S1.2 that

$C_{ij} \le \left(1 - \frac{\epsilon_d}{\epsilon_u}\right)$, since

$$\rho_x^i(A) \ge \left(\frac{\epsilon_d}{\epsilon_u}\right) \frac{\begin{array}{c}\int \mathbb{1}_A(x_u^v) f_u^v(x_u^v \mid k_u^{v(\mathcal{R})}, x_{u-1}^v) g_u^v(y_u^v \mid k_u^{v(\mathcal{R})}, x_u^v) \widetilde{f}_u^v(x_u^v, x_u^{N_u(v)}) \\ \times\, p_{u+1}^{v(\mathcal{R})}(k_{u+1}^{v} \mid k_u, x_u^{v(\mathcal{R})}) \psi^v(dx_u^v)\end{array}}{\begin{array}{c}\int f_u^v(x_u^v \mid k_u^{v(\mathcal{R})}, x_{u-1}^v) g_u^v(y_u^v \mid k_u^{v(\mathcal{R})}, x_u^v) \widetilde{f}_u^v(x_u^v, x_u^{N_u(v)}) \\ \times\, p_{u+1}^{v(\mathcal{R})}(k_{u+1}^{v} \mid k_u, x_u^{v(\mathcal{R})}) \psi^v(dx_u^v)\end{array}};$$

if $v' \ne v$ we have $C_{ij} = 0$. When $u' \notin \{u, u-1, u+1\}$, we have $C_{ij} = 0$.

**Step 2.** When $u = t$, which implies $v \in k_t$. We have $\rho_x^i = \overline{p}_x^i$ and hence $b_i = 0$. Next, we take care of $C_{ij}$.

**Step 2.1.** When $v \in k_{t-1}$, we have

$$\rho_x^i(A) = \frac{\int \mathbb{1}_A(x_t^v) f_t^v(x_t^v \mid k_t^{v(\mathcal{R})}, x_{t-1}^v) g_t^v(y_t^v \mid k_t^{v(\mathcal{R})}, x_t^v) \widetilde{f}_t^v(x_t^v, x_t^{N_t(v)}) \psi^v(dx_t^v)}{\int f_t^v(x_t^v \mid k_t^{v(\mathcal{R})}, x_{t-1}^v) g_t^v(y_t^v \mid k_t^{v(\mathcal{R})}, x_t^v) \widetilde{f}_t^v(x_t^v, x_t^{N_t(v)}) \psi^v(dx_t^v)}.$$

Then if $u' = t-1$ which implies $v' \in k_{t-1}$, when $v' = v$ we have by Theorem S1.2 that $C_{ij} \le \left(1 - \frac{\epsilon_d}{\epsilon_u}\right)$, since

$$\rho_x^i(A) \ge \left(\frac{\epsilon_d}{\epsilon_u}\right) \frac{\int \mathbb{1}_A(x_t^v) g_t^v(y_t^v \mid k_t^{v(\mathcal{R})}, x_t^v) \widetilde{f}_t^v(x_t^v, x_t^{N_t(v)}) \psi^v(dx_t^v)}{\int g_t^v(y_t^v \mid k_t^{v(\mathcal{R})}, x_t^v) \widetilde{f}_t^v(x_t^v, x_t^{N_t(v)}) \psi^v(dx_t^v)};$$

$C_{ij} = 0$ otherwise. If $u' = t$ which implies $v' \in k_t$, when $v' \in N_t(v)$ we have by Theorem S1.2 that $C_{ij} \le \left(1 - \frac{\epsilon_d'}{\epsilon_u'}\right)$, since

$$\rho_x^i(A) \ge \left(\frac{\epsilon_d'}{\epsilon_u'}\right) \frac{\int \mathbb{1}_A(x_t^v) f_t^v(x_t^v \mid k_t^{v(\mathcal{R})}, x_{t-1}^v) g_t^v(y_t^v \mid k_t^{v(\mathcal{R})}, x_t^v) \psi^v(dx_t^v)}{\int f_t^v(x_t^v \mid k_t^{v(\mathcal{R})}, x_{t-1}^v) g_t^v(y_t^v \mid k_t^{v(\mathcal{R})}, x_t^v) \psi^v(dx_t^v)};$$

$C_{ij} = 0$ otherwise. When $u' \notin \{t, t-1\}$, we have $C_{ij} = 0$.

**Step 2.2.** When $v \notin k_{t-1}$ we have

$$\rho_x^i(A) = \frac{\int \mathbb{1}_A(x_t^v) f_t^v(x_t^v \mid k_t^{v(\mathcal{R})}) g_t^v(y_t^v \mid k_t^{v(\mathcal{R})}, x_t^v) \widetilde{f}_t^v(x_t^v, x_t^{N_t(v)}) \psi^v(dx_t^v)}{\int f_t^v(x_t^v \mid k_t^{v(\mathcal{R})}) g_t^v(y_t^v \mid k_t^{v(\mathcal{R})}, x_t^v) \widetilde{f}_t^v(x_t^v, x_t^{N_t(v)}) \psi^v(dx_t^v)}.$$

Then if $u' = t-1$ we have $C_{ij} = 0$. If $u' = t$ which implies $v' \in k_t$, when $v' \in N_t(v)$ we have by Theorem S1.2 that $C_{ij} \le \left(1 - \frac{\epsilon_d'}{\epsilon_u'}\right)$, since

$$\rho_x^i(A) \ge \left(\frac{\epsilon_d'}{\epsilon_u'}\right) \frac{\int \mathbb{1}_A(x_t^v) f_t^v(x_t^v \mid k_t^{v(\mathcal{R})}) g_t^v(y_t^v \mid k_t^{v(\mathcal{R})}, x_t^v) \psi^v(dx_t^v)}{\int f_t^v(x_t^v \mid k_t^{v(\mathcal{R})}) g_t^v(y_t^v \mid k_t^{v(\mathcal{R})}, x_t^v) \psi^v(dx_t^v)};$$

$C_{ij} = 0$ otherwise. When $u' \notin \{t, t-1\}$, we have $C_{ij} = 0$.

**Step 3.** Summing up the above results, for $i = [B_u \cdots B_t]v$ and $j = [B_{u'} \cdots B_t]v'$, we have that

$$\max_{i \in \mathcal{V}^{B_t}} \sum_{j \in \mathcal{V}^{B_t}} e^{\beta_T |d(i) - d(j)|} C_{ij} \tag{S32}$$

$$= \max \left\{ \max_{\substack{i \in \mathcal{V}^{B_t} \\ u \in [t-1]}} \sum_{j \in \mathcal{V}^{B_t}} e^{\beta_T |d(i) - d(j)|} C_{ij}, \ \max_{\substack{i \in \mathcal{V}^{B_t} \\ u = t}} \sum_{j \in \mathcal{V}^{B_t}} e^{\beta_T |d(i) - d(j)|} C_{ij} \right\}.$$

For the first item in equation (S32), we have

$$\max_{\substack{i \in \mathcal{V}^{B_t} \\ u \in [t-1]}} \sum_{j \in \mathcal{V}^{B_t}} e^{\beta_T |d(i) - d(j)|} C_{ij}$$

$$= \max \left\{ \max_{\substack{i \in \mathcal{V}^{B_t} \\ u \in [t-1]}} \sum_{j \in \mathcal{V}^{B_t}} e^{\beta_T |u - u'|} e^{\beta_T d(v, v')} C_{ij} \mathbb{1}_{\{v \in k_u, \ v \notin k_{u-1}, \ v \notin k_{u+1}\}}, \right.$$

$$\max_{\substack{i \in \mathcal{V}^{B_t} \\ u \in [t-1]}} \sum_{j \in \mathcal{V}^{B_t}} e^{\beta_T |u - u'|} e^{\beta_T d(v, v')} C_{ij} \mathbb{1}_{\{v \in k_u, \ v \in k_{u-1}, \ v \notin k_{u+1}\}},$$

$$\max_{\substack{i \in \mathcal{V}^{B_t} \\ u \in [t-1]}} \sum_{j \in \mathcal{V}^{B_t}} e^{\beta_T |u - u'|} e^{\beta_T d(v, v')} C_{ij} \mathbb{1}_{\{v \in k_u, \ v \notin k_{u-1}, \ v \in k_{u+1}\}},$$

$$\left. \max_{\substack{i \in \mathcal{V}^{B_t} \\ u \in [t-1]}} \sum_{j \in \mathcal{V}^{B_t}} e^{\beta_T |u - u'|} e^{\beta_T d(v, v')} C_{ij} \mathbb{1}_{\{v \in k_u, \ v \in k_{u-1}, \ v \in k_{u+1}\}} \right\},$$

where

$$\max_{\substack{i \in \mathcal{V}^{B_t} \\ u \in [t-1]}} \sum_{j \in \mathcal{V}^{B_t}} e^{\beta_T |u - u'|} e^{\beta_T d(v, v')} C_{ij} \mathbb{1}_{\{v \in k_u, \ v \notin k_{u-1}, \ v \notin k_{u+1}\}}$$

$$\leq e^{\beta_T (r + r_T^{\mathcal{R}})} \left( 1 - \frac{\epsilon'_d}{\epsilon'_u} \frac{\kappa_d}{\kappa_u} \right) (\Delta_T + \Delta_T^{\mathcal{R}}),$$

$$\max_{\substack{i \in \mathcal{V}^{B_t} \\ u \in [t-1]}} \sum_{j \in \mathcal{V}^{B_t}} e^{\beta_T |u - u'|} e^{\beta_T d(v, v')} C_{ij} \mathbb{1}_{\{v \in k_u, \ v \in k_{u-1}, \ v \notin k_{u+1}\}}$$

$$\leq e^{\beta_T} \left( 1 - \frac{\epsilon_d}{\epsilon_u} \right) + e^{\beta_T (r + r_T^{\mathcal{R}})} \left( 1 - \frac{\epsilon'_d}{\epsilon'_u} \frac{\kappa_d}{\kappa_u} \right) (\Delta_T + \Delta_T^{\mathcal{R}}),$$

$$\max_{\substack{i \in \mathcal{V}^{B_t} \\ u \in [t-1]}} \sum_{j \in \mathcal{V}^{B_t}} e^{\beta_T |u - u'|} e^{\beta_T d(v, v')} C_{ij} \mathbb{1}_{\{v \in k_u, \ v \notin k_{u-1}, \ v \in k_{u+1}\}}$$

$$\leq e^{\beta_T (r + r_T^{\mathcal{R}})} \left( 1 - \frac{\epsilon'_d}{\epsilon'_u} \frac{\kappa_d}{\kappa_u} \right) (\Delta_T + \Delta_T^{\mathcal{R}}) + e^{\beta_T} \left( 1 - \frac{\epsilon_d}{\epsilon_u} \right),$$

$$\max_{\substack{i \in \mathcal{V}^{B_t} \\ u \in [t-1]}} \sum_{j \in \mathcal{V}^{B_t}} e^{\beta_T |u - u'|} e^{\beta_T d(v, v')} C_{ij} \mathbb{1}_{\{v \in k_u, \ v \in k_{u-1}, \ v \in k_{u+1}\}}$$

$$\leq e^{\beta_T (r + r_T^{\mathcal{R}})} \left( 1 - \frac{\epsilon'_d}{\epsilon'_u} \frac{\kappa_d}{\kappa_u} \right) (\Delta_T + \Delta_T^{\mathcal{R}}) + 2 e^{\beta_T} \left( 1 - \frac{\epsilon_d}{\epsilon_u} \right).$$

Therefore, for the first item in equation (S32),

$$\max_{\substack{i \in \mathcal{V}^{B_t} \\ u \in [t-1]}} \sum_{j \in \mathcal{V}^{B_t}} e^{\beta_T |u - u'|} e^{\beta_T d(v, v')} C_{ij} \tag{S33}$$

$$\leq e^{\beta_T (r + r_T^{\mathcal{R}})} \left( 1 - \frac{\epsilon'_d}{\epsilon'_u} \frac{\kappa_d}{\kappa_u} \right) (\Delta_T + \Delta_T^{\mathcal{R}}) + 2 e^{\beta_T} \left( 1 - \frac{\epsilon_d}{\epsilon_u} \right).$$

For the second item in equation (S32), we have

$$\max_{\substack{i \in \mathcal{V}^{B_t} \\ u = t}} \sum_{j \in \mathcal{V}^{B_t}} e^{\beta_T |d(i) - d(j)|} C_{ij}$$

$$= \max \left\{ \max_{\substack{i \in \mathcal{V}^{B_t} \\ u=t}} \sum_{j \in \mathcal{V}^{B_t}} e^{\beta_T |t-u'|} e^{\beta_T d(v,v')} C_{ij} \mathbb{1}_{\{v \in k_t,\ v \notin k_{t-1}\}}, \right.$$

$$\left. \max_{\substack{i \in \mathcal{V}^{B_t} \\ u=t}} \sum_{j \in \mathcal{V}^{B_t}} e^{\beta_T |t-u'|} e^{\beta_T d(v,v')} C_{ij} \mathbb{1}_{\{v \in k_t,\ v \in k_{t-1}\}} \right\},$$

where

$$\max_{\substack{i \in \mathcal{V}^{B_t} \\ u=t}} \sum_{j \in \mathcal{V}^{B_t}} e^{\beta_T |t-u'|} e^{\beta_T d(v,v')} C_{ij} \mathbb{1}_{\{v \in k_t,\ v \notin k_{t-1}\}} \leq e^{\beta_T r} \left( 1 - \frac{\epsilon'_d}{\epsilon'_u} \right) \Delta_T,$$

$$\max_{\substack{i \in \mathcal{V}^{B_t} \\ u=t}} \sum_{j \in \mathcal{V}^{B_t}} e^{\beta_T |t-u'|} e^{\beta_T d(v,v')} C_{ij} \mathbb{1}_{\{v \in k_t,\ v \in k_{t-1}\}} \leq e^{\beta_T} \left( 1 - \frac{\epsilon_d}{\epsilon_u} \right) + e^{\beta_T r} \left( 1 - \frac{\epsilon'_d}{\epsilon'_u} \right) \Delta_T.$$

Therefore, for the second item in equation (S32),

$$\max_{\substack{i \in \mathcal{V}^{B_t} \\ u=t}} \sum_{j \in \mathcal{V}^{B_t}} e^{\beta_T |d(i)-d(j)|} C_{ij} \leq e^{\beta_T} \left( 1 - \frac{\epsilon_d}{\epsilon_u} \right) + e^{\beta_T r} \left( 1 - \frac{\epsilon'_d}{\epsilon'_u} \right) \Delta_T.$$

Furthermore,

$$\max_{i \in \mathcal{V}^{B_t}} \sum_{j \in \mathcal{V}^{B_t}} e^{\beta_T |d(i)-d(j)|} C_{ij}$$

$$\leq \max \left\{ e^{\beta_T (r+r_T^{\mathcal{R}})} \left( 1 - \frac{\epsilon'_d}{\epsilon'_u} \frac{\kappa_d}{\kappa_u} \right) (\Delta_T + \Delta_T^{\mathcal{R}}) + 2 e^{\beta_T} \left( 1 - \frac{\epsilon_d}{\epsilon_u} \right), \right.$$

$$\left. e^{\beta_T} \left( 1 - \frac{\epsilon_d}{\epsilon_u} \right) + e^{\beta_T r} \left( 1 - \frac{\epsilon'_d}{\epsilon'_u} \right) \Delta_T \right\}$$

$$= e^{\beta_T (r+r_T^{\mathcal{R}})} \left( 1 - \frac{\epsilon'_d}{\epsilon'_u} \frac{\kappa_d}{\kappa_u} \right) (\Delta_T + \Delta_T^{\mathcal{R}}) + 2 e^{\beta_T} \left( 1 - \frac{\epsilon_d}{\epsilon_u} \right).$$

By the definition of $\beta_T$ given in equation (32), we have

$$e^{\beta_T (r+r_T^{\mathcal{R}})} \left( 1 - \frac{\epsilon'_d}{\epsilon'_u} \frac{\kappa_d}{\kappa_u} \right) (\Delta_T + \Delta_T^{\mathcal{R}}) \leq \frac{1}{6} \quad \text{and} \quad e^{\beta_T} \left( 1 - \frac{\epsilon_d}{\epsilon_u} \right) \leq \frac{1}{6}.$$

Thus by Proposition S1.1, we have

$$\max_{i \in \mathcal{V}^{B_t}} \sum_{j \in \mathcal{V}^{B_t}} e^{\beta_T |d(i)-d(j)|} C_{ij} \leq \frac{1}{2}.$$

By Dobrushin comparison theorem (Theorem 3.3) and Theorem S1.3, we obtain

$$\|\rho - \overline{\rho}\|_{[B_t] J} \leq 2 \times \frac{1}{1 - \frac{1}{2}} e^{-\beta_T t} \mathrm{card}(J) = 4 e^{-\beta_T t} \mathrm{card}(J),$$

which completes the proof by equation (S31).  □

In Lemma S1.8, we consider applying $\widetilde{\mathsf{F}}_t \cdots \widetilde{\mathsf{F}}_1$ on two different point masses. Now we extend that result to two different general measures.

**Proposition S1.9.** *Under Assumption 3.1, for any two measures $\mu$ and $\overline{\mu}$, and for $J \subseteq B_t$ and $B_t \in \mathcal{B}(k_t)$,*

$$\left\| \widetilde{\mathsf{F}}_t \cdots \widetilde{\mathsf{F}}_1 \mu - \widetilde{\mathsf{F}}_t \cdots \widetilde{\mathsf{F}}_1 \overline{\mu} \right\|_J \leq 4 e^{-\beta_\mathcal{T} t} \mathrm{card}(J) \left( \frac{\epsilon_u}{\epsilon_d} \frac{\kappa_u}{\kappa_d} \right)^{|\mathcal{B}|^{\widetilde{\mathcal{T}}}} \sum_{\tau \in \mathcal{V}_0^{B_t}} \| \mu^\tau - \overline{\mu}^\tau \|,$$

*Proof.* Recall that $\mathcal{V}_0^{B_t}$ is the set of vertices at time 0 that possibly affact $B_t$, i.e.,

$$\mathcal{V}_0^{B_t} = \left\{ [B_0 \cdots B_t] : B_l \in \mathcal{B}(k_l),\ B_l \cap B_{l+1} \neq \emptyset,\ 0 \leq l < t \right\}$$

and $\mathcal{V}_+^{B_t} = \mathcal{V}^{B_t} \backslash \mathcal{V}_0^{B_t}$ is the set of other vertices. For $i \in \mathcal{V}_+^{B_t}$, define transition

$$\mathcal{T}_i^{B_t} = f_{d(i)}^i(x_{d(i)}^i \mid k_{d(i)}^{i(\mathcal{R})}, x_{d(i)-1}^{k_{d(i)-1} \cap \{i\}}) g_{d(i)}^i(y_{d(i)}^i \mid k_{d(i)}^{i(\mathcal{R})}, x_{d(i)}^i) \widetilde{f}_{d(i)}^i(x_{d(i)}^i, x_{d(i)}^{N_{d(i)}(i)})$$
$$\times \prod_{R \in \mathcal{R}} p_{d(i)}^R(k_{d(i)}^R \mid k_{d(i)-1}, x_{d(i)-1}^{k_{d(i)-1} \cap R}),$$

based on which we define

$$\zeta(A) = \frac{\int \mathbb{1}_A(x_0^{\mathcal{V}_0^{B_t}}) \prod_{i \in \mathcal{V}_+^{B_t}} \mathcal{T}_i^{B_t} \psi^{v(i)}(dx_{d(i)}^i) \mu^{\mathcal{V}_0^{B_t}}(dx_0^{\mathcal{V}_0^{B_t}})}{\int \prod_{i \in \mathcal{V}_+^{B_t}} \mathcal{T}_i^{B_t} \psi^{v(i)}(dx_{d(i)}^i) \mu^{\mathcal{V}_0^{B_t}}(dx_0^{\mathcal{V}_0^{B_t}})},$$

$$\overline{\zeta}(A) = \frac{\int \mathbb{1}_A(x_0^{\mathcal{V}_0^{B_t}}) \prod_{i \in \mathcal{V}_+^{B_t}} \mathcal{T}_i^{B_t} \psi^{v(i)}(dx_{d(i)}^i) \overline{\mu}^{\mathcal{V}_0^{B_t}}(dx_0^{\mathcal{V}_0^{B_t}})}{\int \prod_{i \in \mathcal{V}_+^{B_t}} \mathcal{T}_i^{B_t} \psi^{v(i)}(dx_{d(i)}^i) \overline{\mu}^{\mathcal{V}_0^{B_t}}(dx_0^{\mathcal{V}_0^{B_t}})}.$$

Since,

$$(\widetilde{\mathsf{F}}_t \cdots \widetilde{\mathsf{F}}_1 \mu)(A) = \frac{\int \mathbb{1}_A(x_t^J) \prod_{i \in \mathcal{V}_+^{B_t}} \mathcal{T}_i^{B_t} \psi^{v(i)}(dx_{d(i)}^i) \mu^{\mathcal{V}_0^{B_t}}(dx_0^{\mathcal{V}_0^{B_t}})}{\int \prod_{i \in \mathcal{V}_+^{B_t}} \mathcal{T}_i^{B_t} \psi^{v(i)}(dx_{d(i)}^i) \mu^{\mathcal{V}_0^{B_t}}(dx_0^{\mathcal{V}_0^{B_t}})}$$
$$= \int \frac{\int \mathbb{1}_A(x_t^J) \prod_{i \in \mathcal{V}_+^{B_t}} \mathcal{T}_i^{B_t} \psi^{v(i)}(dx_{d(i)}^i)}{\int \prod_{i \in \mathcal{V}_+^{B_t}} \mathcal{T}_i^{B_t} \psi^{v(i)}(dx_{d(i)}^i)} \zeta(dx_0^{\mathcal{V}_0^{B_t}}), \tag{S34}$$

we have

$$\left\| \widetilde{\mathsf{F}}_t \cdots \widetilde{\mathsf{F}}_1 \mu - \widetilde{\mathsf{F}}_t \cdots \widetilde{\mathsf{F}}_1 \overline{\mu} \right\|_J = 2 \sup_A \left| \int \frac{\int \mathbb{1}_A(x_t^J) \prod_{i \in \mathcal{V}_+^{B_t}} \mathcal{T}_i^{B_t} \psi^{v(i)}(dx_{d(i)}^i)}{\int \prod_{i \in \mathcal{V}_+^{B_t}} \mathcal{T}_i^{B_t} \psi^{v(i)}(dx_{d(i)}^i)} \zeta(dx_0^{\mathcal{V}_0^{B_t}}) \right.$$
$$\left. - \int \frac{\int \mathbb{1}_A(x_t^J) \prod_{i \in \mathcal{V}_+^{B_t}} \mathcal{T}_i^{B_t} \psi^{v(i)}(dx_{d(i)}^i)}{\int \prod_{i \in \mathcal{V}_+^{B_t}} \mathcal{T}_i^{B_t} \psi^{v(i)}(dx_{d(i)}^i)} \overline{\zeta}(dx_0^{\mathcal{V}_0^{B_t}}) \right|$$
$$= \frac{1}{2} \mathrm{osc} \left( \frac{\int \mathbb{1}_A(x_t^J) \prod_{i \in \mathcal{V}_+^{B_t}} \mathcal{T}_i^{B_t} \psi^{v(i)}(dx_{d(i)}^i)}{\int \prod_{i \in \mathcal{V}_+^{B_t}} \mathcal{T}_i^{B_t} \psi^{v(i)}(dx_{d(i)}^i)} \right) \| \zeta - \overline{\zeta} \|, \tag{S35}$$

where we used equation (8.1) in Georgii [2011] that with $\mathrm{osc}(h)$ standing for the oscillation ($h_{\max} - h_{\min}$) of any function $h$,

$$\sup_A |\rho(A) - \rho'(A)| = \sup_h |\rho(h) - \rho'(h)| / \mathrm{osc}(h).$$

Next, since

$$\frac{\int \mathbb{1}_A(x_t^J) \prod_{i \in \mathcal{V}_+^{B_t}} \mathcal{T}_i^{B_t} \psi^{v(i)}(dx_{d(i)}^i)}{\int \prod_{i \in \mathcal{V}_+^{B_t}} \mathcal{T}_i^{B_t} \psi^{v(i)}(dx_{d(i)}^i)}$$

is exactly a filter obtained when the initial condition is a point mass, by Lemma S1.8

$$\left\| \widetilde{\mathsf{F}}_t \cdots \widetilde{\mathsf{F}}_1 \delta_x - \widetilde{\mathsf{F}}_t \cdots \widetilde{\mathsf{F}}_1 \delta_{\overline{x}} \right\|_J \le 4 e^{-\beta_T t} \mathrm{card}(J),$$

we have that

$$\mathrm{osc}\left( \frac{\int \mathbb{1}_A(x_t^J) \prod_{i \in \mathcal{V}_+^{B_t}} \mathcal{T}_i^{B_t} \psi^{v(i)}(dx_{d(i)}^i)}{\int \prod_{i \in \mathcal{V}_+^{B_t}} \mathcal{T}_i^{B_t} \psi^{v(i)}(dx_{d(i)}^i)} \right) \le 2 e^{-\beta_T t} \mathrm{card}(J).$$

Plugging into equation (S35), we have

$$\left\| \widetilde{\mathsf{F}}_t \cdots \widetilde{\mathsf{F}}_1 \mu - \widetilde{\mathsf{F}}_t \cdots \widetilde{\mathsf{F}}_1 \overline{\mu} \right\|_J \le 2 e^{-\beta_T t} \mathrm{card}(J) \| \zeta - \overline{\zeta} \|. \tag{S36}$$

Since

$$(\epsilon_d \kappa_d)^{|\mathcal{B}|_T^\infty} \mathcal{C} \le \prod_{i \in \mathcal{V}_+^{B_t}} \mathcal{T}_i^{B_t} \psi^{v(i)}(dx_{d(i)}^i) \le (\epsilon_u \kappa_u)^{|\mathcal{B}|_T^\infty} \mathcal{C},$$

where $\mathcal{C}$ has no term in $\mathcal{V}_0^{B_t}$ that is defined as below

$$\mathcal{C} = \prod_{i \in \mathcal{V}_+^{B_t}, \, d(i)=1} g_1^i(y_1^i \mid k_1^{i(\mathcal{R})}, x_1^i) \widetilde{f}_1^i(x_1^i, x_1^{N_1(i)}) \psi^{v(i)}(dx_1^i) \prod_{i \in \mathcal{V}_+^{B_t}, \, d(i) \ne 1} \mathcal{T}_i^{B_t} \psi^{v(i)}(dx_{d(i)}^i),$$

we have by Lemma 4.16 of Rebeschini and Van Handel [2015] that

$$\| \zeta - \overline{\zeta} \| \le 2 \left( \frac{\epsilon_u}{\epsilon_d} \frac{\kappa_u}{\kappa_d} \right)^{|\mathcal{B}|_T^\infty} \sum_{\tau \in \mathcal{V}_0^{B_t}} \| \mu^\tau - \overline{\mu}^\tau \|.$$

Plugging into equation (S36), we have

$$\left\| \widetilde{\mathsf{F}}_t \cdots \widetilde{\mathsf{F}}_1 \mu - \widetilde{\mathsf{F}}_t \cdots \widetilde{\mathsf{F}}_1 \overline{\mu} \right\|_J \le 4 e^{-\beta_T t} \mathrm{card}(J) \left( \frac{\epsilon_u}{\epsilon_d} \frac{\kappa_u}{\kappa_d} \right)^{|\mathcal{B}|_T^\infty} \sum_{\tau \in \mathcal{V}_0^{B_t}} \| \mu^\tau - \overline{\mu}^\tau \|,$$

as desired. □

At last, we explore the one-step difference caused by applying $\widetilde{\mathsf{F}}_s$ and $\widehat{\mathsf{F}}_s$.

**Proposition S1.10.** *Under Assumption 3.1, for integer $s \in [T-1]$ and $B_{s+1} \in \mathcal{B}(k_{s+1})$, one has*

$$\left\| \widetilde{\mathsf{F}}_{s+1} \widetilde{\mathsf{F}}_s \widehat{\pi}_{s-1} - \widetilde{\mathsf{F}}_{s+1} \widehat{\mathsf{F}}_s \widehat{\pi}_{s-1} \right\|_{B_{s+1}} \le \frac{16}{\sqrt{N}} \left( \frac{\epsilon_u}{\epsilon_d \epsilon_d'} \right)^{|\mathcal{B}|_T^\infty} \left( \frac{\gamma_u}{\gamma_d} \frac{\epsilon_u'}{\epsilon_d'} \right)^{|\mathcal{B}|_T^\infty + (|\mathcal{B}|_T^\infty)^2} |\mathcal{B}|_T^\infty,$$

*where $|\mathcal{B}|_T^\infty$ is the maximal size of one single cluster in the partition up to time $T$ defined in equation (14).*

*Proof.* For any cluster $B_{s+1} \in \mathcal{B}(k_{s+1})$ and the cluster operator $\mathsf{B}_{s+1}^{B_{s+1}}$ on $B_{s+1}$ at time $s+1$ defined in equation (28), we have by Theorem S1.6 that

$$
\left\| \widetilde{\mathsf{F}}_{s+1} \widetilde{\mathsf{F}}_s \widehat{\pi}_{s-1} - \widetilde{\mathsf{F}}_{s+1} \widehat{\mathsf{F}}_s \widehat{\pi}_{s-1} \right\|_{B_{s+1}}
$$

$$
= \left\| \mathsf{C}_{s+1} \mathsf{B}_{s+1} \mathsf{P}_{s+1} \widetilde{\mathsf{F}}_s \widehat{\pi}_{s-1} - \mathsf{C}_{s+1} \mathsf{B}_{s+1} \mathsf{P}_{s+1} \widehat{\mathsf{F}}_s \widehat{\pi}_{s-1} \right\|_{B_{s+1}}
$$

$$
= \left\| \mathsf{C}_{s+1}^{B_{s+1}} \mathsf{B}_{s+1}^{B_{s+1}} \mathsf{P}_{s+1} \widetilde{\mathsf{F}}_s \widehat{\pi}_{s-1} - \mathsf{C}_{s+1}^{B_{s+1}} \mathsf{B}_{s+1}^{B_{s+1}} \mathsf{P}_{s+1} \widehat{\mathsf{F}}_s \widehat{\pi}_{s-1} \right\|
$$

$$
\leq 2 \left( \frac{\gamma_u}{\gamma_d} \frac{\epsilon_u'}{\epsilon_d'} \right)^{|\mathcal{B}|_T^\infty} \left\| \mathsf{B}_{s+1}^{B_{s+1}} \mathsf{P}_{s+1} \widetilde{\mathsf{F}}_s \widehat{\pi}_{s-1} - \mathsf{B}_{s+1}^{B_{s+1}} \mathsf{P}_{s+1} \widehat{\mathsf{F}}_s \widehat{\pi}_{s-1} \right\|. \tag{S37}
$$

By equation (8.1) in Georgii [2011], we have

$$
\left\| \mathsf{B}_{s+1}^{B_{s+1}} \mathsf{P}_{s+1} \widetilde{\mathsf{F}}_s \widehat{\pi}_{s-1} - \mathsf{B}_{s+1}^{B_{s+1}} \mathsf{P}_{s+1} \widehat{\mathsf{F}}_s \widehat{\pi}_{s-1} \right\|
$$

$$
= \int \left| \frac{ \left( \mathsf{B}_{s+1}^{B_{s+1}} \mathsf{P}_{s+1} \widetilde{\mathsf{F}}_s \widehat{\pi}_{s-1} \right)(dx_{s+1}^{B_{s+1}}) }{ \psi^{B_{s+1}}(dx_{s+1}^{B_{s+1}}) } - \frac{ \left( \mathsf{B}_{s+1}^{B_{s+1}} \mathsf{P}_{s+1} \widehat{\mathsf{F}}_s \widehat{\pi}_{s-1} \right)(dx_{s+1}^{B_{s+1}}) }{ \psi^{B_{s+1}}(dx_{s+1}^{B_{s+1}}) } \right| \psi^{B_{s+1}}(dx_{s+1}^{B_{s+1}}),
$$

where $\psi^{B_{s+1}}(dx_{s+1}^{B_{s+1}})$ is defined in equation (6) and the two items in the above difference are given as follows:

$$
\frac{ \left( \mathsf{B}_{s+1}^{B_{s+1}} \mathsf{P}_{s+1} \widetilde{\mathsf{F}}_s \widehat{\pi}_{s-1} \right)(dx_{s+1}^{B_{s+1}}) }{ \psi^{B_{s+1}}(dx_{s+1}^{B_{s+1}}) } = \frac{ \begin{array}{c} \int \prod_{v \in B_{s+1}} f_{s+1}^v (x_{s+1}^{k_{s+1} \cap \{v\}} \mid k_{s+1}^{v(\mathcal{R})}, x_s^{k_s \cap \{v\}}) \\ \times \prod_{B_s' \in \mathcal{B}(k_s), \, B_s' \cap B_{s+1} \neq \emptyset} \prod_{v' \in B_s'} g_s^{v'} (Y_s^{v'} \mid k_s^{v'(\mathcal{R})}, x_s^{v'}) \\ \times \widetilde{f}_s^{v'} (x_s^{v'}, x_s^{N_s(v')}) \left[ \mathsf{B}^{B_s'} \mathsf{P}_s \widehat{\pi}_{s-1} \right](dx_s^{B_s'}) \end{array} }{ \begin{array}{c} \int \prod_{B_s' \in \mathcal{B}(k_s \cap B_{s+1})} \prod_{v' \in B_s'} g_s^{v'} (Y_s^{v'} \mid k_s^{v'(\mathcal{R})}, x_s^{v'}) \\ \times \widetilde{f}_s^{v'} (x_s^{v'}, x_s^{N_s(v')}) \left[ \mathsf{B}^{B_s'} \mathsf{P}_s \widehat{\pi}_{s-1} \right](dx_s^{B_s'}) \end{array} },
$$

$$
\frac{ \left( \mathsf{B}_{s+1}^{B_{s+1}} \mathsf{P}_{s+1} \widehat{\mathsf{F}}_s \widehat{\pi}_{s-1} \right)(dx_{s+1}^{B_{s+1}}) }{ \psi^{B_{s+1}}(dx_{s+1}^{B_{s+1}}) } = \frac{ \begin{array}{c} \int \prod_{v \in B_{s+1}} f_{s+1}^v (x_{s+1}^{k_{s+1} \cap \{v\}} \mid k_{s+1}^{v(\mathcal{R})}, x_s^{k_s \cap \{v\}}) \\ \times \prod_{B_s' \in \mathcal{B}(k_s), \, B_s' \cap B_{s+1} \neq \emptyset} \prod_{v' \in B_s'} g_s^{v'} (Y_s^{v'} \mid k_s^{v'(\mathcal{R})}, x_s^{v'}) \\ \times \widetilde{f}_s^{v'} (x_s^{v'}, x_s^{N_s(v')}) \left[ \mathsf{B}^{B_s'} \mathsf{S}^N \mathsf{P}_s \widehat{\pi}_{s-1} \right](dx_s^{B_s'}) \end{array} }{ \begin{array}{c} \int \prod_{B_s' \in \mathcal{B}(k_s \cap B_{s+1})} \prod_{v' \in B_s'} g_s^{v'} (Y_s^{v'} \mid k_s^{v'(\mathcal{R})}, x_s^{v'}) \\ \times \widetilde{f}_s^{v'} (x_s^{v'}, x_s^{N_s(v')}) \left[ \mathsf{B}^{B_s'} \mathsf{S}^N \mathsf{P}_s \widehat{\pi}_{s-1} \right](dx_s^{B_s'}) \end{array} }.
$$

By Minkowski's integral inequality,

$$
\left[ \mathbb{E} \left\| \mathsf{B}_{s+1}^{B_{s+1}} \mathsf{P}_{s+1} \widetilde{\mathsf{F}}_s \widehat{\pi}_{s-1} - \mathsf{B}_{s+1}^{B_{s+1}} \mathsf{P}_{s+1} \widehat{\mathsf{F}}_s \widehat{\pi}_{s-1} \right\|^2 \right]^{1/2}
$$

$$
\leq \int \left[ \mathbb{E} \left| \frac{ \left( \mathsf{B}_{s+1}^{B_{s+1}} \mathsf{P}_{s+1} \widetilde{\mathsf{F}}_s \widehat{\pi}_{s-1} \right)(dx_{s+1}^{B_{s+1}}) }{ \psi^{B_{s+1}}(dx_{s+1}^{B_{s+1}}) } \right. \right.
$$

$$-\frac{\left(\mathsf{B}_{s+1}^{B_{s+1}}\mathsf{P}_{s+1}\widehat{\mathsf{F}}_s\widehat{\pi}_{s-1}\right)(dx_{s+1}^{B_{s+1}})}{\psi^{B_{s+1}}(dx_{s+1}^{B_{s+1}})}\Bigg|^2\Bigg]^{1/2}\psi^{B_{s+1}}(dx_{s+1}^{B_{s+1}})$$

$$\leq \psi^{B_{s+1}}(\mathbb{X}^{B_{s+1}})\sup_{x^{B_{s+1}}\in\mathbb{X}^{B_{s+1}}}\Bigg[\mathbb{E}\Bigg|\frac{\left(\mathsf{B}_{s+1}^{B_{s+1}}\mathsf{P}_{s+1}\widetilde{\mathsf{F}}_s\widetilde{\pi}_{s-1}\right)(dx_{s+1}^{B_{s+1}})}{\psi^{B_{s+1}}(dx_{s+1}^{B_{s+1}})}$$

$$-\frac{\left(\mathsf{B}_{s+1}^{B_{s+1}}\mathsf{P}_{s+1}\widehat{\mathsf{F}}_s\widehat{\pi}_{s-1}\right)(dx_{s+1}^{B_{s+1}})}{\psi^{B_{s+1}}(dx_{s+1}^{B_{s+1}})}\Bigg|^2\Bigg]^{1/2}.$$

By Assumption 3.1 and the fact that $f_t(x_t^{k_t}\mid k_t,x_{t-1}^{k_{t-1}})$ defined in equation (8) is a transition density, we have that

$$\epsilon_d\epsilon_d'\psi^v(\mathbb{X}^v)\leq\int f_t^v(x_t^v\mid k_t^{v(\mathcal{R})},x_{t-1}^{k_{t-1}\cap\{v\}})\widetilde{f}_t^v(x_t^v,x_t^{N_t(v)})\psi^v(dx_t^v)=1,$$

which yields

$$\psi^v(\mathbb{X}^v)\leq(\epsilon_d\epsilon_d')^{-1}\quad\text{and}\quad\psi^{B_{s+1}}(\mathbb{X}^{B_{s+1}})\leq(\epsilon_d\epsilon_d')^{-|\mathcal{B}|_T^\infty}.$$

Furthermore, by Assumption 3.1, we have

$$\prod_{v\in B_{s+1}}f_{s+1}^v(x_{s+1}^{k_{s+1}\cap\{v\}}\mid k_{s+1}^{v(\mathcal{R})},x_s^{k_s\cap\{v\}})\leq\epsilon_u^{|\mathcal{B}|_T^\infty}$$

and

$$(\epsilon_d'\gamma_d)^{(|\mathcal{B}|_T^\infty)^2}\leq\prod_{\substack{B_s'\in\mathcal{B}(k_s)\\B_s'\cap B_{s+1}\neq\emptyset}}\prod_{v'\in B_s'}g_s^{v'}(Y_s^{v'}\mid k_s^{v'(\mathcal{R})},x_s^{v'})\widetilde{f}_s^{v'}(x_s^{v'},x_s^{N_s(v')})\leq(\epsilon_u'\gamma_u)^{(|\mathcal{B}|_T^\infty)^2}.$$

Hence,

$$\left[\mathbb{E}\left\|\mathsf{B}_{s+1}^{B_{s+1}}\mathsf{P}_{s+1}\widetilde{\mathsf{F}}_s\widetilde{\pi}_{s-1}-\mathsf{B}_{s+1}^{B_{s+1}}\mathsf{P}_{s+1}\widehat{\mathsf{F}}_s\widehat{\pi}_{s-1}\right\|^2\right]^{1/2}$$

$$\leq 2\left(\frac{\epsilon_u}{\epsilon_d\epsilon_d'}\right)^{|\mathcal{B}|_T^\infty}\left(\frac{\gamma_u}{\gamma_d}\frac{\epsilon_u'}{\epsilon_d'}\right)^{(|\mathcal{B}|_T^\infty)^2}\left\|\bigotimes_{\substack{B_s'\in\mathcal{B}(k_s)\\B_s'\cap B_{s+1}\neq\emptyset}}\mathsf{B}_s^{B_s'}\mathsf{P}_s\widehat{\pi}_{s-1}-\bigotimes_{\substack{B_s'\in\mathcal{B}(k_s)\\B_s'\cap B_{s+1}\neq\emptyset}}\mathsf{B}_s^{B_s'}\mathsf{S}^N\mathsf{P}_s\widehat{\pi}_{s-1}\right\|$$

$$\leq\frac{8}{\sqrt{N}}\left(\frac{\epsilon_u}{\epsilon_d\epsilon_d'}\right)^{|\mathcal{B}|_T^\infty}\left(\frac{\gamma_u}{\gamma_d}\frac{\epsilon_u'}{\epsilon_d'}\right)^{(|\mathcal{B}|_T^\infty)^2}|\mathcal{B}|_T^\infty,$$

where the first inequality is by Theorem S1.6 and Assumption 3.1, and the second inequality is by Corollary 4.21 of Rebeschini and Van Handel [2015]. Now, plugging the above result into equation (S37) yields that

$$\left\|\widetilde{\mathsf{F}}_{s+1}\widetilde{\mathsf{F}}_s\widehat{\pi}_{s-1}-\widetilde{\mathsf{F}}_{s+1}\widehat{\mathsf{F}}_s\widehat{\pi}_{s-1}\right\|_{B_{s+1}}\leq 2\left(\frac{\gamma_u}{\gamma_d}\frac{\epsilon_u'}{\epsilon_d'}\right)^{|\mathcal{B}|_T^\infty}\frac{8}{\sqrt{N}}\left(\frac{\epsilon_u}{\epsilon_d\epsilon_d'}\right)^{|\mathcal{B}|_T^\infty}\left(\frac{\gamma_u}{\gamma_d}\frac{\epsilon_u'}{\epsilon_d'}\right)^{(|\mathcal{B}|_T^\infty)^2}|\mathcal{B}|_T^\infty$$

$$=\frac{16}{\sqrt{N}}\left(\frac{\epsilon_u}{\epsilon_d\epsilon_d'}\right)^{|\mathcal{B}|_T^\infty}\left(\frac{\gamma_u}{\gamma_d}\frac{\epsilon_u'}{\epsilon_d'}\right)^{|\mathcal{B}|_T^\infty+(|\mathcal{B}|_T^\infty)^2}|\mathcal{B}|_T^\infty.$$

$\square$

## S2. Proofs of the main theorems

In this section, we provide the proofs for Theorems 3.2 and 3.4.

### S2.1. Proof of Theorem 3.2

Since we can write $\pi_T$ and $\widetilde{\pi}_T$ in a recursive way as follows:

$$\pi_T = \mathsf{F}_T \mathsf{F}_{T-1} \cdots \mathsf{F}_{s+1} \mathsf{F}_s \mathsf{F}_{s-1} \cdots \mathsf{F}_1 \pi_0,$$

$$\widetilde{\pi}_T = \widetilde{\mathsf{F}}_T \widetilde{\mathsf{F}}_{T-1} \cdots \widetilde{\mathsf{F}}_{s+1} \widetilde{\mathsf{F}}_s \widetilde{\mathsf{F}}_{s-1} \cdots \widetilde{\mathsf{F}}_1 \widetilde{\pi}_0,$$

where $\pi_0 = \widetilde{\pi}_0$, we can bound $\|\widetilde{\pi}_T - \pi_T\|_J$ by means of error decomposition

$$\|\widetilde{\pi}_T - \pi_T\|_J \leq \sum_{s=1}^{T} \|\mathsf{F}_T \cdots \mathsf{F}_{s+1} \widetilde{\mathsf{F}}_s \widetilde{\pi}_{s-1} - \mathsf{F}_T \cdots \mathsf{F}_{s+1} \mathsf{F}_s \widetilde{\pi}_{s-1}\|_J. \tag{S38}$$

For $s \in [T-1]$ in equation (S38), we have that for every $J \subseteq B_T$ and $B_T \in \mathcal{B}(k_T)$,

$$
\begin{aligned}
&\left\| \mathsf{F}_T \cdots \mathsf{F}_{s+1} \mathsf{F}_s \widetilde{\pi}_{s-1} - \mathsf{F}_T \cdots \mathsf{F}_{s+1} \widetilde{\mathsf{F}}_s \widetilde{\pi}_{s-1} \right\|_J \\
&\leq 2 e^{-\beta_T(T-s)} \sum_{v \in J} \max_{v' \in k_s} e^{-\beta_T d(v,v')} \sup_{x_s^{k_s}, x_{s+1}^{k_{s+1}} \in \mathcal{X}} \left\| (\widetilde{\mathsf{F}}_s \widetilde{\pi}_{s-1})_{\chi_s, \chi_{s+1}}^{v'} - (\mathsf{F}_s \widetilde{\pi}_{s-1})_{\chi_s, \chi_{s+1}}^{v'} \right\| \\
&= 2 e^{-\beta_T(T-s)} \sum_{v \in J} \max_{v' \in B_s', B_s' \in \mathcal{B}(k_s)} e^{-\beta_T d(v,v')} \sup_{x_s^{k_s}, x_{s+1}^{k_{s+1}} \in \mathcal{X}} \left\| (\widetilde{\mathsf{F}}_s \widetilde{\pi}_{s-1})_{\chi_s, \chi_{s+1}}^{v'} - (\mathsf{F}_s \widetilde{\pi}_{s-1})_{\chi_s, \chi_{s+1}}^{v'} \right\| \\
&\leq 8 e^{-\beta_T(T-s)} \sum_{v \in J} \max_{v' \in B_s', B_s' \in \mathcal{B}(k_s)} e^{-\beta_T} e^{-\beta_T d(v,v')} e^{-\beta_T d(v', \partial B_s')} \left( 1 - \frac{\epsilon_d}{\epsilon_u} \right) \\
&\leq 8 e^{-\beta_T(T-s)} \sum_{v \in J} \max_{v' \in B_s', B_s' \in \mathcal{B}(k_s)} e^{-\beta_T} e^{-\beta_T d(v, \partial B_s')} \left( 1 - \frac{\epsilon_d}{\epsilon_u} \right) \\
&\leq 8 e^{-\beta_T(T-s+1)} \max_{B_s' \in \mathcal{B}(k_s)} e^{-\beta_T d(J, \partial B_s')} \left( 1 - \frac{\epsilon_d}{\epsilon_u} \right) \mathrm{card}(J), 
\end{aligned}
\tag{S39}
$$

where the first inequality is by Proposition S1.5, the second inequality is by Proposition S1.7, the third inequality is by the triangle inequality, and the last inequality is by the definition of $d(J, B_s')$ in equation (12). Next, when $s = T$ in equation (S38), by Proposition S1.4, we have for every $B_T \in \mathcal{B}(k_T)$ and $J \subseteq B_T$,

$$\left\| \mathsf{F}_T \widetilde{\pi}_{T-1} - \widetilde{\mathsf{F}}_T \widetilde{\pi}_{T-1} \right\|_J \leq 4 \left( 1 - \frac{\epsilon_d}{\epsilon_u} \right) e^{-\beta_T} e^{-\beta_T d(J, \partial B_T)} \mathrm{card}(J). \tag{S40}$$

Plugging equations (S39) and (S40) into equation (S38), we have

$$
\begin{aligned}
\|\widetilde{\pi}_T - \pi_T\|_J &\leq \sum_{s=1}^{T-1} 8 e^{-\beta_T(T-s+1)} \max_{B_s' \in \mathcal{B}(k_s)} e^{-\beta_T d(J, \partial B_s')} \left( 1 - \frac{\epsilon_d}{\epsilon_u} \right) \mathrm{card}(J) \\
&\quad + 4 \left( 1 - \frac{\epsilon_d}{\epsilon_u} \right) e^{-\beta_T} e^{-\beta_T d(J, \partial B_T)} \mathrm{card}(J) \\
&< \sum_{s=1}^{T} 8 e^{-\beta_T(T-s+1)} \left( 1 - \frac{\epsilon_d}{\epsilon_u} \right) \mathrm{card}(J) \max_{B_s' \in \mathcal{B}(k_s)} e^{-\beta_T d(J, \partial B_s')}
\end{aligned}
$$

$$< 8 \frac{e^{-\beta_T}}{1 - e^{-\beta_T}} \left(1 - \frac{\epsilon_d}{\epsilon_u}\right) \operatorname{card}(J) \left[\max_{s \in [T]} \max_{B'_s \in \mathcal{B}(k_s)} e^{-\beta_T d(J, \partial B'_s)}\right]$$

as desired.

## S2.2. Proof of Theorem 3.4

Since we can write $\widetilde{\pi}_T$ and $\widehat{\pi}_T$ in a recursive way as follows:

$$\widetilde{\pi}_T = \widetilde{\mathsf{F}}_T \widetilde{\mathsf{F}}_{T-1} \cdots \widetilde{\mathsf{F}}_{s+1} \widetilde{\mathsf{F}}_s \widetilde{\mathsf{F}}_{s-1} \cdots \widetilde{\mathsf{F}}_1 \widetilde{\pi}_0,$$
$$\widehat{\pi}_T = \widehat{\mathsf{F}}_T \widehat{\mathsf{F}}_{T-1} \cdots \widehat{\mathsf{F}}_{s+1} \widehat{\mathsf{F}}_s \widehat{\mathsf{F}}_{s-1} \cdots \widehat{\mathsf{F}}_1 \widehat{\pi}_0,$$

where $\widehat{\pi}_0 = \widetilde{\pi}_0$, we can bound $\|\|\widetilde{\pi}_T - \widehat{\pi}_T\|\|_J$ by means of error decomposition

$$\|\|\widetilde{\pi}_T - \widehat{\pi}_T\|\|_J \leq \sum_{s=1}^{T} \left\|\left\| \widetilde{\mathsf{F}}_T \cdots \widetilde{\mathsf{F}}_{s+1} \widetilde{\mathsf{F}}_s \widehat{\pi}_{s-1} - \widetilde{\mathsf{F}}_T \cdots \widetilde{\mathsf{F}}_{s+1} \widehat{\mathsf{F}}_s \widehat{\pi}_{s-1} \right\|\right\|_J. \tag{S41}$$

In Proposition S1.9, we obtained that for any two measures $\mu$ and $\overline{\mu}$, and for $J \subseteq B_T$ and $B_T \in \mathcal{B}(k_T)$,

$$\left\|\widetilde{\mathsf{F}}_T \cdots \widetilde{\mathsf{F}}_1 \mu - \widetilde{\mathsf{F}}_T \cdots \widetilde{\mathsf{F}}_1 \overline{\mu}\right\|_J \leq 4 e^{-\beta_T T} \operatorname{card}(J) \left(\frac{\epsilon_u}{\epsilon_d} \frac{\kappa_u}{\kappa_d}\right)^{|\mathcal{B}|_T^\infty} \sum_{\tau \in \mathcal{V}_0^{B_T}} \|\mu^\tau - \overline{\mu}^\tau\|,$$

where $\mathcal{V}_0^{B_T}$ is defined in equation (S30) standing for the set of clusters at time 0 that could possibly affact $B_T$. Recall that $|\mathcal{B}|_T^\infty$ defined in equation (14) denotes the maximal size of one single cluster up to time $T$. For each layer of $T$ layers, the branching factor of one cluster at most $|\mathcal{B}|_T^\infty$. Hence, we have

$$\mathbb{E}\left[\left\|\widetilde{\mathsf{F}}_T \cdots \widetilde{\mathsf{F}}_1 \mu - \widetilde{\mathsf{F}}_T \cdots \widetilde{\mathsf{F}}_1 \overline{\mu}\right\|_J^2\right]^{1/2}$$
$$\leq 4 \left(\frac{\epsilon_u}{\epsilon_d} \frac{\kappa_u}{\kappa_d}\right)^{|\mathcal{B}|_T^\infty} e^{-\beta_T T} \left(|\mathcal{B}|_T^\infty\right)^T \operatorname{card}(J) \max_{B_0 \in \mathcal{B}(k_0)} \left[\mathbb{E}\|\mu - \overline{\mu}\|_{B_0}^2\right]^{1/2}.$$

Noting that the above bound holds uniformly in the sequence of $Y$, we could generalize the above result to the following for $s \in [T-2]$:

$$\mathbb{E}\left[\left\|\widetilde{\mathsf{F}}_T \cdots \widetilde{\mathsf{F}}_{s+2} \mu - \widetilde{\mathsf{F}}_T \cdots \widetilde{\mathsf{F}}_{s+2} \overline{\mu}\right\|_J^2\right]^{1/2}$$
$$\leq 4 \left(\frac{\epsilon_u}{\epsilon_d} \frac{\kappa_u}{\kappa_d}\right)^{|\mathcal{B}|_T^\infty} e^{-\beta_T (T-s-1)} \left(|\mathcal{B}|_T^\infty\right)^{(T-s-1)} \operatorname{card}(J) \max_{B_{s+1} \in \mathcal{B}(k_{s+1})} \mathbb{E}\left[\|\mu - \overline{\mu}\|_{B_{s+1}}^2\right]^{1/2}.$$

Therefore, for $s \in [T-2]$,

$$\left\|\left\|\widetilde{\mathsf{F}}_T \cdots \widetilde{\mathsf{F}}_{s+2} \widetilde{\mathsf{F}}_{s+1} \widetilde{\mathsf{F}}_s \widehat{\pi}_{s-1} - \widetilde{\mathsf{F}}_T \cdots \widetilde{\mathsf{F}}_{s+2} \widetilde{\mathsf{F}}_{s+1} \widehat{\mathsf{F}}_s \widehat{\pi}_{s-1}\right\|\right\|_J$$
$$\leq \mathbb{E}\left[\left\|\left\|\widetilde{\mathsf{F}}_T \cdots \widetilde{\mathsf{F}}_{s+2} (\widetilde{\mathsf{F}}_{s+1} \widetilde{\mathsf{F}}_s \widehat{\pi}_{s-1}) - \widetilde{\mathsf{F}}_T \cdots \widetilde{\mathsf{F}}_{s+2} (\widetilde{\mathsf{F}}_{s+1} \widehat{\mathsf{F}}_s \widehat{\pi}_{s-1})\right\|\right\|_J^2\right]^{1/2}$$
$$\leq 4 \left(\frac{\epsilon_u}{\epsilon_d} \frac{\kappa_u}{\kappa_d}\right)^{|\mathcal{B}|_T^\infty} \exp\left(-\beta_T (T-s-1)\right) \left(|\mathcal{B}|_T^\infty\right)^{(T-s-1)} \operatorname{card}(J)$$

$$\times \max_{B_{s+1} \in \mathcal{B}(k_{s+1})} \mathbb{E}\left[\|\widetilde{\mathsf{F}}_{s+1}\widetilde{\mathsf{F}}_s\widehat{\pi}_{s-1} - \widetilde{\mathsf{F}}_{s+1}\widehat{\mathsf{F}}_s\widehat{\pi}_{s-1}\|^2_{B_{s+1}}\right]^{1/2}$$

$$\leq 4\left(\frac{\epsilon_u \, \kappa_u}{\epsilon_d \, \kappa_d}\right)^{|\mathcal{B}|^\infty_T} \exp\left(-(\beta_T - \log(|\mathcal{B}|^\infty_T))(T-s-1)\right)\mathrm{card}(J)$$

$$\times \frac{16}{\sqrt{N}}\left(\frac{\epsilon_u}{\epsilon_d \epsilon'_d}\right)^{|\mathcal{B}|^\infty_T}\left(\frac{\gamma_u \, \epsilon'_u}{\gamma_d \, \epsilon'_d}\right)^{|\mathcal{B}|^\infty_T + (|\mathcal{B}|^\infty_T)^2}|\mathcal{B}|^\infty_T$$

$$= \frac{64}{\sqrt{N}}\left(\frac{\epsilon_u^2 \kappa_u}{\epsilon_d^2 \epsilon'_d \kappa_d}\right)^{|\mathcal{B}|^\infty_T}\left(\frac{\gamma_u \, \epsilon'_u}{\gamma_d \, \epsilon'_d}\right)^{|\mathcal{B}|^\infty_T + (|\mathcal{B}|^\infty_T)^2}|\mathcal{B}|^\infty_T \mathrm{card}(J)$$

$$\times \exp\left(-(\beta_T - \log(|\mathcal{B}|^\infty_T))(T-s-1)\right). \tag{S42}$$

Here, Proposition S1.10 is used in obtaining the last inequality above, and it provides the result for $s = T-1$ case as follows:

$$\left\|\left\|\widetilde{\mathsf{F}}_T\widetilde{\mathsf{F}}_{T-1}\widehat{\pi}_{T-2} - \widetilde{\mathsf{F}}_T\widehat{\mathsf{F}}_{T-1}\widehat{\pi}_{T-2}\right\|\right\|_J \leq \max_{B_T \in \mathcal{B}(k_T)}\left[\mathbb{E}\left\|\widetilde{\mathsf{F}}_T\widetilde{\mathsf{F}}_{T-1}\widehat{\pi}_{T-2} - \widetilde{\mathsf{F}}_T\widehat{\mathsf{F}}_{T-1}\widehat{\pi}_{T-2}\right\|^2_{B_T}\right]^{1/2}$$

$$\leq \frac{16}{\sqrt{N}}\left(\frac{\epsilon_u}{\epsilon_d \epsilon'_d}\right)^{|\mathcal{B}|^\infty_T}\left(\frac{\gamma_u \, \epsilon'_u}{\gamma_d \, \epsilon'_d}\right)^{|\mathcal{B}|^\infty_T + (|\mathcal{B}|^\infty_T)^2}|\mathcal{B}|^\infty_T. \tag{S43}$$

At last, by Theorem S1.6 and a standard Monte Carlo analysis, we have

$$\left\|\left\|\widetilde{\mathsf{F}}_T\widehat{\pi}_{T-1} - \widehat{\mathsf{F}}_T\widehat{\pi}_{T-1}\right\|\right\|_{B_T} = \left\|\left\|\mathsf{C}_T^{B_T}\mathsf{B}^{B_T}\mathsf{P}_T\widehat{\pi}_{T-1} - \mathsf{C}_T^{B_T}\mathsf{B}^{B_T}\mathsf{S}^N\mathsf{P}_T\widehat{\pi}_{T-1}\right\|\right\|$$

$$\leq 2\left(\frac{\epsilon'_u \gamma_u}{\epsilon'_d \gamma_d}\right)^{|\mathcal{B}|^\infty_T}\left\|\left\|\mathsf{P}_T\widehat{\pi}_{T-1} - \mathsf{S}^N\mathsf{P}_T\widehat{\pi}_{T-1}\right\|\right\|$$

$$\leq \frac{2}{\sqrt{N}}\left(\frac{\epsilon'_u \gamma_u}{\epsilon'_d \gamma_d}\right)^{|\mathcal{B}|^\infty_T}. \tag{S44}$$

Now, plugging equations (S42)-(S44) into equation (S41), we complete the proof as follows:

$$\|\|\widetilde{\pi}_T - \widehat{\pi}_T\|\|_J \leq \sum_{s=1}^{T-2}\frac{64}{\sqrt{N}}\left(\frac{\epsilon_u^2 \kappa_u}{\epsilon_d^2 \epsilon'_d \kappa_d}\right)^{|\mathcal{B}|^\infty_T}\left(\frac{\gamma_u \, \epsilon'_u}{\gamma_d \, \epsilon'_d}\right)^{|\mathcal{B}|^\infty_T + (|\mathcal{B}|^\infty_T)^2}|\mathcal{B}|^\infty_T \mathrm{card}(J)$$

$$\times \exp\left(-(\beta_T - \log(|\mathcal{B}|^\infty_T))(T-s-1)\right)$$

$$+ \frac{16}{\sqrt{N}}\left(\frac{\epsilon_u}{\epsilon_d \epsilon'_d}\right)^{|\mathcal{B}|^\infty_T}\left(\frac{\gamma_u \, \epsilon'_u}{\gamma_d \, \epsilon'_d}\right)^{|\mathcal{B}|^\infty_T + (|\mathcal{B}|^\infty_T)^2}|\mathcal{B}|^\infty_T + \frac{2}{\sqrt{N}}\left(\frac{\epsilon'_u \gamma_u}{\epsilon'_d \gamma_d}\right)^{|\mathcal{B}|^\infty_T}$$

$$< \sum_{s=1}^{T-2}\frac{64}{\sqrt{N}}\left(\frac{\epsilon_u^2 \kappa_u}{\epsilon_d^2 \epsilon'_d \kappa_d}\right)^{|\mathcal{B}|^\infty_T}\left(\frac{\gamma_u \, \epsilon'_u}{\gamma_d \, \epsilon'_d}\right)^{|\mathcal{B}|^\infty_T + (|\mathcal{B}|^\infty_T)^2}|\mathcal{B}|^\infty_T \mathrm{card}(J)$$

$$\times \exp\left(-(\beta_T - \log(|\mathcal{B}|^\infty_T))(T-s-1)\right)$$

$$+ \frac{64}{\sqrt{N}}\left(\frac{\epsilon_u^2 \kappa_u}{\epsilon_d^2 \epsilon'_d \kappa_d}\right)^{|\mathcal{B}|^\infty_T}\left(\frac{\gamma_u \, \epsilon'_u}{\gamma_d \, \epsilon'_d}\right)^{|\mathcal{B}|^\infty_T + (|\mathcal{B}|^\infty_T)^2}|\mathcal{B}|^\infty_T \mathrm{card}(J).$$

By the definition of $\beta_T$ given in (32), and the last condition (31) in Assumption 3.1, we have

$$\beta_T - \log(|\mathcal{B}|^\infty_T) > 0,$$

because of which,

$$\|\|\widetilde{\pi}_T - \widehat{\pi}_T\|\|_J < \frac{64}{\sqrt{N}} \left(\frac{\epsilon_u^2 \kappa_u}{\epsilon_d^2 \epsilon_d' \kappa_d}\right)^{|\mathcal{B}|_T^\infty} \left(\frac{\gamma_u}{\gamma_d}\frac{\epsilon_u'}{\epsilon_d'}\right)^{|\mathcal{B}|_T^\infty + (|\mathcal{B}|_T^\infty)^2} \frac{|\mathcal{B}|_T^\infty \mathrm{card}(J)}{1 - \exp\left(-(\beta_T - \log(|\mathcal{B}|_T^\infty))\right)}.$$

## References

Hans-Otto Georgii. *Gibbs measures and phase transitions*, volume 9. Walter de Gruyter, 2011.

Patrick Rebeschini and Ramon Van Handel. Can local particle filters beat the curse of dimensionality? *The Annals of Applied Probability*, 25(5):2809–2866, 2015.


\end{document}